\newif\ifpreprint\preprintfalse
\newif\ifnotpreprint
\preprinttrue

\ifpreprint
    \notpreprintfalse
    \documentclass{article}
    \usepackage{arxiv}
\else
    \notpreprinttrue
    \documentclass[ruler]{CUP-JNL-EDS}%
\fi

\usepackage[utf8]{inputenc}

\ifpreprint
    \usepackage{authblk}

    \renewcommand{\headeright}{}
    \renewcommand{\undertitle}{}
\fi
\usepackage{booktabs}
\usepackage{paralist} %
\usepackage{xspace}
\usepackage{booktabs} %
\usepackage{graphicx}
\usepackage{mathtools}
\usepackage{colortbl}
\usepackage{siunitx}
\usepackage{xcolor}
\usepackage{hyperref}
\usepackage{cleveref}
\usepackage{anyfontsize} %
\usepackage{subcaption}
\usepackage{arydshln}
\usepackage{colortbl}
\usepackage{algorithm}
\usepackage{algpseudocode}
\usepackage{placeins}
 \usepackage{lscape}

\usepackage{latexsym}
\usepackage{multicol,multirow}
\usepackage{amsmath,amssymb,amsfonts}
\usepackage{mathrsfs}
\usepackage{amsthm}
\usepackage{rotating}
\usepackage{appendix}
\usepackage[authoryear]{natbib}
\usepackage{ifpdf}
\usepackage[T1]{fontenc}
\usepackage{times}
\usepackage{newtxmath}
\usepackage{textcomp}%

\let\originalleft\left
\let\originalright\right
\renewcommand{\left}{\mathopen{}\mathclose\bgroup\originalleft}
\renewcommand{\right}{\aftergroup\egroup\originalright}

\usepackage[acronyms, shortcuts, nonumberlist]{glossaries}
\glsdisablehyper\newacronym{CNN}{CNN}{Convolutional Neural Network}
\newacronym{GNN}{GNN}{Graph Neural Network}
\newacronym{MPNN}{MPNN}{Message Passing Neural Network}
\newacronym{MLP}{MLP}{MultiLayer Perceptron}
\newacronym{ViT}{ViT}{Vision Transformer}

\newacronym{RMSE}{RMSE}{Root Mean Squared Error}
\newacronym{MSE}{MSE}{Mean Squared Error}
\newacronym{NMAE}{NMAE}{Normalized Mean Absolute Error}
\newacronym{ME}{ME}{Mean Error}
\newacronym{STDEV_ERR}{STDEV\_ERR}{Standard Deviation of the Error}
\newacronym{LSD}{LSD}{Log Spectral Distance}
\newacronym{FSS}{FSS}{Fractions Skill Score}
\newacronym{SAL}{SAL}{Structure Amplitude Location}
\newacronym{ETS}{ETS}{Equitable Threat Score}
\newacronym{FBI}{FBI}{Frequency Bias Index}

\newacronym{MLWP}{MLWP}{Machine Learning Weather Prediction}
\newacronym{ML}{ML}{Machine Learning} %

\newacronym{GPU}{GPU}{Graphics Processing Unit}

\newacronym{DANRA}{DANRA}{DANish ReAnalysis}
\newacronym{DMI}{DMI}{Danish Meteorological Institute}
\newacronym{COSMO}{COSMO}{Consortium for Small-scale Modeling}
\newacronym{KENDA}{KENDA}{Kilometre-scale Ensemble Data Assimilation}
\newacronym{MCH}{MCH}{Federal Office for Meteorology and Climatology MeteoSwiss}
\newacronym{ERA5}{ERA5}{ECMWF ReAnalysis v5}

\newacronym{IFS}{IFS}{Integrated Forecasting System}
\newacronym{GFS}{GFS}{Global Forecasting System}

\newacronym{LAM}{LAM}{Limited Area Model}
\newacronym{NWP}{NWP}{Numerical Weather Prediction}
\newacronym{SYNOP}{SYNOP}{Surface Synoptic Observations}
\newacronym{DA}{DA}{Data Assimilation}
\newacronym{CRS}{CRS}{Coordinate Reference System}

\ifnotpreprint
    \makeglossaries
\fi

\usepackage{amssymb} %
\usepackage{amsmath}
\usepackage{bm} %

\newcommand{\eq}[2][my_equation]{\begin{equation}\label{eq:#1}#2\end{equation}}

\newcommand{\alse}[2][my_eq]{\begin{subequations}\label{eq:#1}\begin{align}#2\end{align}\end{subequations}}

\newcommand{\qm}[1]{``#1''}

\newcommand{\loss}{\mathcal{L}}

\newcommand{\set}[1]{\{ #1 \}}
\newcommand{\setsize}[1]{\left| #1 \right|}

\newcommand{\vect}[1]{\mathbf{#1}} %

\DeclareMathOperator{\var}{Var}

\newcommand{\numsamples}{N} %
\newcommand{\fclen}{K} %

\newcommand{\wvar}[1]{\texttt{#1}}

\newcommand{\nnmodel}{g}
\newcommand{\wstate}{\vect{x}}
\newcommand{\predstate}{\vect{\hat{x}}}

\newcommand{\iforcing}{\vect{f}}
\newcommand{\istatic}{\vect{c}}
\newcommand{\bforcing}{\vect{f}^B}
\newcommand{\bstatic}{\vect{c}^B}

\newcommand{\interior}{\Omega}
\newcommand{\boundary}{\Omega^B}

\newcommand{\interiorpoints}{\mathbb{G}}

\newcommand{\statevars}{\mathbb{V}}
\newcommand{\invtimediff}{\lambda}
\newcommand{\varweight}{w}

\newcommand{\gtmedges}{\mathcal{E}_{g2m}} 
\newcommand{\mtgedges}{\mathcal{E}_{m2g}} 
 
\newcommand{\edgelen}{\ell} %

\newcommand{\danraplots}[3]{
    \begin{figure}
        \centering
        \begin{subfigure}[b]{0.5\textwidth}
            \centering
            \includegraphics[width=\textwidth]{graphics/design_experiments/danra/#1/t2m_rmse.pdf}
            \caption{\SI{2}{m} temperature (\wvar{2t})}
            \label{fig:#2_t2m}
        \end{subfigure}%
        \hfill%
        \begin{subfigure}[b]{0.5\textwidth}
            \centering
            \includegraphics[width=\textwidth]{graphics/design_experiments/danra/#1/wv10m_rmse.pdf}
            \caption{\SI{10}{m} wind}
            \label{fig:#2_wv10m}
        \end{subfigure}%
        \hfill%
        \begin{subfigure}[b]{0.5\textwidth}
            \centering
            \includegraphics[width=\textwidth]{graphics/design_experiments/danra/#1/lwavr0m_rmse.pdf}
            \caption{Surface net longwave radiation (\wvar{athb\_s})}
            \label{fig:#2_athb_s}
        \end{subfigure}%
        \begin{subfigure}[b]{0.5\textwidth}
            \centering
            \includegraphics[width=\textwidth]{graphics/design_experiments/danra/#1/z600_rmse.pdf}
            \caption{Geopotential at \SI{600}{\hecto\pascal} (\wvar{z600})}
            \label{fig:#2_z600}
        \end{subfigure}%
        \caption{#3}
        \label{fig:#2}
    \end{figure}
}
\newcommand{\cosmoplots}[3]{
    \begin{figure}
        \centering
        \begin{subfigure}[b]{0.5\textwidth}
            \centering
            \includegraphics[width=\textwidth]{graphics/design_experiments/cosmo/#1/T_2M_rmse.pdf}
            \caption{\SI{2}{m} temperature (\wvar{2t})}
            \label{fig:#2_t2m}
        \end{subfigure}%
        \hfill%
        \begin{subfigure}[b]{0.5\textwidth}
            \centering
            \includegraphics[width=\textwidth]{graphics/design_experiments/cosmo/#1/wv10m_rmse.pdf}
            \caption{\SI{10}{m} wind}
            \label{fig:#2_vw10m}
        \end{subfigure}%
        \hfill%
        \begin{subfigure}[b]{0.5\textwidth}
            \centering
            \includegraphics[width=\textwidth]{graphics/design_experiments/cosmo/#1/ATHB_S_rmse.pdf}
            \caption{Surface net longwave radiation (\wvar{athb\_s})}
            \label{fig:#2_athb_s}
        \end{subfigure}%
        \begin{subfigure}[b]{0.5\textwidth}
            \centering
            \includegraphics[width=\textwidth]{graphics/design_experiments/cosmo/#1/TOT_PREC_rmse.pdf}
            \caption{Precipitation (\wvar{tp01})}
            \label{fig:#2_tp}
        \end{subfigure}%
        \caption{#3}
        \label{fig:#2}
    \end{figure}
}

\ifnotpreprint
    
    \jname{Environmental Data Science}
    \jyear{2020}
    \jvol{4}
    \jdoi{10.1017/eds.2020.xx}
\fi

\newcommand{\papertitle}{Building Machine Learning Limited Area Models: Kilometer-Scale Weather Forecasting in Realistic Settings}
\ifpreprint
    \renewcommand{\papertitle}{Building Machine Learning Limited Area Models: Kilometer-Scale Weather Forecasting in\\ Realistic Settings}
\fi
\newcommand{\papertitleshort}{Building Machine Learning Limited Area Models}

\begin{document}

\ifpreprint
    \title{\papertitle}
    \renewcommand{\shorttitle}{\papertitleshort}
    \author[*,1,2]{Simon Adamov}
    \author[*,\textdagger,3]{Joel Oskarsson}
    \author[4]{Leif Denby}
    \author[5]{Tomas Landelius}
    \author[4]{Kasper Hintz}
    \author[4]{Simon Christiansen}
    \author[6]{Irene Schicker}
    \author[1]{Carlos Osuna}
    \author[3]{Fredrik Lindsten}
    \author[1]{Oliver Fuhrer}
    \author[7]{Sebastian Schemm}
    \affil[1]{Federal Office for Meteorology and Climatology MeteoSwiss}
    \affil[2]{Institute of Atmospheric and Climate Science, Department of Environmental Science, ETH Zurich}
    \affil[3]{Division of Statistics and Machine Learning, Department of Computer and Information Science, Linköping University}
    \affil[4]{Danish Meteorological Institute}
    \affil[5]{Swedish Meteorological and Hydrological Institute}
    \affil[6]{GeoSphere Austria}
    \affil[7]{Department of Applied Mathematics and Theoretical Physics, University of Cambridge, Cambridge, UK}
    {
        \renewcommand{\thefootnote}{\fnsymbol{footnote}}
        \footnotetext[1]{These authors contributed equally.}
        \footnotetext[2]{Corresponding author. Email: \texttt{joel.oskarsson@outlook.com}}
    }
    \date{\vspace{-2em}\today\vspace{-2em}} %
    \maketitle
    \section*{Abstract}
    Machine learning is revolutionizing global weather forecasting, with models that efficiently produce highly accurate forecasts.
Apart from global forecasting there is also a large value in high-resolution regional weather forecasts, focusing on accurate simulations of the atmosphere for a limited area.
Initial attempts have been made to use machine learning for such limited area scenarios, but these experiments do not consider realistic forecasting settings and do not investigate the many design choices involved.
We present a framework for building kilometer-scale machine learning limited area models with boundary conditions imposed through a flexible boundary forcing method.
This enables boundary conditions defined either from reanalysis or operational forecast data.
Our approach employs specialized graph constructions with rectangular and triangular meshes, along with multi-step rollout training strategies to improve temporal consistency. 
We perform systematic evaluation of different design choices, including the boundary width, graph construction and boundary forcing integration.
Models are evaluated across both a Danish and a Swiss domain, two regions that exhibit different orographical characteristics.
Verification is performed against both gridded analysis data and in-situ observations, including a case study for the storm Ciara in February 2020.
Both models achieve skillful predictions across a wide range of variables, with our Swiss model outperforming the numerical weather prediction baseline for key surface variables.
With their substantially lower computational cost, our findings demonstrate great potential for machine learning limited area models in the future of regional weather forecasting.

\fi

\section{Introduction}
\gls{MLWP} has become a key area of research due to its ability to complement or potentially replace conventional \gls{NWP} \citep{panguweather,Ben-Bouallegue_2023_rise,aifs,gencast,graph_dop,graph_efm,graphcast,keisler}.
\glsunset{ML} %
\gls{MLWP} methods, specifically based on neural networks, have shown to produce forecasts that are more accurate for certain variables.
At the same time, these methods are also much more computationally efficient than physics-based \gls{NWP}.
This creates new possibilities for what is possible for weather forecasting, both in terms of accurate day-to-day information \citep{nwp_at_crossroads} and possibilities to predict extreme weather events \citep{ai_extreme_weather_climate}.

Despite important progress in global-scale \gls{NWP}, several small-scale phenomena remain insufficiently resolved.
For example, the daily cycle of summertime convection in mountainous regions or the dynamics of sea-breeze circulations near coastlines cannot be accurately modeled with global model configurations \citep{lean_hectometric_nodate}.
Dedicated \glspl{LAM} offer a solution by targeting mesoscale, or even microscale, processes. 
Considering only a specific region allows for modeling at higher spatial resolutions at reduced computational costs with rapid-refresh forecast-assimilation cycles \citep{Zhang_2024_LAM-global}.
Given regional data, such \glspl{LAM} additionally allow for tailoring the forecasting system to properties and interests of the specific area.
A challenge of local approaches is, however, the presence of boundary conditions at the border of the region.
In physics-based \gls{NWP} these boundary conditions are derived from larger-scale reanalysis datasets or global forecasts, which must be handled with particular care to ensure the physical correctness and numerical stability of predictions \citep{Zhang_2024_LAM-global}.
On one hand it has been shown that \glspl{LAM} can produce high quality simulations even with boundary information limited to a few relevant fields on a coarse scale \citep{Denis_2003_LAM-LBC-sens}.
On the other hand, the inevitable mismatch of the interior state and boundary conditions leads to artificial transients.
As the effect of these is largely reduced after some model integration, there is a connection between the size of the \gls{LAM} domain and the boundary conditions. 
Hence a wider \gls{LAM} domain generally leads to better results because of the remoteness of the boundary \citep{Davies_2017_LAM-resolution}.

\gls{MLWP} offers intriguing opportunities for relieving some of the challenges of \gls{NWP} \glspl{LAM} and pushing this modeling paradigm forward.
There have been a small number of attempts at building such \gls{MLWP} \glspl{LAM}, which show initial encouraging results \citep{neural_lam,graph_efm,storm_cast,yinglong,diffusion_lam}.
These methods build on methodologies developed for global \gls{MLWP} with some adaptations to handle the limited area setting.
In particular, the integration of boundary conditions through additional boundary forcing inputs is a crucial addition to these models.
\gls{ML} models offer new possibilities in terms of integrating this boundary condition information, as they do not have the same physical constraints as \gls{NWP} \glspl{LAM}.
As we show in this work, in \gls{MLWP} the boundary information can contain different atmospheric variables, at a different temporal resolution, compared to the \gls{LAM}.

Initial \gls{ML} \glspl{LAM} have shown promising results, but do not consider realistic forecasting settings \citep{neural_lam,yinglong,diffusion_lam}.
These works use forecasts from existing physics-based \glspl{LAM} for boundary forcing.
This means that they describe a somewhat idealized setting and do not capture the model behavior in an operational setting, where the \gls{ML} \gls{LAM} would be run without the support of an auxiliary regional \gls{NWP} forecast.
As \gls{ML} \glspl{LAM} are being created for different regions, there is also a substantial interest in better understanding the different design choices involved in building these models.
While initial works show examples of successful approaches, they do not systematically investigate different design choices related to the models \citep{graph_efm,storm_cast}.

While multiple types of model architectures have been used for \gls{MLWP} \citep{graphcast, panguweather, fourcastnet}, of particular interest to this work is graph-based methods \citep{keisler,graphcast,gencast,graph_efm,aifs}.
Graph-based \gls{MLWP} is a flexible framework that leverages \glspl{GNN} to move past traditional latitude--longitude grid layouts in modeling atmospheric processes.
These methods can additionally be leveraged to capture both large-scale circulations and microscale processes with computational efficiency.
In this work we build on graph-based \gls{MLWP} and show that these models offer a unique advantage for integration of irregular domain boundaries, an advantage largely unexplored in previous works on graph-based \glspl{LAM} \citep{graph_efm}.

Verification of regional weather prediction models, whether physics-based or data-driven, demands rigorous quantitative and qualitative methodologies.
We employ validation against in-situ observations and gridded data, showing how well the model handles real-world complexity at relevant spatial and temporal scales \citep{Ben-Bouallegue_2023_rise,Bremnes_2023_Evaluation}.
Comparisons against operational \gls{NWP} \glspl{LAM} then allow for assessing model skill.
As \gls{MLWP} models do not have the built-in physical constraints of traditional \gls{NWP} models, of particular interest for these are physical consistency checks.
We tackle this through specific distributional metrics for important diagnostic fields, such as precipitation.

In this work, we propose a practical training and evaluation methodology for \gls{ML} \glspl{LAM} using \glspl{GNN}.
Building on existing works on graph-based modeling \citep{graphcast,neural_lam}, we describe how these models can be used for \gls{LAM} forecasting under realistic conditions in terms of data-availability.
The proposed framework is evaluated by training and verification of \gls{ML} \gls{LAM} models for two geographical regions.
Our main contributions are:
\begin{enumerate}
    \item We demonstrate a framework for training and evaluating \gls{ML} \glspl{LAM} in a setting with realistic boundary forcing.
          By using separate encoding components in the graph-based framework we allow for boundary data of arbitrary temporal resolution, selection of variables and spatial gridding.
    \item We empirically investigate different design choices associated with graph-based \gls{MLWP} for \glspl{LAM}.
          We compare rectangular and triangular mesh graphs in multi-scale and hierarchical configurations.
          This is the first large scale systematic comparison of different graph constellations used in \gls{MLWP}.
    \item We also investigate how to define the boundary area.
        Configurations with different boundary widths are evaluated and we explore options for overlapping the boundary and forecasting region.
    \item The general framework is evaluated across two geographic regions, one centered on Denmark and one on Switzerland.
          Models are trained on these different regional datasets and evaluated with boundary forcing coming from the global \gls{IFS}.
    \item
          The forecast accuracy and physical coherence of the models is extensively verified against both high-resolution gridded analyses and point-wise observational data.
          Our Swiss model achieves competitive or better performance compared to operational \gls{NWP} forecasts for multiple important variables.
          Forecasts from the Danish model show competitive performance on station observations in Denmark.
          Both models are capable of rapidly producing full high-resolution forecasts in just one minute.

    \item 
    To further understand the ability of high-resolution \gls{ML} \glspl{LAM} to model convective systems we conduct a case study for the storm Ciara in February 2020.
    The \gls{ML} models show some tendencies to underforecast the most extreme values, but overall transport the storm conditions accurately across and within domain boundaries.

\end{enumerate}

\section{Related work} \label{sec:related_work}

\paragraph{Graph-based \gls{MLWP}}
Pioneering work on global \gls{MLWP} using \glspl{GNN} has enabled scalable forecasting, balancing computational demands with the need to resolve fine-scale atmospheric processes \citep{keisler,graphcast,gencast,aifs,graph_efm,stretch_grid,aifs_crps}. 
The \glspl{GNN} used in these models include Interaction Networks \citep{interaction_nets, keisler, graphcast, graph_efm}, as we use in this paper, and sparse transformer constructions \citep{gencast, aifs, aifs_crps}.
The graph-based approach has also been used in initial proof-of-concept \gls{ML} \gls{LAM} models \citep{neural_lam, graph_efm}.
By efficiently embedding regional domains into graph representations, these methods facilitate high-resolution localized forecasts.

\paragraph{\gls{ML} \glspl{LAM}}
Regional \gls{MLWP} models were initially used in nowcasting regimes, where boundary conditions play a limited role due to the short lead times \citep{metnet,skilful_precip_nowcasting,bihlo_gan_model}.
The first \gls{ML} \gls{LAM} models to explicitly handle boundary conditions and allow for autoregressive rollouts to longer lead times were developed by \cite{neural_lam}.
They trained graph-based surrogate models as fast emulators of an existing \gls{NWP} system, operating on a Nordic domain at \SI{10}{\kilo\metre} resolution.
\cite{neural_lam} handle boundary conditions by replacing the model predictions with ground truth data along the boundaries in each autoregressive forecasting step.
Another early deterministic \gls{ML} \gls{LAM} model is the YingLong model by \cite{yinglong}, which is based on an adaptive Fourier neural operator architecture \citep{afno}.
YingLong produces forecasts over the southeast US at \SI{3}{\kilo\metre} resolution and up to \SI{12}{\hour} ahead.
Boundary conditions in YingLong are handled similarly as by \cite{neural_lam}, by replacing the values in a boundary region.
The new boundary values are however created by smoothly blending the forecast from the \gls{ML} model with those from an \gls{NWP} forecast.
These two approaches both rely on boundary conditions from an existing \gls{NWP} \gls{LAM} model.
Therefore they do not represent a fully realistic setting, where an \gls{ML} \gls{LAM} model runs with only boundary information from a global model.
We build on the ideas from these works, but extend them to such a realistic setting.

\paragraph{Stretch-grid models}
One approach to regional \gls{MLWP}, that is different from the \gls{LAM} setup, involves stretch-grid models \citep{stretch_grid}. 
This method utilizes a global \gls{GNN} model with increased resolution specifically over the region of interest, initialized with both global and high-resolution regional analyses. 
By employing a stretched grid, the model aims to achieve seamless transition of weather systems between global and regional domains without explicit boundary treatment. 
While stretch-grid models target forecasting for a specific region, they still require simulating the full global atmospheric state.
We further discuss the practical differences between \gls{ML} \glspl{LAM} and stretch-grid models in \cref{sec:discussion_future}.
 
\paragraph{Using foundation models for regional forecasting}
Another direction of research explores adapting earth system foundation models for localized weather forecasting. 
\cite{efficient_localized_adaptation} experiment with fine-tuning ClimaX \citep{climax}, a pre-trained transformer model for climate and weather, 
for forecasting a region covering the Middle East and North Africa.
To address the computational challenges associated with training large models, they utilize Low Rank Adaptation \citep{lora} for parameter-efficient fine-tuning.
\cite{efficient_localized_adaptation} describe no treatment of boundary conditions in their regional adaption.

\paragraph{Statistical downscaling}
An alternative to running a regional weather forecasting model is to apply statistical downscaling techniques to a global forecast, increasing its resolution, correcting biases and adapting the forecast to fine-scaled orographic features \citep{downscaling_review, comparing_statistical_downscaling}.
There has been great success in employing \gls{ML} methods for this problem \citep{corrdiff, downscaling_st_diff, ai_weather_as_downscaling}.
Statistical downscaling approaches generally do not have to actually simulate atmospheric dynamics over time, as opposed to a regional forecasting model.
The downscaling operation happens at a snapshot in time, and only requires a coarse-scaled global forecast as input.
\gls{LAM} models are more complex as they start from a high resolution initial condition, and simulate the dynamics with the global forecast as a forcing.
This is more computationally demanding, but has the benefit of integrating information from the regional initial condition.
By incorporating additional observations into this initialization, through regional \gls{DA} setups, \gls{LAM} forecasts can be more accurate than a downscaled global one.
Additionally, actually simulating dynamics at a finer resolution allows \gls{LAM} models to accurately represent atmospheric processes that are not resolved in the global forecast.

\paragraph{Probabilistic \gls{MLWP}}
Within global \gls{MLWP} there is an increasing interest in probabilistic models, capable of generating ensemble forecasts \citep{swinvrnn, gencast, graph_efm, archesweather, aifs_crps}.
There are also similar developments for \gls{ML} \glspl{LAM}, targeting regional high-resolution ensemble forecasting \citep{graph_efm, storm_cast, diffusion_lam}.
Based on a latent variable approach, \cite{graph_efm} developed the Graph-EFM model which can be used for both global and \gls{LAM} ensemble forecasting.
Also diffusion models \citep{denoising_diffusion,edms} have been applied to the \gls{LAM} setting \citep{storm_cast, diffusion_lam}.
The StormCast diffusion model of \cite{storm_cast} targets the convection-allowing \SI{3}{\kilo\metre} scale and includes forcing from the \gls{GFS}.
Their model relies on a U-Net as a diffusion backbone, meaning that the global forecast has to be re-gridded to the same resolution as the \gls{LAM} area.
In StormCast the input from the global forecast is given for the full \gls{LAM} region, rather than only along the boundary.
This makes the model operate as a hybrid approach between a \gls{LAM} and downscaling model, a possibility that we also investigate in this work.
\cite{diffusion_lam} consider a diffusion-based \gls{ML} \gls{LAM} with forcing only along the area boundary, but instead investigate the possibility of including boundary conditions from both past and future time steps as input.
A common challenge in all these probabilistic \gls{ML} \gls{LAM} models has been to achieve sufficient spread in the ensemble forecasts.

\paragraph{Exploring design choices}
Most previous works in \gls{MLWP} only briefly discuss and investigate design choices leading to their final model \citep{graphcast, fourcastnet,panguweather}.
One work that tries to shed some light on these choices, in the context of global \gls{MLWP}, is \cite{exploring_mlwp_design_space}.
They examine factors such as architectural choices, pre-training strategies, loss functions, and multi-step fine-tuning.
Although \glspl{LAM} introduce unique challenges related to boundary conditions and smaller spatial scales, the insights gained from global-scale studies
serve as an initial guide also for our study.
While \cite{exploring_mlwp_design_space} focus on many factors relevant for general \gls{MLWP}, we specifically investigate design choices central to graph-based and \gls{LAM} models.
One notable finding of \cite{exploring_mlwp_design_space} is that regular grid architectures outperform grid-invariant models under comparable compute limits.
While our framework uses grid-invariant graph models, we argue that this grid invariance is especially suitable in the \gls{LAM} setting.
As we further motivate in \cref{sec:boundary_forcing}, the graph-based approach is particularly advantageous when we deal with multiple heterogeneous data sources, in the form of interior and boundary data.
\cite{exploring_mlwp_design_space} also discuss the possibility of combining grid-invariant encoders and decoders with regular grid neural network layers, which could be a useful future direction also for \gls{ML} \gls{LAM} models.

\section{\texorpdfstring{\acrfull{MLWP}}{Machine Learning Weather Prediction}}

\begin{figure}
    \centering
    \includegraphics[width=\linewidth]{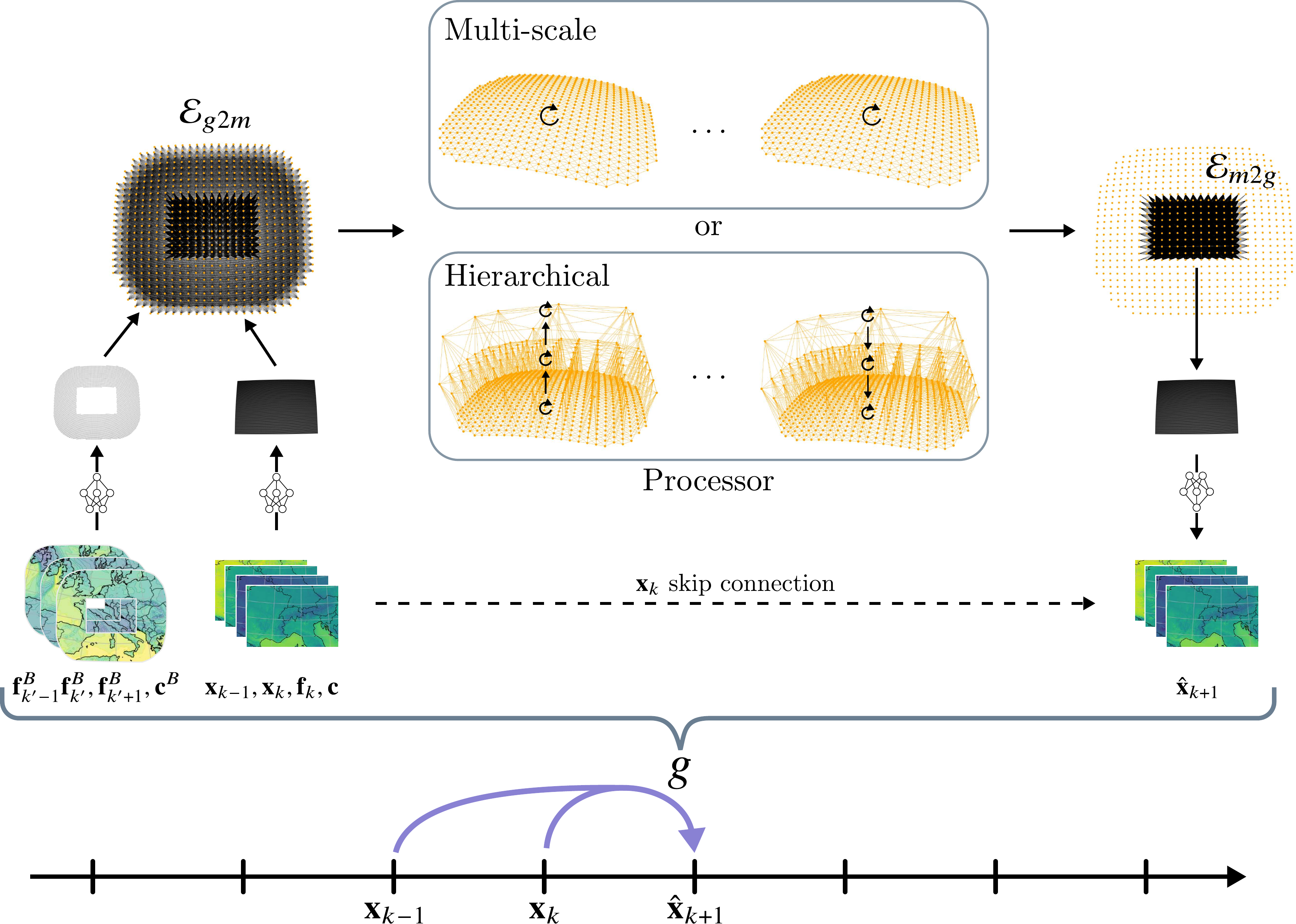}
    \caption{
    Overview of the graph-based \gls{MLWP} approach applied in our \gls{LAM} setting.
    The dual \acrshort{MLP} encoders for interior and boundary inputs are introduced in this work.
    }
    \label{fig:graph_model_overview}
\end{figure}

Highly capable \gls{MLWP} models have been created using a number of different methods from the \gls{ML} literature.
In this work we build on the graph-based \gls{MLWP} framework \citep{keisler, graphcast} and in particular the adaptation to \gls{LAM} settings from \cite{neural_lam}.
This choice is motivated by the flexibility of graph-based \gls{MLWP}, allowing for handling arbitrary spatial grids and resolutions.
In this section we give an introduction to graph-based \gls{MLWP} and then proceed to define our \gls{LAM} forecasting problem.

\subsection{Graph-based \texorpdfstring{\gls{MLWP}}{MLWP}}
In the graph-based framework a \gls{GNN} model \citep{encode_process_decode, mpnn, interaction_nets} $\nnmodel$ is learned to autoregressively predict the next atmospheric state $\wstate_{k+1}$ at time $t_{k+1}$, given $\wstate_k$ at time $t_k$ and $\wstate_{k-1}$ at time $t_{k-1}$.
Each state contains the values of multiple atmospheric variables modeled in a grid over some spatial region or globally.
In \gls{MLWP} it is common practice to give the model the two previous states \citep{graphcast, graph_efm, aurora}. 
This allows for capturing first-order dynamics at time $t_k$.
The model $\nnmodel$ is defined as an encode-process-decode sequence \citep{encode_process_decode}.
Using \gls{GNN} layers, the original gridded data is encoded to a mesh graph, where a number of processing steps are carried out, and the results are finally decoded back to the grid to produce a prediction \citep{keisler}.
This is illustrated in \cref{fig:graph_model_overview}.
The encoding to a mesh graph constitutes a spatial dimensionality reduction from the high-resolution input grid to a smaller set of mesh nodes.
Working on this mesh graph rather than the original grid is crucial for the computational efficiency of the model.
The exact construction of the mesh graph is an important design choice \citep{graphcast,graph_efm}, which we discuss further for our \gls{LAM} setting in \cref{sec:graph_construction}.

The first step in the model $\nnmodel$ is to encode the model inputs.
At each grid point $i$ the model takes as input the previous states $\wstate_{k,i}$ and $\wstate_{k-1,i}$, but also forcing inputs $\iforcing_{k,i}$ and static inputs $\istatic_i$.
These additional inputs $\iforcing_{k,i}$ and $\istatic_i$ are pre-computed quantities related to the time and location, and should help the model to accurately capture the evolution of the atmospheric state.
Using a \gls{MLP} applied independently at each grid point, all these inputs are encoded to a joint latent representation.
The grid-to-mesh \gls{GNN} is then applied to map all of these latent representations from the grid nodes to the mesh graph.
This \gls{GNN} operates on a directed edge set $\gtmedges$ with edges starting in the grid points and ending in the mesh nodes.
By message passing over $\gtmedges$ the latent representations in the grid nodes are mapped to the mesh graph and aggregated to new representations in each mesh node.
These new representations are then the starting point for the processor step.

The processor is the core part of the model that maps the encoded state from time $t_k$ to $t_{k+1}$.
This is achieved by applying a sequence of flexible \gls{GNN} layers to update the node and edge representations in the mesh graph.
In this work we consider two types of processors:
\begin{itemize}
    \item The multi-scale processor of \cite{graphcast} works with a single mesh graph, applying a sequence of \gls{GNN} layers all operating on the same graph.
    \item The hierarchical processor from \cite{neural_lam} instead operates over multiple levels in a hierarchical mesh graph.
    Higher levels in the hierarchy contain fewer nodes and longer edges.
    One layer of the processor consists of multiple \gls{GNN} layers, sequentially performing message passing both within each level and between them.
\end{itemize}
The specific \gls{GNN} layers used in both types of processors are Interaction Networks \citep{interaction_nets}.

After the processor, the final node representations are decoded back to the grid to produce the prediction of $\wstate_{k+1}$.
This decoding happens in an analogous way to the encoding, through a mesh-to-grid \gls{GNN} operating over an edge set $\mtgedges$ with edges from the mesh nodes to the grid points.
The result of this decoding is additionally added to the previous state $\wstate_k$ to produce the final prediction.
This skip connection in principle means that the model is predicting the residual between the states $\wstate_{k+1}$ and $\wstate_{k}$ \citep{graphcast, graph_efm}.

While one forward pass through the model $\nnmodel$ only takes us one time step forward, the model can be iteratively applied to its own predictions in order to unroll a longer forecast.
It is common to train these models first on single-step prediction, and then fine-tune on longer rollouts \citep{keisler, graphcast, graph_efm, aifs}.

\subsection{\texorpdfstring{\gls{LAM}}{LAM} problem statement}
\label{sec:lam_problem_statement}
The limited area modeling problem focuses on weather prediction over a bounded geographical region $\interior$, where atmospheric states evolve according to both internal dynamics and external forcing. 
We refer to this region as the interior.
An important external forcing, and the key difference of \glspl{LAM} compared to global models, is the influence of boundary conditions on $\interior$ \citep{fundamentals_of_nwp}.
In an \gls{ML} \gls{LAM} we need to inform the model of the atmospheric state also in a boundary area $\boundary$ around $\interior$.
The problem of interest is to forecast a sequence $\set{\wstate_{k}}_{k=1}^\fclen$ of atmospheric states, given:
\begin{itemize}
    \item Initial states $\wstate_{-1}, \wstate_{0}$ within the domain $\interior$. 
    In an operational setting $\wstate_0$ and $\wstate_{-1}$ could be retrieved from different iterations of a running \gls{DA} cycle.
    \item Forcing inputs $\set{\iforcing_{k}}_{k=1}^\fclen$ for the domain $\interior$. 
    \item Boundary forcing $\set{\bforcing_{k'}}_{k'=0}^{\fclen'}$ from the boundary region $\boundary$.
    This boundary forcing contains the values of atmospheric variables over $\boundary$, coming from a global or larger regional forecast.
    It can also include pre-computed forcings similar to $\iforcing_k$.
    Note that we here assume that $\bforcing_{k'}$ for all time points is available already when starting the \gls{LAM} forecast.
    With the advent of global \gls{MLWP} the time to actually produce global forecasts is shrinking, and while there is still a need to wait for global \gls{DA} this can also be expected to reduce through \gls{ML} \gls{DA} methods \citep{diffda, towards_self_contained_mlwp} and forecasting models directly initialized from observations \citep{transformer_dop, graph_dop}.
    We generally do not assume that the boundary forcing has the same time steps as the \gls{LAM} model, meaning that times $t'_0, \dots, t'_{\fclen'}$ and $t_0, \dots, t_{\fclen}$ might not line up.
    For simplicity we do however assume that the boundary forcing has longer or equal time steps than the interior, which is the common situation in practice.
    \item Static features $\istatic$ for $\interior$ and $\bstatic$ for $\boundary$.
    These are static over time and relate to the terrain of the two regions.
\end{itemize}

Following this problem formulation, the goal is to train \gls{ML} \gls{LAM} models with improved local details, while leveraging global model guidance through boundary forcing.
The restriction to a limited domain $\interior$ allows for high-resolution modeling at significantly lower computational costs compared to global high-resolution forecasting. 
Note that this problem formulation differs from that of stretch-grid models \citep{stretch_grid}, that need to represent the atmospheric state over the entire global domain. 
This is not true in this \gls{LAM} formulation, where boundary forcing can be restricted to only what is necessary to obtain good quality forecasts for the interior domain.

\section{Building \texorpdfstring{\gls{ML} \gls{LAM}}{ML LAM} models}
The two key challenges in applying graph-based \gls{MLWP} to the \gls{LAM} setting is how to incorporate the boundary forcing in a useful way and how to construct the mesh graph.
To build \gls{ML} \gls{LAM} models that are applicable in realistic forecasting settings, we here outline a flexible framework for these modeling choices.

\subsection{Boundary forcing}
\label{sec:boundary_forcing}

\begin{figure}[b]
    \centering
    \begin{subfigure}[b]{0.33\textwidth}
        \centering
        \includegraphics[width=\textwidth]{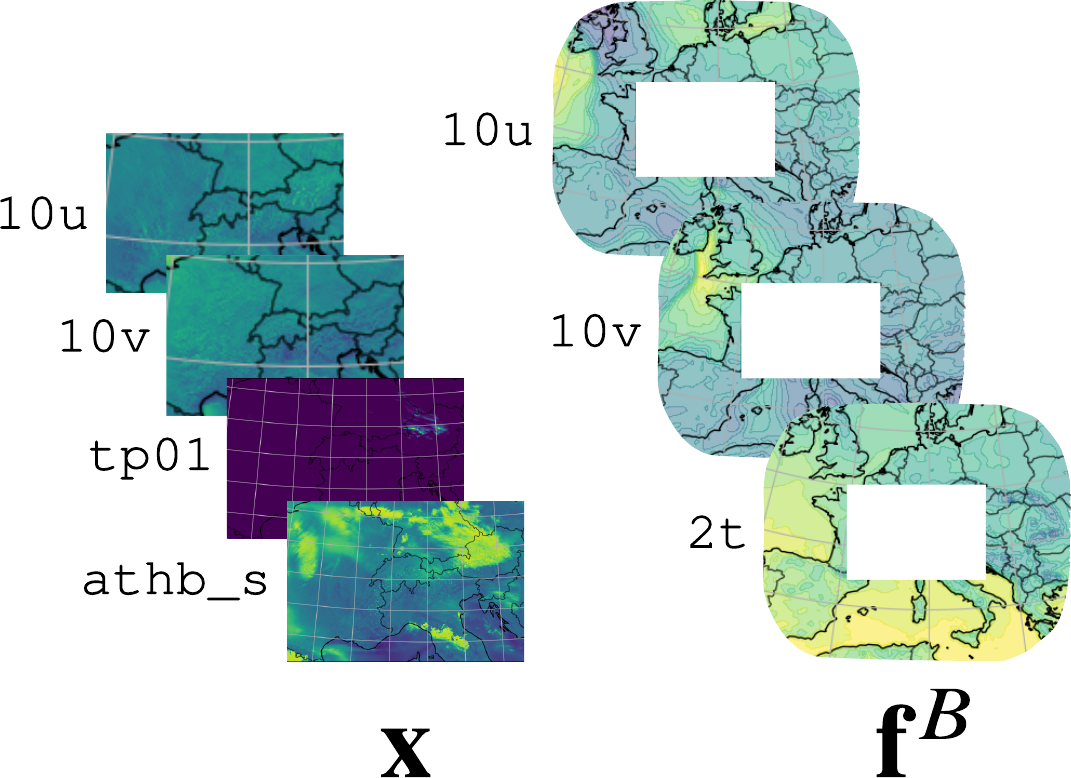}
        \caption{Different atmospheric variables}
    \end{subfigure}%
    \hfill%
    \begin{subfigure}[b]{0.33\textwidth}
        \centering
        \includegraphics[width=.8\textwidth]{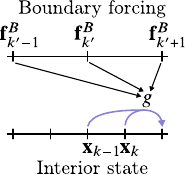}
        \caption{Different length of time steps}
    \end{subfigure}%
    \hfill%
    \begin{subfigure}[b]{0.33\textwidth}
        \centering
        \includegraphics[width=\textwidth]{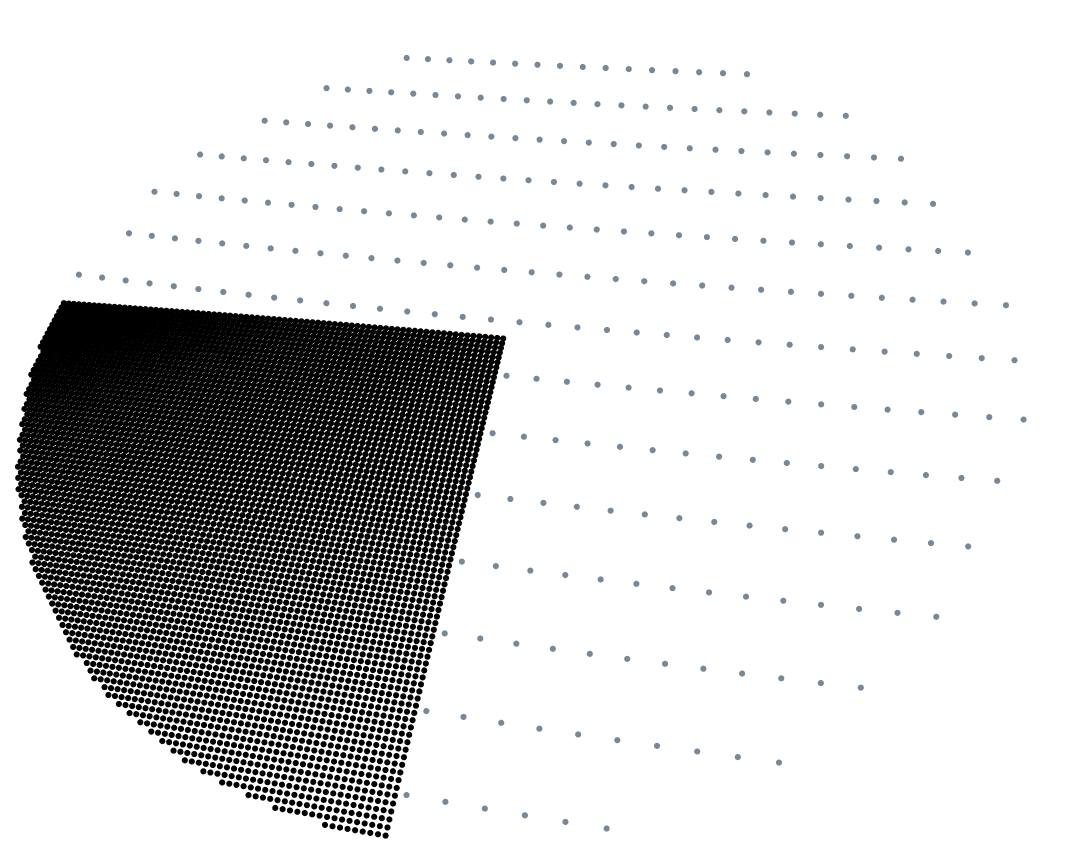}
        \caption{Different grid layouts}
    \end{subfigure}%
    \caption{Desiderata for the boundary forcing framework.}
    \label{fig:boundary_forcing_desiderata}
\end{figure}

Limited area models require careful treatment of boundary conditions to maintain forecast accuracy and physical validity.
While previous works consider simplified settings \citep{neural_lam, yinglong}, where boundary forcing already comes from an \gls{NWP} \gls{LAM}, moving to a fully realistic \gls{LAM} setting introduces additional challenges.
We aim for a flexible boundary forcing approach, that can apply to a diverse range of \gls{LAM} scenarios.
As regional and global data can have different structure and resolution, it is particularly useful with an approach that does not require the boundary forcing to be on the exact same format as the interior state data.
As a desiderata, we aim for a framework that allows for interior state and boundary forcing with:
\begin{enumerate}
    \item Different atmospheric variables,
    \item Different length of time steps,
    \item Different spatial resolution and grid layouts.
\end{enumerate}
We illustrate these points in \cref{fig:boundary_forcing_desiderata}.
This makes the \gls{ML} \gls{LAM} maximally applicable to diverse \gls{LAM} scenarios, without having to re-configure existing \gls{NWP} systems or re-running expensive simulations.
While differences in spatial and temporal resolution can be tackled through regridding and interpolation, such computations do not add any information that can be useful to the model.
Instead, this artificially expands the data size, increasing computational cost.
We follow a philosophy of feeding the \gls{ML} model with minimally processed data, and instead letting it learn any useful correlations across coarse spatial or temporal representations.

\paragraph{Different atmospheric variables}
To tackle point 1 above, our model architecture employs dual encoder \glspl{MLP}, mapping to a joint latent space (see \cref{fig:graph_model_overview}).
At time $t_k$ the interior \gls{MLP} takes as input $\wstate_{k,i}, \wstate_{k-1,i}, \iforcing_{k,i}, \istatic_{k,i}$ at an interior grid point $i$.
The boundary \gls{MLP} instead takes as input three time steps of boundary forcing $\bforcing_{k'-1,i'} \bforcing_{k',i'}, \bforcing_{k'+1,i'}$ and static features $\bstatic_{i'}$ for a grid point $i'$ in the boundary region.
Note here that variables present at multiple vertical levels in the atmosphere are simply stacked with all other variables in the state and forcing vectors. 
The input at grid point $i$ or $i'$ thus contains values for the surface and the full atmospheric column above it.
The boundary time step $k'$ is here always chosen as the last step in the boundary time series such that $ t'_{k'} \leq t_k$, meaning that we use the last boundary forcing that is from before or the same time as the interior.
These separate \glspl{MLP} allow the model to handle boundary data with a different set of variables from the interior.
The boundary \gls{MLP} processes variables coming from a global model forecast (in this work \gls{IFS} or the \gls{ERA5} dataset) and projects it into a latent space compatible with the interior representation.

\paragraph{Future boundary forcing}
Including multiple time steps of boundary forcing as input at time $t_k$ is motivated similarly as the inclusion of two initial states, to give information about dynamics.
However, we specifically choose to also include $\bforcing_{k'+1}$, which is boundary forcing from the future time $t'_{k' + 1}$.
This is possible by making use of the assumption that the full boundary forcing time series is available already when the \gls{LAM} forecast is started.
As argued also by \cite{diffusion_lam}, boundary information from the future should clearly be useful for the model.
This choice is also crucial for producing \gls{LAM} forecasts that are consistent across the transition from the interior to the boundary area, without sudden discontinuities.
If the model is only fed $\bforcing_{k'}$, as in previous work \citep{neural_lam, yinglong, storm_cast}, it can not learn how to make a prediction that is consistent with boundary values at the future prediction time.
Instead the model has to keep correcting any boundary discontinuities as it is unrolled over multiple time steps.
By feeding also the future boundary forcing $\bforcing_{k'+1}$ to the model, it is tasked with directly producing a prediction consistent with the boundary values.
While we in principle could include boundary forcing from more than three time steps, values from times $t'_{k' - 1}, t'_{k'}$ and $t'_{k'+1}$ are sufficient for estimating second-order dynamics and the boundary state at the prediction time.

\paragraph{Different length of time steps}
With boundary forcing at different time steps than the interior state, point 2 in our desiderata, the time difference $t_k - t'_{k'}$ between interior time $t_k$ and boundary time $t'_{k'}$ can vary depending on $k$.
For example, with global data at \SI{6}{\hour} time steps and interior data at \SI{1}{\hour} steps this difference is between 0 and \SI{5}{\hour}.
The time difference between $t_k$ and boundary forcing at times $t'_{k' - 1}$ and $t'_{k' + 1}$ varies in similar ways.
The age of the boundary information should be an important factor for the model to take into consideration.
Information about these time differences does reach the model indirectly through time of day forcing features in $\iforcing$ and $\bforcing$.
However, we also hypothesize that including features that more directly describe this time difference information could be useful.
One way to achieve this is by encoding the time differences directly and including these as forcing features.
We experiment with sinusoidal encodings \citep{attention_all_you_need, transformer_ts_survey} for this in \cref{sec:exp_boundary_forcing}.

\paragraph{Different spatial resolution and grid layouts}
The final requirement in our desiderata concerns having boundary forcing at a different spatial resolution and grid layout.
The graph-based framework is inherently suitable for handling this irregularity, as it makes no assumptions about inputs on regular grids, as opposed to methods based on \glspl{CNN} or \glspl{ViT} \citep{storm_cast, diffusion_lam, yinglong}.
We only need to construct the edges $\gtmedges$ and $\mtgedges$ that map between the grid points and mesh graph in a suitable ways, including both the grid points in the interior and boundary.
The exact construction of these edge sets is discussed in details below in \cref{sec:graph_construction}.

\paragraph{Overlapping boundary}
An intriguing prospect of this flexible boundary forcing framework is to loosen the definition of the boundary region $\boundary$ as something that lays outside the interior $\interior$.
Instead of enforcing disjoint sets $\interior \cap \boundary = \varnothing$, we consider the possibility of including boundary forcing\footnote{For consistency we still refer to this as the \qm{boundary region} and \qm{boundary forcing} also in the overlapping setting, even though $\boundary$ is no longer restricted to lay outside the boundary of $\interior$.} also within the interior region as $\interior \subset \boundary$.
Such overlap is also considered by \cite{storm_cast}, but in a simpler setting with regular grids and $\boundary = \interior$.
With overlapping boundary forcing, the \gls{LAM} forecasting can also incorporate aspects of downscaling.
The model can both produce forecasts by simulating physics forward from the state at the previous time step, and through learning to re-grid variables in the coarser boundary forcing to higher resolution within the interior.
We experiment with this overlapping setup in \cref{sec:exp_boundary_forcing} and there further discuss these different model behaviors.

\subsection{Graph construction}
\label{sec:graph_construction}
The construction of the mesh graph defines how the model handles spatial relationships, and is thus an important design choice.
As the graph connectivity is independent of the grid layout of the data, there is a large design space of possible mesh graphs.
We consider here both multi-scale mesh graphs \citep{graphcast} and hierarchical mesh graphs \citep{neural_lam}.
For both types the construction procedure starts by building a number of mesh levels.
In the hierarchical graphs these levels are then kept separately, and additional edges added in-between \citep{neural_lam}.
In the multi-scale graphs the levels are instead collapsed into one graph, taking the union of their edge sets \citep{graphcast}.
To be able to perform this collapsing, nodes at the different levels are required to line up at some positions, which is generally guaranteed by construction.
It should be noted that the levels in the mesh graphs do not relate to vertical levels in the atmosphere, as variables from the full atmospheric column is already encoded to a joint latent vector before the mapping to the mesh graph.
While nodes and edges within each level can be laid out in an infinite number of ways, we here focus on two geometrically motivated procedures: rectangular and triangular graphs.

\begin{algorithm}[tbp]
\caption{Rectangular graph construction}
\label{alg:rect_graph}
\begin{algorithmic}[1]
    \For{each mesh level}
        \State Determine the node spacing, with higher levels having larger spacing. 
        \State Lay out nodes in regular rows and columns, covering both the interior and boundary. %
        \State Connect each node to its closest neighbors, horizontally, vertically and diagonally. \Comment{\cref{fig:rect_nodes_edges}}
    \EndFor \Comment{\cref{fig:rect_levels}}
    \If{multi-scale graph} %
        \State Combine all mesh levels into single graph. 
    \ElsIf{hierarchical graph} 
        \For{each mesh level, except bottom level}
            \State Add edges to and from the level below.
        \EndFor \Comment{\cref{fig:rect_hierarchical}}
    \EndIf
\end{algorithmic}
\end{algorithm}

\begin{figure}[tbp]
    \centering
    \begin{subfigure}[t]{0.33\textwidth}
        \centering
        \includegraphics[width=\textwidth]{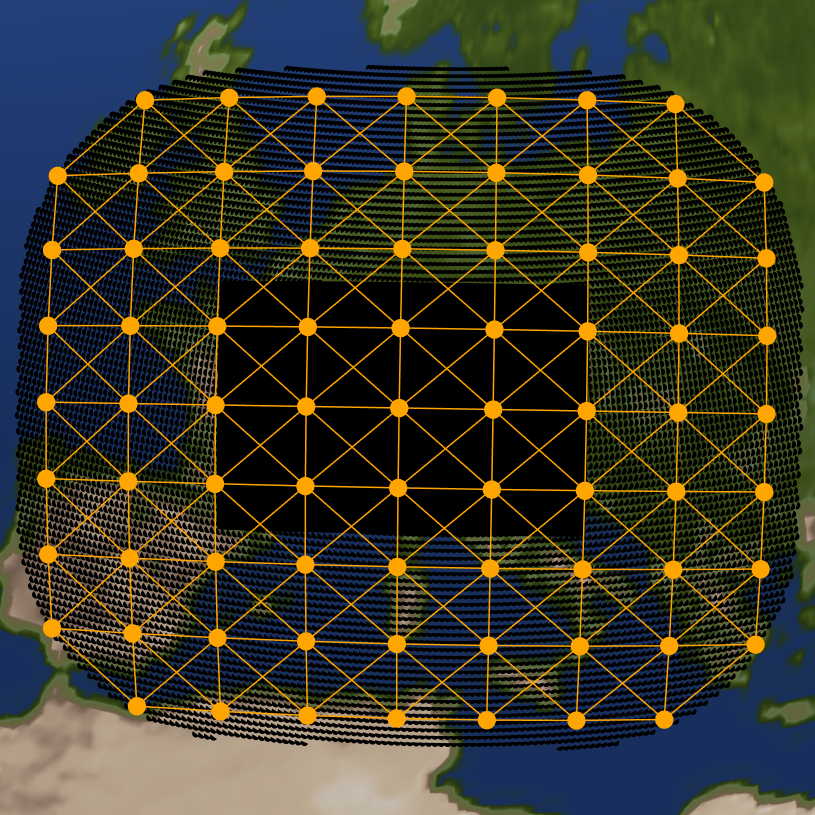}
        \caption{Node and edge layout}
        \label{fig:rect_nodes_edges}
    \end{subfigure}%
    \hfill%
    \begin{subfigure}[t]{0.33\textwidth}
        \centering
        \includegraphics[width=\textwidth]{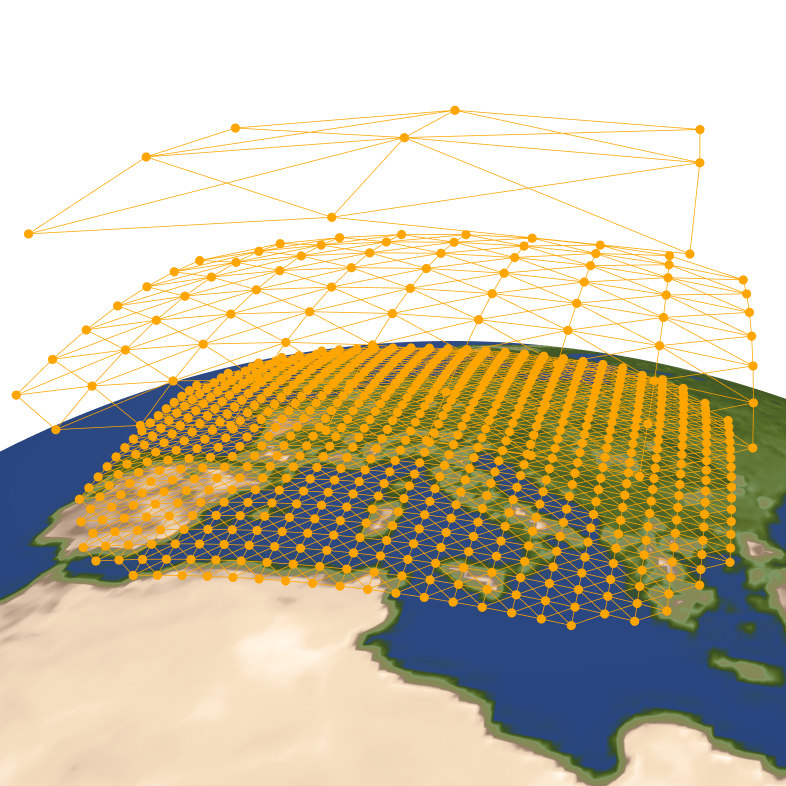}
        \caption{Mesh levels}
        \label{fig:rect_levels}
    \end{subfigure}%
    \hfill%
    \begin{subfigure}[t]{0.33\textwidth}
        \centering
        \includegraphics[width=\textwidth]{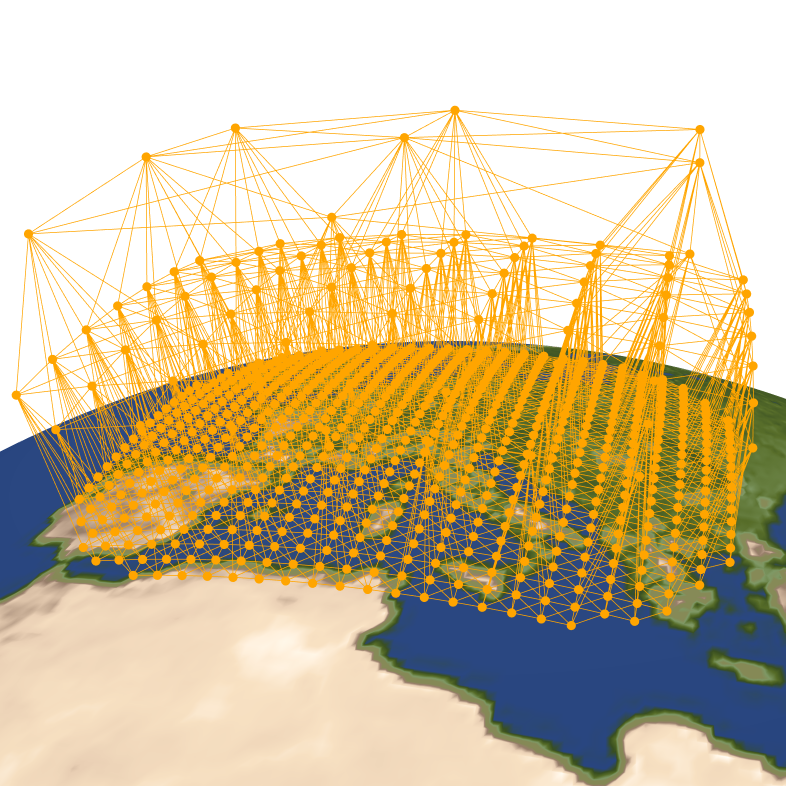}
        \caption{Hierarchical mesh graph}
        \label{fig:rect_hierarchical}
    \end{subfigure}%
    \caption[Illustration of different steps in the rectangular graph creation process.]{
        Illustration of different steps in the rectangular graph creation process. 
        \begin{inparaenum}[(a)]
            \item Nodes are laid out in a regular grid and connected to neighbors horizontally, vertically and diagonally.
            \item The set of mesh levels, with different number of nodes. Note that the vertical position of mesh levels is purely for visualization purposes and does not relate to vertical atmospheric levels. All mesh levels should be conceptually viewed as laying flat on the surface of the earth.
            \item 3-level hierarchical mesh graph, including edges connecting the different levels. Each node is connected to a set of $3 \times 3$ nodes at the level below, if present.
        \end{inparaenum}
    }
    \label{fig:rect_graphs}
\end{figure}

\paragraph{Rectangular graphs}
Rectangular graphs, or more verbosely quadrilateral graphs \citep{neural_lam}, are constructed within the two-dimensional \gls{CRS} where the interior data is laid out.
For example, in our \gls{DANRA} setting, this \gls{CRS} is defined by a Lambert conformal conic projection.
Rectangular graphs are built in this two-dimensional space following the high-level procedure described in \cref{alg:rect_graph} and illustrated in \cref{fig:rect_graphs}.
The nodes are placed only within the convex hull of all grid points.
For the multi-scale graphs node spacings are determined such that node positions align at multiple levels, enabling the combinations of all levels into one graph.
We do not make this restriction in the hierarchical graphs.
For further details we refer to \cite{neural_lam}.

\paragraph{Triangular graphs}
Triangular graphs, or more verbosely icosahedral graphs \citep{graphcast, graph_efm, aifs}, are constructed by iteratively subdividing an icosahedron built around the earth.
They are thus constructed in three-dimensional space on the surface of earth.
As opposed to rectangular graphs, this means that the node positions respect the spherical geometry of earth.
While this is likely not crucial for \gls{LAM} models operating on smaller regions, it becomes relevant as the size of the interior and boundary grows.
Triangular mesh graphs were originally proposed for global \gls{MLWP} models \citep{graphcast} and we here introduce a way to adapt these also to the \gls{LAM} setting.
This adaptation is done by subsetting nodes and edges to those within the spherical convex hull of all grid points.
The high-level procedure for constructing triangular mesh graphs is described in \cref{alg:tri_graph} and \cref{fig:tri_graphs}.
Note that the subsetting to the convex hull is done after all splitting of triangles.
This is important, as we want to keep smaller triangles in the mesh graph even if the larger triangles they originated from do not fully fit within the convex hull.
While the iterative subdivision process splits triangles in many iterations, we only keep a few of the finest meshes as mesh levels for the final graph.
Most of the meshes from this splitting process would have very few nodes when restricted to the convex hull, and are not relevant to include in the final mesh hierarchy.
For further details we refer to \cite{graphcast} for the multi-scale version and \cite{graph_efm} for the hierarchical adaptation.

\begin{algorithm}[tbp]
\caption{Triangular graph construction}
\label{alg:tri_graph}
\begin{algorithmic}[1]
    \State Construct icosahedron around earth. \Comment{\cref{fig:tri_ico}}
    \For{number of splits}
        \State Split each triangle in the previous mesh into 4 new ones. \Comment{\cref{fig:tri_splitting}}
    \EndFor
    \State Keep only the finest meshes, equal to the number of levels.
    \For{each mesh level}
        \State Subset mesh to spherical convex hull of all grid points. \Comment{\cref{fig:tri_chull}}
    \EndFor
    \If{multi-scale graph}
        \State Combine all mesh levels into single graph.
    \ElsIf{hierarchical graph} 
        \For{each mesh level, except bottom level}
            \State Add edges to and from the level below.
        \EndFor \Comment{\cref{fig:tri_hierarchical}}
    \EndIf
\end{algorithmic}
\end{algorithm}

\begin{figure}[tbp]
    \centering
    \begin{subfigure}[t]{0.33\textwidth}
        \centering
        \includegraphics[width=\textwidth]{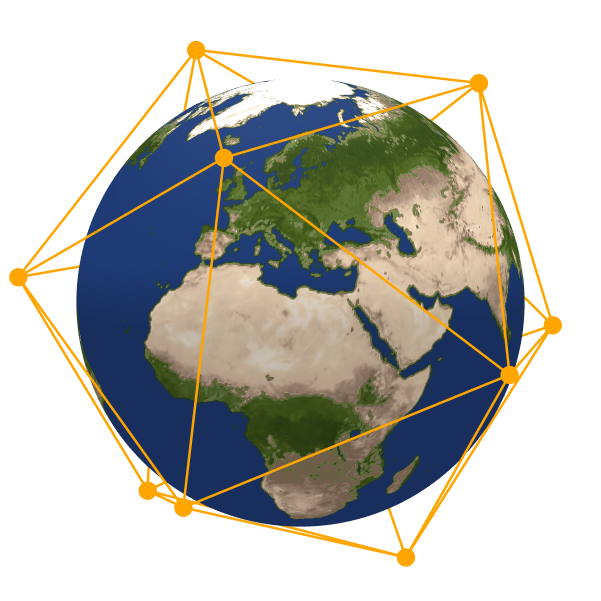}
        \caption{Global icosahedron}
        \label{fig:tri_ico}
    \end{subfigure}%
    \hspace{.1\textwidth}
    \begin{subfigure}[t]{0.33\textwidth}
        \centering
        \includegraphics[width=.9\textwidth]{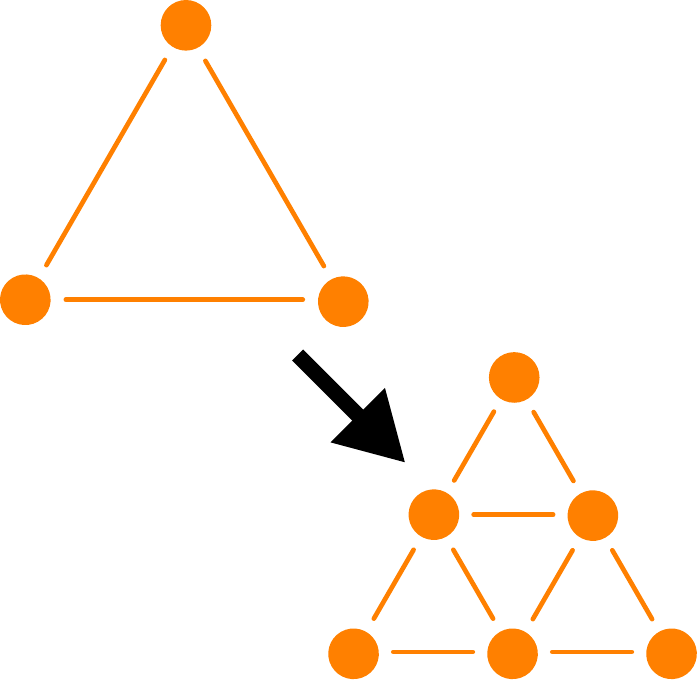}
        \caption{Triangle splitting}
        \label{fig:tri_splitting}
    \end{subfigure}\\
    \begin{subfigure}[t]{0.33\textwidth}
        \centering
        \includegraphics[width=\textwidth]{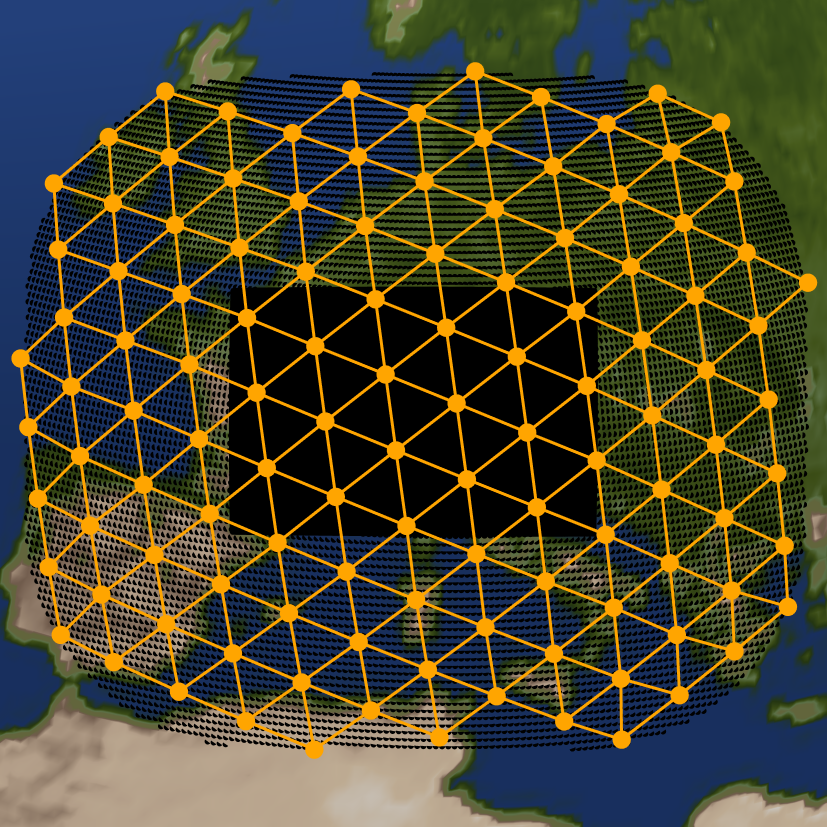}
        \caption{Node and edge layout}
        \label{fig:tri_chull}
    \end{subfigure}%
    \hspace{.1\textwidth}
    \begin{subfigure}[t]{0.33\textwidth}
        \centering
        \includegraphics[width=\textwidth]{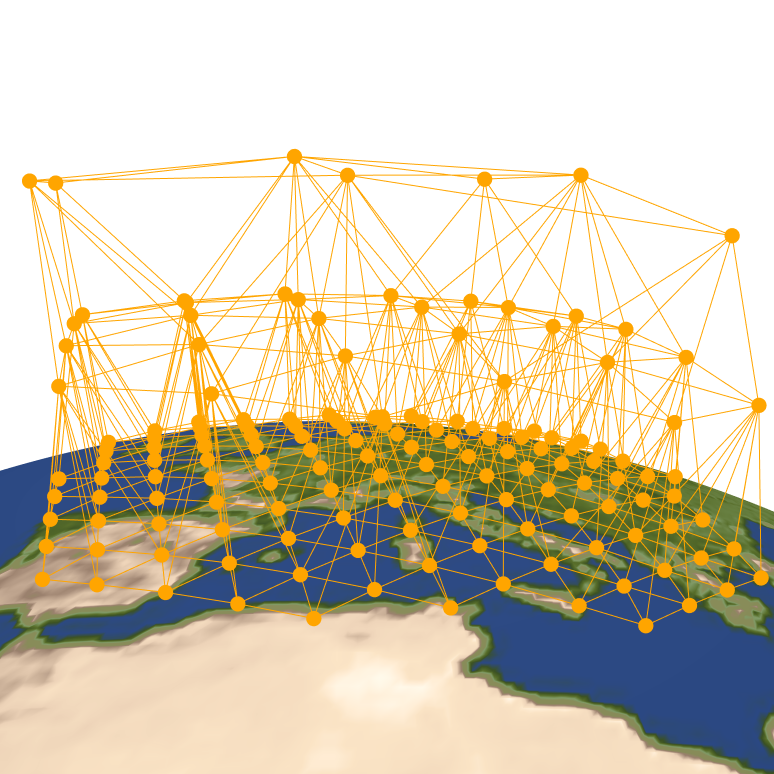}
        \caption{Hierarchical graph}
        \label{fig:tri_hierarchical}
    \end{subfigure}%
    \caption[Illustration of different steps in the triangular graph creation process.]{
        Illustration of different steps in the triangular graph creation process. 
        \begin{inparaenum}[(a)]
            \item The global icosahedron that the subdividing process starts from.
            \item Triangle splitting method. Each triangle of 3 nodes and 3 edges is split into 6 nodes and 9 edges.
            \item Nodes and edges in triangular graph, subset to the convex hull of all grid points.
            \item 3-level hierarchical mesh graph, including edges connecting the different levels. Each node is connected to 7 nodes at the level below, if present.
        \end{inparaenum}
    }
    \label{fig:tri_graphs}
\end{figure}

\paragraph{Connecting grid and mesh}
Once the mesh graph has been created, also the edge sets $\gtmedges$ and $\mtgedges$ have to be constructed to connect the grid points and mesh nodes.
For the hierarchical graphs, only the bottom level is connected to the grid through these edges.
When constructing the edge sets, grid points in the interior and boundary are treated separately.
This is necessary both due to possible differences in grid layout, but also due to their different roles in the model.
The $\gtmedges$ edges, that encode the grid input to the mesh graph, are created based on a radius around each point.
Each grid point is connected to every mesh node within this given radius.
Separate radii are used for interior grid points and boundary grid points.
These radii are chosen such that information is encoded from a sufficiently large neighborhood around each mesh node and to make sure that all nodes are connected and receive information from the model input.
The $\mtgedges$ edges, that decode from the mesh back to the grid for prediction, only connect to the interior grid points.
This is sufficient, as we do not make predictions in the boundary region.
Each interior grid points is in $\mtgedges$ connected to the mesh graph by looking up the polygon (rectangle or triangle) in the mesh graph that contains its coordinates.
Edges are then created from the corners (4 or 3 mesh nodes, respectively) of this polygon to the grid point.
This ensures that each grid point receives the spatially most relevant information for prediction.
Some further details about graph-related input features and the exact graph configurations used throughout our experiments are given in \cref{sec:graph_details}.

\subsection{Model training}
\label{sec:model_training}

\paragraph{Loss function}
We train our models by minimizing a \gls{MSE} loss function averaged over space and forecast length.
As we are purely interested in accurately forecasting the interior region, the boundary does not enter the loss formulation.
The exact loss for one forecast used for training is
\eq[loss]{
    \loss_{\fclen}
    =
    \frac{1}{
        \fclen
        \setsize{\interiorpoints}
    }
    \sum_{k=1}^{\fclen}
    \sum_{i \in \interiorpoints}
    \sum_{j \in \statevars}
    \varweight_j \invtimediff_j
    \left(\predstate_{k,i,j} - \wstate_{k,i,j}\right)^2
}
where
\begin{itemize}
    \item $\fclen$ is the number of forecast steps rolled out during training,
    \item $\interiorpoints \subset \interior$ is the discrete set of interior grid points,
    \item $\statevars$ is the set of variables that we forecast,
    \item $\invtimediff_j = \var\left[\wstate_{k+1,i,j} - \wstate_{k,i,j}\right]^{-1}$ is the inverse variance of time differences for variable $j$, a weighting which accounts for variables varying at different temporal scales \citep{graphcast},
    \item $\varweight_j$ is a manually chosen variable-specific weighting,
    \item $\predstate_{k,i,j}$ is the prediction from the model and $\wstate_{k,i,j}$ the corresponding ground truth value.
\end{itemize}
Training a deterministic model using the \gls{MSE} loss means that we are optimizing the model to produce forecasts that have the lowest error on average.
While this is a reasonable objective when producing a single deterministic forecast, it also comes with side-effects such as spatially smooth fields \citep{graph_efm, storm_cast, gencast}.
We discuss some perspectives on this in more detail in \cref{sec:discussion_limitations}.

\paragraph{Variable weighting}
To balance the influence of different variables during training, they are commonly weighted using per-variable weights $\varweight_j$ \citep{graphcast, neural_lam, aifs}.
While the overall framework allows for choosing these weights arbitrarily, in our experiments we choose to weight all variables uniformly.
Specifically, we weight each surface variable equal to the total weight assigned to a variable present at multiple vertical levels.
For example, the surface variable \SI{2}{m} temperature is given weight 1, whereas the vertical temperature, that is modeled at 8 pressure levels, is given weight $1 / 8$ per level.
Previous works have assigned lower weight to levels higher in the atmosphere \citep{graphcast, neural_lam}, steering the model to prioritize vertical levels closer to the surface.
While we still assign relatively high weights to surface variables, we do not prioritize among vertical levels.
This should make the models more applicable to a variety of use cases, such as within aviation, where accurate forecasts also of variables higher up in the atmosphere is important \citep{Mazzarella_2022_lam_aviation}. 
Reducing the weight with height would be appropriate for levels close to the model top of the underlying \gls{NWP} model, for example below \SI{5}{\hecto\pascal}.
At such levels the \gls{NWP} models apply strong Rayleigh damping to remove vertically propagating gravity waves. 
However, as our training data does not extend above the \SI{100}{\hecto\pascal} level it is not motivated to down-weight higher levels for this reason.

\paragraph{Training curriculum}
Unrolling forecasts during training of deterministic \gls{MLWP} models has proven important to achieve skillful and stable forecasting \citep{graphcast, graph_efm, fourcastnet}.
We follow a training curriculum of:
\begin{enumerate}
    \item Pre-training on $\fclen = 1$ single-step prediction,
    \item Fine-tuning on $\fclen > 1$ steps of rolled out forecasts.
\end{enumerate}
The fine-tuning helps the model adapt to the distributional shift of the input being a model prediction from the previous time rather than an initial condition from the dataset.
This induces stability also for longer rollouts during evaluation.
A challenge of the fine-tuning is however the larger computational cost and memory footprint, as each training step involves $\fclen$ forward passes through the model.
As it is often desirable to scale up the model size until only a single forward pass can fit into \gls{GPU} memory at once, the memory requirement is especially problematic.
To still enable fine-tuning on rollouts, we follow previous works \citep{graphcast, aifs, aurora} 
and make use of gradient checkpointing \citep{optimal_grad_checkpointing, grad_checkpointing} in-between each step $k$.
This trades off compute for a memory footprint near-constant in $\fclen$.

\subsection{Open source implementation}
\label{sec:open_source}
To promote reproducibility and enable other researchers to build on our work, we provide a complete open source implementation of our modeling framework.
We hope this will enable and accelerate the development of \gls{ML} \glspl{LAM} for more regions.
The implementation is split across three main repositories\footnote{
    For the experiments in this work we have made some additions to the code on top of the public main branches.
    Tagged releases of each package that exactly reproduces this paper can be found at 
    \begin{itemize}
        \item \url{https://github.com/joeloskarsson/neural-lam-dev/releases/tag/building-ml-lams},
        \item \url{https://github.com/sadamov/mllam-data-prep/releases/tag/building-ml-lams}
        \item \url{https://github.com/joeloskarsson/weather-model-graphs/releases/tag/building-ml-lams}
    \end{itemize}
    or alternatively in one archive at \url{https://doi.org/10.5281/zenodo.15161033}.
    Features from this work that have proven useful are being worked into the main branch of each repository. 
}:
\begin{itemize}
    \item \texttt{neural-lam} (\url{https://github.com/mllam/neural-lam}): Core implementation of the graph-based \gls{ML} \gls{LAM} architectures, including handling of boundary forcing (\cref{sec:boundary_forcing}), simple model verification, triangular graph construction and pipelines for data loading and model training.

    \item \texttt{mllam-data-prep} (\url{https://github.com/mllam/mllam-data-prep}):
          Preprocessing pipelines for converting meteorological datasets to formats that are optimized for efficient data loading during model training.
          The implementation uses \texttt{Xarray} \citep{xarray} for ingesting and processing datasets, and the training-optimized output is stored in compressed \texttt{Zarr} format \citep{zarr}.

    \item \texttt{weather-model-graphs} (\url{https://github.com/mllam/weather-model-graphs}): Construction of the rectangular graphs used with the models, including multi-scale and hierarchical graphs.
\end{itemize}
The implementation emphasizes modularity between different components, disentangling preprocessing, graph construction and model training.
The code has been designed for adaptation to new domains and data sources, avoiding design choices that restrict the usage to one specific \gls{LAM} setting.
To get started training models using the overall software framework, the only requirement is interior and boundary data in a format that can be read using \texttt{Xarray} \citep{xarray}. %
This open source approach enables validation of results and fosters international collaboration within \gls{MLWP} research.

\section{Experiments}
\label{sec:experiments}

\begin{figure}[tbp]
    \centering
    \begin{subfigure}[b]{0.5\textwidth}
        \centering
        \includegraphics[width=\textwidth]{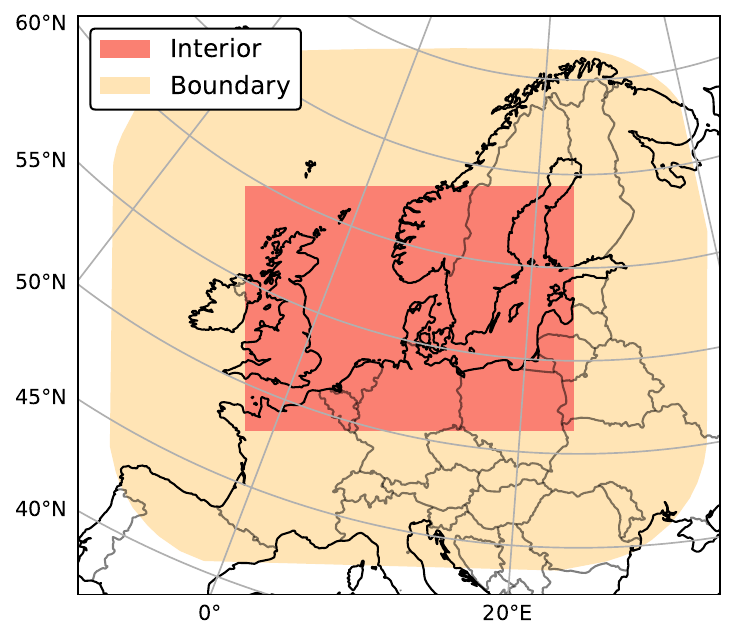}
        \caption{\gls{DANRA}}
    \end{subfigure}%
    \hfill%
    \begin{subfigure}[b]{0.5\textwidth}
        \centering
        \includegraphics[width=\textwidth]{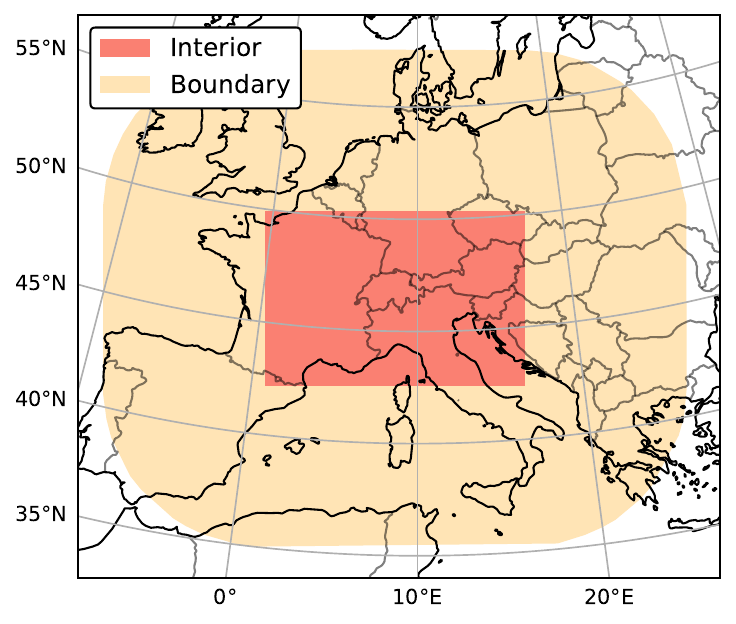}
        \caption{\gls{COSMO}}
    \end{subfigure}%
    \caption{The regions being modeled in our two \gls{LAM} settings. The boundary region shown here has width \SI{800}{\kilo\metre}. Note that while the boundary area can be substantially larger than the interior, the spatial resolution of the boundary is much lower, so it consists of fewer grid points.}
    \label{fig:model_regions}
\end{figure}

To understand key aspects of the proposed \gls{ML} \gls{LAM} framework, we carry out extensive experiments over two geographic regions:
one region covering northern Europe, centered on Denmark
and one region covering central Europe, centered on Switzerland.
These two regions are shown, together with example boundary setups, in \cref{fig:model_regions}.

This experiment section is structured as follows:
We first introduce the datasets and experimental setups considered (\crefrange{sec:data}{sec:metrics}).
The first experiments then investigate different design choices in the building of \gls{ML} \gls{LAM} models.
This includes choices related to integration of boundary forcing (\cref{sec:exp_boundary_forcing}) and graph construction (\cref{sec:exp_graph_construction}).
In \cref{sec:extra_finetuning} we additionally showcase the impact of different fine-tuning options.
Using the insights from these design studies, we then choose our final models for each region.
These models then undergo extensive verification in a realistic forecasting setting with boundary forcing from the global \gls{IFS} model.
The verification combines gridded evaluation against reanalysis/analysis data (\cref{sec:eval_gridded}) with point-wise station observations (\cref{sec:eval_station}), ensuring both broad-scale and localized accuracy assessment.
The evaluation strategy spans traditional meteorological metrics and specialized verification techniques for high-impact weather events.
We finally showcase example forecasts from both models for storm Ciara in February 2020 (\cref{sec:ciara_case_study}).

For the design studies we start from a standard model configuration and then adjust different aspects of the design.
This standard configuration is motivated by previous work \citep{neural_lam, graph_efm, graphcast} and computational considerations.
The setup uses an \SI{800}{\kilo\metre} boundary that does not overlap with the interior region.
The graph used is a rectangular hierarchical graphs with 3 levels.
We treat the dimensionality of latent vector representations (in nodes and edges) as the main parameter for model scaling.
This choice is crucial, as it determines both the model flexibility and the \gls{GPU} memory requirement.
We maximize this hyperparameter under the constraint of fitting one full autoregressive model step in memory during training.
The resulting settings are a latent dimensionality of 300 in the processor module and 150 for the encoding and decoding stages. 
As described in \cref{sec:model_training}, models are first pre-trained on single-step prediction and then fine-tuned using $\fclen = 4$ step rollout.
For further hyperparameter details, see \cref{sec:hyperparameter_details}.
In the design studies we focus on \gls{ERA5} boundary forcing both during training and evaluation, to isolate the impact of modeling choices.
We consider forecasts up to \SI{48}{\hour}, to include both early lead times with strong dependence on the initial state and a later period with more influence from the boundary forcing.

\subsection{Data}
\label{sec:data}
To clearly understand different design choices and quantify the performance of \gls{ML} \gls{LAM} models we consider an extensive set of data for training and verification.
A summary of all datasets used in this study is given in \cref{tab:datasets}.
Our design experiments are evaluated on the validation split, which covers 2018-11-05 -- 2019-10-22 for \gls{DANRA} and 2019-09-30 -- 2019-10-30 for \gls{COSMO}.
The final verification is carried out for the test period 2019-10-30 -- 2020-10-29 for both models.
Remaining data, from before the validation period, is all used for training.
During training we start forecasts from all time points in the datasets, but during evaluation we only consider 00 and 12 UTC as initialization times.
\begin{table}[tbp]
\centering
\caption{Summary of datasets used in experiments. While the original versions of this data is available for longer time periods, the length here denotes the subset used for this study. The length of the \gls{ERA5} subset used matches the training data for each model.}
\label{tab:datasets}
\begin{tabular}{@{}llllll@{}}
\toprule
\textbf{Dataset} & \textbf{Type} & \textbf{Use} & \textbf{Resolution} & \textbf{Time step} & \textbf{Length} \\ \midrule
COSMO-2/KENDA & Analysis & \multirow{2}{*}{\centering Training} & \SI{2.2}{\kilo\meter} & \SI{1}{\hour} & 4 years \\
DANRA & Reanalysis & & \SI{2.5}{\kilo\meter} & \SI{3}{\hour} & 21 years \\ \midrule
ERA5 & Reanalysis & \multirow{2}{*}{\centering Boundary} & \SI{0.25}{\degree} & \SI{6}{\hour} & 4/21 years \\
IFS HRES & Forecasts & & \SI{0.25}{\degree} & \SI{6}{\hour} & 2 years \\ \midrule
MCH SYNOP & Observations & \multirow{4}{*}{\centering Verification} & \centering 158 Stations & \centering 1 h & 1 year \\
DMI SYNOP & Observations & & \centering 50 Stations & \centering 1 h & 1 year \\
COSMO-E & Forecast & & \SI{2.2}{\kilo\meter} & \SI{1}{\hour} & 1 year \\
DANRA Forecasts & Forecast & & \SI{2.5}{\kilo\meter} & \SI{3}{\hour} & 1 year \\ \bottomrule
\end{tabular}
\end{table}

\paragraph{\gls{DANRA}}
The DANRA dataset is a regional reanalysis produced by the \gls{DMI} \citep{danra}.
The full dataset covers the years 1990--2020, and we use data from year 2000 onward.
Our split contains both surface variables and variables at pressure levels, all at \SI{3}{\hour} time steps.
DANRA has a spatial resolution of \SI{2.5}{\kilo\metre} and covers a $1472 \times $ \SI{1972}{\kilo\metre} area ($589 \times 789$ grid).
During training the boundary forcing for the \gls{ML} model is provided from the \gls{ERA5} global reanalysis \citep{ERA5}.
We use a version of \gls{ERA5} provided through Weatherbench 2 \citep{weatherbench2}, with \SI{6}{h} time steps and latitude-longitude gridding at \SI{0.25}{\degree} resolution.
Note that this boundary data has a different spatial and temporal resolution than the \gls{DANRA} data for the interior.
Our flexible methodology for integrating boundary forcing enables direct combination of such heterogeneous datasets.
During verification we instead run the model under realistic boundary conditions, coming from a forecast.
For this we use forecasts from the global \gls{IFS} model \citep{ifs}, at the same resolutions as \gls{ERA5}.
In our verification we include \gls{SYNOP} observations from 50 stations in Denmark as ground truth measurements.
We additionally consider a collection of \SI{18}{h} \gls{NWP} forecasts of surface variables for the \gls{DANRA} domain as a baseline.
These forecasts were created with the operational Harmonie \gls{NWP} forecasting system \citep{GleesonHarmonie24} and initialized from the operational analysis.
The \gls{DANRA} fields, used to initialize the \gls{ML} model, contain some additional information from \gls{SYNOP} stations compared to this operational analysis.
While this additional information is believed to have minor impact on the model performance, it should still be considered when interpreting the results.

\paragraph{\gls{COSMO}}
The \gls{COSMO} data consists of the operational analysis from \gls{MCH} \gls{KENDA} system, covering the Alpine region with Switzerland at its center.
This data has hourly temporal resolution and covers an area of approximately $860 \times 1280$ km with grid resolution \SI{2.2}{km} ($390 \times 582$ grid).
The domain uses a rotated pole projection with pole longitude \SI{190}{\degree} and pole latitude \SI{43}{\degree}.
Boundary forcing in the \gls{COSMO} setting uses the same \gls{ERA5} and \gls{IFS} data setup as DANRA.
For verification of the \gls{COSMO} \gls{ML} model we use \gls{SYNOP} observations in Switzerland, retrieved from 158 stations in the \gls{MCH} automatic monitoring network.
To compare the \gls{ML} model with an \gls{NWP} baseline we use a set of forecasts called COSMO-E, containing surface variables.
These are operational forecasts, initialized from the mean state of an ensemble \gls{DA} process. The \gls{ML} model however, is initialized from the deterministic analysis state (the control member). This should be considered when interpreting results. The comparison against ground truth can be seen as unfair to the NWP-model as the \gls{ML}-model was trained and initialized by the same data considered ground truth. The comparison against station based observations, however, is completely fair.

\paragraph{Variables}
\newcommand{\varsepline}{\arrayrulecolor{lightgray}\hdashline[5pt/10pt]}
\setlength\extrarowheight{1pt}

\begin{table}[tbp]
\centering
\caption{Variables used in the the \gls{DANRA} and \gls{COSMO} experimental setups. The \gls{ERA5}/\gls{IFS} columns describe the boundary forcing $\bforcing$ for each setup. Variables on specific vertical levels are abbreviated also using the corresponding level, for example geopotential at \SI{600}{\hecto\pascal} is abbreviated \wvar{z600}.}
\label{tab:variables}
\begin{tabular}{@{}lclcccc@{}}
\toprule
\textbf{} & \textbf{} & \textbf{} & \multicolumn{2}{c}{\textbf{DANRA Setup}} & \multicolumn{2}{c}{\textbf{COSMO Setup}} \\ \cmidrule(l){4-5} \cmidrule(l){6-7}
\textbf{Variable} & \textbf{Unit} & \textbf{Abbrev.} & DANRA & \begin{tabular}[c]{@{}c@{}}ERA5 \\ / IFS\end{tabular} & COSMO & \begin{tabular}[c]{@{}c@{}}ERA5 \\ / IFS\end{tabular} \\ \midrule
\textbf{Surface variables} &  &  &  &  &  &  \\ \midrule
\SI{10}{\meter} u-component of wind & \si{\metre\per\second} & \wvar{10u} & \checkmark & \checkmark & \checkmark & \checkmark \\ \varsepline
\SI{10}{\meter} v-component of wind & \si{\metre\per\second} & \wvar{10v} & \checkmark & \checkmark & \checkmark & \checkmark \\ \varsepline
\SI{2}{\meter} temperature & \si{\kelvin} & \wvar{2t} & \checkmark & \checkmark & \checkmark & \checkmark \\ \varsepline
Mean sea level pressure & \si{\pascal} & \wvar{msl} & \checkmark & \checkmark & \checkmark & \checkmark \\ \varsepline
Surface pressure & \si{\pascal} & \wvar{sp} & \checkmark & \checkmark & \checkmark & \checkmark \\ \varsepline
Net short wave radiation flux & \si{\watt\per\metre\squared} & \wvar{asob\_s} & \checkmark &  & \checkmark &  \\ \varsepline
Net long wave radiation flux & \si{\watt\per\metre\squared} & \wvar{athb\_s} & \checkmark &  & \checkmark &  \\ \varsepline
Sensible heat net flux & \si{\watt\per\metre\squared} & \wvar{ashfl\_s} &  &  & \checkmark &  \\ \varsepline
Total precipitation in \SI{1}{h} & \si{\metre} & \wvar{tp01} &  &  & \checkmark &  \\ \varsepline
Total precipitation in \SI{6}{h} & \si{\metre} & \wvar{tp06} &  &  &  & \checkmark \\ \midrule
\textbf{Variables on vertical levels} &  &  &  &  &  &  \\ \midrule
Geopotential & \si{\meter\squared\per\second\squared} & \wvar{z} & \checkmark & \checkmark &  & \checkmark \\ \varsepline
Pressure & \si{\pascal} & \wvar{pres} &  &  & \checkmark &  \\ \varsepline
Temperature & \si{\kelvin} & \wvar{t} & \checkmark & \checkmark & \checkmark & \checkmark \\ \varsepline
u-component of wind & \si{\metre\per\second} & \wvar{u} & \checkmark & \checkmark & \checkmark & \checkmark \\ \varsepline
v-component of wind & \si{\metre\per\second} & \wvar{v} & \checkmark & \checkmark & \checkmark & \checkmark \\ \varsepline
Vertical velocity & \si{\pascal\per\second} & \wvar{w} & \checkmark & \checkmark & \checkmark & \checkmark \\ \varsepline
Relative humidity & - & \wvar{r} & \checkmark &  & \checkmark &  \\ \varsepline
Specific humidity & \si{\kilo\gram\per\kilo\gram} & \wvar{q} &  & \checkmark &  & \checkmark \\ \midrule
\textbf{Forcing} &  &  & \multicolumn{1}{l}{} & \multicolumn{1}{l}{} & \multicolumn{1}{l}{} & \multicolumn{1}{l}{} \\ \midrule
Top of atmosphere radiation flux & \si{\watt\per\metre\squared} & \wvar{toarf} & \checkmark & \checkmark & \checkmark & \checkmark \\ \varsepline
$\sin$-encoded time of day & - & \wvar{sin\_d} & \checkmark & \checkmark & \checkmark & \checkmark \\ \varsepline
$\cos$-encoded time of day & - & \wvar{cos\_d} & \checkmark & \checkmark & \checkmark & \checkmark \\ \varsepline
$\sin$-encoded day of year & - & \wvar{sin\_y} & \checkmark & \checkmark & \checkmark & \checkmark \\ \varsepline
$\cos$-encoded day of year & - & \wvar{cos\_y} & \checkmark & \checkmark & \checkmark & \checkmark \\ \midrule
\textbf{Static fields} &  &  & \multicolumn{1}{l}{} & \multicolumn{1}{l}{} & \multicolumn{1}{l}{} & \multicolumn{1}{l}{} \\ \midrule
Land-sea mask & - & \wvar{lsm} & \checkmark & \checkmark & \checkmark & \checkmark \\ \varsepline
Surface geopotential (orography) & \si{\meter\squared\per\second\squared} & \wvar{oro} & \checkmark & \checkmark & \checkmark & \checkmark \\ \arrayrulecolor{black}\bottomrule
\end{tabular}
\end{table}

\setlength\extrarowheight{0pt}

We choose to model a selection of variables both at the surface and at vertical levels in the atmosphere.
The full list of variables is given in \cref{tab:variables}.
This selection was motivated by both meteorological relevance and data availability.
The variables were also chosen to largely match between the two settings, with some exceptions.
The main exception to this is that the \gls{DANRA} vertical variables are modeled on pressure levels (100, 200, 400, 600, 700, 850, 925 and \SI{1000}{\hecto\pascal}), whereas the \gls{COSMO} vertical variables are on the original model levels based on terrain following sigma coordinates \citep{numerical_weather_and_clim_pred}.
Vertical variables in the boundary forcing uses the same pressure levels as \gls{DANRA} for both setups.
The specific subset of 8 model levels in the \gls{COSMO} setup was chosen to on average be close to the pressure levels of the boundary forcing.
It should be noted that because of our flexible boundary encoding framework, working with different vertical levels in the interior and boundary does not create any modeling problems.
All these vertical variables are mapped into joint latent spaces, and the model is tasked with learning their interactions.
Similarly, this flexibility also allows for using entirely different variables in the interior and boundary, such as relative humidity in the interior and specific humidity in the boundary forcing.
We also forecast variables, specifically surface radiation fluxes, that are only present in the interior datasets.

\paragraph{Forcing and static fields}
In addition to the forecasted variables in the state $\wstate$, \cref{tab:variables} also contains forcing $\iforcing$ and static fields $\istatic$ given as input to the model.
All the forcing fields are derived purely from coordinates and timestamps.
Following previous work \citep{graphcast,neural_lam} we compute each vector of forcing features from a window of three timestamps, so $\iforcing_k$ contains features computed based on times $t_{k-1}$, $t_{k}$ and $t_{k+1}$.
This forcing was selected to help the model capture diurnal and seasonal variations.
The static fields include a land-sea mask and orography, encoding information about the terrain.
Worth noting is that we do not include spatial coordinates as a direct inputs to the model, in contrast to some previous works \citep{neural_lam,graphcast,aifs}.
Including coordinates as inputs explicitly breaks the spatial equivariance of the model.
We prefer to keep this equivariance, to guarantee that we simulate the same physics everywhere.
This could in particular be useful for later performing transfer learning to a new spatial region.

\subsection{Efficiency, training times and hardware}
\label{sec:time_hardware}
Our models are capable of rapidly producing long forecasts using only a single \gls{GPU}.
Producing a 5 day forecast takes
1 minute on a GH200 chip for our final \gls{COSMO} model and 
50 seconds on an A100 \gls{GPU} for our final \gls{DANRA} model.
This is a massive speedup, using comparatively modest hardware, in relation to existing \gls{NWP} \glspl{LAM}.

Training \gls{MLWP} models, especially on high-resolution data, requires the use of state-of-the-art \gls{GPU}clusters.
With high-resolution data a key constraint is often the \gls{GPU} memory.
We here work with \glspl{GPU} with memory $\geq$ \SI{80}{\giga\byte}, make use of reduced numerical precision through \texttt{bfloat16} computations \citep{bfloat16} and use a batch size of 1 per \gls{GPU}.
In \cref{tab:compute_resources} we give a complete overview of our computational setup and training time, both for a single model and for all experiments in the paper.
We note that the \gls{COSMO} models demand substantially more fine-tuning, which we relate to the shorter time steps used.
\begin{table}[tbp]
    \centering
    \caption{
    Computational resource utilization for model training across \gls{DANRA} and COSMO datasets, including training epochs for each phase. Both configurations demonstrate efficient scaling across distributed compute infrastructure, with consistent performance characteristics for pre-training and fine-tuning phases.
    \textsuperscript{*}While our standard \gls{COSMO} setup uses fine-tuning with 4 rollout steps, we also experiment with up to 12 steps in \cref{sec:extra_finetuning}.
    }
    \begin{tabular}{@{}lcc@{}}
        \toprule
        \textbf{Configuration}       & \textbf{DANRA}   & \textbf{COSMO} \\
        \midrule
        Dataset size (time steps)    & 54,896           & 33,660         \\
        GPU Type                     & NVIDIA A100/H100 & NVIDIA GH200   \\
        Total GPUs per run           & 16               & 256 (H100s)  \\
        \midrule
        \textbf{Pre-training Phase}  &                  &                \\
        \midrule
        Epochs                       & 80               & 200            \\
        Autoregressive rollout steps & 1                & 1              \\
        Average training time [h]    & 144              & 12             \\
        Total GPU hours              & 2,304             & 3,072          \\
        \midrule
        \textbf{Fine-tuning Phase}   &                  &                \\
        \midrule
        Epochs                       & 3                & 50             \\
        Autoregressive rollout steps & 4                & 4\textsuperscript{*}\\
        Average training time [h]    & 36               & 14             \\
        Total GPU hours              & 576              & 3,584          \\
        \midrule
        \textbf{Number of Trainings} &                  &                \\
        \midrule
        Pre-training                 & 12                & 9              \\
        Fine-tuning                  & 12                & 12             \\
        \midrule
        \textbf{Total GPU-hours}     & 34,560                & 70,656         \\
        \bottomrule
    \end{tabular}
    \label{tab:compute_resources}
\end{table}

\subsection{Metrics}
\label{sec:metrics}
Our verification employs multiple complementary metrics to assess different aspects of forecast quality. 
Metrics are calculated using the Scores library \citep{scores}, or for simple calculations directly using Xarray \citep{xarray} or NumPy \citep{numpy}.
Below we define the different metrics used.
We typically average over $\numsamples$ forecasts initialized at different times, and over all interior grid points in the set $\interiorpoints$.
Metrics in relation to sparse station observations are computed the same, but with $\interiorpoints$ then made up of only the station locations.

\paragraph{RMSE} For our model design experiments we mainly consider \gls{RMSE} for a few highly interesting variables.
The \gls{RMSE} metric for a specific variable $y$ and lead time $t_k$ is defined as
\eq[rmse_def]{
    \text{RMSE}_k = \sqrt{\frac{1}{\numsamples \setsize{\interiorpoints}}\sum_{i=1}^{\numsamples} \sum_{j \in \interiorpoints} \left(\hat{y}_{k,j}^{(i)} - y_{k,j}^{(i)}\right)^2}
}
where $\hat{y}_{k,j}^{(i)}$ is the forecasted value at grid point $j$ and lead time $t_k$ in forecast $i$ and $y_{k,j}^{(i)}$ the corresponding ground truth value.
For wind fields we compute \gls{RMSE} based on the vector difference as
\eq[rmse_wind_def]{
    \text{RMSE}_k^{\text{wind}}
    =
    \sqrt{
        \frac{1}{\numsamples \setsize{\interiorpoints}}
        \sum_{i=1}^{\numsamples}
        \sum_{j \in \interiorpoints}
        \left(\hat{u}_{k,j}^{(i)} - u_{k,j}^{(i)}\right)^2 + \left(\hat{v}_{k,j}^{(i)} - v_{k,j}^{(i)}\right)^2
    }
}
where $\hat{u},u$ and $\hat{v},v$ are specifically the u and v wind component variables.

\paragraph{Normalized MAE} 
For the 3D variables, on vertical levels, we use the \gls{NMAE}, defined
\alse[nmae_def]{
    \text{NMAE}_k
    &= 
    \frac{1}{\numsamples \setsize{\interiorpoints}}\sum_{i=1}^{\numsamples} \sum_{j \in \interiorpoints} 
    \frac{
        \left|\hat{y}_{k,j}^{(i)} - y_{k,j}^{(i)}\right|
    }{
    \Delta_k
    }\\ %
    \Delta_k
    &=
    \max_{i,j} y_{k,j}^{(i)} 
    - 
    \min_{i,j} y_{k,j}^{(i)} 
}
where $\Delta_k$ represents the range of the ground truth values.
This normalization expresses the error as a percentage of the variable's range, making it particularly useful for comparing errors across vertical levels where variables can have substantially different scales. %

\paragraph{ME (Mean Error or Bias)} 
The \gls{ME}
\begin{equation}
    \text{ME}_k = 
    \frac{1}{\numsamples \setsize{\interiorpoints}}
    \sum_{i=1}^{\numsamples} \sum_{j \in \interiorpoints} 
    \left(\hat{y}_{k,j}^{(i)} - y_{k,j}^{(i)}\right),
\end{equation}
also known as bias, measures the average difference between the predicted and true values:
The \gls{ME} provides insight into systematic over- or under-prediction by the model. A positive \gls{ME} indicates a tendency to overforecast, while a negative \gls{ME} suggests underforecasting. Ideally, the \gls{ME} should be close to zero.

\paragraph{STDEV\_ERR (Standard Deviation of the Error)} 
The \gls{STDEV_ERR} measures the spread or variability of the errors around the mean error
\begin{equation}
    \text{STDEV\_ERR}_k = \sqrt{\frac{1}{\numsamples \setsize{\interiorpoints}-1}\sum_{i=1}^{\numsamples} \sum_{j \in \interiorpoints} \left[\left(\hat{y}_{k,j}^{(i)} - y_{k,j}^{(i)}\right) - \text{ME}_k\right]^2}.
\end{equation}

\paragraph{Threshold-based metrics} 
For evaluating categorical forecasts, we use the \gls{ETS} and \gls{FBI}.
These metrics are based on set threshold values, and count the number of times the ground truth data and forecast exceeds these.
The \gls{ETS} is defined as
\begin{equation}
    \text{ETS}_k = \frac{H-H_r}{H+M+F-H_r}
\end{equation}
where $H$ represents hits (both forecast and data exceed threshold), $M$ misses (only data exceeds threshold), $F$ false alarms (only model exceeds threshold), and $H_r$ the number of hits of a random forecast \citep{calculating_ets}.
The score ranges from $-\frac{1}{3}$ to 1, with 0 representing skill comparable to a random guess.

For a threshold $\tau$, the \gls{FBI} is defined
\begin{equation}
    \text{FBI}_k
    = 
    \frac{
        \sum_{i=1}^{\numsamples} \sum_{j \in \interiorpoints} 
        \mathbb{I}\left(\hat{y}_{k,j}^{(i)} \geq \tau\right)
    }{
        \sum_{i=1}^{\numsamples} \sum_{j \in \interiorpoints} 
        \mathbb{I}\left(y_{k,j}^{(i)} \geq \tau\right)
    }
\end{equation}
where $\mathbb{I}(\cdot)$ is the indicator function.
Standard thresholds are used for precipitation (0.1, 1, 5 mm/h) and wind speed (2.5, 5, 10 m/s), corresponding to meteorologically significant levels.

\paragraph{Precipitation SAL score} For spatial verification of precipitation, we employ the \gls{SAL} score.
\gls{SAL} combines three components as $|S| + |A| + |L|$,
where $S \in [-2, 2]$, $A \in [-2, 2]$ and $L \in [0, 2]$, giving a total range of $[0,6]$ for the combined metric.
Values close to 0 are better, indicating good alignment between forecast and data.
The different components evaluate the structure, amplitude and location of rain clouds.
For details on the computation of these components we refer to
\cite{SAL}.
We calculate SAL using the PySteps implementation \citep{pysteps}.

\paragraph{\acrfull{LSD}}
For verification of the spatial scales represented by the models we compute the spectra of forecasts in \cref{sec:eval_gridded}.
To quantitatively measure deviations from the spectra of the ground truth data we use \gls{LSD}
\begin{equation}
    \text{LSD} 
    =
    \sqrt{
    \frac{1}{\fclen\numsamples\Lambda}
    \sum_{k=1}^{\fclen}\sum_{i=1}^{\numsamples} 
        \sum_{\lambda=1}^{\Lambda} 
        \left[
            \log_{10}S\left(\hat{y}_{k}^{(i)}\right)_\lambda
            - 
            \log_{10}S\left(y_{k}^{(i)}\right)_\lambda
        \right]^2
    }
\end{equation}
where $S(y)$ is the spectra of the field $y$ and $\lambda$ denotes the specific wavenumber, with $\Lambda$ the total range of the spectra.

\subsection{Integrating boundary forcing}
\label{sec:exp_boundary_forcing}
The incorporation of boundary forcing is crucial for \gls{ML} \gls{LAM} models to retain skill over time.
We thus devote substantial effort in this section to empirically investigating open questions related to the boundary region and the integration of boundary forcing in the model.

\paragraph{Boundary width}
To comprehensively study the impact of boundary region size we train models using five different boundary widths: 0, 400, 800, 1200, and \SI{1600}{\kilo\metre}.
The \SI{0}{\kilo\metre} case serves as an edge case to evaluate model behavior without explicit boundary forcing, while the maximum width of 1600 km tests the benefits of extended spatial context. 
This range of distances was motivated by how far different meteorological features can travel during one time step.
The maximum distance considered corresponds to approximately twice the distance traversable by the fastest meteorological features in our domain, which is 750 km per 3-hour time step in \gls{DANRA}.

\cosmoplots{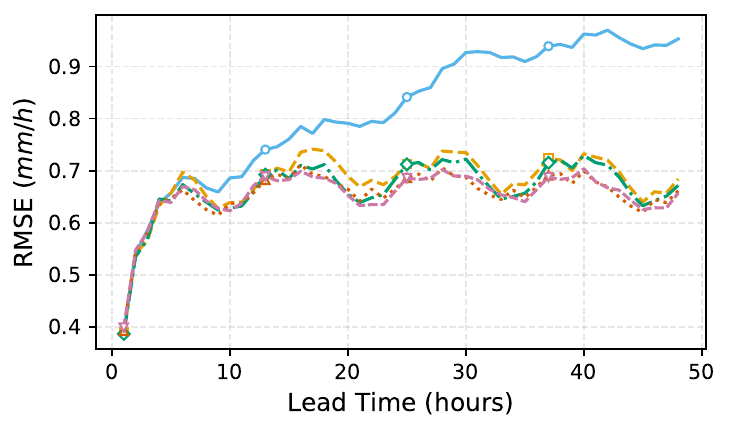}{cosmo_boundary_width_experiment}{
    \gls{RMSE} on the validation set for \gls{COSMO} models with different boundary widths. The boundary forcing from all models comes from \gls{ERA5}.
}

In \cref{fig:cosmo_boundary_width_experiment} we show results for models trained for the \gls{COSMO} domain and with boundary areas of different width.
We first note that removing the boundary forcing completely significantly degrades model performance. After a few hours the model clearly diverges from the ground truth, missing the information coming from the boundary region.
This is expected, as the model has no way of predicting the atmospheric state  propagating into the region.

When considering the different boundary widths, our experiments reveal that even a \SI{400}{\kilo\metre} boundary zone captures enough relevant meteorological phenomena to be useful for the model. 
We note no significant difference in performance between all the different boundary widths included in \cref{fig:cosmo_boundary_width_experiment}.
This does imply that the models do not make crucial use of fast-moving features, propagating into the interior region over a single time step.
While these results are for the \gls{COSMO} setup with the shorter \SI{1}{\hour} step, in \cref{sec:extra_boundary_width} we show similar results for the \gls{DANRA} setup.
Also in that setting, with \SI{3}{\hour} time steps, the \SI{400}{\kilo\metre} boundary is sufficient.
Overall we conclude that the boundary forcing is crucial for model performance, but the width of the boundary region does not play a large role.
While the sufficiency of a smaller boundary region is positive for practical applications, it should be noted that the boundary size does not have a major impact on the computational cost of the model.
This is because of our flexible boundary forcing framework, where the boundary data can remain at its native, coarser, resolution.
Thus the boundary consists of very few grid points, contributing little to both data storage and \gls{GPU} memory.
While we could consider finding the smallest boundary width $\leq \SI{400}{\kilo\metre}$ that is still sufficient, this is of little practical interest due to the minor computational contribution of the boundary.

\paragraph{Overlapping boundary region}
Given the freedom of our framework in choosing the boundary region, we here experiment with the possibility of having a boundary overlapping with the interior \gls{LAM} area.
An overlapping boundary region $\boundary$ is defined as the set of points within the interior $\interior$ and all points at a distance $\leq$ \SI{800}{\kilo\metre} from it.
As opposed to \cite{storm_cast}, we consider boundary forcing both overlapping the interior and extending beyond it.
This is well-motivated, as the meteorological features entering the interior domain during one time step are located outside of it at the initialization time.
Note that the boundary forcing $\bforcing$ located within the interior region is defined in different grid points than the state $\wstate$, still at the original coarser resolution.
The information from both of these sets of grid points is combined through the mapping to the mesh graph using $\gtmedges$.

\cosmoplots{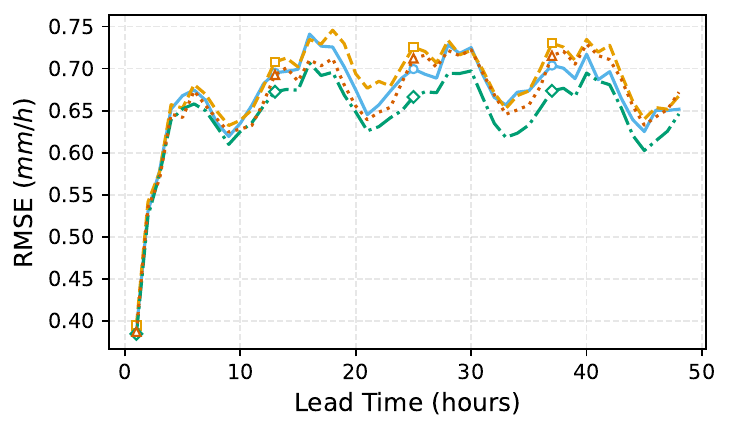}{cosmo_overlapping_experiment}{
    \gls{RMSE} on the validation set for \gls{COSMO} models with and without boundary forcing overlapping the forecasting area (interior). Results are shown for settings with boundary forcing either from \gls{ERA5} or \gls{IFS}.
}

\danraplots{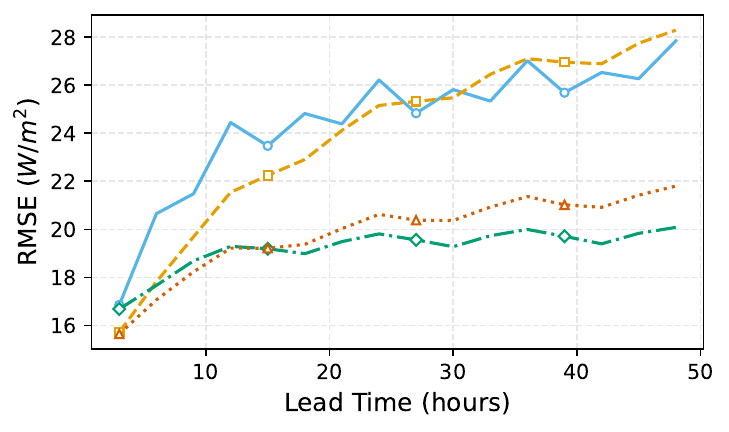}{danra_overlapping_experiment}{
    \gls{RMSE} on the validation set for \gls{DANRA} models with and without boundary forcing overlapping the forecasting area (interior). Results are shown for settings with boundary forcing either from \gls{ERA5} or \gls{IFS}.
}

While we generally only focus on boundary forcing from \gls{ERA5} in these design studies, we here additionally consider model performance also using \gls{IFS} boundary forcing, as this is especially interesting for the overlapping boundary setting.
We present results for the \gls{COSMO} setting in \cref{fig:cosmo_overlapping_experiment} and for the \gls{DANRA} setting in \cref{fig:danra_overlapping_experiment}.

When comparing models forced by \gls{ERA5} boundaries, allowing overlap between boundary and interior regions improves model performance across variables. 
The overlap enables the model to effectively downscale global weather patterns, while simultaneously simulating regional dynamics. 
Thus the model operates as a form of hybrid between statistical downscaling and \gls{LAM} forecasting.
Note that this intriguing configuration is something made possible by our \gls{ML} framework, and would be highly challenging to achieve with physics-based \gls{NWP}.
The impact varies by variable, with some meteorological fields benefiting more from this downscaling capability than others.
Improvements from the overlapping setup are also small for shorter lead times, when the most important information comes from the initial condition rather than the boundary forcing.

It should come as no surprise that a hybrid downscaling model keeps being skillful forever as long as we feed it with \gls{ERA5} boundary forcing.
However, it is interesting to consider the model performance as we swap out the boundary forcing to \gls{IFS} forecasts, which themselves have an error increasing with lead time.
For the \gls{COSMO} models in \cref{fig:cosmo_overlapping_experiment} we see higher errors with \gls{IFS} boundaries compared to \gls{ERA5} forcing, but the overlapping setup is still significantly better.
With the short \SI{1}{\hour} time steps, this overlapping configuration helps ground the model in the large-scale atmospheric state, reducing error accumulation throughout the many autoregressive steps.
In the \gls{DANRA} case, in \cref{fig:danra_overlapping_experiment}, the switch to \gls{IFS} boundaries has a major impact, largely removing the performance gap between the models with and without overlapping boundary.
In this \gls{IFS} setting, the \gls{DANRA} model without overlap even gives lower errors at short lead times.
This seems to be because of an over-reliance on the boundary forcing in the overlapping model.
Instead of focusing on simulating dynamics from the well-estimated high-resolution initial condition it mainly downscales the global forecast, which is a less accurate representation of the atmosphere in the interior domain.

\Cref{fig:danra_overlapping_experiment} also uncovers a peculiar oscillating error in the overlapping \gls{DANRA} model with \gls{IFS} boundary forcing.
To investigate this closer, we note that the green line (overlap, ERA5 forcing) in \cref{fig:danra_overlapping_experiment_wv10m,fig:danra_overlapping_experiment_z600} is also somewhat jagged, but in the opposite direction. 
This means that when the error of the model is high when forced by IFS, it is low when forced by ERA5. 
This points to an over-reliance on the boundary forcing at lead times \SI{6}{\hour}, \SI{12}{\hour}, \dots, but less at \SI{3}{\hour}, \SI{9}{\hour}, \dots.
The first set of lead times corresponds to those where the boundary forcing input $\bforcing_{k'+1}$ temporally aligns with the forecast time $t_{k+1}$ (recall that the boundary forcing has \SI{6}{\hour} time steps, whereas the \gls{DANRA} model takes \SI{3}{\hour} steps).
This indicates that the model at these times focuses mainly on downscaling the overlapping boundary forcing. 
Consequently, at the other lead times (\SI{3}{\hour}, \SI{9}{\hour}, \dots) the model would then rather be focusing on actually simulating the atmospheric dynamics forward. 
This does however not explain why the error drops at these lead times, when the \gls{IFS} boundary forcing is used.
By visually inspecting the forecasts from this model, we have also noted that they tend to lack coherence over time.
What seems to be happening is that the model at lead times \SI{3}{\hour}, \SI{9}{\hour}, \dots is rather simulating the dynamics forward \SI{6}{h} from the state $\wstate_{k-1}$, instead of using the (mainly downscaled) previous state $\wstate_k$.
There is essentially two forecasting processes going on at the same time, one that is mostly downscaling and following the boundary forcing and one that is mainly simulating dynamics with \SI{6}{\hour} time steps. 
This odd model behavior does explain the jagged error lines in \cref{fig:danra_overlapping_experiment}.
The reason why this behavior does not appear in the COSMO setting is likely because of the difference in time step pairing (\SI{1}{\hour} and \SI{6}{\hour}, compared to \SI{3}{\hour} and \SI{6}{\hour} in \gls{DANRA}).
Taking 5 steps in-between each alignment of $\bforcing_{k'+1}$ and $t_{k+1}$ should make the model more focused on actually simulating dynamics, making this dual-mode behavior less of an optimal solution.
While we here make no specific changes to the model in the overlapping boundary setup, these issues point towards possibilities for further model improvements.
We further discuss directions for such future work in \cref{sec:discussion_future}.

\paragraph{Boundary forcing time}
Our \gls{ML}\gls{LAM} framework allows to include any number of boundary forcing time steps in the past and the future of the current forecast time.
This opens up great flexibility, but also requires additional capabilities from the model.
We here investigate some of the design choices related to temporal aspects of the boundary forcing.
To understand these, we train models without the future boundary forcing $\bforcing_{k'+1}$ as input and models with explicit sinusoidal encodings of the time difference between interior and boundary.
The results from these experiments are shown in \cref{fig:cosmo_boundary_handling_experiment} for the \gls{COSMO} setup and in \cref{sec:extra_boundary_handling} for \gls{DANRA}.
As the conclusions for the \gls{COSMO} and \gls{DANRA} here closely align we focus on only the \gls{COSMO} setup in the main paper.
Note that we here again consider our standard setup, which does not have overlapping boundary forcing.

\cosmoplots{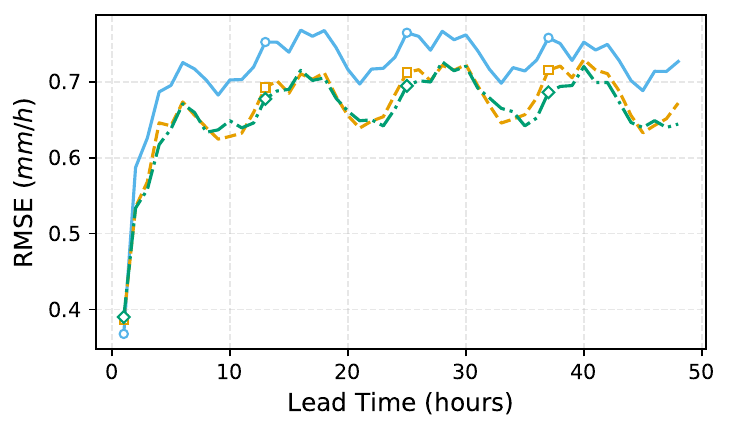}{cosmo_boundary_handling_experiment}{
    \gls{RMSE} on the validation set for \gls{COSMO} models with different configurations for the boundary forcing.
    For the model with future boundary (our standard model), $\bforcing_{k' + 1}$ is included in its input, whereas for the model with no future boundary it is not.
    All boundary forcing here comes from \gls{ERA5}.
}

From \cref{fig:cosmo_boundary_handling_experiment} we note that removing the future boundary forcing leads to a significant degradation in model performance. 
Having both past and future boundary forcing allows the model to internally interpolate between the two, improving forecast quality.
The possibility of including future boundary information was also considered by \cite{diffusion_lam} in a less realistic setting, drawing the same conclusion.
Still, most \gls{ML} \gls{LAM} models so far have not made use of this possibility \citep{neural_lam,storm_cast,yinglong}.

As we allow for interior states and boundary forcing with different time steps, the difference $t_k - t'_{k'}$ between interior time $t_k$ and boundary time $t'_{k'}$ can vary depending on $k$.
As noted in our investigation of overlapping boundary forcing, the relationship between these time points can be highly important for the model.
It is thus necessary that the model is able to clearly learn the valid time of the different input features.
We experiment with explicitly including the time differences $(t_k - t'_{k'-1}), (t_k - t'_{k'}), (t_k - t'_{k'+1})$ as inputs, as part of the boundary forcing.
Before being fed to the model, these times are encoded as sinusoidal features \citep{attention_all_you_need, transformer_ts_survey}.
\Cref{fig:cosmo_boundary_handling_experiment} shows the error of our standard model and a model trained including this additional time embedding. 
The results show that the time embedding does not significantly improve model performance. 
The model seems able to learn the temporal dependencies between the boundary and interior data without this additional complexity. 
As these temporal relationships can already be derived from the time of day and day of year forcing features, this appears sufficient to capture the temporal dependencies.
Hence it is not beneficial to include this time-embedding to the model.

\subsection{Graph design}
\label{sec:exp_graph_construction}
A key choice in our \gls{ML} \gls{LAM} framework is how to construct the mesh graph used for the processor component in the model. 
To shed some light on this, we here perform an extensive evaluation by training models using different configurations of graphs.
As described in \cref{sec:graph_construction}, we consider both rectangular and triangular graphs, as well as multi-scale and hierarchical configurations.
This is the first large scale comprehensive comparison of all these different graph constellations used in \gls{MLWP}, with previous works comparing only some of the options \citep{graph_efm, graphcast}.
It should be noted that while we mainly discuss the multi-scale and hierarchical graphs as different graph constructs, these are necessarily always used with different \gls{GNN} processors.
So when considering the difference between these graph constructs, we are also considering the difference between the processor \glspl{GNN} tailored to working with them.

Our main graph investigation is carried out for the \gls{DANRA} setup, with a smaller study for the \gls{COSMO} domain given in \cref{sec:extra_graph_design}. 
For rectangular graph configurations, we consider hierarchical graphs with 2--4 levels and multi-scale graphs where 3 or 4 levels have been combined to make up the final mesh.
For triangular graphs we focus on 3 and 4 level hierarchical versions and a multi-scale variant with 3 levels combined.
More details about the different configurations are given in \cref{sec:graph_details}.
We trained a model also with a single level rectangular graph, but as the forecasts from this model were significantly worse than any of the others we do not include it in this comparison.
This poor performance of single mesh level models is in line with previous findings in the literature \citep{neural_lam, graphcast}.

\danraplots{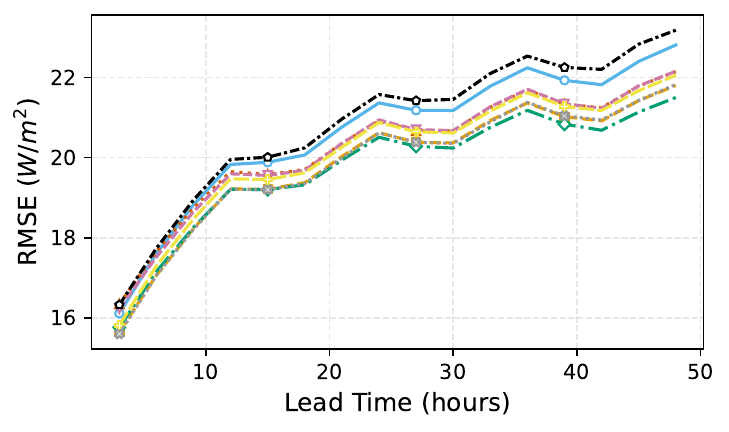}{danra_graph_experiment}{
    \gls{RMSE} on validation set for \gls{DANRA} models trained using different graph configurations.
    We consider Rectangular (Rect.) and Triangular (Tri.) graphs, both in Hierarchical (Hi.) and Multi-Scale (M.S) setups with different number of levels (lev.). 
    Recall that in multi-scale graphs all the levels are collapsed into one final mesh graph.
}

The main results, for all different kinds of graphs, are shown in \cref{fig:danra_graph_experiment}.
We note that including more levels is generally beneficial, both for rectangular and triangular graphs, and for hierarchical and multi-scale setups.
For the rectangular graphs there is a significant reduction of error when going from the 2 level to the 3 level hierarchical graph, and a smaller improvement also when going from 3 to 4.

The mesh graph layout, rectangular or triangular, seems to play a minor role for the hierarchical graphs.
Focusing on the 3 level multi-scale versions however, the triangular one performs substantially worse.
Inspecting forecasts from this model, they also show extensive graph-related artifacts in many fields.
We hypothesize that this poor performance relates to edge lengths, as the length of the longest edges grows slower with the number of included levels in the triangular graphs than the rectangular.
The 3-level hierarchical triangular model still seems able to compensate for shorter edges, perhaps because of its level-specific \gls{GNN} parameters.

Focusing on rectangular graphs, when comparing hierarchical and multi-scale models we see that the performance difference is small between these arguably quite different architectures.
This again points to the fact that the most important design dimension here is the inclusion of longer edges, capable of transporting information long spatial distances in each \gls{GNN} layer.
If sufficiently long edges are included, other options are less important.
Still, for some variables such as the \SI{10}{\metre} wind in \cref{fig:danra_graph_experiment_wv10m}, there is a clear advantage to the hierarchical configurations with more levels.

One explanation for the superior performance of the hierarchical graph processor is that this type of model has more flexibility than the multi-scale processor.
By not sharing \gls{GNN} layer parameters between the different mesh graph levels the hierarchical models can further adapt to the data.
Through reduced parameter sharing the hierarchical processors get more learnable parameters, but without directly increasing requirements on the limiting \gls{GPU} memory resource.
However, the addition of edges in-between levels does introduce the need for storing additional representation vectors, having some memory footprint.
This impact of this is however quite minor, as the memory use is dominated by edges in the finest mesh level.

\subsection{Evaluation against gridded data}
\label{sec:eval_gridded}

With the insights from our design studies, we choose the final models for each domain to put through additional in-depth verification on the test splits of the data.
Based on insights from \cref{sec:exp_boundary_forcing} these final models use an \SI{800}{\kilo\metre} boundary and include future boundary forcing $\bforcing_{k'+1}$ as input.
The final \gls{COSMO} model is the version with overlapping boundary, operating in a hybrid downscaling and \gls{LAM} fashion, as we saw substantial advantages of this setup for the Swiss domain.
For the final \gls{DANRA} model we do not use the overlapping boundary setup, as this showed no performance improvement when forced with \gls{IFS} forecasts.
Both models use rectangular hierarchical graphs with the maximum number of levels from \cref{sec:exp_graph_construction}.

Throughout the verification we compare our \gls{ML} models with \gls{NWP} forecasts for the two domains.
As described in \cref{sec:data} there are some limitations to these \gls{NWP} forecasts, that should be kept in mind.
In particular, the \gls{NWP} forecasts include only surface variables and for the \gls{DANRA} domain only up to \SI{18}{\hour}.
In the Danish setting we thus focus on shorter lead times ($\leq$ \SI{48}{\hour}). %
For both domains there is a general emphasis on surface fields, both due to this being the only fields available in baseline \gls{NWP} forecasts and because of practical relevance.

As we have access to long \gls{NWP} forecasts for the \gls{COSMO} domains we consider lead times up to \SI{120}{\hour} for the verification.
To achieve skillful performance to these longer lead times, despite the many \SI{1}{\hour} autoregressive steps, we found it beneficial to fine-tune the final \gls{COSMO} model further.
We include an additional stage of unrolled training with $\fclen = 12$ steps before verification (see \cref{sec:extra_finetuning} for more details).

\begin{figure}[tbp]
    \centering
    \begin{subfigure}[b]{0.5\textwidth}
        \centering
        \includegraphics[width=\textwidth]{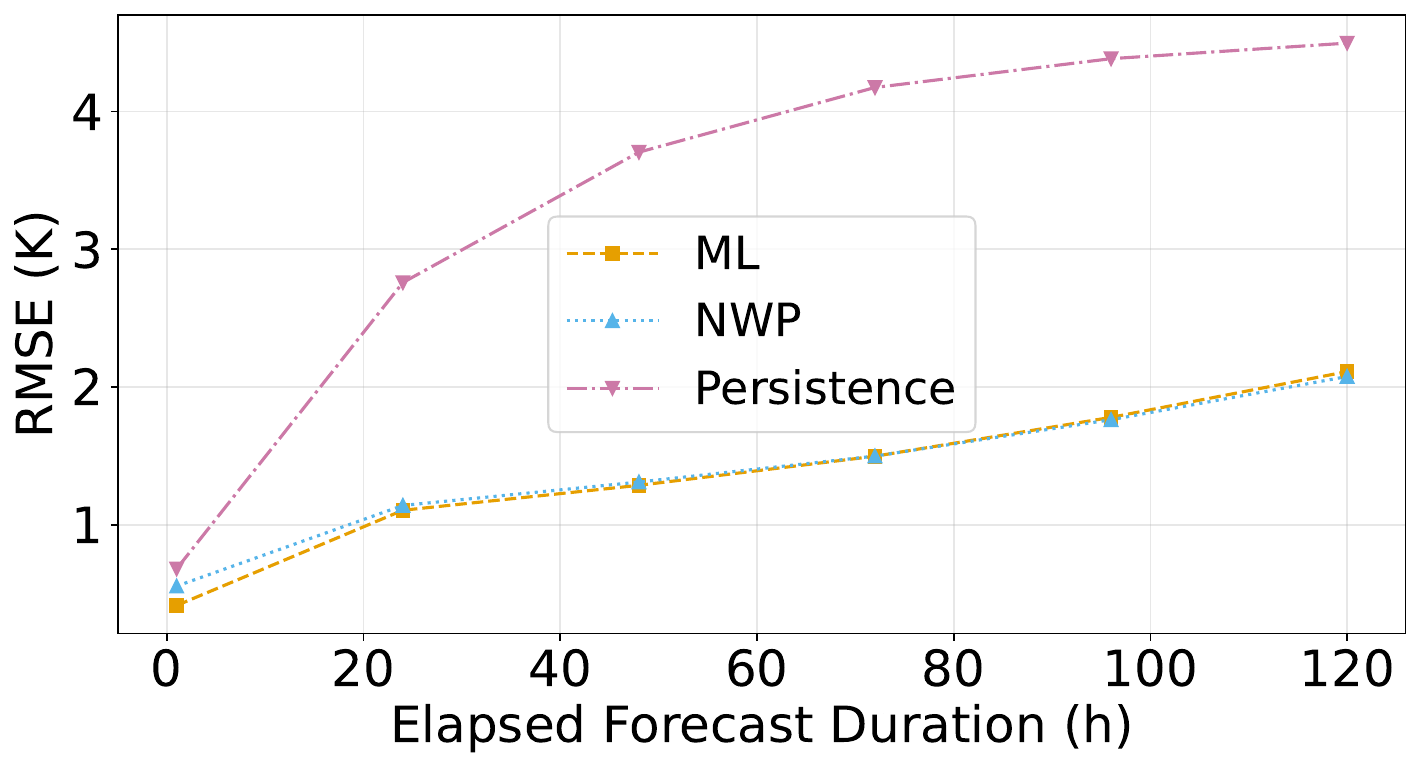}
        \caption{\SI{2}{m} temperature (\wvar{2t})}
    \label{fig:cosmo_verif_gridded_rmse_t2m}
    \end{subfigure}%
    \hfill%
    \begin{subfigure}[b]{0.5\textwidth}
        \centering
        \includegraphics[width=\textwidth]{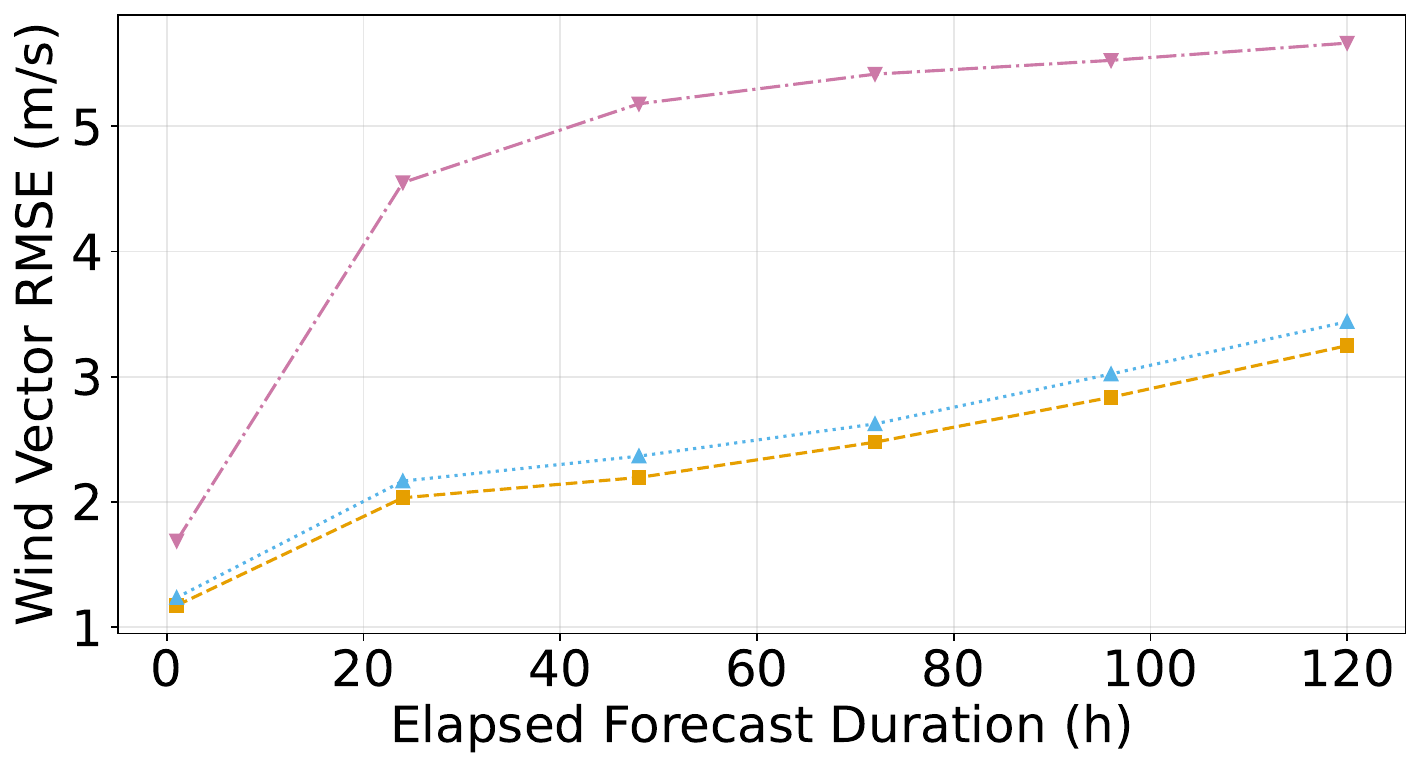}
        \caption{\SI{10}{m} wind}
    \label{fig:cosmo_verif_gridded_rmse_wv10m}
    \end{subfigure}%
    \hfill%
    \begin{subfigure}[b]{0.5\textwidth}
        \centering
        \includegraphics[width=\textwidth]{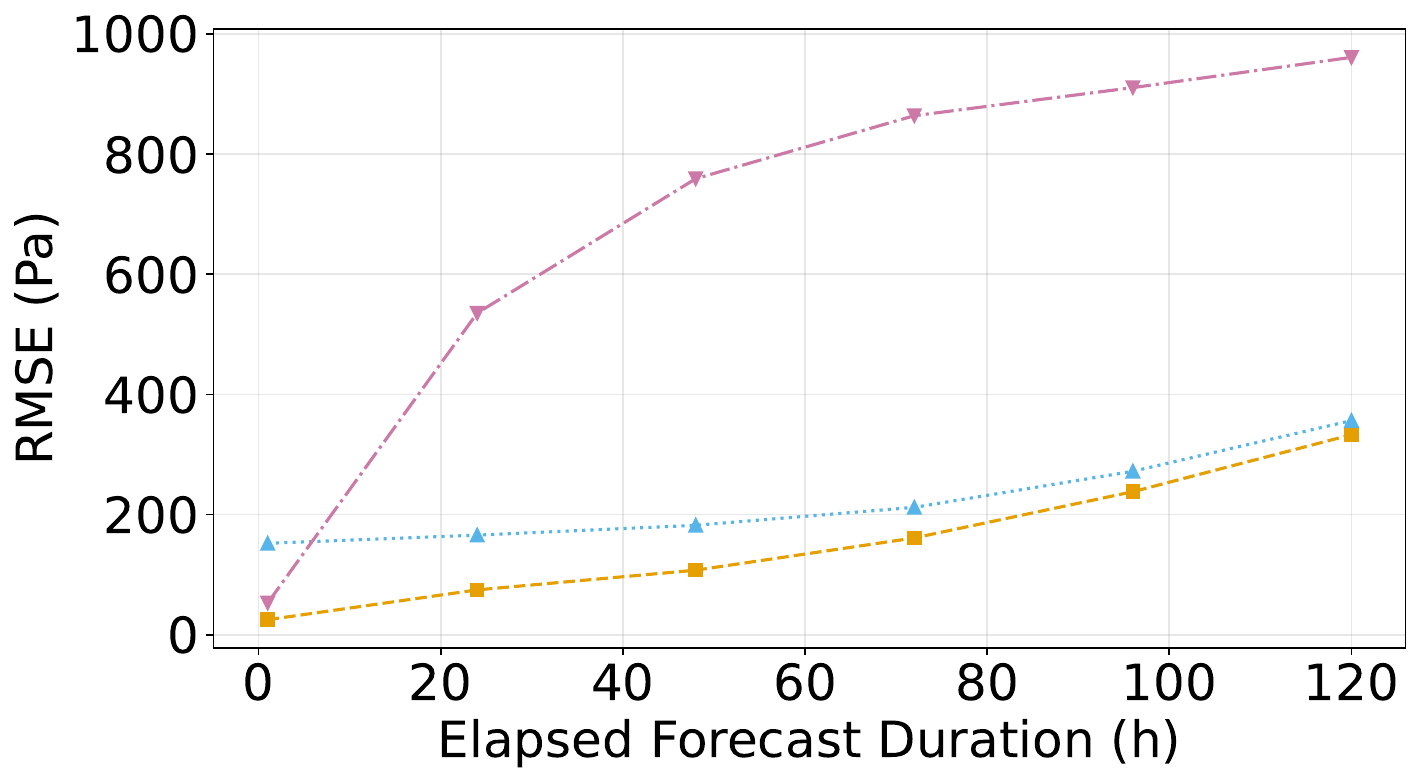}
        \caption{Surface pressure (\wvar{sp})}
    \label{fig:cosmo_verif_gridded_rmse_sp}
    \end{subfigure}%
    \begin{subfigure}[b]{0.5\textwidth}
        \centering
        \includegraphics[width=\textwidth]{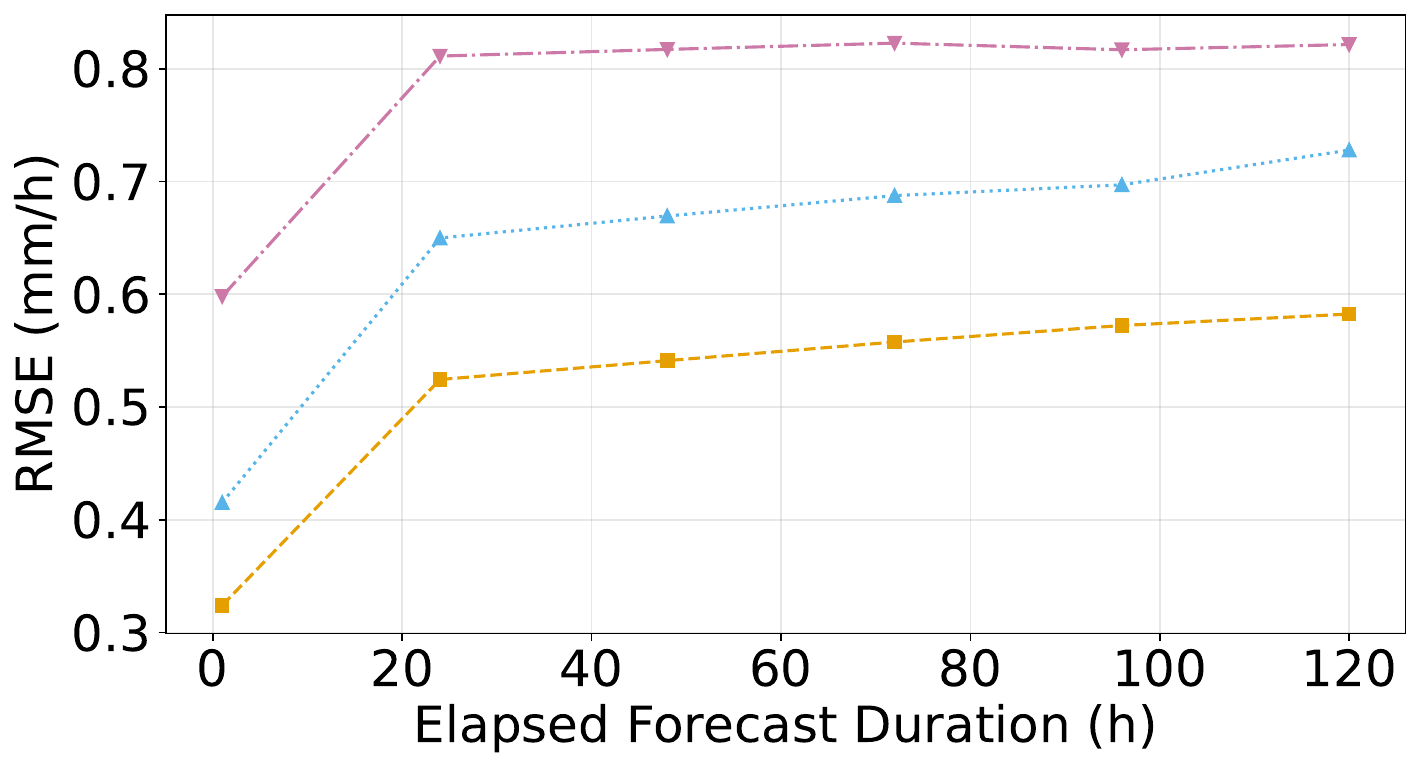}
        \caption{Precipitation (\wvar{tp01})}
    \label{fig:cosmo_verif_gridded_rmse_tp}
    \end{subfigure}%
    \caption{
        \gls{RMSE} for the \gls{COSMO} models compared to the gridded analysis.
        We show \glspl{RMSE} for a subset of lead times, meaning that the diurnal cycle is not visible.
    }
    \label{fig:cosmo_verif_gridded_rmse}
\end{figure}

\paragraph{\glspl{RMSE}}
To get an overview of model performance, we show \gls{RMSE} values of forecasts compared against the gridded \gls{COSMO} analysis in \cref{fig:cosmo_verif_gridded_rmse}.
The error is plotted along the elapsed forecast duration, corresponding to different lead times.
As a reference we also include the error of a persistence baseline, predicting the value of the initial condition $\wstate_0$ at all forecast times.
For all the surface variables shown, the \gls{COSMO} \gls{ML} model shows impressive performance on this average metric.
The \gls{ML} model is outperforming the \gls{NWP} baseline on all variables, except \SI{2}{m} temperature where the errors are similar.
While precipitation \gls{RMSE} values seems especially low for the \gls{ML} model, this is not the most suitable metric for this variable.
We devote specific effort, employing more well-motivated metrics, to investigation precipitation performance later in our verification.

\begin{figure}[tbp]
    \centering
    \begin{subfigure}[b]{0.5\textwidth}
        \centering
        \includegraphics[width=\textwidth]{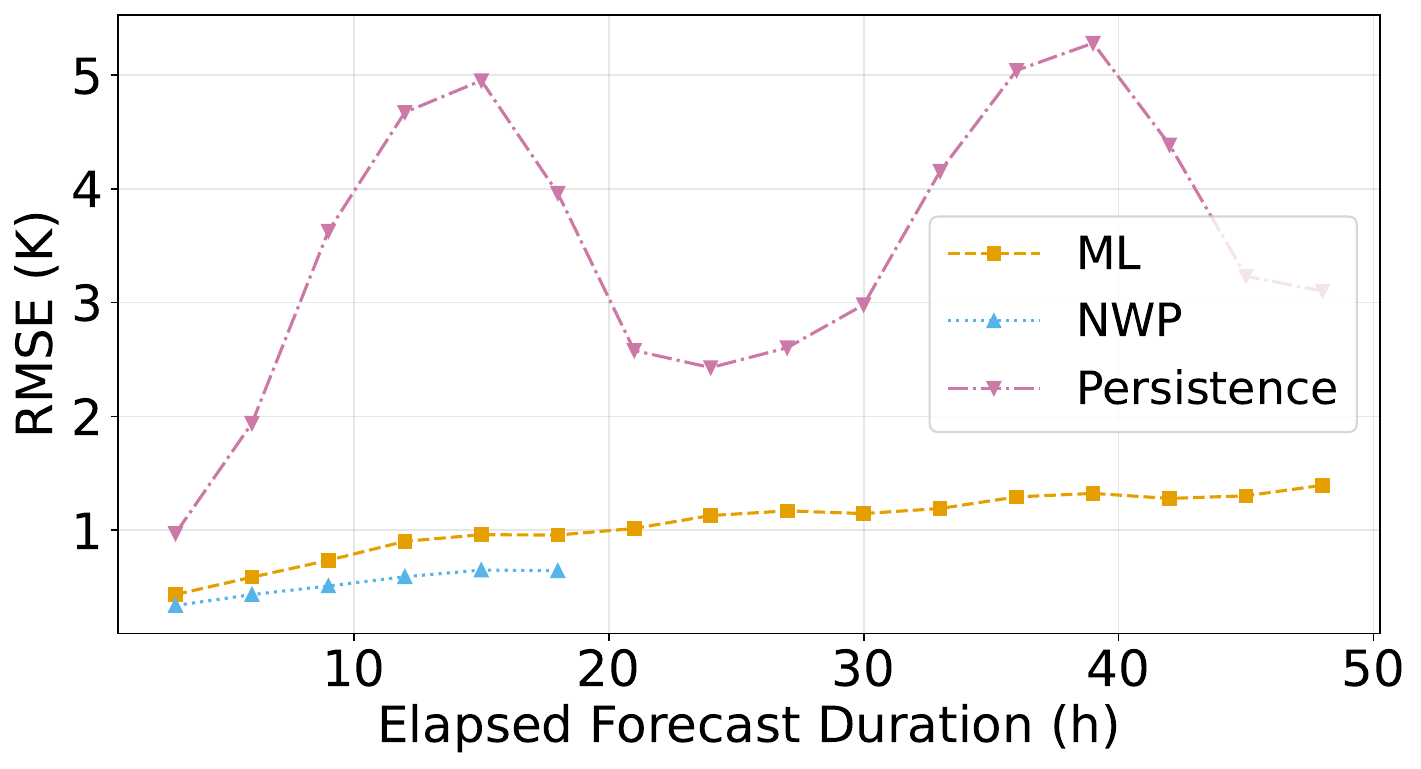}
        \caption{\SI{2}{m} temperature (\wvar{2t})}
    \label{fig:danra_verif_gridded_rmse_t2m}
    \end{subfigure}%
    \hfill%
    \begin{subfigure}[b]{0.5\textwidth}
        \centering
        \includegraphics[width=\textwidth]{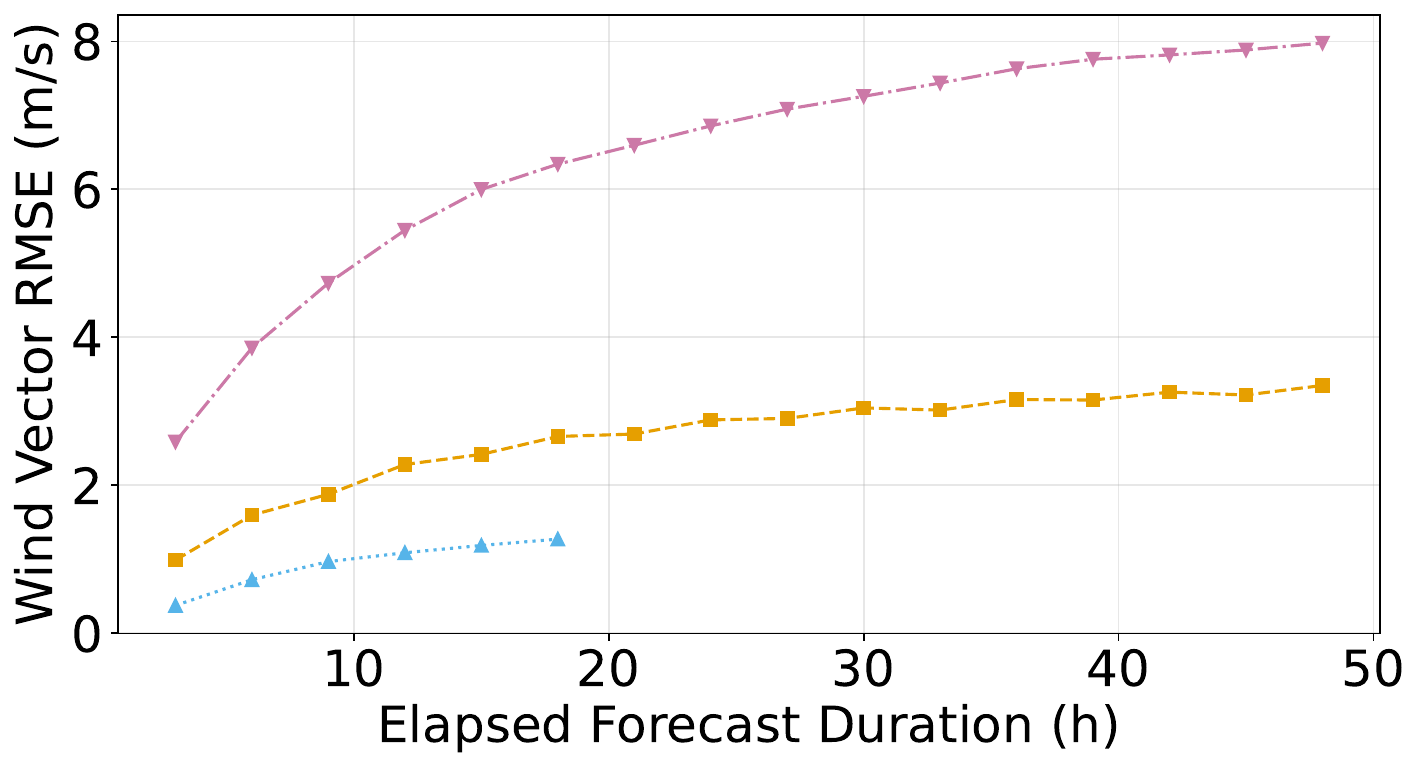}
        \caption{\SI{10}{m} wind}
        \label{fig:danra_verif_gridded_rmse_wv10m}
    \end{subfigure}%
    \caption{
        \gls{RMSE} for the \gls{DANRA} models compared to the gridded reanalysis.
    }
    \label{fig:danra_verif_gridded_rmse}
\end{figure}

\Cref{fig:danra_verif_gridded_rmse} shows \glspl{RMSE} of surface variables for the \gls{DANRA} models compared to the gridded reanalysis.
While our \gls{ML} model clearly produces skillful forecasts, its accuracy does not reach all the way to compete with the \gls{NWP} baseline.
Both models however demonstrate a similar rate of error growth, and our \gls{ML} model is stable to roll out for many days.
In \cref{sec:extra_eval_gridded} we additionally consider the gridded \gls{RMSE} computed for a sub-region covering only Denmark.
For this smaller region, which sits at the center of the \gls{DANRA} domain, the \gls{ML} model is more competitive and achieves errors not far from the \gls{NWP} system.

\paragraph{Mean error maps}

To better understand where in the domain each model has potential weaknesses, we look at mean error maps across lead time.
In \cref{fig:cosmo_verif_gridded_error_map_temp} we note that for temperature the \gls{COSMO} \gls{ML} model has much lower biases in the \SI{1}{\hour} prediction.
However, when considering longer forecast durations the \gls{NWP} model shows lower biases. 
There are two main reasons for the deteriorating performance of the \gls{ML} model.
First, there are negative biases in areas north and south of the Alps, which could be related to inversion and overestimated cold pool formation.
Second, there are positive biases in the South-Eastern corner of the domain, relating to the boundary forcing. 
Similar bias can be seen along other parts of the boundary over the Mediterranean sea, indicating that the model struggles with incorporating boundary information from the south.
It should be noted that our approach does not explicitly target having smooth boundary transitions, and as the training loss is only calculated in the interior, this is not guaranteed.
Worth considering is also that the operational analysis, which the \gls{COSMO} model is trained on, contains non-physical boundary grid-cells which are used to introduce boundary information into the \gls{NWP}-forecast. 
The \gls{ML}-model has learned to largely ignore these artifacts, but they might still affect the training process.
Another interesting observation in \cref{fig:cosmo_verif_gridded_error_map_temp} is that the \gls{NWP} model suffers from a small, but consistent warm bias over the Alps. 
The \gls{ML} model on the other hand does not seem to be affected by this systematic elevation bias.

\begin{figure}[tbp]
    \centering
    \includegraphics[width=.76\textwidth]{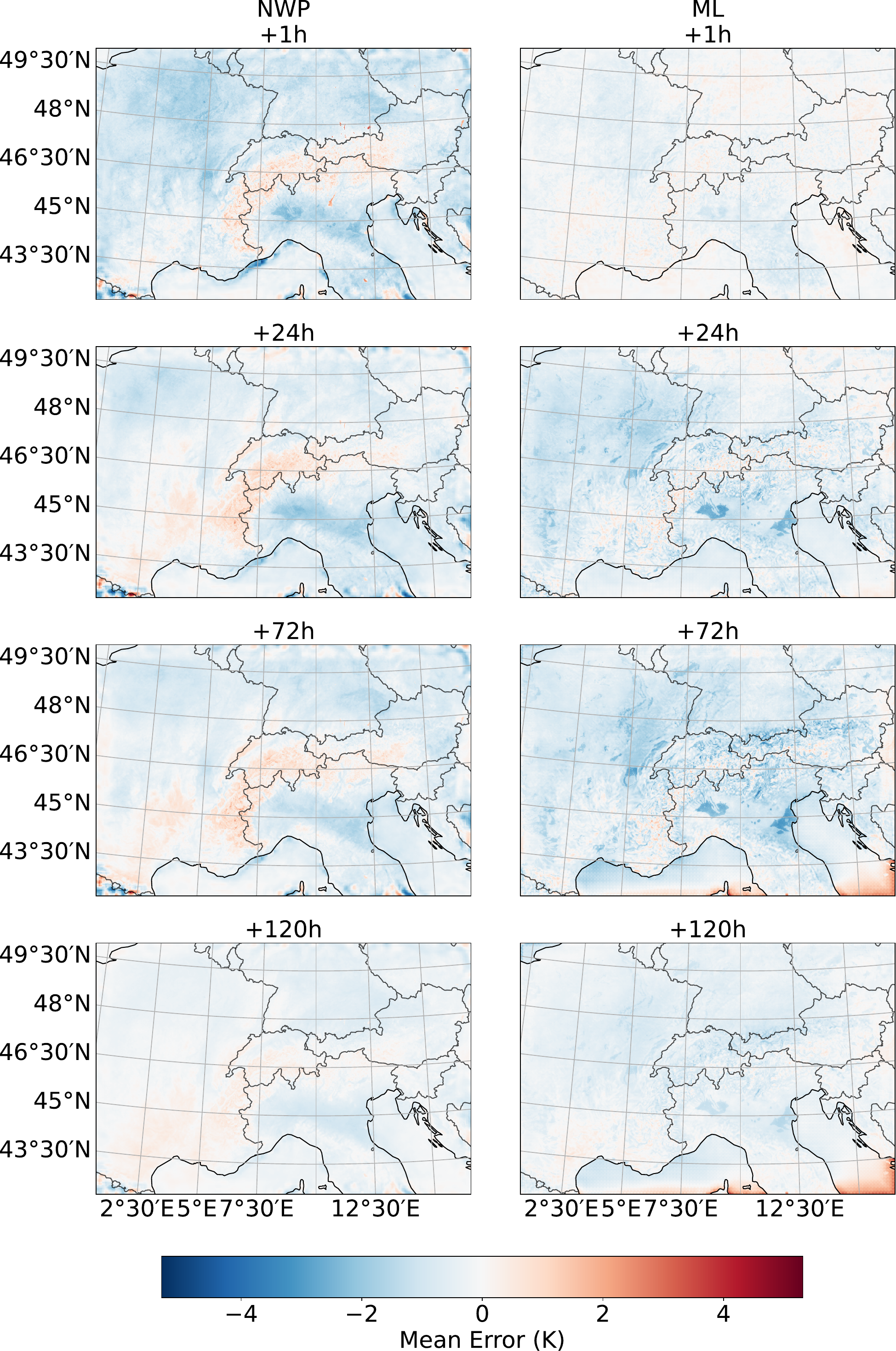}
    \caption{
        Mean error maps of \SI{2}{m} temperature (\wvar{2t}) for the \gls{COSMO} models over the full test year period.
    }
    \label{fig:cosmo_verif_gridded_error_map_temp}
\end{figure}

For the wind in \cref{fig:cosmo_verif_gridded_error_map_wind}, the error is composed differently. While the \gls{ML} model has an overall lower bias, smaller turbulences over the orography are clearly visible in the error.
This suggests that the model is smoothing out the forecasts too strongly, in particular gravity waves.
For the \gls{NWP} model on the other hand we can clearly see large biases related to land-sea processes. The model is generally overestimating winds over the Adriatic sea in the south-east, while underestimating them over the rest of the Mediterranean Sea in the south.

\begin{figure}[tbp]
    \centering
    \includegraphics[width=.8\textwidth]{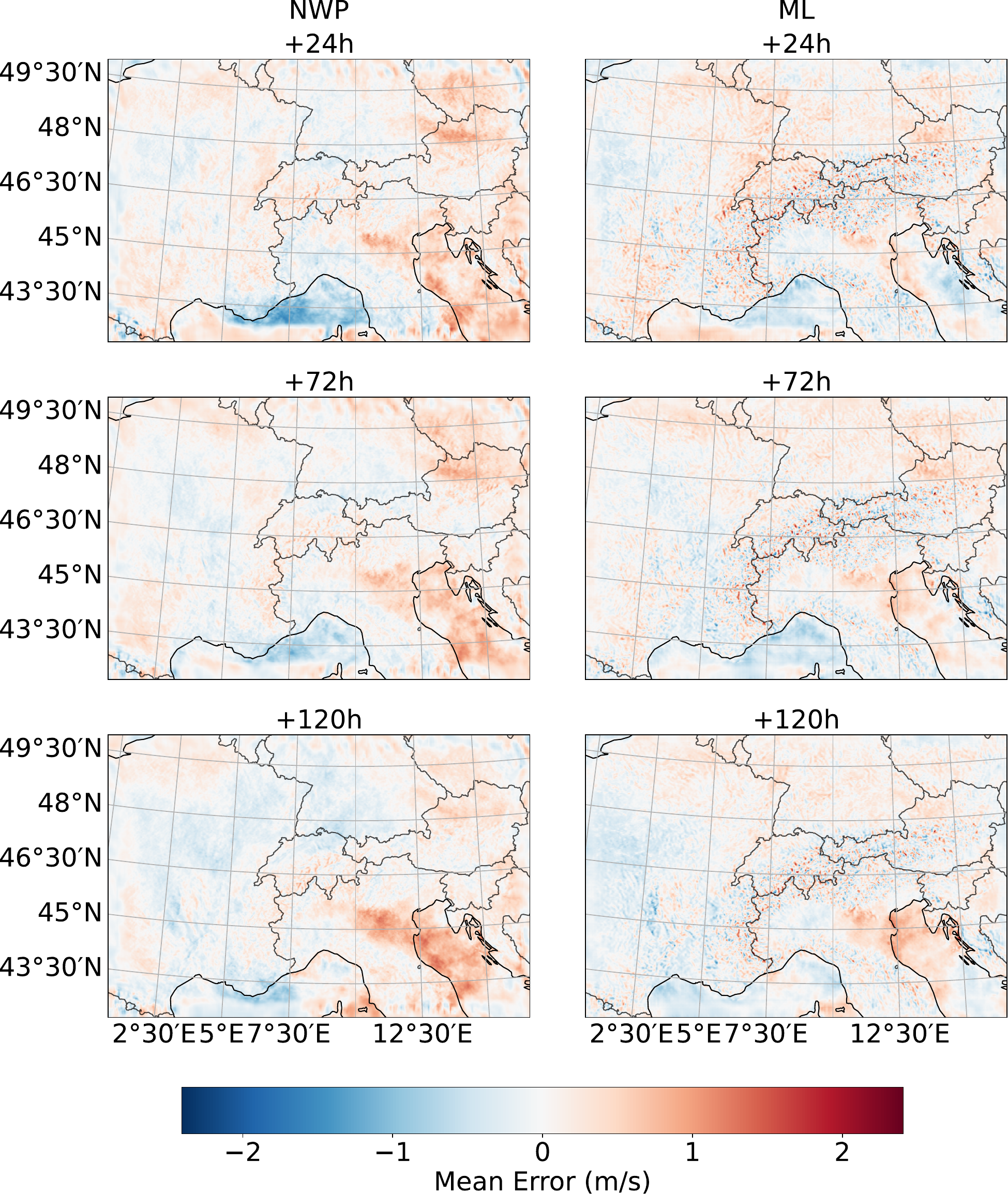}
    \caption{
        Mean error maps of \SI{10}{m} wind u-component (\wvar{10u}) for the \gls{COSMO} models over the full test year period.
    }
    \label{fig:cosmo_verif_gridded_error_map_wind}
\end{figure}

\begin{figure}[tbp]
    \centering
    \includegraphics[width=0.8\textwidth]{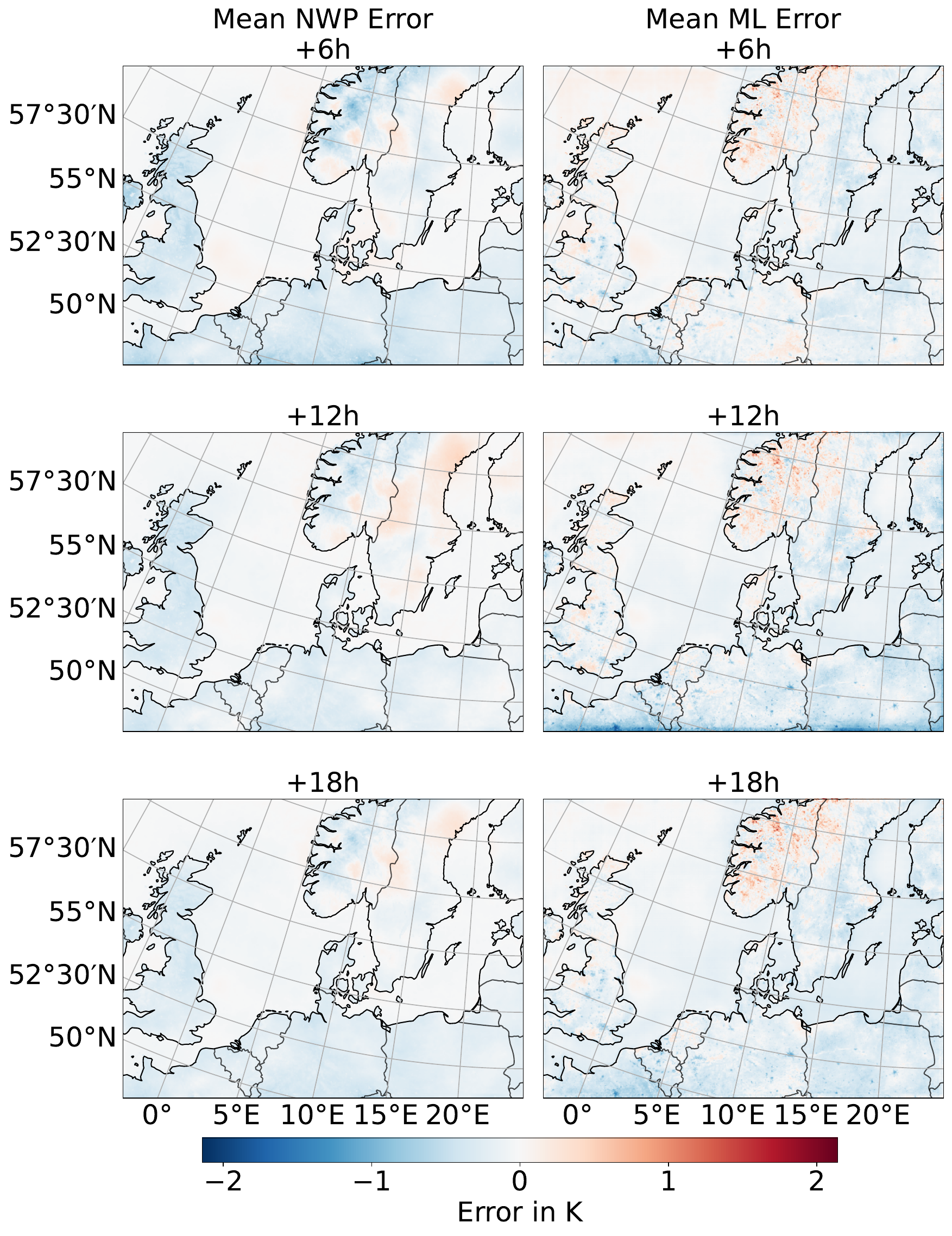}
    \caption{
        Mean error maps of \SI{2}{m} temperature (\wvar{2t}) for the \gls{DANRA} models over the full test year period.
    }
    \label{fig:danra_verif_gridded_error_map_temp}
\end{figure}

In \cref{fig:danra_verif_gridded_error_map_temp} we show mean error maps of \SI{2}{\meter} temperature for the \gls{DANRA} models.
Both models show a general cold bias over most of the domain.
Interestingly the \gls{DANRA} \gls{NWP} model has a general cold bias over the Scandinavian Mountains in Norway, whereas the \gls{ML} model shows areas both too warm and too cold in the region.
The error maps for the \gls{ML} model highlights also a failure to accurately capture the temperature in many urban centers.
This manifests as small areas of cold bias throughout Germany, Poland and the UK.
While we do not observe this in the \gls{COSMO} model, it does indicate that the \gls{ML} models could benefit from additional information about land use.
Such inputs could be included as static fields to the \gls{LAM} models.

\Cref{fig:danra_verif_gridded_error_map_wind} shows the mean error for the u-component of \SI{10}{\meter} wind.
While biases are generally small, we see that the \gls{ML} model struggles with finer orography, especially in the Norwegian mountains.
This matches the problems with capturing smaller turbulences also observed in the \gls{COSMO} model, and again relates to forecast smoothness.
It is however encouraging that these biases do no increase over time, meaning that they do not clearly result in compound errors, exacerbated by the autoregressive steps of the model.
The mitigation of such effects is an important role of the fine-tuning on rolled out forecast.

For the vertical variables, with lower loss weighting, we observe in some fields biases that can be related back to the graph structure used in the \gls{DANRA} model.
Some examples of this are shown in \cref{sec:extra_eval_gridded}.
This seems to most strongly impact the smoother fields.

\begin{figure}[tbp]
    \centering
    \includegraphics[width=0.8\textwidth]{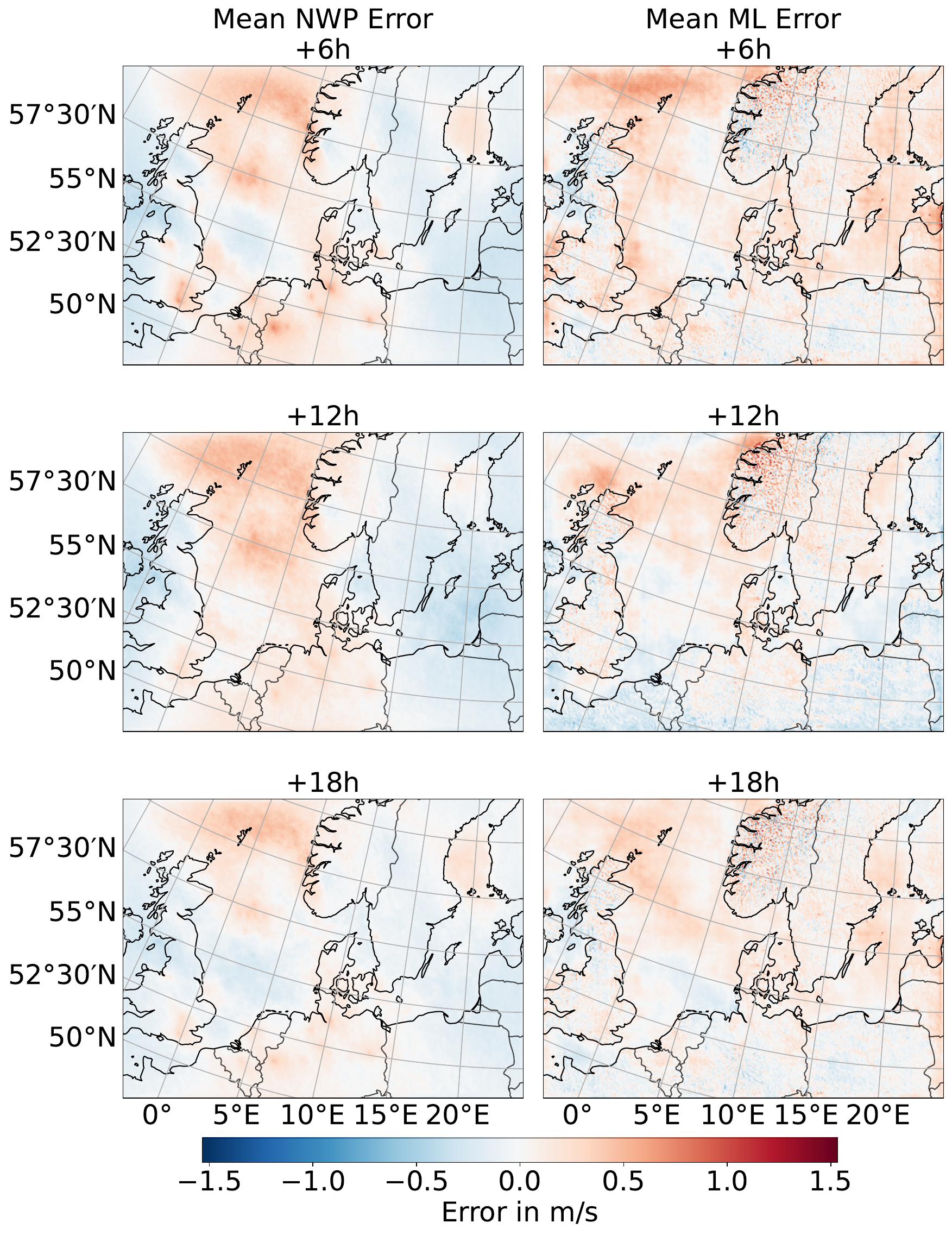}
    \caption{
        Mean error maps of \SI{10}{m} wind u-component (\wvar{10u}) for the \gls{DANRA} models over the full test year period.
    }
    \label{fig:danra_verif_gridded_error_map_wind}
\end{figure}

\paragraph{3D atmospheric variables}
While 3-dimensional fields are not available for the \gls{NWP} models, we can still uncover much about the ability of the \gls{ML} model to capture the full atmospheric state, including its vertical motion. 
In \cref{fig:cosmo_verif_gridded_vertical} we study the error distribution across vertical levels for different lead times and 4 different variables.
Results for all variables are shown in \cref{sec:extra_eval_gridded}.
For most variables the error clearly increases with lead time across all vertical levels.
For wind, temperature and pressure the error is higher towards the top of the atmosphere. 
This might be related to the variable weighting. %
To achieve low errors on the highly weighted surface fields, it should be necessary for the model to also accurately capture the dynamics in vertical levels close to  the surface.
This could lead to the model focusing more on the close-to-surface levels, where we note that the error is lower, and less on higher vertical fields.

Comparing the normalized errors with each other the vertical velocity shows the lowest errors across all vertical levels. 
This is a positive sign that the model is able to capture vertical motion accurately in the atmosphere, even in a convection-resolving resolution of < 3 km. 
This is very important for thunderstorm formation and convective rain clouds. 
Physical processes related to orography, such as  near the Alps, also heavily depend on vertical motion. 
Relative humidity on the other hand has the highest relative errors, especially in the middle of the atmosphere. 
Given the decreasing specific humidity with height, the relative humidity can change rapidly. 
We do not want the model to learn this, as it is a mathematical artifact rather than a physical one.
Limitations and challenges around the modeling of humidity are discussed further in \cref{sec:discussion_limitations}.

\begin{figure}[tbp]
    \centering
    \begin{subfigure}[b]{0.5\textwidth}
        \centering
        \includegraphics[width=\textwidth]{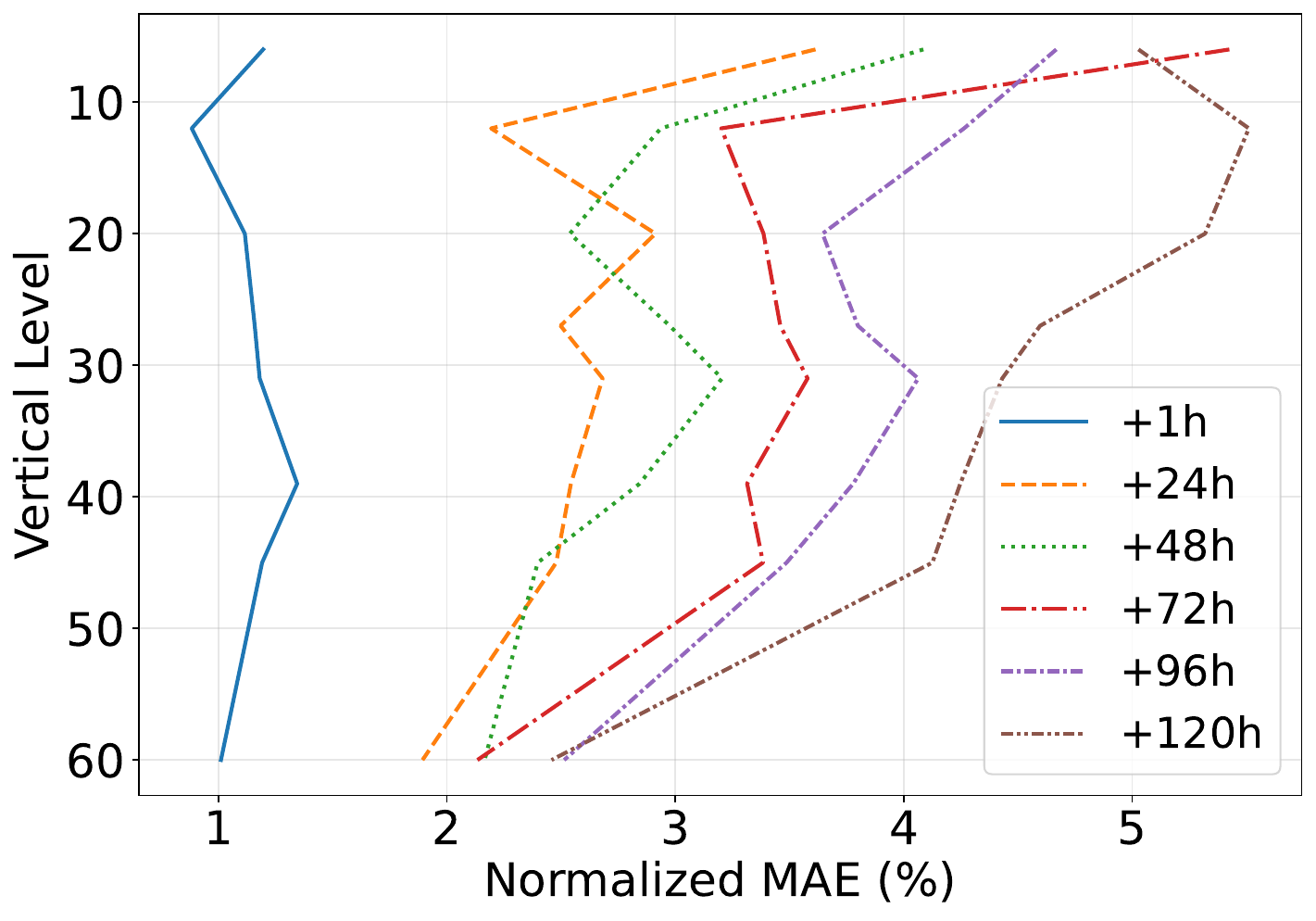}
        \caption{Wind u-component (\wvar{u})}
    \end{subfigure}%
    \begin{subfigure}[b]{0.5\textwidth}
        \centering
        \includegraphics[width=\textwidth]{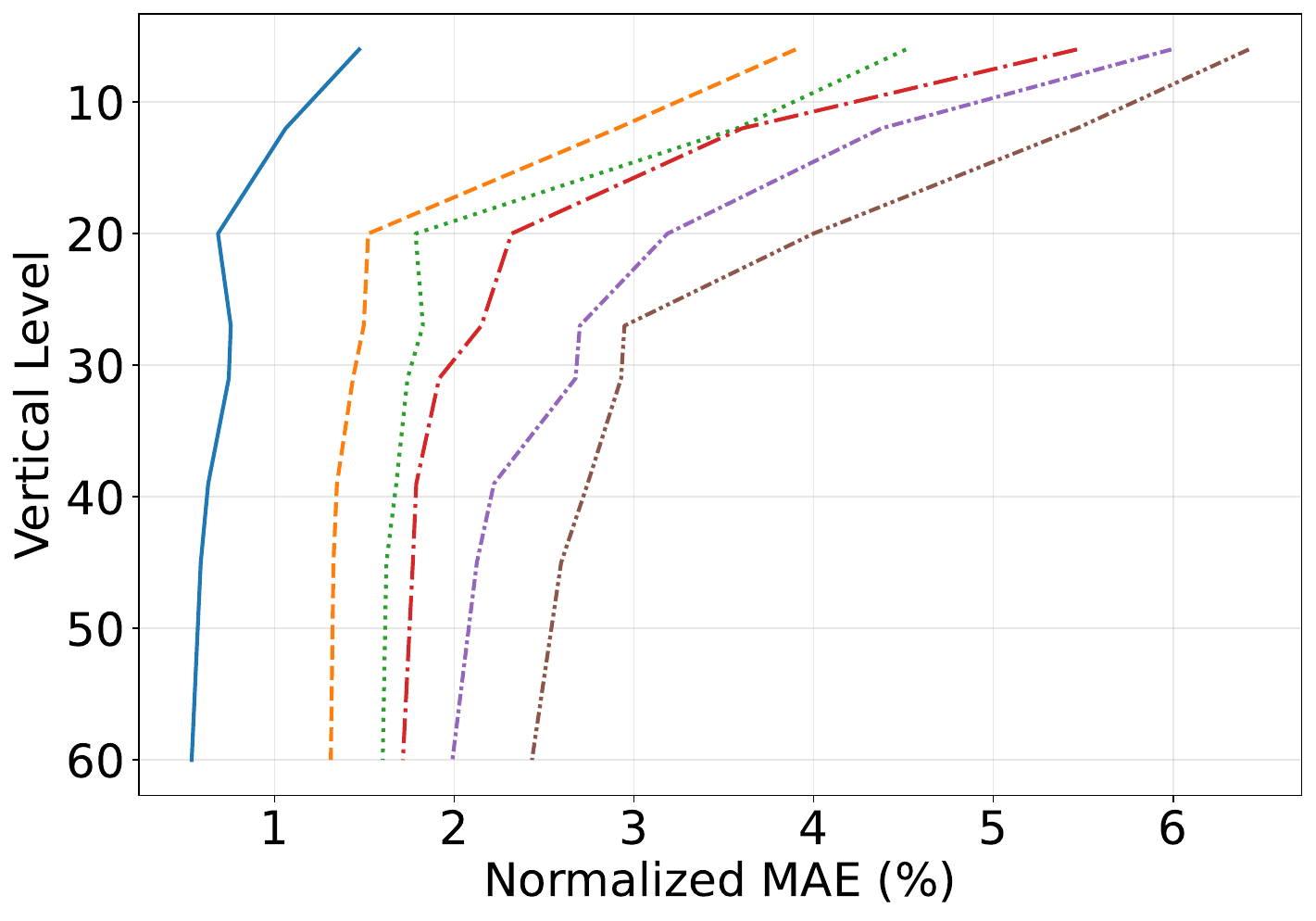}
        \caption{Temperature (\wvar{t})}
    \end{subfigure}%
    \hfill%
    \begin{subfigure}[b]{0.5\textwidth}
        \centering
        \includegraphics[width=\textwidth]{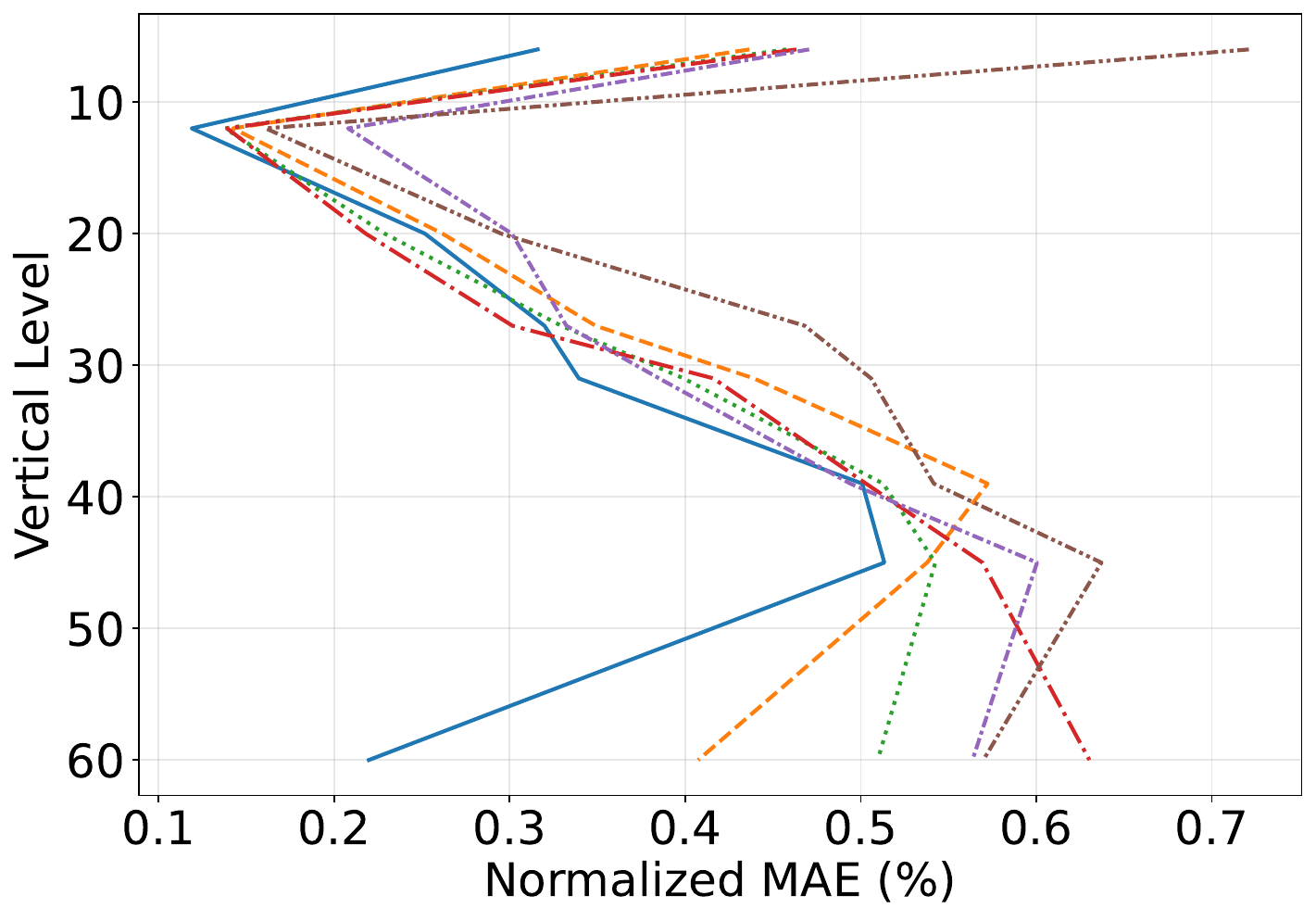}
        \caption{Vertical velocity (\wvar{w})}
    \end{subfigure}%
    \hfill%
    \begin{subfigure}[b]{0.5\textwidth}
        \centering
        \includegraphics[width=\textwidth]{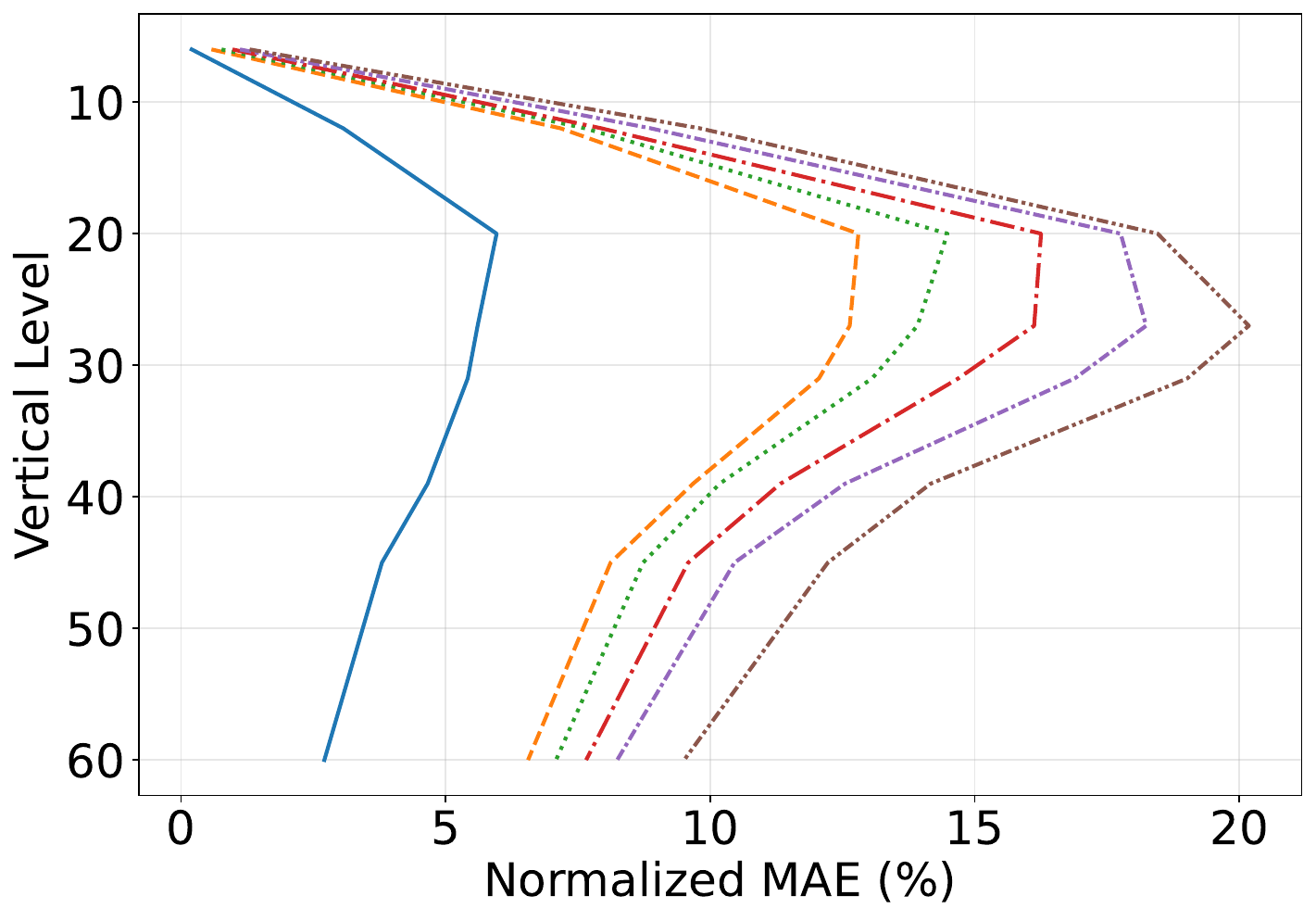}
        \caption{Relative humidity (\wvar{r})}
    \end{subfigure}%
    \caption{
        Vertical profiles of the \gls{NMAE} for the \gls{COSMO} \gls{ML} model.
        Higher numbered vertical model levels are closer to the ground.
    }
    \label{fig:cosmo_verif_gridded_vertical}
\end{figure}

In \cref{fig:danra_verif_gridded_vertical} we show the error of the \gls{DANRA} \gls{ML} model along vertical profiles for the wind u-component and temperature, with results for all variables in \cref{fig:danra_verif_gridded_vertical_full} in \cref{sec:extra_eval_gridded}.
For the \gls{DANRA} setting there is less of a tendency of lower errors close to the surface, with only temperature clearly showing this pattern.
Much of the \gls{DANRA} domain is lightly impacted by orography, which could make it less crucial to capture dynamics at lower vertical levels with high accuracy for predicting the heavily weighted surface fields.
Also for this model the vertical velocity has the lowest \gls{NMAE} and relative humidity has the highest, which can again be attributed to the challenges of humidity modeling.

\begin{figure}[tbp]
    \centering
    \begin{subfigure}[b]{0.5\textwidth}
        \centering
        \includegraphics[width=\textwidth]{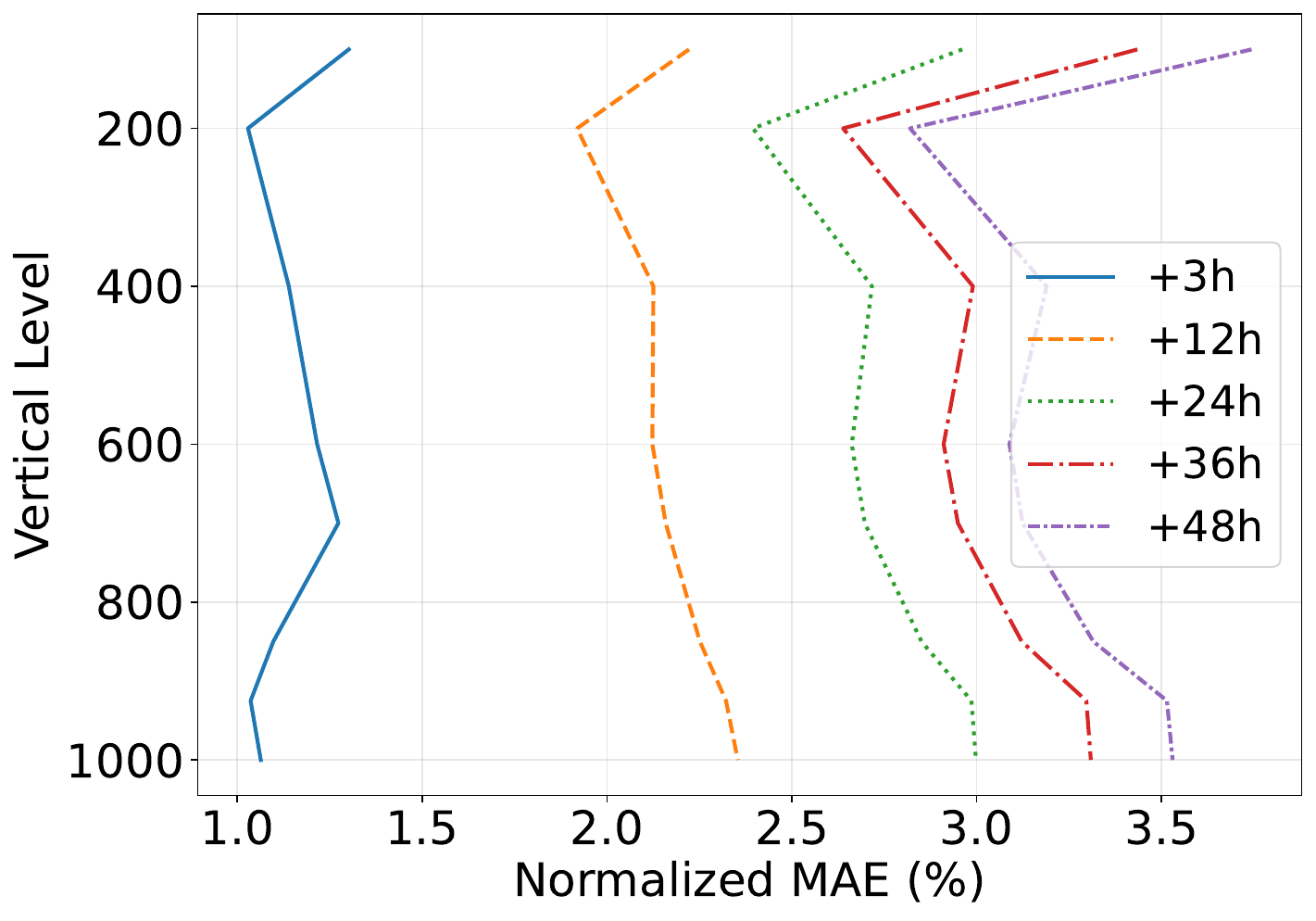}
        \caption{Wind u-component (\wvar{u})}
    \end{subfigure}%
    \begin{subfigure}[b]{0.5\textwidth}
        \centering
        \includegraphics[width=\textwidth]{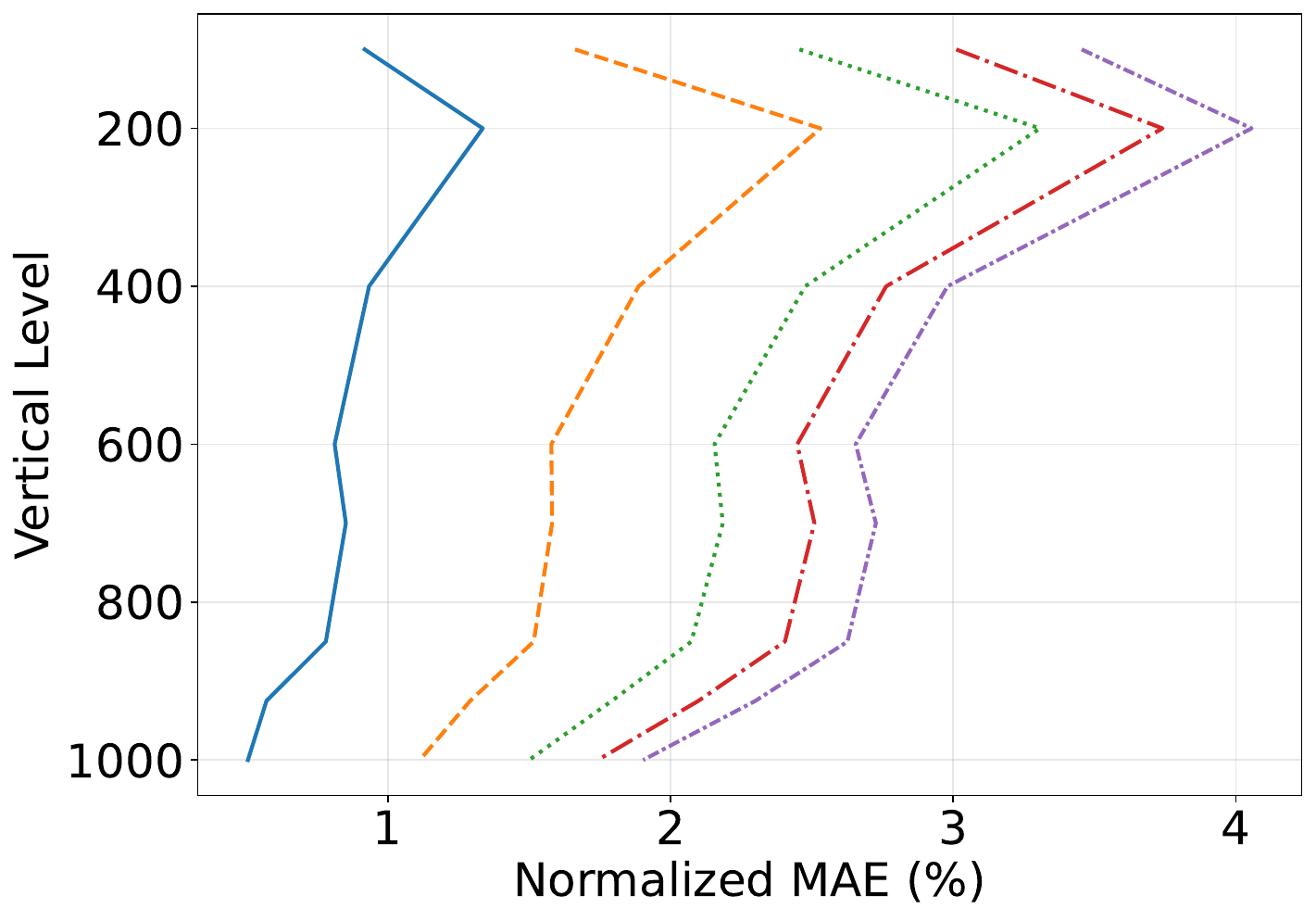}
        \caption{Temperature (\wvar{t})}
    \end{subfigure}%
    \caption{
        Vertical profiles of the \gls{NMAE} for the \gls{DANRA} \gls{ML} model.
        The vertical coordinate constitutes pressure levels.
    }
    \label{fig:danra_verif_gridded_vertical}
\end{figure}

\paragraph{Spatial verification of precipitation}
To quantify the capabilities of the \gls{COSMO} \gls{ML} model for forecasting precipitation we look at the \gls{SAL} metric in \cref{fig:cosmo_verif_gridded_sal}.
For both the \gls{ML} and \gls{NWP} models this metric shows good performance for precipitation forecasting.
The errors in amplitude and location are close to 0 for both models. 
The structure however, is almost perfect in the \gls{NWP} model, whereas the \gls{ML} model predicts a more diffuse structure. 
This is due to the fact that deterministic \gls{ML} models tend to smooth out high-frequency features. 
We later observe this directly in forecasts in \cref{fig:cosmo_verif_case_study_precipitation}.
As the \gls{DANRA} model is not trained to forecast precipitation we do not compute any \gls{SAL} scores for it.

\begin{figure}[tbp]
    \centering
    \hspace{2cm}
    \includegraphics[width=0.7\textwidth]{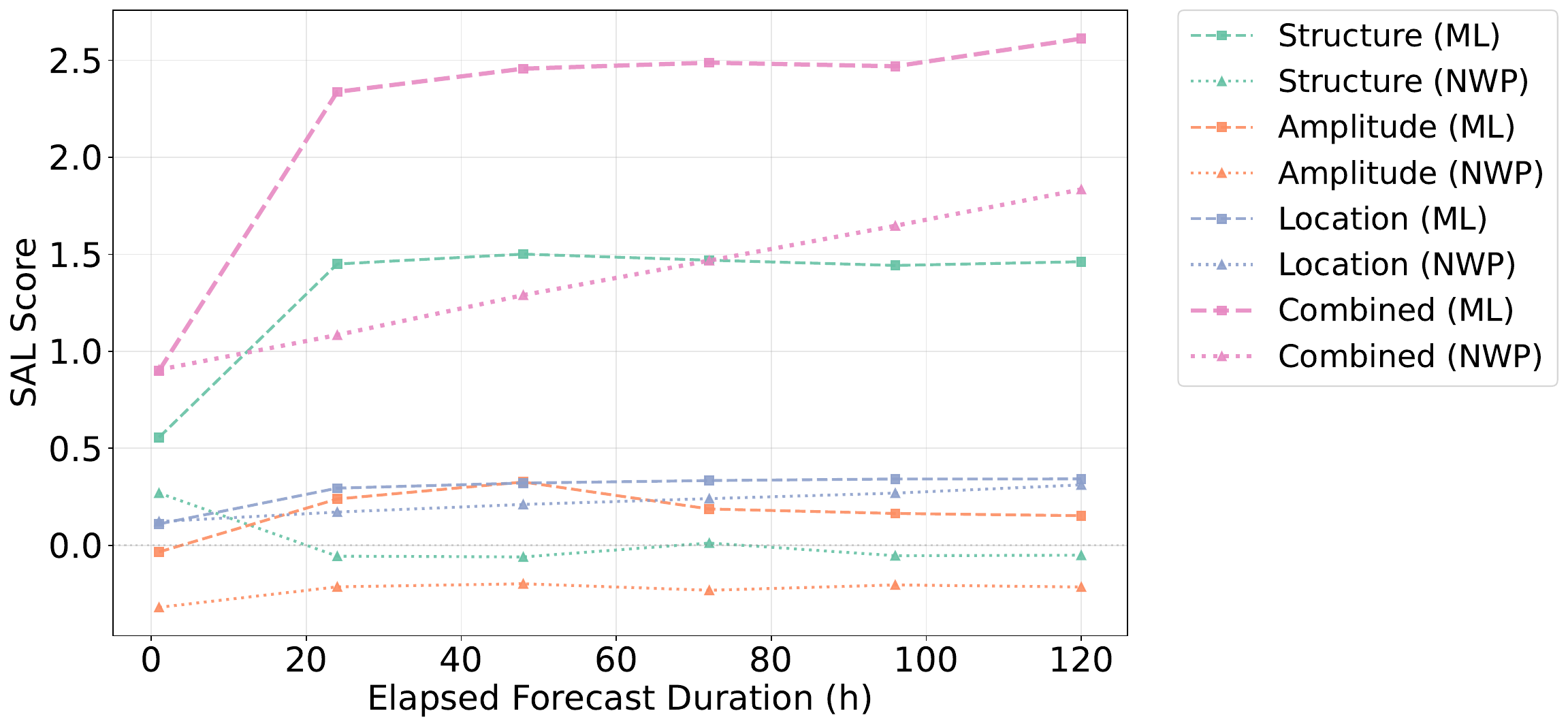}
    \caption{
        \gls{SAL} scores for precipitation (\wvar{tp01}) forecasts the \gls{COSMO} models.
    }
    \label{fig:cosmo_verif_gridded_sal}
\end{figure}

\paragraph{Energy spectra analysis}
We compute kinetic energy spectra for different surface variables to examine the ability of the model to represent atmospheric motion across different spatial scales. 
This analysis reveals how well the models capture both large-scale flow patterns and smaller-scale features, particularly important near complex orography. 
In \cref{fig:cosmo_verif_gridded_energy_spectra} we observe that smaller-scale features, corresponding to  higher wavenumbers, for wind and precipitation are underestimated in the \gls{COSMO} \gls{ML} model. 
This is a direct consequence of deterministic \gls{MSE}-based training, which tends to smooth out high-frequency features \citep{graph_efm, Selz_2025}.
This is especially true for fields with high variability, like precipitation, or high advection and turbulence like wind.
Otherwise, both models show similar energy spectra to the ground truth data, indicating that they are able to capture the different scales of motion and energy flux in the atmosphere.
In \cref{sec:extra_eval_gridded} we additionally look at the evolution of the spectral distance over lead time for specific wavenumbers.

\begin{figure}[tbp]
    \centering
    \begin{subfigure}[b]{0.5\textwidth}
        \centering
        \includegraphics[width=\textwidth]{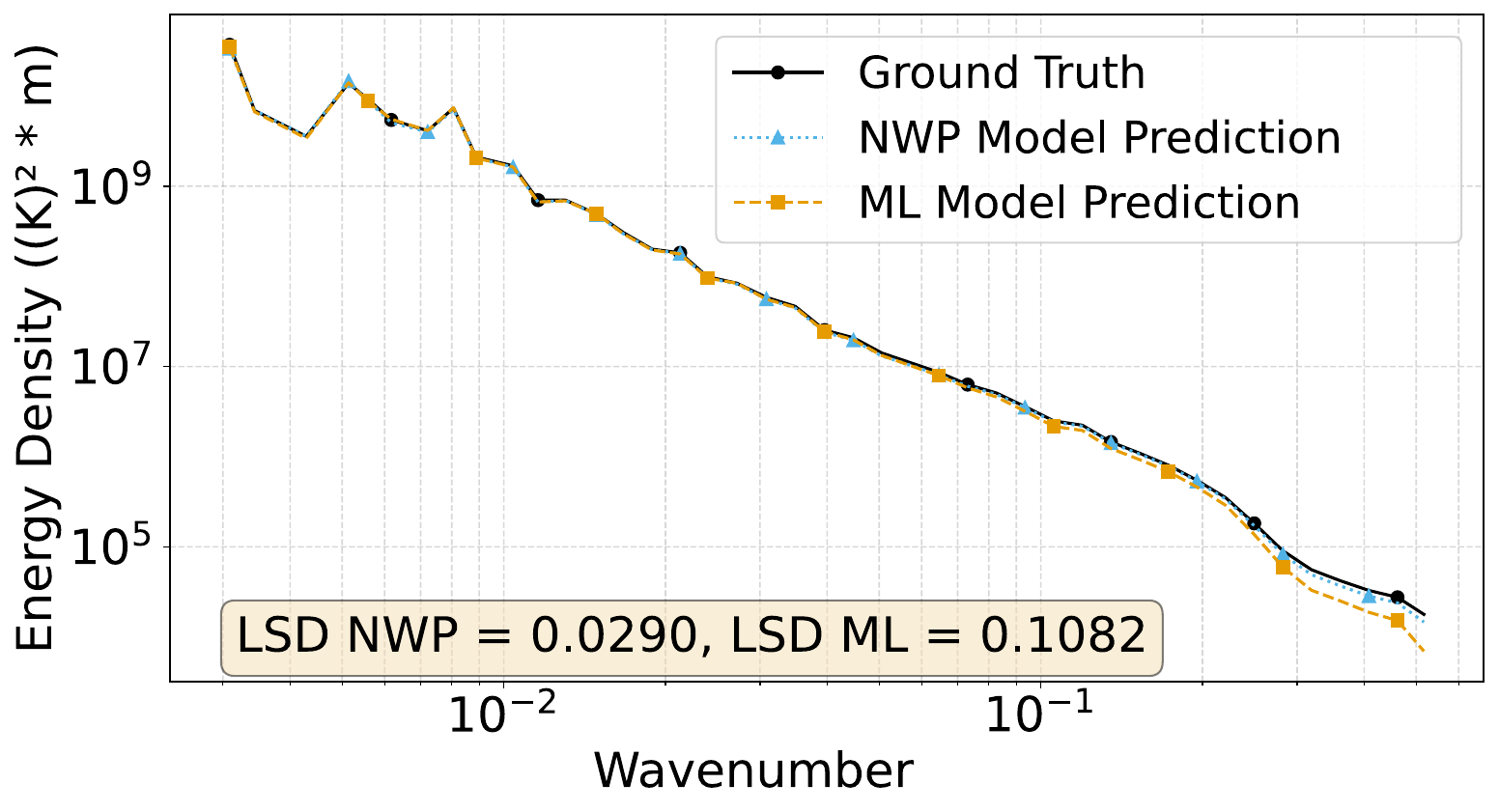}
        \caption{\SI{2}{m} temperature (\wvar{2t})}
    \end{subfigure}%
    \begin{subfigure}[b]{0.5\textwidth}
        \centering
        \includegraphics[width=\textwidth]{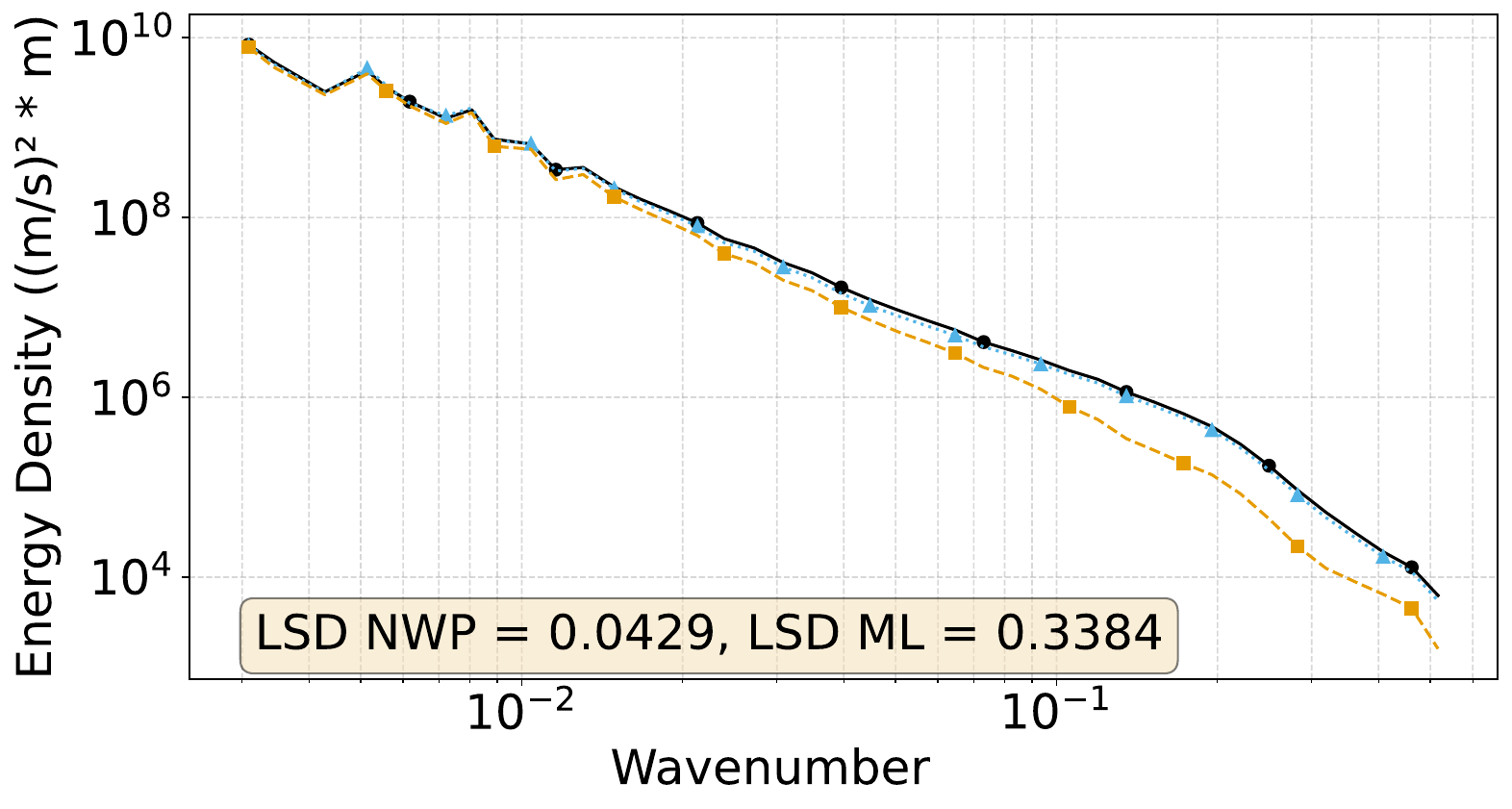}
        \caption{\SI{10}{m} wind-component (\wvar{10u})}
    \end{subfigure}%
    \hfill%
    \begin{subfigure}[b]{0.5\textwidth}
        \centering
        \includegraphics[width=\textwidth]{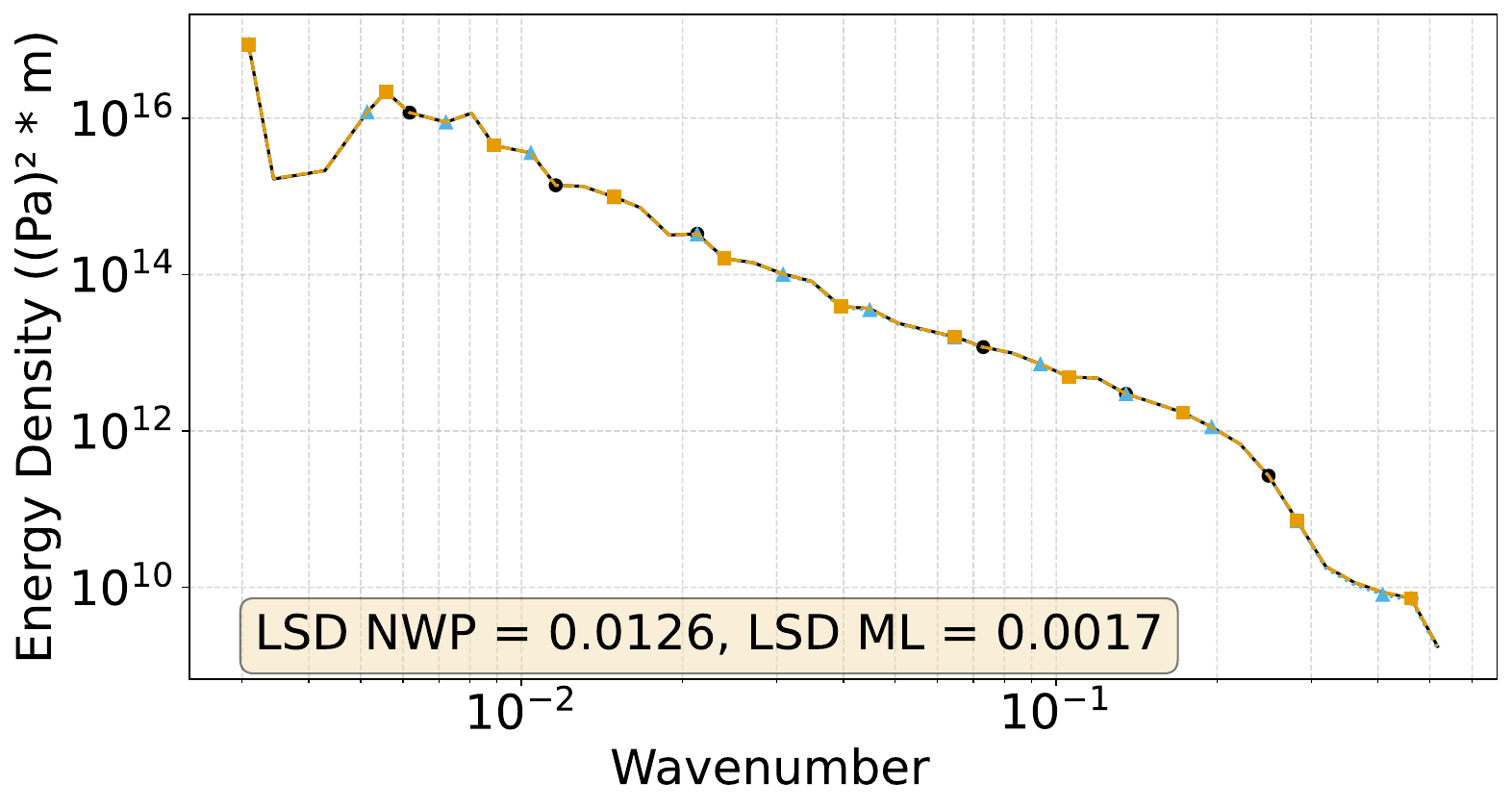}
        \caption{Surface pressure (\wvar{sp})}
    \end{subfigure}%
    \begin{subfigure}[b]{0.5\textwidth}
        \centering
        \includegraphics[width=\textwidth]{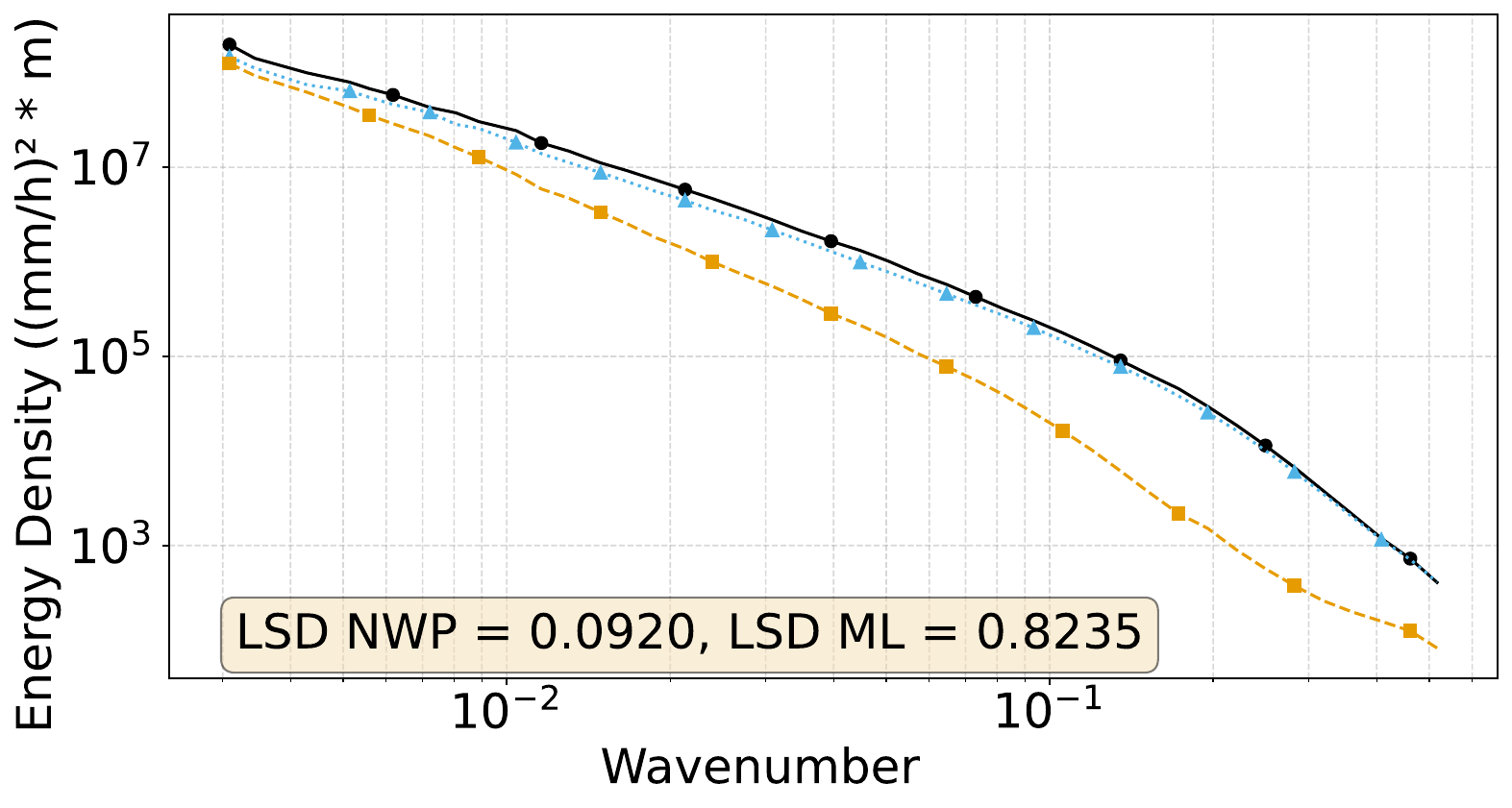}
        \caption{Precipitation (\wvar{tp01})}
    \end{subfigure}%
    \caption{
        Energy spectra of surface variables for the \gls{COSMO} models and ground truth data, averaged over lead times. 
        In each plot we display the \gls{LSD} for both models, compared to the spectra of the ground truth data.
        The unit of the wavenumber is $(\text{number of grid points})^{-1}$, meaning for example that wavenumber $10^{-1}$ corresponds to wavelength $10 \times 2.2 = \SI{22}{km}$.
    }
    \label{fig:cosmo_verif_gridded_energy_spectra}
\end{figure}

\begin{figure}[tbp]
    \centering
    \begin{subfigure}[b]{0.5\textwidth}
        \centering
        \includegraphics[width=\textwidth]{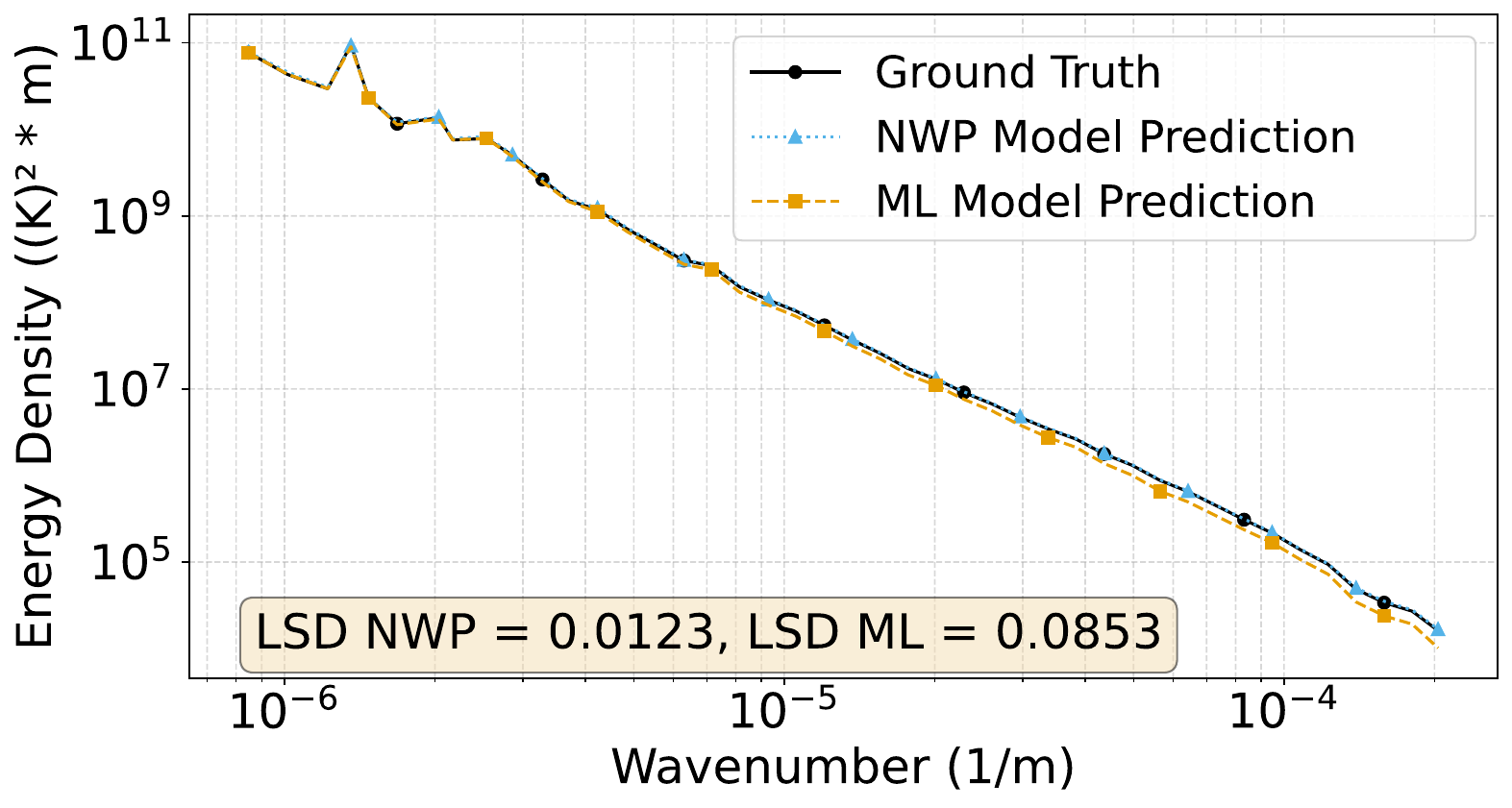}
        \caption{\SI{2}{m} temperature (\wvar{2t})}
    \end{subfigure}%
    \begin{subfigure}[b]{0.5\textwidth}
        \centering
        \includegraphics[width=\textwidth]{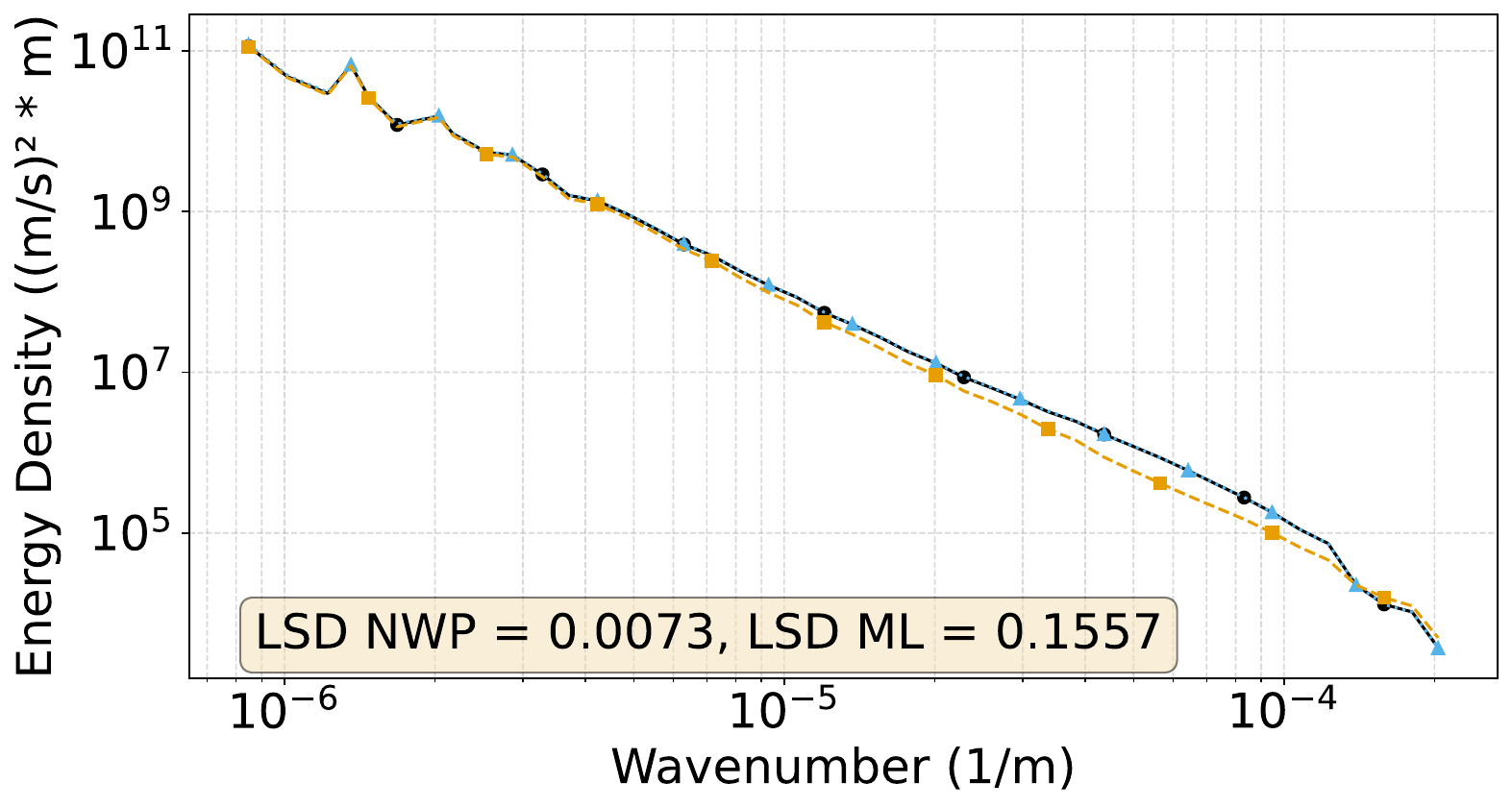}
        \caption{\SI{10}{m} wind-component (\wvar{10u})}
    \end{subfigure}%
    \caption{
        Energy spectra of surface variables for the \gls{DANRA} models and ground truth data, averaged over lead times. 
    }
    \label{fig:danra_verif_gridded_energy_spectra}
\end{figure}

\Cref{fig:danra_verif_gridded_energy_spectra} shows the spectra of the \gls{DANRA} models for a selection of surface variables.
We see similar trends as for the \gls{COSMO} model, with a small spectral difference for the smoother temperature field, and a more substantial deviation in the wind spectra.
Additional spectra from the \gls{DANRA} model, for vertical variables, are shown in \cref{sec:extra_eval_gridded}.

\subsection{Evaluation against station observations}
\label{sec:eval_station}
To provide crucial real-world validation of model performance we here perform point-wise verifications against surface station observations.
We generally focus on key meteorological parameters of high importance for downstream use cases: \SI{2}{m} temperature (\wvar{2t}) and 10-meter wind (\wvar{10u}, \wvar{10v}).
For the \gls{COSMO} model we also consider surface pressure (\wvar{sp}) and precipitation (\wvar{tp01}).
The verification is performed against quality-controlled \gls{SYNOP} observations. 
For the \gls{DANRA} models these are stations in Denmark from the \gls{DMI} network and for the \gls{COSMO} model \gls{MCH} stations in Switzerland are used (data details are given in \cref{sec:data}).
When considering station observations the different geographical features of the two domains are further highlighted.
Many stations in the Danish domain are close to the coast, sitting in flat areas next to the sea.
The Swiss stations are instead far from the sea, with many located in Alpine areas of complex terrain.
These stations are of particular interest, as they measure how well the model represents orographic features.

\begin{figure}[tbp]
    \centering
    \begin{subfigure}[b]{0.5\textwidth}
        \centering
        \includegraphics[width=\textwidth]{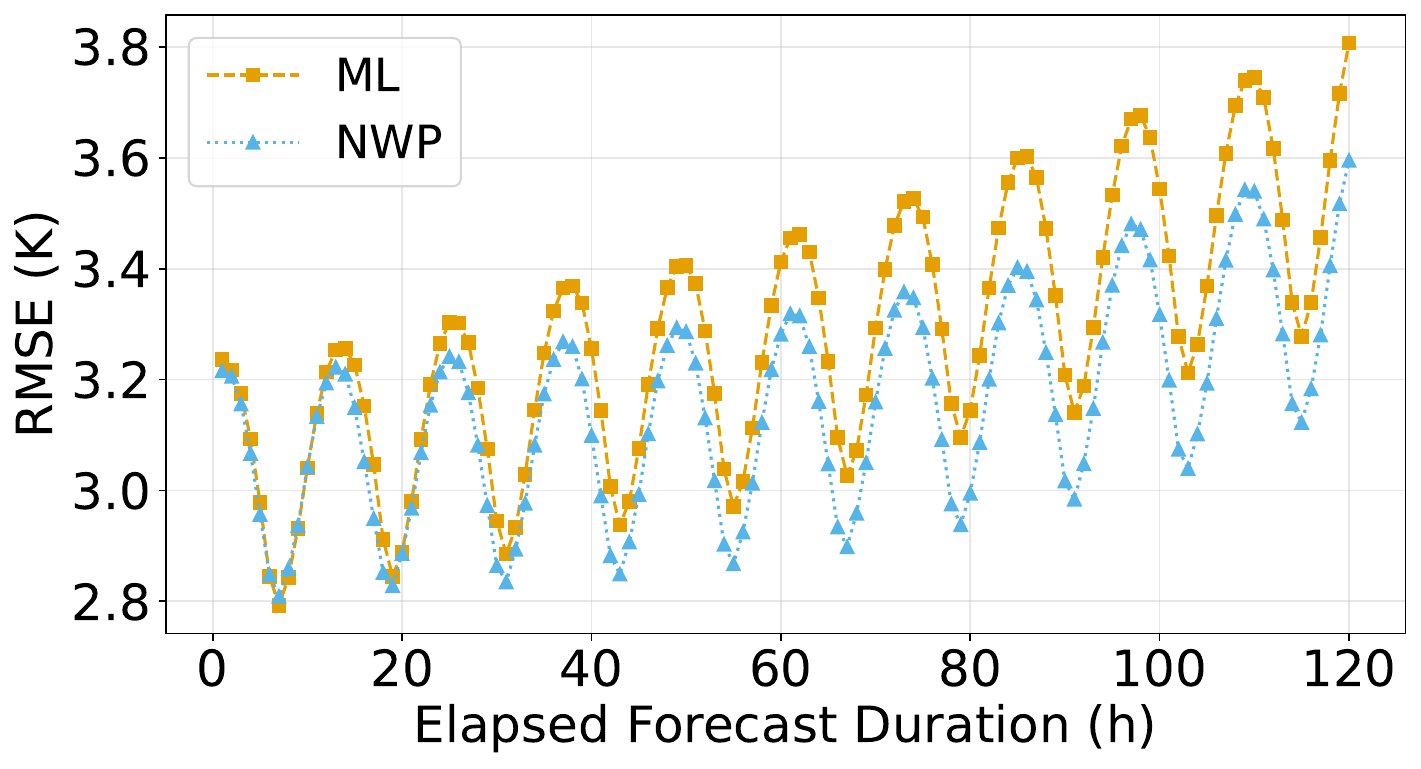}
        \caption{\SI{2}{m} temperature (\wvar{2t})}
    \label{fig:cosmo_verif_sparse_rmse_t2m}
    \end{subfigure}%
    \hfill%
    \begin{subfigure}[b]{0.5\textwidth}
        \centering
        \includegraphics[width=\textwidth]{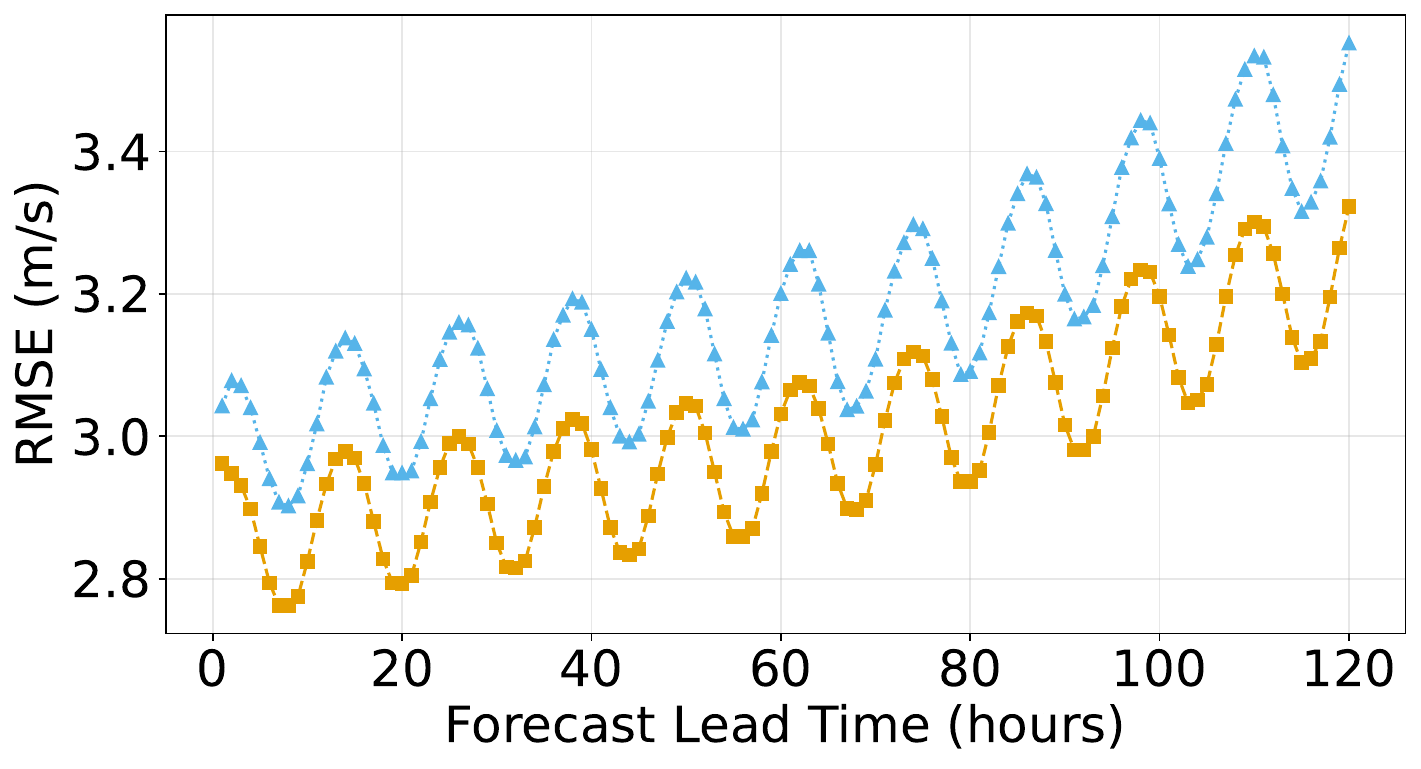}
        \caption{\SI{10}{m} wind}
    \label{fig:cosmo_verif_sparse_rmse_wv10m}
    \end{subfigure}%
    \hfill%
    \begin{subfigure}[b]{0.5\textwidth}
        \centering
        \includegraphics[width=\textwidth]{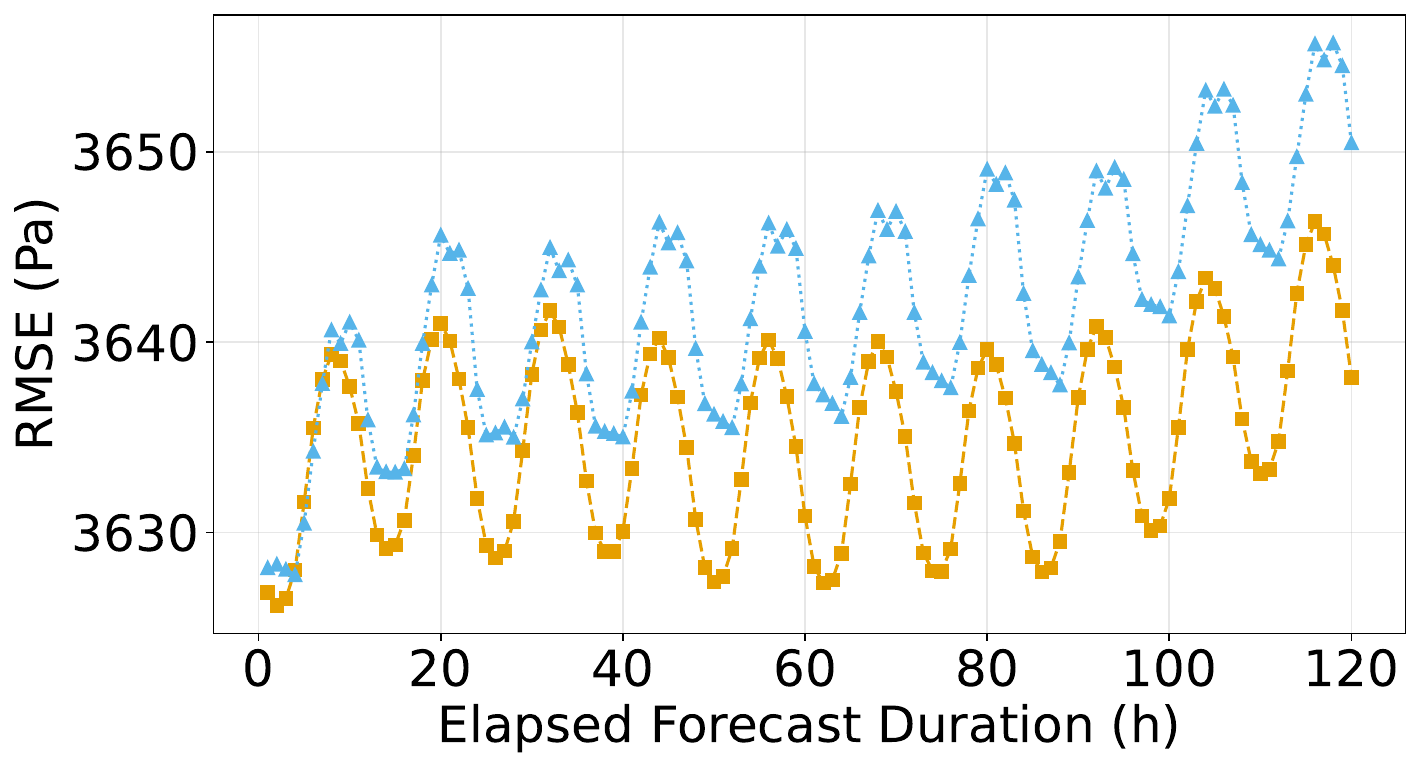}
        \caption{Surface pressure (\wvar{sp})}
    \label{fig:cosmo_verif_sparse_rmse_sp}
    \end{subfigure}%
    \begin{subfigure}[b]{0.5\textwidth}
        \centering
        \includegraphics[width=\textwidth]{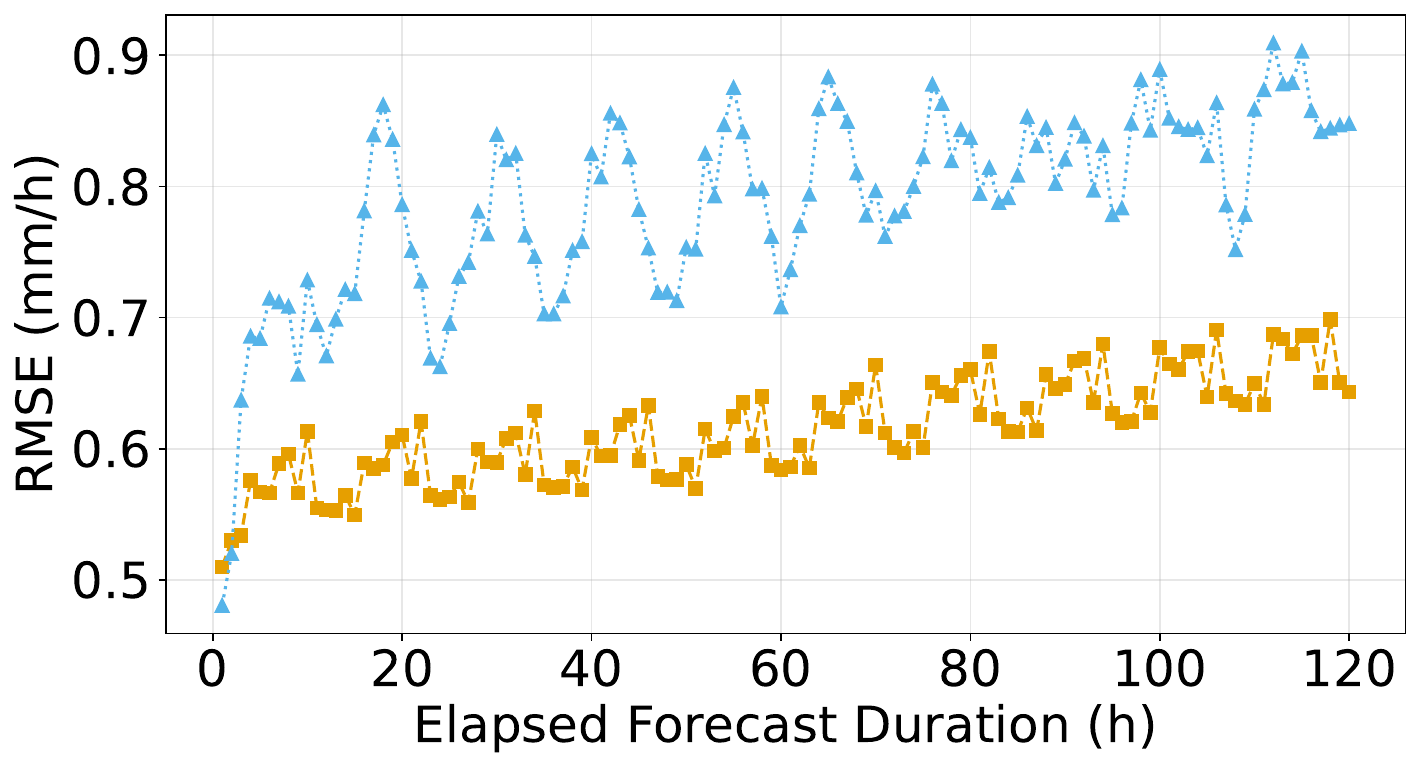}
        \caption{Precipitation (\wvar{tp01})}
    \label{fig:cosmo_verif_sparse_rmse_tp}
    \end{subfigure}%
    \caption{
        \gls{RMSE} for the \gls{COSMO} models compared to station observations.
        The diurnal cycle has a strong impact on all variables, here manifesting with a \SI{12}{\hour} periodicity due to the averaging of forecasts initialized at 00 and 12 UTC.
    }
    \label{fig:cosmo_verif_sparse_rmse}
\end{figure}

\paragraph{\glspl{RMSE}}
Looking at \gls{RMSE} values in \cref{fig:cosmo_verif_sparse_rmse}, the \gls{COSMO} \gls{ML} \gls{LAM} model outperforms the \gls{NWP} model on all variables except temperature. 
This is a remarkable result for a \gls{ML} model on \SI{2.2}{\kilo\metre} and hourly resolution over complex orography. 
It should however be noted that, since the \gls{ML} model was trained to optimize \gls{MSE}, we should expect it to perform well on this metric.
The more blurry \gls{ML} forecasts have an advantage in terms of \gls{RMSE}, as they avoid the double penalty effect \citep{Selz_2025,Ben-Bouallegue_2023_rise}.
The magnitude of the error for all variables is dominated by the impact of the diurnal cycle, which has a heavy influence on all considered variables in this region of complex orography.
This cycle shows up with a \SI{12}{h} period in \cref{fig:cosmo_verif_sparse_rmse} due to the averaging of forecasts initialized at 00 and 12 UTC.
The \gls{COSMO} \gls{ML} model has a lower or comparable error growth to the \gls{NWP} model.
This is impressive, considering the fact that the \gls{COSMO} model predicts over 120 autoregressive steps.
However, it shows the impact of the overlapping boundary forcing, grounding the model in the global \gls{IFS} forecast and preventing the error from growing unconstrained.
For temperature in \cref{fig:cosmo_verif_sparse_rmse_t2m} the error of both models is comparable in the first \SI{12}{\hour}, but the \gls{NWP} model performs better for longer forecast times.
In \cref{sec:extra_eval_station} we decompose the temperature error into bias and standard deviation, finding that the difference in \gls{RMSE} can be mainly attributed to a cold bias in the \gls{ML} model.
The precipitation \gls{RMSE} in \cref{fig:cosmo_verif_sparse_rmse_tp} is lower for the \gls{ML} model after the first few hours.
This is most likely related to a tendency to simply predict less precipitation than the \gls{NWP} model, which avoids the double penalty effect.
We later consider more informative metrics for precipitation.

\begin{figure}[tbp]
    \centering
    \begin{subfigure}[b]{0.5\textwidth}
        \centering
        \includegraphics[width=\textwidth]{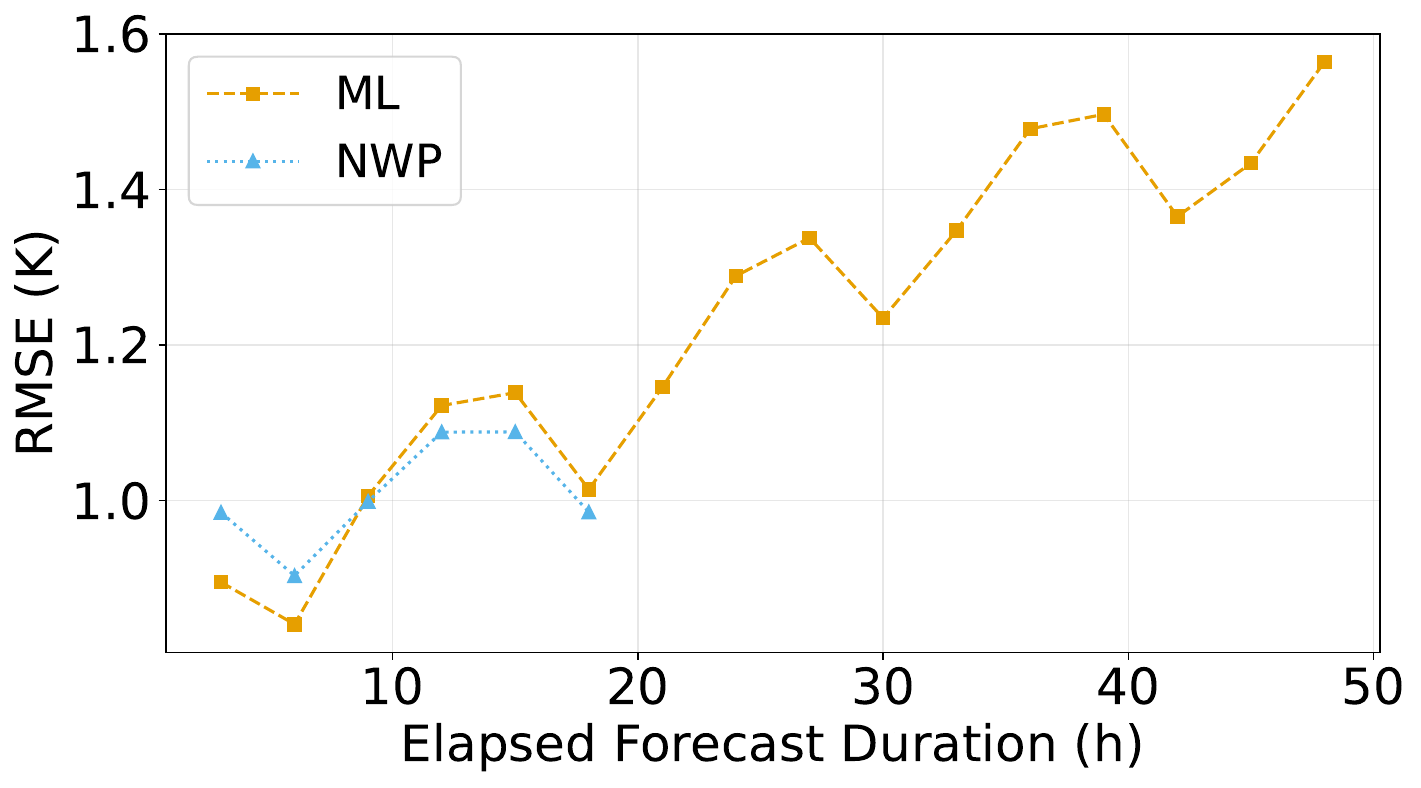}
        \caption{\SI{2}{m} temperature (\wvar{2t})}
    \label{fig:danra_verif_sparse_rmse_t2m}
    \end{subfigure}%
    \hfill%
    \begin{subfigure}[b]{0.5\textwidth}
        \centering
        \includegraphics[width=\textwidth]{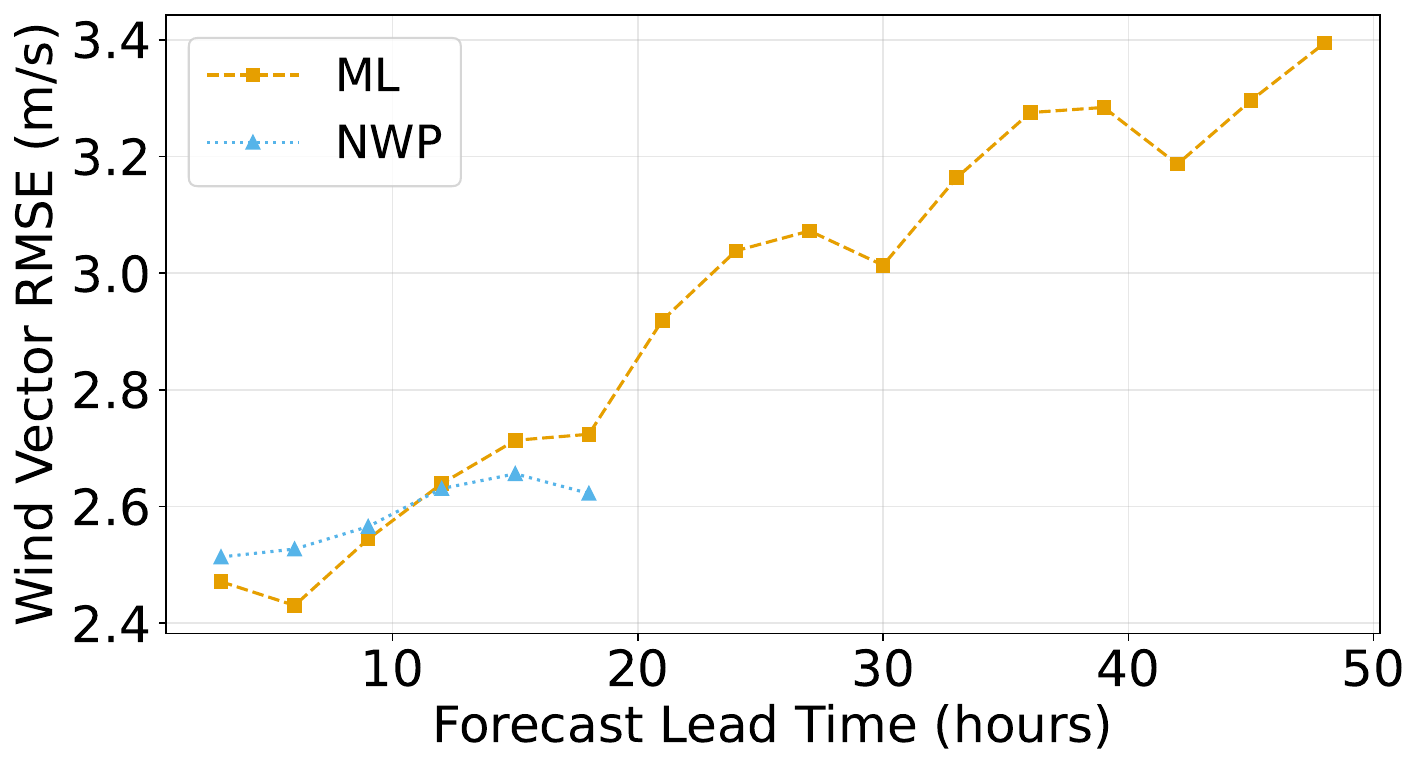}
        \caption{\SI{10}{m} wind}
    \label{fig:danra_verif_sparse_rmse_wv10m}
    \end{subfigure}%
    \caption{
        \gls{RMSE} for the \gls{DANRA} models compared to station observations.
    }
    \label{fig:danra_verif_sparse_rmse}
\end{figure}

Looking at the \gls{DANRA} setting, we show \gls{RMSE} values relative the Danish stations in \cref{fig:danra_verif_sparse_rmse}.
While we only have comparable \gls{NWP} forecasts for the first \SI{18}{\hour}, the results from the \gls{ML} model here look very promising.
For both temperature and wind, the \gls{ML} model achieves somewhat lower errors than the \gls{NWP} counterpart for the first \SI{12}{\hour}.
While the error growth past this point could be lower for the \gls{NWP} model, it is hard to get any clear indications from the short forecasts and comparably few observations.
An important consideration for these results is that there is a slight difference in the initial conditions used for the two models, with a few additional observations included in the \gls{DANRA} fields that the \gls{ML} model is started from.
This could partially or fully explain the difference in performance at short lead times.

While the \gls{DANRA} \gls{ML} model did not fully match the performance of the \gls{NWP} baseline in our gridded verification, compared here to the station observations the errors seem much more similar.
One reason for this discrepancy is that we here consider only Denmark, rather than the full \gls{DANRA} domain.
In \cref{sec:extra_eval_gridded} we show that when considering a region only around Denmark the gridded \gls{RMSE} is more similar for the two models.
We should also expect both models to inherit some of the biases, in terms of the exact physics configuration, from the \gls{DANRA} reanalysis process.
The \gls{ML} model gets these biases from training on the data, while the \gls{NWP} model is directly running the exact same configuration.
These biases should lead to lower errors when comparing to the \gls{DANRA} reanalysis as ground truth, but not be helpful when comparing to station observations.
If these biases are stronger in the \gls{NWP} model than the \gls{ML} one, this should lead to lower errors in the gridded reanalysis but not in \cref{fig:danra_verif_sparse_rmse}.

\begin{figure}[tbp]
    \centering
    \includegraphics[width=0.7\textwidth]{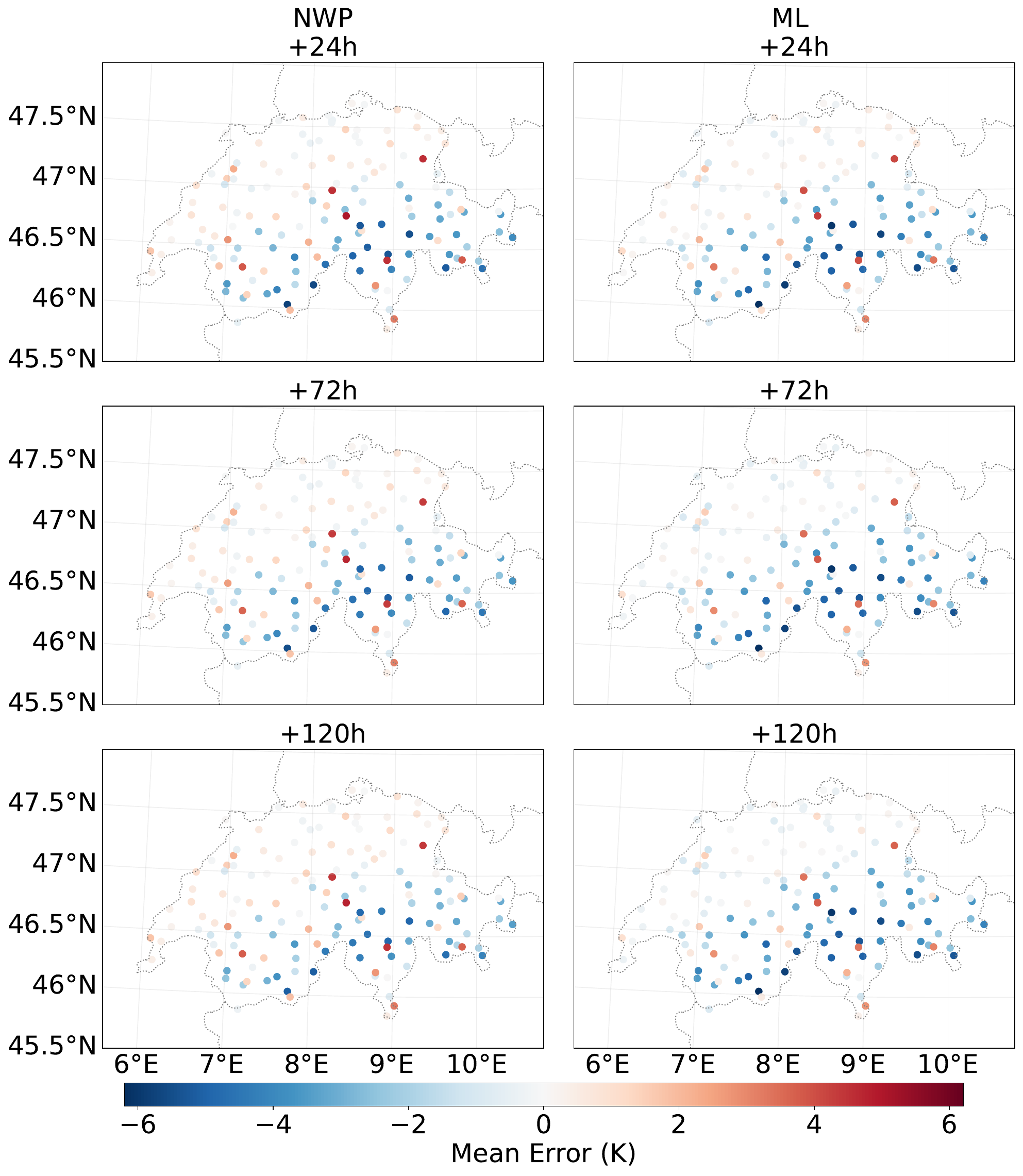}
    \caption{
        Mean error map for \SI{2}{m} temperature (\wvar{2t}) for the \gls{COSMO} models over the full test year period.
    }
    \label{fig:cosmo_verif_sparse_error_map_temp}
\end{figure}

\paragraph{Mean error maps}
In \cref{fig:cosmo_verif_sparse_error_map_temp} we show the mean error maps for temperature over the Swiss domain.
We see that both models share nearly identical biases across the Swiss stations. 
This trend also holds for the other surface variables.
While the error in the central midlands is low, the error in the Alpine stations are larger and mostly negative. 
The more pronounced cold bias of the \gls{ML} model is visible across the Alpine ridge from Wallis to Graubünden.
A possible reason for this bias is that the \gls{ML} model is failing to fully emulate small-scale thermal processes in the Alpine valleys. 
Additionally, for higher lead times the \gls{ML} model might rely strongly on downscaling of the overlapping boundary forcing.
This could lead to a stronger cold bias in the Alpine region, where the boundary forcing is less representative of the local conditions.

\begin{figure}[tbp]
    \centering
    \includegraphics[width=0.7\textwidth]{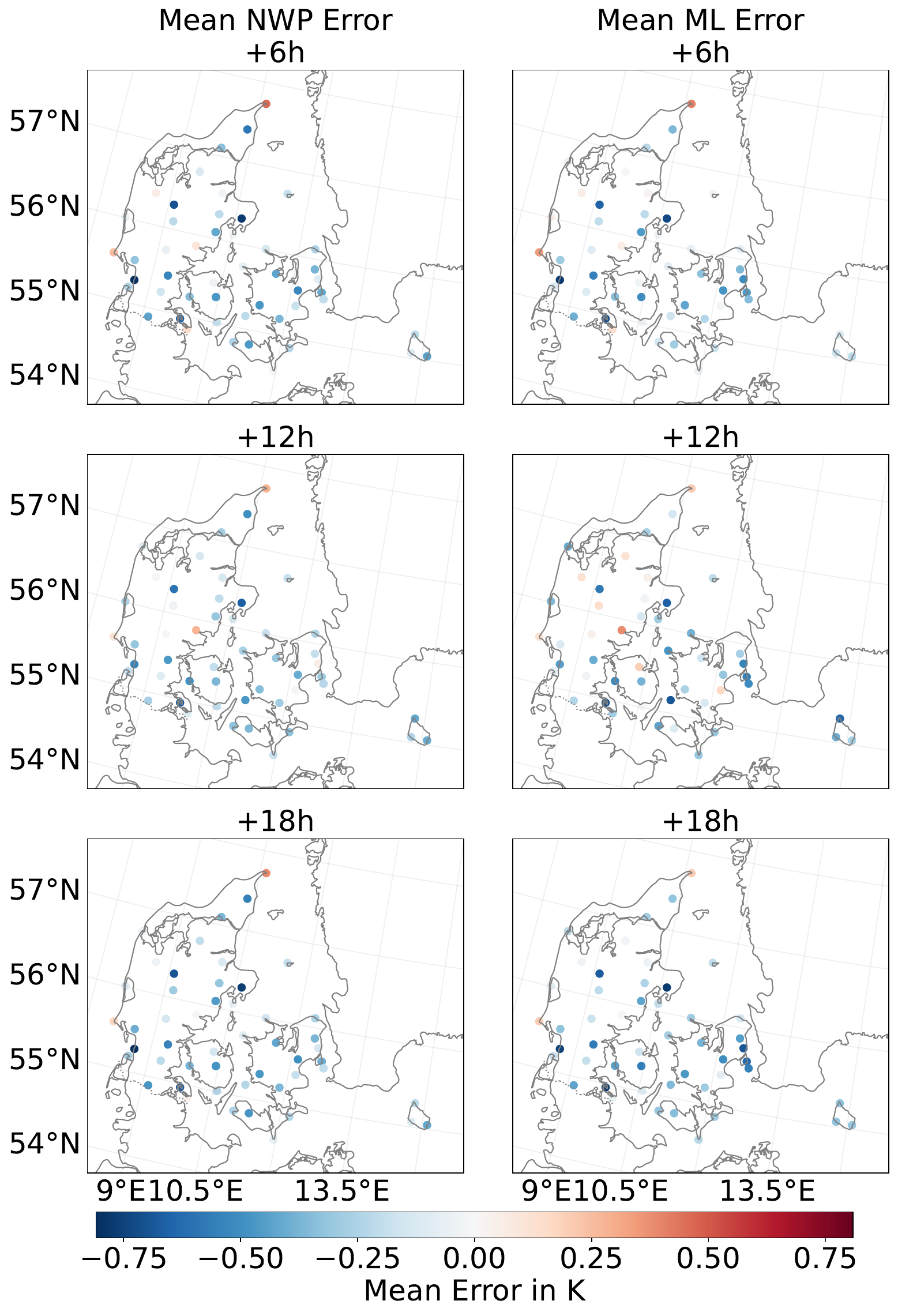}
    \caption{
        Mean error map for \SI{2}{m} temperature (\wvar{2t}) for the \gls{DANRA} models over the full test year period.
    }
    \label{fig:danra_verif_sparse_error_map_temp}
\end{figure}

\Cref{fig:danra_verif_sparse_error_map_temp} shows maps over the mean error of the \gls{DANRA} models for stations in Denmark.
As in the Swiss domain, the biases are largely shared between the \gls{NWP} and \gls{ML} models.
It is clear that the \gls{ML} models inherit some of the \gls{NWP} modeling errors from the training data, or from the overall modeling setup.
There are still some minor differences in \cref{fig:danra_verif_sparse_error_map_temp} worth pointing out.
The \gls{ML} model interestingly has a much stronger cold bias in Eastern Zealand, around Copenhagen.
On the other hand, the \gls{NWP} model exhibits fairly strong biases for the northernmost stations, which are less severe in forecasts from the \gls{ML} model.
Additional mean error maps for both domains are given in \cref{sec:extra_eval_station}.

\paragraph{Categorical scores}
For metrics like precipitation and wind, it is common to use categorical scores like \gls{ETS} and \gls{FBI}.
These measure the ability of the models to capture rare high-magnitude events, potentially with increased societal-impact.
For wind we compute the scores for the u and v components individually. 
These scores are shown in \cref{fig:cosmo_verif_sparse_categorical} for the \gls{COSMO} models.
From the \gls{FBI} results we see that the \gls{ML} model is underforecasting across all wind and the higher two precipitation categories. For the lowest precipitation category the \gls{ML} model is instead overforecasting.
This is most likely a direct consequence of training a deterministic model on \gls{MSE}. 
Such a model will predict smoother fields and therefore underforecast more extreme events, matching our observations for the energy spectra in \cref{fig:cosmo_verif_gridded_energy_spectra}. 
The \gls{NWP} model is underforecasting for all wind categories and the highest precipitation category. 
Across the board the \gls{NWP} model shows lower frequency biases than the \gls{ML} model.
Looking at the \gls{ETS} however, thereby taking into account accurate predictions by pure chance and spatiotemporal information, the differences between the two models are smaller.
The \gls{ML} model still underforecasts the highest threshold category for both wind and precipitation.

\begin{figure}[tbp]
    \centering
    \begin{subfigure}[b]{0.5\textwidth}
        \centering
        \includegraphics[width=\textwidth]{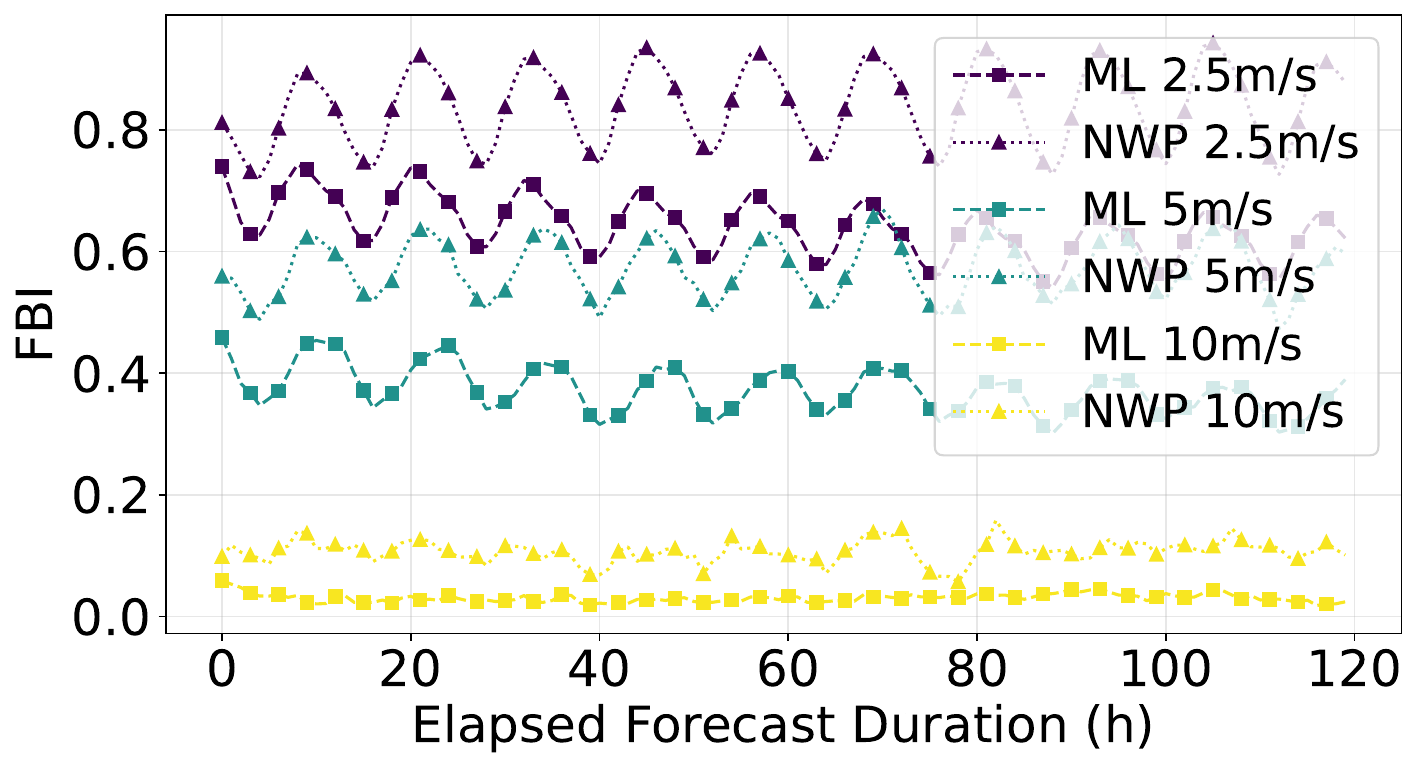}
        \caption{Wind u-component (\wvar{10u}) \gls{FBI}}
    \end{subfigure}%
    \begin{subfigure}[b]{0.5\textwidth}
        \centering
        \includegraphics[width=\textwidth]{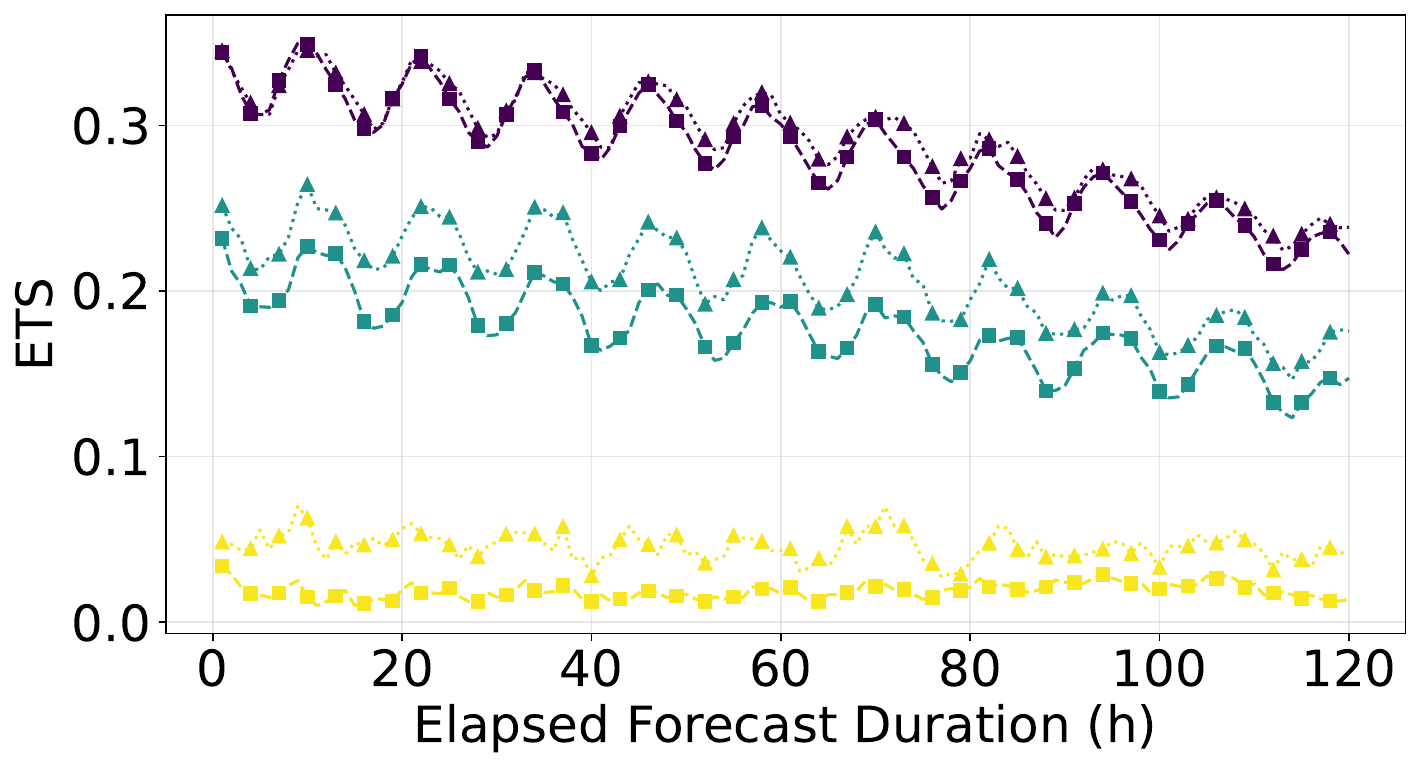}
        \caption{Wind u-component (\wvar{10u}) \gls{ETS}}
    \end{subfigure}%
        \hfill%
    \begin{subfigure}[b]{0.5\textwidth}
        \centering
        \includegraphics[width=\textwidth]{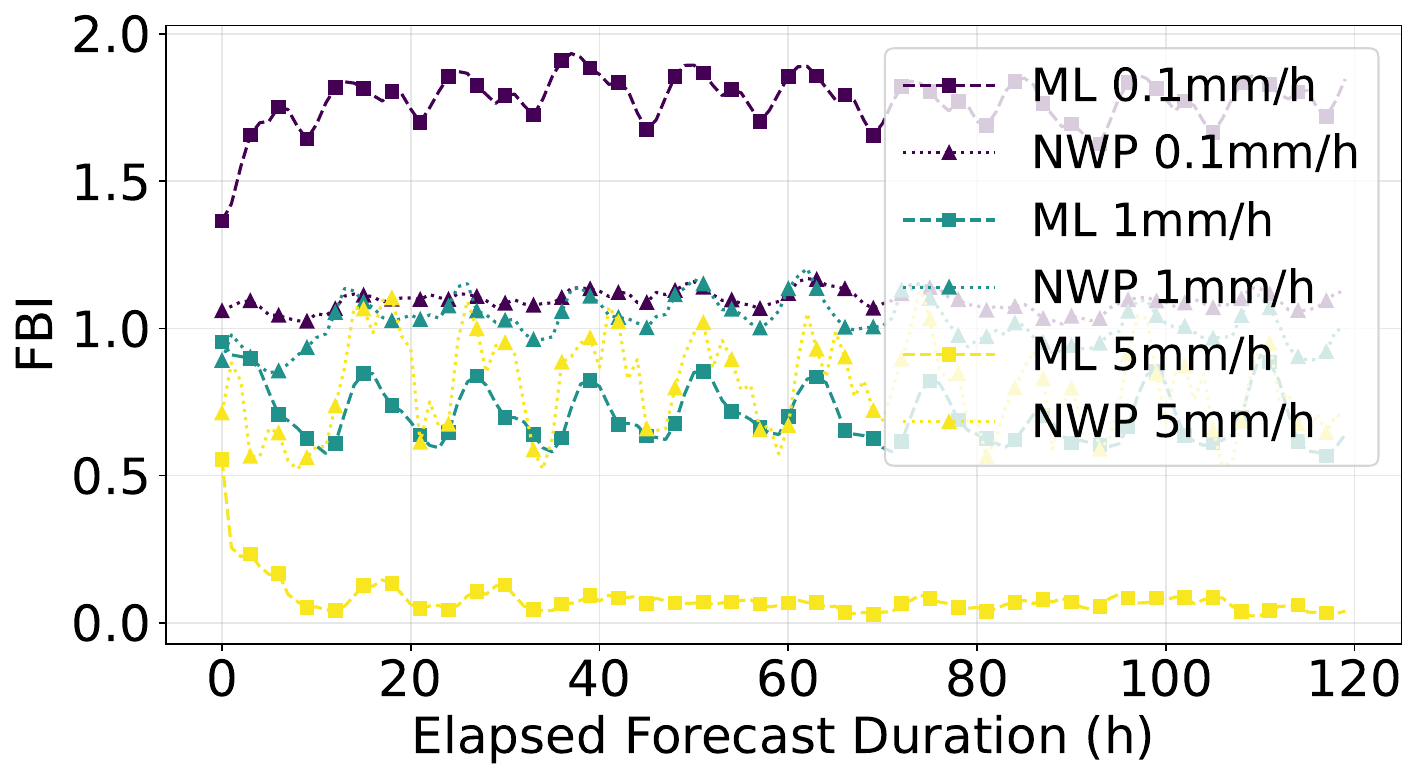}
        \caption{Precipitation (\wvar{tp01}) \gls{FBI}}
    \end{subfigure}%
    \begin{subfigure}[b]{0.5\textwidth}
        \centering
        \includegraphics[width=\textwidth]{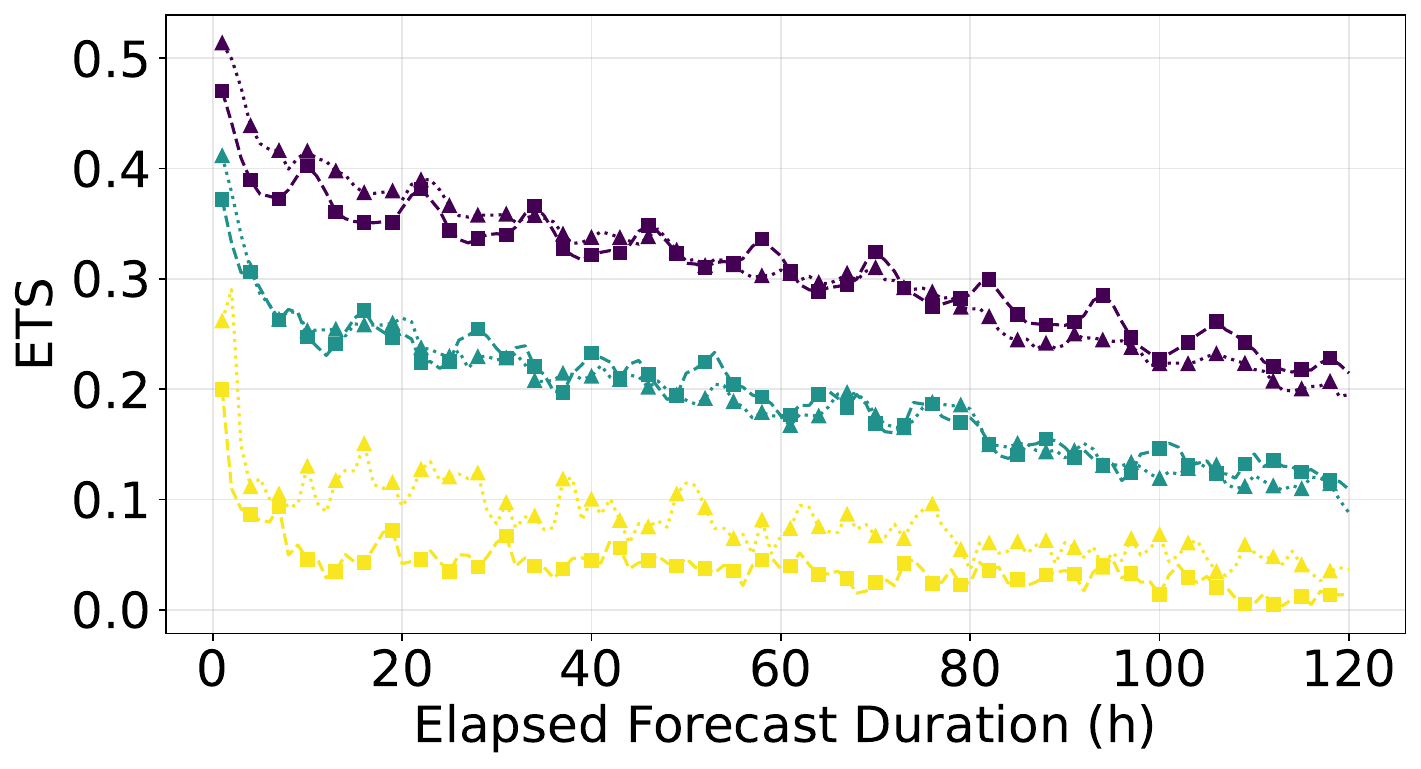}
        \caption{Precipitation (\wvar{tp01}) \gls{ETS}}
    \end{subfigure}%
    \caption{
        \gls{ETS} and \gls{FBI} scores for the \gls{COSMO} models, compared to station observations.
        We show values only for the u-component of wind, as the corresponding plots for the v-component show the same patterns.
    }
    \label{fig:cosmo_verif_sparse_categorical}
\end{figure}

Categorical scores for wind are shown in \cref{fig:danra_verif_sparse_categorical} for the \gls{DANRA} models.
Across the board the metrics are highly similar for the \gls{ML} and \gls{NWP} models.
The \gls{FBI} of the u-component in \cref{fig:danra_verif_sparse_categorical_u_fbi} shows that both models overforecast, but only slightly.
This eastward wind component is important, as it captures wind impacting Denmark from the North Sea.
The v-component in \cref{fig:danra_verif_sparse_categorical_v_fbi} instead shows underforecasting, especially at the highest threshold.
Compared to the \gls{COSMO} model, this underforecasting is however far less severe, especially for winds under \SI{5}{\meter\per\second}.
\gls{ETS} values are similar for both models, except for the highest threshold of v-component values, where we observe a decreasing trend with \gls{ETS} for the \gls{ML} model.
Overall these scores indicate an ability of the \gls{DANRA} \gls{ML} model to capture much of the stronger winds over Denmark.

\begin{figure}[tbp]
    \centering
    \begin{subfigure}[b]{0.5\textwidth}
        \centering
        \includegraphics[width=\textwidth]{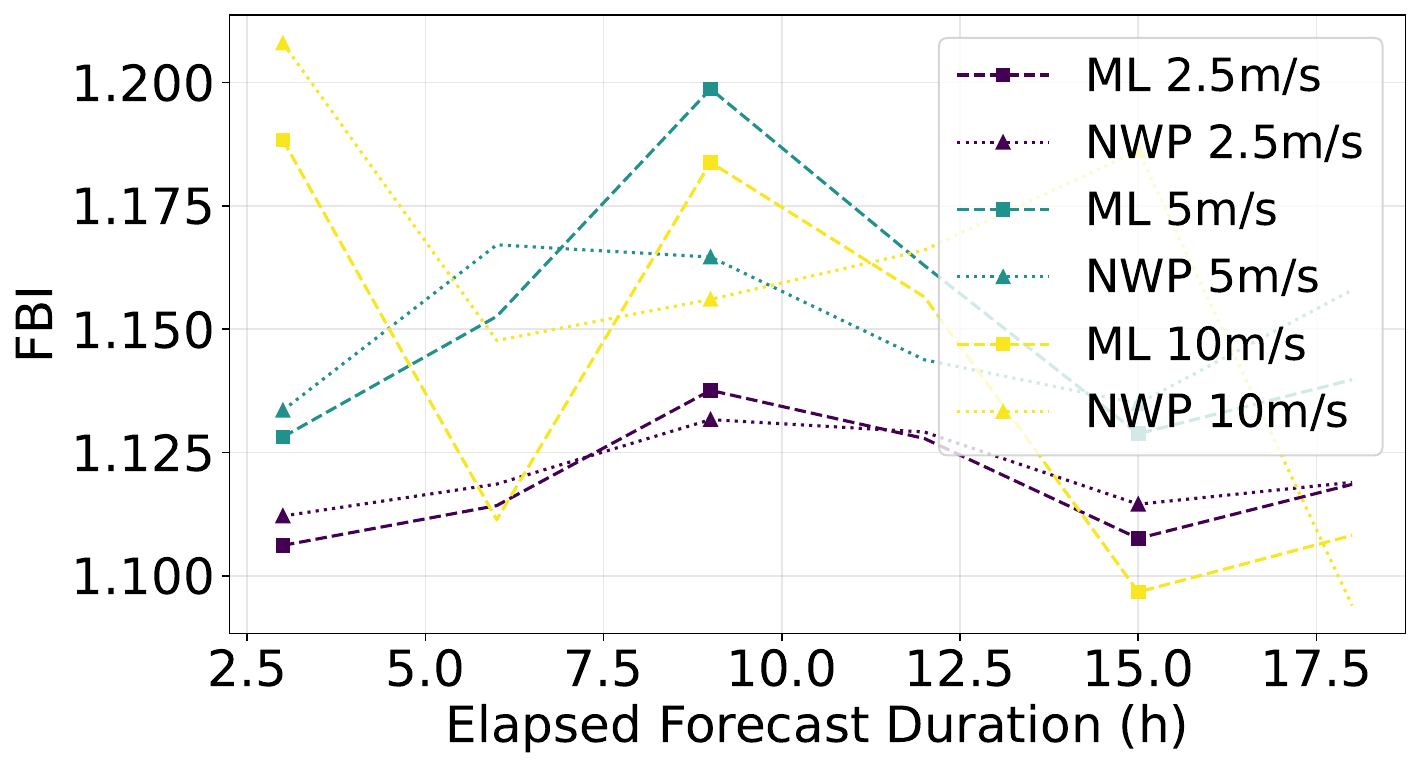}
        \caption{Wind u-component (\wvar{10u}) \gls{FBI}}
        \label{fig:danra_verif_sparse_categorical_u_fbi}
    \end{subfigure}%
    \begin{subfigure}[b]{0.5\textwidth}
        \centering
        \includegraphics[width=\textwidth]{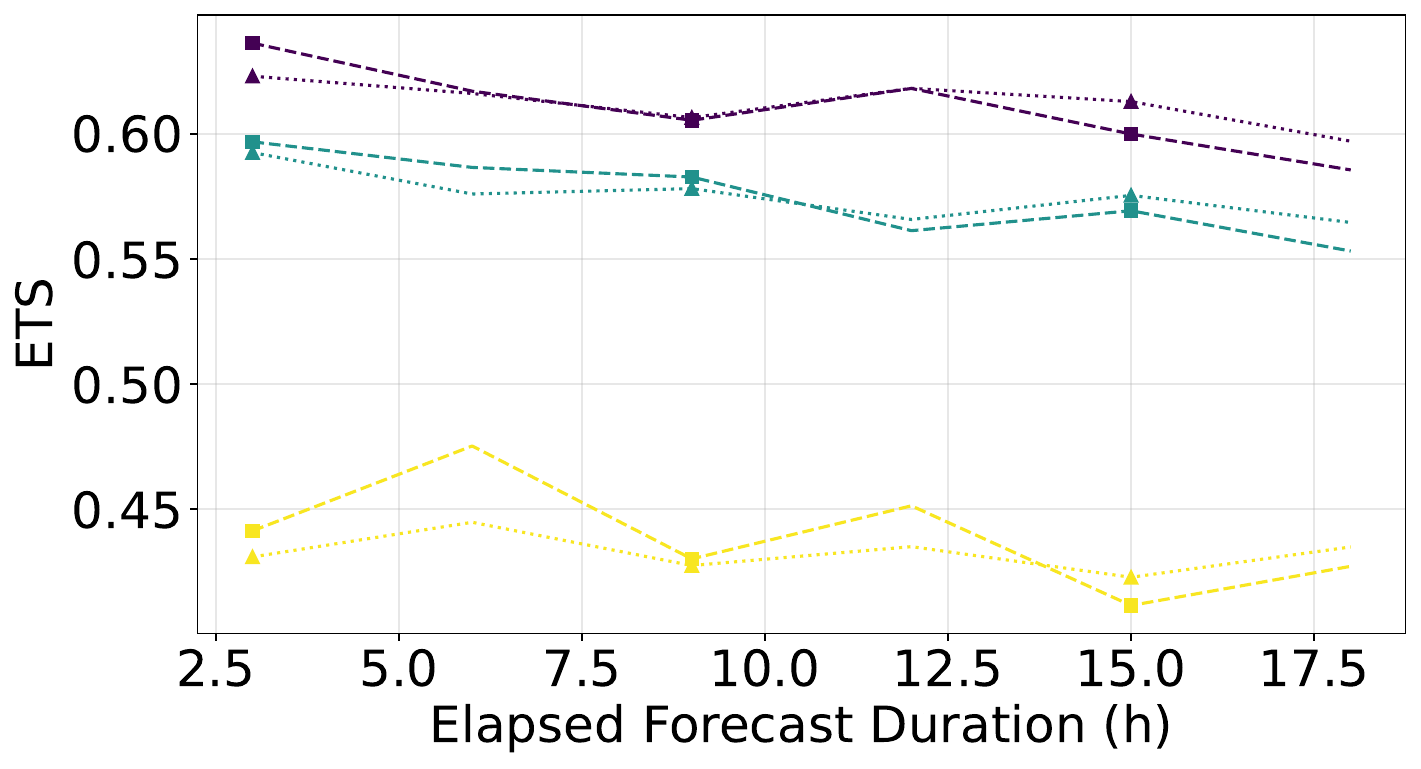}
        \caption{Wind u-component (\wvar{10u}) \gls{ETS}}
    \end{subfigure}%
    \hfill%
    \begin{subfigure}[b]{0.5\textwidth}
        \centering
        \includegraphics[width=\textwidth]{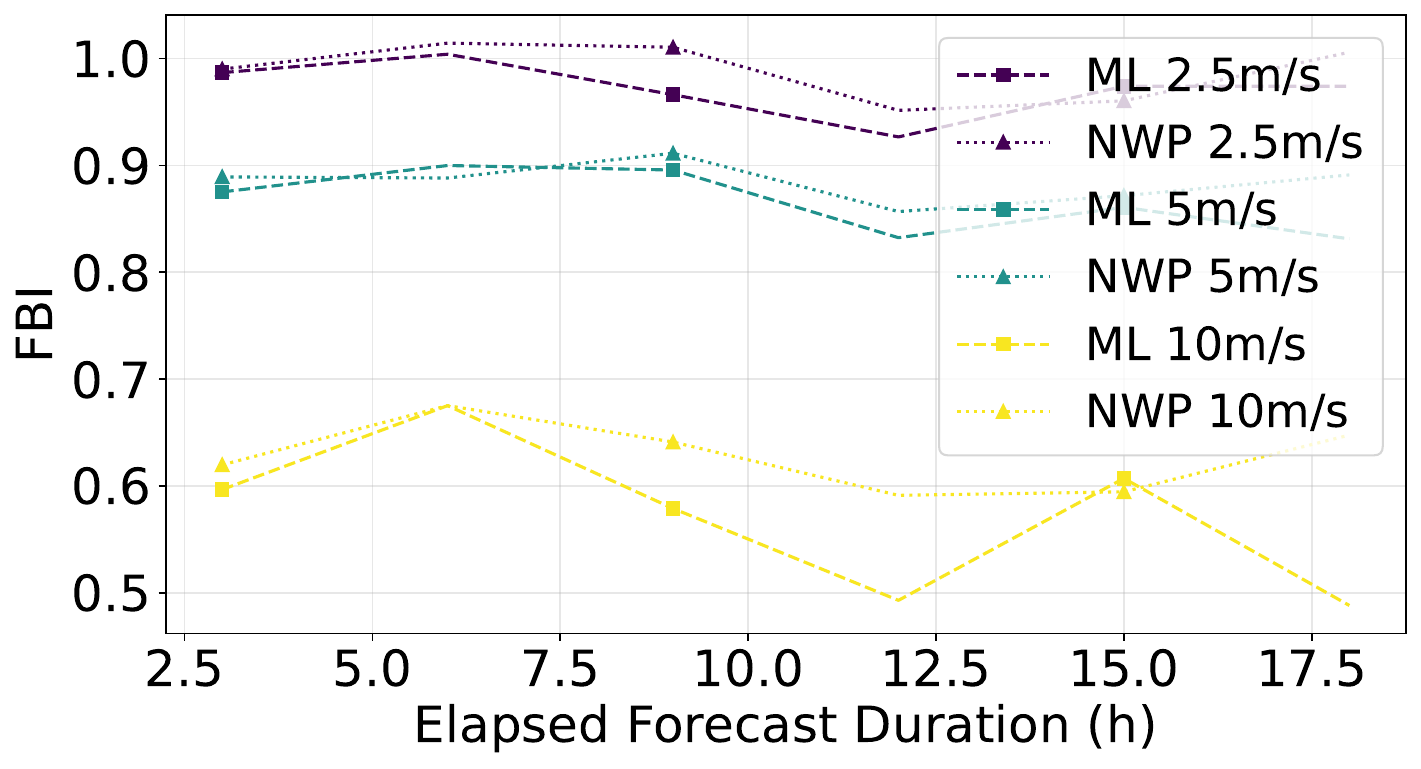}
        \caption{Wind v-component (\wvar{10v}) \gls{FBI}}
        \label{fig:danra_verif_sparse_categorical_v_fbi}
    \end{subfigure}%
    \begin{subfigure}[b]{0.5\textwidth}
        \centering
        \includegraphics[width=\textwidth]{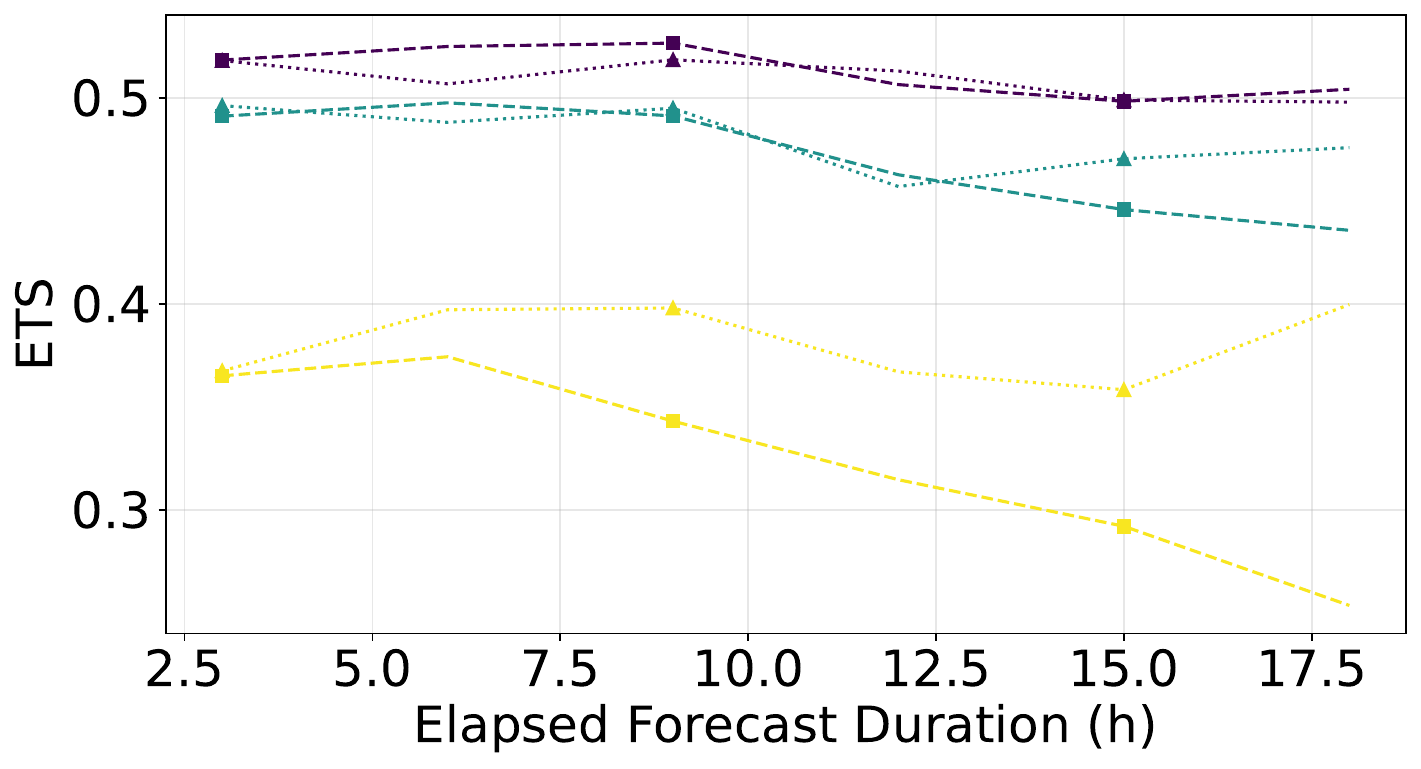}
        \caption{Wind v-component (\wvar{10v}) \gls{ETS}}
    \end{subfigure}%
    \caption{
        \gls{ETS} and \gls{FBI} scores for the \gls{DANRA} models, compared to station observations.
    }
    \label{fig:danra_verif_sparse_categorical}
\end{figure}

\paragraph{Distributional alignment}
To understand if the models are capable of representing the full distribution of values for different variables, in \cref{sec:extra_eval_station} we inspect histograms of predictions from the models.
By comparing such histograms to data both from the sparse station observations and the gridded datasets we are able to detect distributional shifts in the values produced by the models.
For the \gls{COSMO} setting we observe close distributional alignment to the gridded analysis data for both the \gls{ML} and \gls{NWP} models.
However, the distributions of forecast values interpolated to station locations differs somewhat from that of the observations.
This can be related to the interpolation method used, and the complex orography of the Swiss domain.
We perform a similar investigation for the \gls{DANRA} setting.
For that domain, with simpler orography, model distributions align well both with the gridded reanalysis and station observations.

\subsection{Case study: storm Ciara complex}
\label{sec:ciara_case_study}
In February 2020, a sequence of powerful extratropical cyclones created exceptional meteorological conditions across Europe.
These storms provide an ideal test case for the boundary condition handling and multi-scale weather representation capabilities of our \gls{ML} \glspl{LAM} . 
This storm sequence began with Ciara (called Sabine in Germany and neighboring countries, and Elsa in Norway), which developed over the North Atlantic on February 8--9.
After formation it traversed the British Isles, Northern Europe, and eventually reaching the Alpine regions and Scandinavia by February 10-11 \citep{ciara2023, Haeseler2020Sabine}. 
Just days later, Storm Dennis formed over the Atlantic, rapidly deepening to a remarkable \SI{920}{\hecto\pascal} by February 15, affecting the UK and Ireland before sweeping across Northern Europe through February 17 \citep{MetOffice2020Dennis}. 
These storms were dynamically linked through the interaction of the North Atlantic jet stream and exhibited explosive cyclogenesis.

This storm complex had substantial societal impact across the British Isles and central Europe.
The most severe impacts occurred in South Wales, Herefordshire, Worcestershire, and Shropshire (UK), where major flooding developed after Dennis delivered rainfall onto ground already saturated by Ciara just days earlier. In Germany, North Rhine-Westphalia and Lower Saxony reported significant structural damage, while Alpine regions faced a combination of high winds and heavy snowfall. The 9-day period from February 8-16 saw many western UK areas receive their entire February monthly average precipitation, with some locations recording 150-200\% of normal values.

To understand the abilities of the \gls{ML} \gls{LAM} models to capture extreme weather conditions, we produce forecasts for storm Ciara using both the \gls{COSMO} and \gls{DANRA} models.
From a modeling perspective, the Ciara-Dennis complex presents several compelling challenges. 
Both systems featured deep low-pressure centers with intense pressure gradients, well-defined frontal boundaries, and rapid transitions across domain boundaries.
Additionally, the impact of the storms varied substantially with local orography, with mountainous regions experiencing significantly higher wind speeds and orographically enhanced precipitation, testing especially the ability of the \gls{COSMO} model to represent sub-grid scale processes in complex terrain.

\ifpreprint
    \newcommand{\includeciaraplot}[1]{\includegraphics[width=0.86\textwidth]{#1}}
\fi
\begin{figure}[tbp]
    \centering
    \includeciaraplot{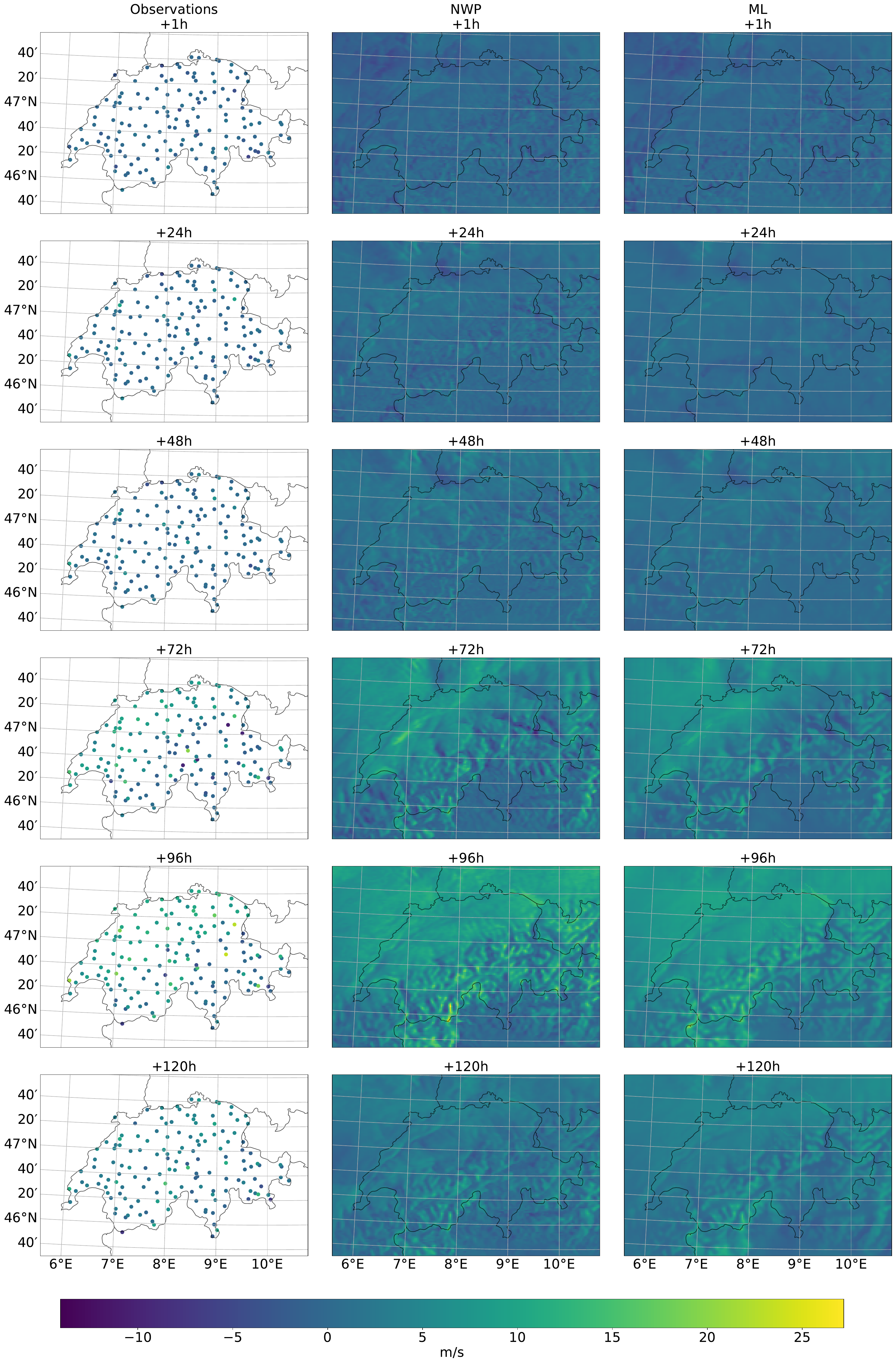}
    \caption{
        Comparison of wind u-component (\wvar{10u}) between station measurements, \gls{NWP} model and \gls{ML} model for the storm Ciara over Switzerland. 
        Forecasts are initialized at 2020-02-07T00 UTC.
    }
    \label{fig:cosmo_verif_case_study_wind_u}
\end{figure}

\begin{figure}[tbp]
    \centering
    \includeciaraplot{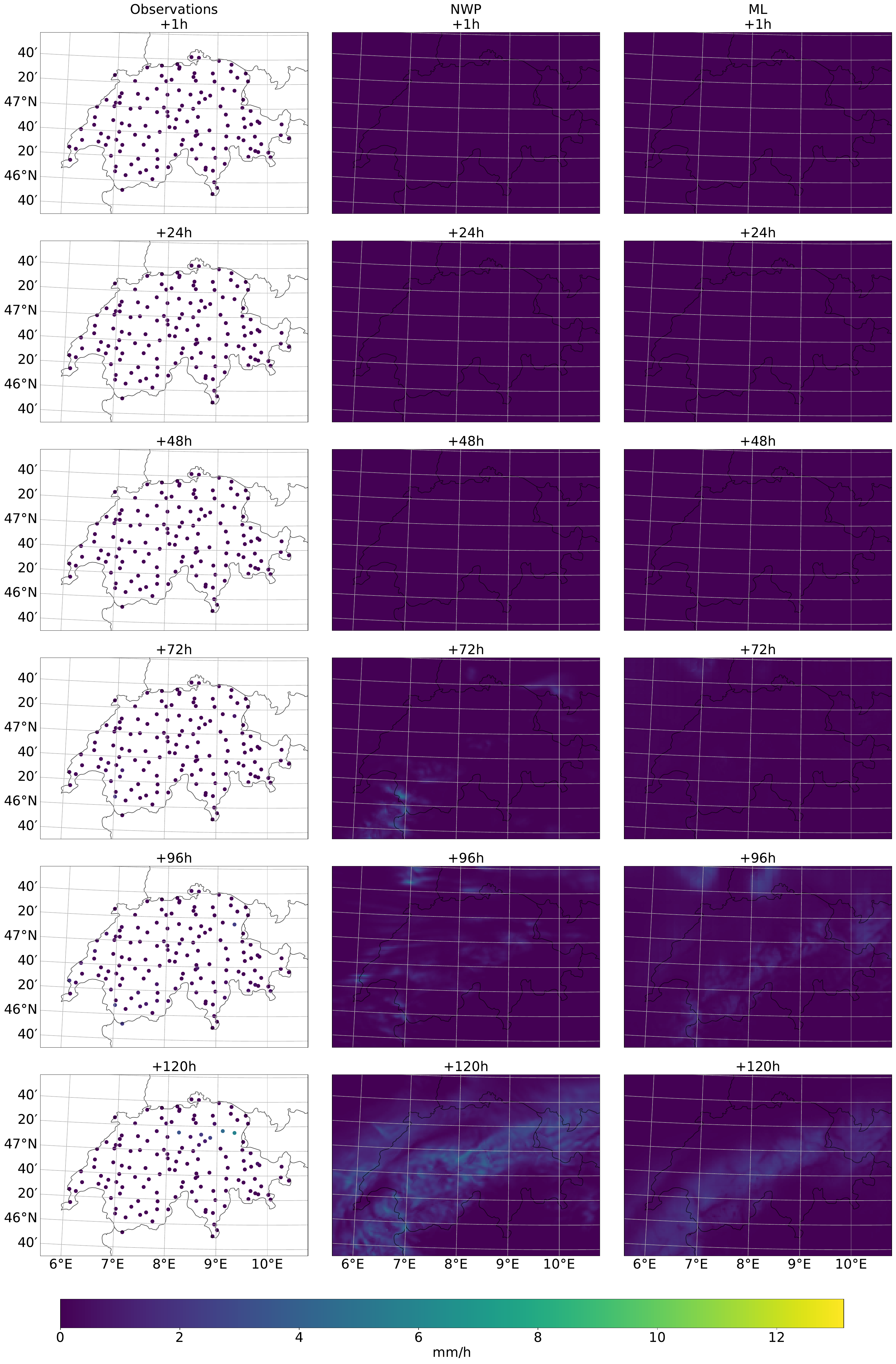}
    \caption{
        Comparison of precipitation (\wvar{tp01}) between station measurements, \gls{NWP} model and \gls{ML} model for the storm Ciara over Switzerland.
        Forecasts are initialized at 2020-02-07T00 UTC.
    }
    \label{fig:cosmo_verif_case_study_precipitation}
\end{figure}

In \cref{fig:cosmo_verif_case_study_wind_u} we show the wind u-component, between station measurements and the \gls{COSMO} \gls{NWP} and \gls{ML} models.
The forecasts are initialized at 2020-02-07T00 UTC, before storm Ciara had entered the \gls{LAM} domain.
Even after three days of forecasting, when the storm has a large impact over Switzerland, both models manage to capture the higher wind speeds in alpine valleys.
This shows the ability of the \gls{ML} model to accurately propagate weather patterns into the domain, even under extraordinary conditions.
The \gls{ML} model however shows a certain tendency to underforecast some of the highest wind speeds, in comparison to the \gls{NWP} model.
This is consistent with the categorical scores for the \gls{COSMO} model in \cref{sec:eval_station} and can be observed by the smoother looking fields.
A similar observation can be made for the rain patterns in \cref{fig:cosmo_verif_case_study_precipitation}, \SI{120}{\hour} into the forecast.
While both models predict the precipitation reaching the affected stations in the north-west of Switzerland, the precipitation fields of the \gls{ML} model are more smooth and do not take as extreme values.

\begin{figure}[tbp]
    \centering
    \includegraphics[width=\textwidth]{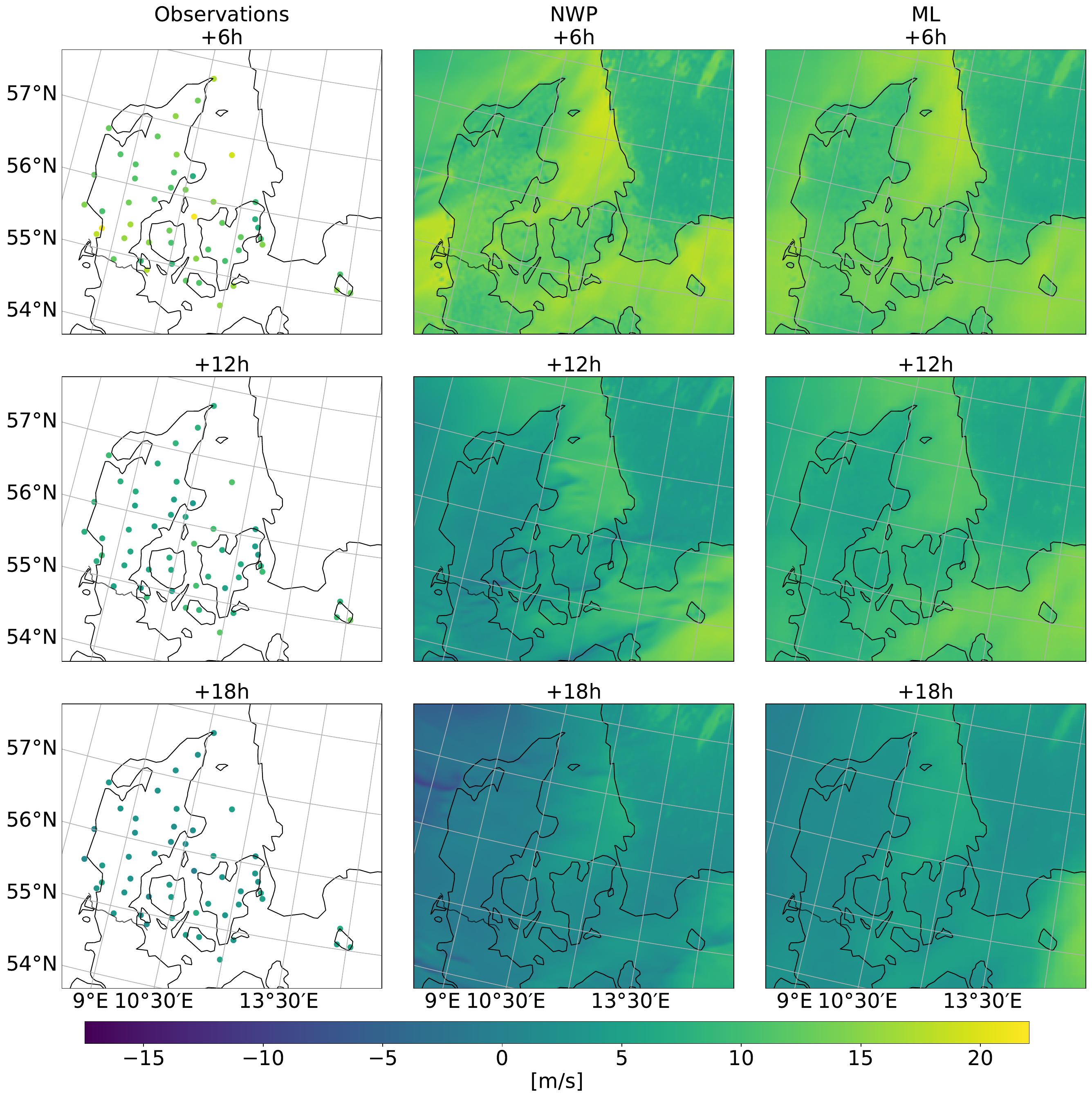}
    \caption{
        Comparison of wind v-component (\wvar{10v}) between station measurements, \gls{NWP} model and \gls{ML} model for the storm Ciara over Denmark.
        Forecasts are initialized at 2020-02-09T12 UTC.
    }
    \label{fig:danra_verif_case_study_wind_v}
\end{figure}

As we focus on shorter forecast lengths for the \gls{DANRA} model, we consider forecasts initialized at 2020-02-09T12 UTC in \cref{fig:danra_verif_case_study_wind_v}.
Both models do a good job in capturing the high wind station measurements after \SI{6}{\hour}.
These models are however initialized with the storm already in the domain, but it is encouraging that the \gls{ML} model manages to accurately simulate the atmospheric dynamics to bring the extreme conditions to Denmark.
Also for later lead times, when the storm has largely passed, the \gls{ML} model does a good job at matching the station observations.
Similar to the \gls{COSMO} setting, the forecasts from the \gls{DANRA} \gls{ML} model are notably more smooth in comparison to the \gls{NWP} forecasts.
While still capturing the larger-scale dynamics, the \gls{ML} model does not reproduce smaller fronts without a clear connection to orography or the coast.
As there is great uncertainty in these smaller scale features, the model has instead learned to average them out.
It is however encouraging that despite this smoothness, the most extreme predicted values of the \gls{ML} model do not fall far from the observations, still indicating usefulness in high-impact scenarios.

\FloatBarrier
\section{Discussion}

\subsection{Challenges and limitations}
\label{sec:discussion_limitations}
We here discuss some of the limitations of this study and challenges in the proposed \gls{ML} \gls{LAM} framework.

\paragraph{Focus on specific model class}
Our work focuses specifically on graph-based \gls{LAM} models, and do not consider other types of neural network architectures.
While we in \cref{sec:boundary_forcing} make the argument that the flexibility enabled by graph-based models is especially suitable for this problem, there is definitely merit to tailoring also other architectures for \gls{LAM} forecasting.
In particular, regular grid architectures such as \glspl{CNN} \citep{storm_cast,diffusion_lam} or \glspl{ViT} \citep{yinglong} can have computational benefits on modern \glspl{GPU}, compared to \glspl{GNN} using many sparse operations.
These do however not fulfill the flexibility criteria of our desiderata in \cref{sec:boundary_forcing} without modifications.

Another drawback of the architecture is that we for some fields observe graph-related artifacts in predictions and in the model biases.
This is a problem observed also in earlier works for both rectangular \citep{neural_lam,graph_efm} and triangular graphs \citep{keisler}.
The problem does however seem to diminish as models are scaled up \citep{graph_efm}, indicating that flexible enough models can learn to counteract these.

Our investigation is similarly only focused on the \gls{LAM} formulation of regional forecasting, as outlined in \cref{sec:lam_problem_statement}.
Empirical comparisons between different approaches to regional forecasting, such as the stretch-grid method of \cite{stretch_grid}, are not the focus of this work and were deemed outside our scope

\paragraph{Deterministic models}
We focus in this work specifically on deterministic \gls{MLWP} models, as opposed to probabilistic models generating ensemble forecasts \citep{graph_efm,storm_cast,diffusion_lam}.
This delimitation reduces the already high training costs, enabling more extensive experimentation around design choices.
Still, we believe that many of our findings in \cref{sec:exp_boundary_forcing,sec:exp_graph_construction} should generalize also to probabilistic \gls{ML} \gls{LAM} models.

A direct consequence of training deterministic models with an \gls{MSE} loss is however that forecast fields become overly smooth \citep{Selz_2025}.
This is a recurring observation throughout our verification, as well as in previous literature \citep{graphcast,fourcastnet,aifs}.
It should be noted that this smoothing effect is not some unfortunate side-effect, but arguably the model solving exactly the problem it is being optimized for.
As we are training a statistical model for the forecasting problem, we have to deal with the inherent uncertainty in predictions.
There are mainly three ways of doing this:
\begin{enumerate}
    \item Aim for a single prediction minimizing the average error (\gls{MSE}).
    This minimizer is the mean of the predictive distribution, averaging out the uncertainty and necessarily producing smooth fields \citep{Selz_2025}.
    \item Aim for a prediction minimizing the average error, but under some constraints regularizing the model to output less smooth fields \citep{fourier_amplitude_loss,fixing_double_penalty}.
    Such regularization is possible, but it is generally not clear what this optimal regularized solution is, and if all such solutions necessarily correspond to fields that appear physically realistic.
    As something different than the \gls{MSE} is being optimized, one can also expect some cost to the average error.
    \item Move to probabilistic models, producing samples from the predictive distributions rather than a single forecast.
    Such ensemble forecasting approaches do however come with substantial complexity, and if a single forecast is desired in the end, this is anyhow the smooth ensemble mean.
\end{enumerate}
While we have chosen option 1 here, the best tradeoff depends highly on the downstream uses of the forecasts.

\paragraph{Choice of atmospheric variables and vertical levels}
Our selection of atmospheric variables and forcing inputs was decided based on a few preliminary model trainings, but we do not systematically search for an optimal subset of variables or vertical levels. 
While this would be interesting, it is also a challenging task due to the many interdependencies among different variables.

We found modeling of moisture-related variables, total precipitation and relative humidity, specifically challenging.
While we handle precipitation similar to all other variables, this is often not suitable as it is rarely assimilated.
High-quality precipitation fields for training, coherent with the rest of an analysis dataset, are also hard to produce.
This is for example something we lack for the \gls{DANRA} dataset.
Precipitation is additionally one of the fields most impacted by the double-penalty effect of \gls{MSE} training, especially motivating probabilistic approaches.

The challenges in forecasting humidity relates more to the specific modeling setup.
To model relative humidity, we restrict the model output to $[0, 1]$ by applying sigmoid activation functions.
While this prevents unphysical outputs, the approach does seem suboptimal for the model training when combined with the \gls{MSE} minimization. 
We observe that for some high moisture regions the model struggles to reduce the high moisture values, and does not accurately transport the moist air.
This could likely be improved by a loss function tailored for the $[0,1]$ domain, but a simpler alternative is to just directly model specific humidity.
    
\paragraph{Choice of boundary forcing}
For the majority of the design studies \gls{ERA5} has been used as boundary forcing for both the DANRA and COSMO domains. 
While using ERA5 as a best possible state of the global atmosphere in the design studies is valid, such reanalysis products are not available in a realistic setting.
We therefore switch to \gls{IFS} boundary forcing for the final verification, to keep this true to operational conditions.
The model has never seen \gls{IFS} data during training or fine-tuning. 
Unsurprisingly, results show that the model performs worse when using \gls{IFS} as boundary forcing.
This is expected, as the \gls{IFS} forecasts come with their own biases and errors, not present in \gls{ERA5}.
This is especially true when the latest available forecast from \gls{IFS} started a long time ago.
While we have not observed instabilities or forecast deterioration when swapping the \gls{ERA5} boundary to \gls{IFS}, this still constitutes a distributional shift in the input data.
With this in mind, exposing the model to \gls{IFS} boundary forcing already during fine-tuning could improve the model performance. 
We leave any detailed investigation of training curriculums for different boundary forcing setups for future work.
While conceptually similar to using \gls{IFS}, it would also be interesting to take the boundary forcing from a global \gls{MLWP} model.

\paragraph{Design studies and verification}
There are a number of limitations related to the experiments in \cref{sec:experiments}.
In our design studies we perform independent investigations for a number of different modeling choices, treated as orthogonal.
While this is a common assumption, and necessary to make such studies feasible, it is likely an oversimplification.
For example, the best graph configuration might depend on whether overlapping boundary forcing is used or not.

One limitation to the station verification is the simplistic interpolation method used, based on a weighted 4-nearest-neighbors approach.
This lacks adjustment to station height, which especially impacts the interpolation over the mountainous Swiss terrain.
While this affects both the \gls{ML} and \gls{NWP} models the same, we note in \cref{sec:eval_station} that this results in some distributional shifts between the model forecasts and station observations.

Another challenge is achieving a fully fair comparison between the \gls{NWP} and \gls{ML} models.
The biggest caveat to these comparisons is the differences in initial conditions (ensemble \gls{DA} mean vs control member for \gls{COSMO} and operational analysis vs  reanalysis for \gls{DANRA}, see \cref{sec:data}). 
While we believe this has a minor impact, it is hard to quantify and the discrepancy should be taken into account when interpreting the results.

The limited availability of comparable \gls{NWP} forecasts means that much of the verification is restricted to surface variables and, for \gls{DANRA}, short lead times.
As only subsets of operational forecasts are often archived, compiling a set of comparable \gls{NWP} forecasts can be a substantial challenge.

\subsection{Future prospects of \texorpdfstring{\gls{ML} \gls{LAM}}{ML LAM} models} 
\label{sec:discussion_future}
The capabilities demonstrated in this work point to many exciting future opportunities for kilometer-scale \gls{ML} \glspl{LAM}.
We here discuss some of the possible extensions of these models, as well as their future use cases.

\paragraph{Ensemble forecasting and probabilistic models}
We have limited the scope of our work to deterministic models, but extensions of our framework to ensemble forecasting are of great interest.
To achieve the important uncertainty quantification provided by ensemble forecasting, our models could be used with ensemble \gls{DA}, starting forecasts from different initial conditions \citep{Bulte_2024_Uncertainty}.
Our framework is also directly compatible with multi-model ensemble methods \citep{dlwp_ensemble,calibration_of_large_neurwp}, capturing also uncertainty from the model training.
Extensions to full probabilistic \gls{ML} \gls{LAM} models would both tackle forecast smoothness and enable ensemble forecasting.
While there are initial exciting works in this direction, they do not consider realistic boundary settings \citep{graph_efm, diffusion_lam} or produce only short forecasts \citep{storm_cast}.
Our flexible boundary forcing framework could be combined with these approaches to build ensemble models based on diffusion or latent variable formulations.

\paragraph{Temporal resolution}
We are in this work pushing \gls{MLWP} to higher spatial resolutions, but while our 3 and \SI{1}{h} time steps are shorter than those typically used in global \gls{MLWP} \citep{graphcast, panguweather, gencast}, this is still a coarse temporal resolution for many applications.
For example, the EU energy sector now trades in \SI{15}{\minute} intervals \citep{EU2019_943}.
Simply taking shorter time steps in the \gls{ML} models is not a good option, as this leads to high error accumulation throughout the forecast \citep{panguweather,arci}.
A better solution is to unroll forecasts with longer time steps and interpolate to intermediate times, either through a separate \gls{ML} model \citep{modulated_afno_interpolation} or by building in this capability in the forecasting model itself \citep{arci}.

\paragraph{Overlapping boundary forcing}
For the \gls{COSMO} \gls{ML} model, with \SI{1}{\hour} time steps, we saw great benefits in the overlapping boundary configuration.
However, in \cref{sec:exp_boundary_forcing} we observed that the \gls{DANRA} model, with \SI{3}{\hour} time steps, struggled to balance the forecasting and downscaling capabilities in this hybrid setup.
Further tailoring the \gls{LAM} models for the overlapping boundary setting is an interesting direction to investigate in future work.
In particular, it should be further explored how to effectively simulate dynamics under the soft constraints of the boundary information, avoiding over-reliance on the boundary forcing.

\paragraph{Learning directly from observations}
A major limitation to current \gls{MLWP} models is still the need for gridded initial conditions from a \gls{DA} process.
Recent developments in the field \citep{graph_dop,transformer_dop,aardvark,fuxi_weather} aim to move past this dependence on traditional \gls{DA} systems, initializing models directly from observations.
Development of such models also in the \gls{LAM} setting is highly interesting, as much added value of regional models can come from integration of additional local observations.

\paragraph{\glspl{LAM} vs stretch-grid models}
For producing high-resolution regional \gls{MLWP} forecasts, there are two main approaches: \glspl{LAM} and stretch-grid models \citep{stretch_grid}.
We discuss some of their strengths and weaknesses here, but also note that further experiments to understand and contrast these approaches is an interesting avenue for future work.
\gls{ML} \gls{LAM} models are fully defined on a sub-region of earth's surface, whereas stretch-grid models are global with a higher resolution over a specific region.
Both methods require both regional high-resolution data and more coarse global data during training, with \gls{LAM} models only needing global data within the small boundary region.

To produce a forecast from an \gls{ML} \gls{LAM} model, we first need a global (or larger-domain) forecast to use as boundary forcing.
This can come from any source providing such forecast products, or simply running a global \gls{ML} model from a global initial condition.
The \gls{LAM} model then takes the boundary forcing and a regional initial condition as input to roll out its forecast.
The stretch-grid approach instead starts directly from both a regional and a global initial condition, producing a forecast both for the full globe and the region of interest.
This allows for feedback effects between the high-resolution domain and the rest of the globe, with meteorological features moving seamlessly in and out from the domain.
However, this also closely entangles the global and regional forecasting capabilities.
The information available around the region of interest in a stretch-grid model is only as accurate as the global forecast capabilities of that model.
For a \gls{LAM} model on the other hand, the global forecast comes from a different model, meaning that we can swap out the source of this boundary forcing data as more accurate global models become available.
The caveat to this is however that a too large distributional shift in the boundary forcing features would likely hurt the \gls{LAM} model, and could require fine-tuning using this new boundary data.
In a stretch-grid model, for the usefulness of regional boundary information to match that of a \gls{LAM} model, its global forecasting capabilities must match those of a non-stretch-grid global model.
So as a side effect of accurate regional modeling, there is a requirement to also learn a highly skillful global model.
This can be viewed both as an additional benefit or as a wasteful byproduct, and its justifiability should depend on the domain size and available computational resources.

One possible benefit of the stretch-grid approach is the possibility of global patterns to impact the regional domain.
During training, if parameters are shared between the global and regional parts of the model it should be possible to more quickly learn the atmospheric dynamics to be simulated by both.
On the other hand, for the high-resolution regional domain we want to capture additional convective processes that are not resolved on the global scale.
The model thus also needs to operate differently on the global scale and regionally, and some loss weighting is needed to achieve this \citep{stretch_grid}.
This is not a consideration for our \gls{ML} \gls{LAM} models, as these only forecast the interior domain.
Similarly, a stretch-grid approach is locked in to using the same time step length for the global and regional modeling.
Our \gls{LAM} framework allows for making use of global forecast with long time steps, avoiding large error accumulation, while still taking shorter steps in the \gls{LAM} model.
As we observe with our \gls{COSMO} model, boundary forcing can constrain the error accumulation to allow for time steps as short as \SI{1}{\hour}.
Having access to global weather patterns in each forecasting step of stretch-grid models is still substantially more information than in the boundary of our \gls{LAM} models.
However, our findings that large boundaries do not improve model performance calls into question the usefulness of this information to produce accurate regional forecasts.

Looking at computational considerations, the stretch-grid model requires keeping both the model state representing global atmospheric conditions and the state of the high-resolution domain in memory at the same time.
\gls{LAM} setups can handle these sequentially, potentially first running a global \gls{ML} forecast and then the \gls{LAM}.
Taking \gls{GPU} memory as the main limiting factor, running the models sequentially means that the global and regional components do not have to compete for this resource. %
This is less of a consideration in resource-rich settings, where model-parallelism can be employed to enable scaling past the memory of a single \gls{GPU} \citep{stretch_grid}, but becomes relevant in settings with limited resources.
Also when moving to probabilistic approaches, such as diffusion models \citep{storm_cast,diffda}, the computational costs grow manifold, putting larger emphasis on optimizing memory and compute usage.
At the end of the day this choice between stretch-grid and \gls{LAM} models is in many contexts a practical one, if the option to run a single global and regional model is preferable and affordable over separate global and regional forecasts.

\paragraph{Cross-domain generalization}
Through their low computational cost, \gls{ML} \glspl{LAM} can enable possibilities for new actors to run high-resolution weather forecasting models, possibly for new regions.
Our modeling framework and open source implementation help make this possible.
However, a major hurdle is still the need for large amounts of high-resolution training data.
Current models, including ours, are trained and evaluated for one specific region only.
An important direction to make regional \gls{MLWP} accessible to many is to develop \gls{LAM} models that can generalize to new domains. 
This could be done either through transfer learning, where models are fine-tuned on small amounts of data from the new domain, or optimally by training models that can immediately be deployed for an arbitrary domain. 
Such cross-domain \gls{LAM} models relate closely to the many efforts on building earth system foundation models \citep{climax, atmorep, aurora}, but this specific possibility remains unexplored.

\paragraph{Accurate and energy-efficient forecasting}
As shown in this work, regional \gls{MLWP} has the possibility of producing faster and more accurate forecasts.
Improvements to forecast accuracy is important for industries like renewable energy and transportation, and can under extreme conditions help save lives and reduce property damage.

Forecasts from our \gls{ML} models run in only a minute, allowing for rapidly producing new timely forecasts.
To make the most use of this new capability, the \gls{ML} models must be backed up by a \gls{DA} system of similar speed.
While the \gls{DA} systems of today do not satisfy this, encouraging developments in \gls{ML} \gls{DA} here hold promise \citep{diffda, towards_self_contained_mlwp}.
The situation is similar for boundary forcing from global models, where we already now have access to many good global \gls{MLWP} configurations \citep{graphcast,aifs,aurora}.

Apart from speed, the \gls{ML} models also come with a heavy reduction in the energy consumption needed for the forecasting process.
This reduction is somewhat offset by the energy cost of model training, but as this is done much more rarely than forecasting there are still substantial energy savings \citep{fourcastnet}.
A research project such as this also has a larger environmental impact due to the energy consumption of training multiple large models for our design study.
We hope however that by sharing our findings with the scientific community we reduce the need for re-running many such experiments in future modeling efforts.

\section{Conclusion}
In this work we have proposed a framework for building graph-based \gls{ML} \glspl{LAM} for regional weather forecasting.
The framework includes flexible methods for integration of boundary conditions either from reanalysis datasets or other forecasting models.
This flexibility has been demonstrated by training \glspl{LAM} for two different domains, with different geographic conditions and available data.
We have conducted a thorough investigation of different design choices in our framework, shedding light on boundary widths, graph constructions and overlapping boundary forcing configurations.
Graphs with long spatial edges were shown to be crucial for model performance, and in particular hierarchical graphs showed some advantages.
In the Swiss setting we saw great benefits of overlapping the boundary forcing with the interior domain, essentially creating a hybrid downscaling and \gls{LAM} forecasting model.
Our final models have gone through extensive verification against both gridded data and in-situ observations, illuminating their strengths and weaknesses.
Both \gls{ML} models achieve competitive scores on standard metrics.
The Swiss model especially outperforms the \gls{NWP} baseline on important surface variables.
Forecasts from the deterministic \gls{ML} models are however smooth, exhibiting low energy levels for higher wavenumbers.
The models still produce highly useful forecasts even under extreme conditions, as indicated by our case study for storm Ciara.

Our experiments demonstrate that there are great opportunities in building \gls{ML} \gls{LAM} models.
The general modeling framework and open source implementation is developed to make this possible for a range of different regions and datasets.
We hope also that the results from our design studies can help inform developments of similar \gls{ML} \gls{LAM} models, but also be of use in extended modeling paradigms such as probabilistic models \citep{graph_efm,storm_cast,diffusion_lam} and direct-from-observations forecasting \citep{graph_dop,transformer_dop}.
Our efforts show that \gls{ML} enables accurate high-resolution regional forecasts, at a minuscule computational cost in comparison to existing \gls{NWP} systems.
This creates entirely new possibilities not just for operational forecasting centers, but also researchers and downstream users of forecast products.

\ifpreprint
    \section*{Data availability}
    Our code is released as a set of open source repositories, see \cref{sec:open_source}.
Model checkpoints from our final models are openly available\footnote{\url{https://doi.org/10.5281/zenodo.15131837}} under a CC BY 4.0 license.
A 2-months sample of the \gls{COSMO} dataset has been uploaded to the ETH Research Collection for reproducibility \citep{cosmo_data}.
For a complete 5 year record of the dataset please contact the authors.
The COSMO-E forecasts and station observations used for the Swiss verification are part of the \gls{MCH} archive and not publicly accessible.
With the transition to the ICON model, the COSMO forecasts are considered a legacy dataset and not accessible via a public API.
\gls{MCH} makes the ICON-based successor data publicly available\footnote{ICON data will be public under the Swiss Open Government Data initiative at \url{https://data.geo.admin.ch/browser/} starting May 2025.}, which is highly recommended for future use in similar studies.
\ifpreprint
    A publication further describing the DANRA data is currently being drafted, and the full dataset will be released with it.
\else
    The DANRA dataset is openly available at (link), and the corresponding NWP forecasts at (link). A publication further describing the DANRA dataset is currently being drafted.
\fi
The Danish station measurements used for verification are publicly available from the \gls{DMI} open data\footnote{\url{https://opendatadocs.dmi.govcloud.dk/en/DMIOpenData}} under the CC BY 4.0 license.
All boundary forcing data (\gls{ERA5} and \gls{IFS}) are from the WeatherBench 2 public archive \citep{weatherbench2}.

    \section*{Acknowledgments and funding}
    We would like to thank
the whole MeteoSwiss ML-Team for many fruitful discussions on data curation, model verification and training.
We especially thank Francesco Zanetta (MCH) for help with data handling.
We thank also the DMI ML-Team for help with data handling and many open-source contributions to the codebase,
Dana Grund (ETH) and Tobias Selz (LMU) for theoretical discussions about ML-Emulators and error growth,
Erik Larsson (LiU) and Martin Andrae (LiU) for useful \gls{MLWP} discussions
and
Reto Knutti (ETH) for contributing to the review of this manuscript.
This work was supported as part of the Swiss AI Initiative by a grant from the Swiss National Supercomputing Centre (CSCS) under project ID a01 on Alps.
Our computations were enabled by the Berzelius resource provided by the Knut and Alice Wallenberg Foundation at the National Supercomputer Centre.
We also acknowledge the Danish Centre for AI Innovation (DCAI) for providing access to the AI supercomputer Gefion Super Pod, enabling the development of our models.
    Fredrik Lindsten and Joel Oskarsson acknowledge funding from the Swedish Research Council (project no: 2020-04122, 2024-05011),
the Wallenberg AI, Autonomous Systems and Software Program (WASP) funded by the Knut and Alice Wallenberg Foundation,
and
the Excellence Center at Linköping--Lund in Information Technology (ELLIIT).
Tomas Landelius work was funded by the JPP ERA-Net SES Focus Initiative on Digitalisation of the Energy System (“EnerDigit”). This initiative has received funding from the European Union's Horizon 2020 research and innovation programme under grant agreements no. 646039, 775970 and 883973.
Sebastian Schemm acknowledges funding from the European Research Council (ERC) project grant 848698.

    \bibliography{references}

    \newpage
    \appendix
    \section{Hyperparameter details}
\label{sec:hyperparameter_details}
In all models we use 4 processor layers.
Following the convention in \cite{graph_efm}, we define one layer of the hierarchical processor as one complete pass through the mesh graph, either from the bottom level up or from the top level down.
We train the models using the AdamW optimizer \citep{adamw} with a learning rate of 0.001 during the single-step pre-training.
For the \gls{DANRA} models we use learning rate 0.0001 for fine-tuning on $\fclen=4$ step rollouts.
We observed higher learning rates leading to worse convergence.
For the \gls{COSMO} models we use the higher 0.001 learning rate also during fine-tuning, which did not show any convergence problems.
The \gls{COSMO} fine-tuning was however impacted by an implementation issue where a learning rate of 0.0001 was used for one epoch, before switching to 0.001.
This behavior was the same for all models and we believe this 1 epoch of lower learning rate to have had neglectable impact on the training process.
For the additional fine-tuning of the final \gls{COSMO} model, described in \cref{sec:extra_finetuning} and \cref{fig:cosmo_finetuning_strategy_experiment}, we diverge from this setup slightly and use a lower learning rate of 0.0001 also for the \gls{COSMO} model.

\section{Graph details}
\label{sec:graph_details}
We here give some further details on the graph creation processes and give more in-depth descriptions of the graphs used in experiments in \cref{sec:graph_construction}.

\paragraph{Triangular graph creation details}
Triangular graphs have some heterogeneity in their connectivity around the corners of the original global icosahedron. 
To avoid this impacting the \gls{LAM} models we rotate the icosahedron to line up the center of one of its faces with the center of the interior region.
This generally means that the entire interior fits within one of the triangles making up the icosahedron.
For all our triangular graphs the icosahedron goes through 9 steps of triangle splitting, resulting in edges of length $\approx \SI{16.5}{\kilo\metre}$ in the finest mesh.

\paragraph{Static graph features}
Together with the graph structure itself, the graphs also come with static node and edge features.
In the models these features are embedded using \glspl{MLP} to give the initial node and edge representations that the processor then updates.
Our specific graph features are based on previous work \citep{graphcast,neural_lam} and represent relative positions and distances.
For nodes these features are their coordinates.
Edge features contain the vector difference between the receiver and sender nodes, as well as the length of the edge.
These features are all normalized by dividing by the length of the longest edge. 
Graph features can either be computed in the 2-dimensional \gls{CRS} of the regional data or in a 3-dimensional coordinate system.
For the rectangular graphs, which are constructed in the 2-dimensional \gls{CRS}, we use this coordinate system also for the graph features.  
Global triangular graphs have earlier been used with graph features computed from 3-dimensional cartesian coordinates \citep{graphcast,graph_efm}.
However, when using such features for the \gls{DANRA} model with triangular graphs we experienced issues during training.
With the 3-dimensional graph features the model did not seem able to make use of the graph, likely failing to capture the spatial relationships correctly.
We thus use the 2-dimensional graph features also for the \gls{DANRA} triangular graphs.
In the \gls{COSMO} setting we did not observe this issue, and the model was able to learn graph relationships equally well using the 2- or 3-dimensional features.
Results from the triangular model in \cref{fig:cosmo_graph_design_experiment} are with the 3-dimensional features.

\paragraph{Grid-mesh connectivity radius}
For constructing the $\gtmedges$ edges a radius parameter has to be specified, determining which grid points and mesh nodes to connect.
This should be selected such that all grid points have some connection to the mesh graph, and each mesh node some connection to the grid.
The radius is set separately when connecting grid points in the interior and boundary regions.
In the interior we use a radius of $0.51 \edgelen$ for the rectangular graphs and $0.6 \edgelen$ for the triangular ones.
The constant $\edgelen$ is here the length of the longest edge in the finest level of the mesh graph. %
These values narrowly guarantee that all grid points get connected to the graph.
Note that due to the different geometries of the rectangular and triangular graphs it is motivated to set these to somewhat different values.
For the boundary region, with substantially fewer grid points, larger values are used.
We here set the radius to $1.6 \edgelen$ for the rectangular graphs and $1.85 \edgelen$ for triangular.

\paragraph{Details of graphs used in experiments}
In \cref{tab:danra_graph_details,tab:cosmo_graph_details} we list the exact value of $\edgelen$ and the number of nodes and edges in each graph configuration.
\Cref{tab:danra_graph_details} features details for all graphs used for the \gls{DANRA} setting in \cref{sec:exp_graph_construction}.
\Cref{tab:cosmo_graph_details} contain details for the \gls{COSMO} graphs compared in \cref{fig:cosmo_graph_design_experiment}.
Note that for the rectangular graphs there are small differences in the value of $\edgelen$ due to exactly how the mesh nodes are laid out.
When constructing rectangular graphs, we set a minimum mesh node distance (diagonal edges of length \SI{17.7}{km}). We then try to find a node spacing that allows for creating all graph levels and covers the full area, while staying as close to the minimum distance as possible.
The graphs illustrated in \cref{fig:rect_graphs,fig:tri_graphs} are not the exact ones used in experiments, but only created for visualization purposes. 
The exact graphs described in \cref{tab:danra_graph_details,tab:cosmo_graph_details} have too many nodes and edges to allow for clear illustrations in figures. 

\ifpreprint
    \begin{table}[tbp]
\centering
\caption{
Details about the \gls{COSMO} graphs used in our graph design study (\cref{fig:cosmo_graph_design_experiment}).
}
\label{tab:cosmo_graph_details}
\begin{tabular}{@{}llS[table-format=7.0]S[table-format=7.0]@{}}
\toprule
 &  & \multicolumn{1}{c}{\textbf{Rectangular}} & \multicolumn{1}{c}{\textbf{Triangular}} \\ 
 \cmidrule(lr){3-3} \cmidrule(lr){4-4}
 &  & \multicolumn{1}{c}{\textbf{Hierarchical}} & \multicolumn{1}{c}{\textbf{Hierarchical}} \\ 
 \cmidrule(lr){3-3} \cmidrule(lr){4-4}
 & \textbf{Levels} & 3 & 3 \\ \midrule
 & $\edgelen$ & {\SI{16.1}{\kilo\metre}} & {\SI{16.5}{\kilo\metre}} \\ \midrule
\multirow{3}{*}{\textbf{Mesh nodes}} & Level 1 & 50075 & 35426 \\
 & Level 2 & 5569 & 8858 \\
 & Level 3 & 624 & 2210 \\ \midrule
 & \textbf{Total nodes} & 56268 & 46494 \\ \midrule
\multirow{3}{*}{\textbf{Intra-level edges}} & Level 1 & 398126 & 210528 \\
 & Level 2 & 43726 & 52137 \\
 & Level 3 & 5569 & 12756 \\ \midrule
\multirow{2}{*}{\textbf{Up edges}} & Level 1 $\rightarrow$ 2 & 50075 & 61661 \\
 & Level 2 $\rightarrow$ 3 & 5569 & 15319 \\ \midrule
\multirow{2}{*}{\textbf{Down edges}} & Level 2 $\rightarrow$ 1 & 50075 & 61661 \\
 & Level 3 $\rightarrow$ 2 & 5569 & 15319 \\ \midrule
\multirow{2}{*}{\textbf{\begin{tabular}[c]{@{}l@{}}Grid-mesh\\ edges\end{tabular}}} & $\setsize{\gtmedges}$ & 535184 & 518116 \\
 & $\setsize{\mtgedges}$\rule{0pt}{.4cm} & 907920 & 680940 \\ \midrule
 & \textbf{Total edges} & 2057457 & 1705417 \\ \bottomrule
\end{tabular}
\end{table}

    \begin{landscape}
\begin{table}[p]
\centering
\caption{
Details about the \gls{DANRA} graphs used in our graph design study (\cref{sec:exp_graph_construction}).
Note that in multi-scale (M.S.) graphs we collapse all initial levels into a single graph with more edges.
We list the multi-scale graphs on the same row as level 1 of hierarchical graphs.
Multi-scale graphs do not have up or down edges.
}
\label{tab:danra_graph_details}
\begin{tabular}{@{}llS[table-format=7.0]S[table-format=7.0]S[table-format=7.0]S[table-format=7.0]S[table-format=7.0]S[table-format=7.0]S[table-format=7.0]S[table-format=7.0]@{}}
\toprule
 &  & \multicolumn{5}{c}{\textbf{Rectangular}} & \multicolumn{3}{c}{\textbf{Triangular}} \\ 
 \cmidrule(lr){3-7} \cmidrule(lr){8-10}
 &  & \multicolumn{3}{c}{\textbf{Hierarchical}} & \multicolumn{2}{c}{\textbf{Multi-scale}} & \multicolumn{2}{c}{\textbf{Hierarchical}} & \multicolumn{1}{c}{\textbf{Multi-scale}} \\ 
 \cmidrule(lr){3-5} \cmidrule(lr){6-7} \cmidrule(lr){8-9} \cmidrule(lr){10-10}
 & \textbf{Levels} & 2 & 3 & 4 & 3 & 4 & 3 & 4 & 3 \\ \midrule
 & $\edgelen$ & {\SI{17.8}{\kilo\metre}} & {\SI{18.1}{\kilo\metre}} & {\SI{18.4}{\kilo\metre}}& {\SI{18.1}{\kilo\metre}}& {\SI{18.4}{\kilo\metre}} & {\SI{16.5}{\kilo\metre}} & {\SI{16.5}{\kilo\metre}} & {\SI{16.5}{\kilo\metre}} \\ \midrule
\multirow{4}{*}{\textbf{Mesh nodes}} & Level 1 / M.S. & 65891 & 63721 & 61670 & 63721 & 61670 & 49509 & 49509 & 49509 \\
 & Level 2 & 7332 & 7093 & 6866 &  &  & 12379 & 12379 &  \\
 & Level 3 &  & 789 & 764 &  &  & 3094 & 3094 &  \\
 & Level 4 &  &  & 86 &  &  &  & 773 &  \\ \midrule
 & \textbf{Total nodes} & 73223 & 71603 & 69386 & 63721 & 61670 & 64982 & 65755 & 49509 \\ \midrule
\multirow{4}{*}{\textbf{Intra-level edges}} & Level 1 / M.S. & 524252 & 506940 & 490580 & 568742 & 550974 & 294519 & 294519 & 385464 \\
 & Level 2 & 57696 & 55802 & 54002 &  &  & 73011 & 73011 &  \\
 & Level 3 &  & 6000 & 5806 &  &  & 17934 & 17934 &  \\
 & Level 4 &  &  & 586 &  &  &  & 4323 &  \\ \midrule
\multirow{3}{*}{\textbf{Up edges}} & Level 1 $\rightarrow$ 2 & 65891 & 63721 & 61670 &  &  & 86224 & 86224 &  \\
 & Level 2 $\rightarrow$ 3 &  & 7093 & 6866 &  &  & 21450 & 21450 &  \\
 & Level 3 $\rightarrow$ 4 &  &  & 764 &  &  &  & 5307 &  \\ \midrule
\multirow{3}{*}{\textbf{Down edges}} & Level 2 $\rightarrow$ 1 & 65891 & 63721 & 61670 &  &  & 86224 & 86224 &  \\
 & Level 3 $\rightarrow$ 2 &  & 7093 & 6866 &  &  & 21450 & 21450 &  \\
 & Level 4 $\rightarrow$ 3 &  &  & 764 &  &  &  & 5307 &  \\ \midrule
\multirow{2}{*}{\textbf{\begin{tabular}[c]{@{}l@{}}Grid-mesh\\ edges\end{tabular}}} & $\setsize{\gtmedges}$ & 1046067 & 1045868 & 1046597 & 1045868 & 1046597 & 885886 & 885886 & 885886 \\
 & $\setsize{\mtgedges}$\rule{0pt}{.4cm} & 1858884 & 1858884 & 1858884 & 1858884 & 1858884 & 1394163 & 1394163 & 1394163 \\ \midrule
 & \textbf{Total edges} & 3684572 & 3685936 & 3664355 & 3473494 & 3456455 & 2988535 & 3008779 & 2665513 \\ \bottomrule
\end{tabular}
\end{table}
\end{landscape}

\else
    
    \begin{table}[tbp]
\centering
\caption{
Details about the \gls{COSMO} graphs used in our graph design study (\cref{fig:cosmo_graph_design_experiment}).
}
\label{tab:cosmo_graph_details}
\begin{tabular}{@{}llS[table-format=7.0]S[table-format=7.0]@{}}
\toprule
 &  & \multicolumn{1}{c}{\textbf{Rectangular}} & \multicolumn{1}{c}{\textbf{Triangular}} \\ 
 \cmidrule(lr){3-3} \cmidrule(lr){4-4}
 &  & \multicolumn{1}{c}{\textbf{Hierarchical}} & \multicolumn{1}{c}{\textbf{Hierarchical}} \\ 
 \cmidrule(lr){3-3} \cmidrule(lr){4-4}
 & \textbf{Levels} & 3 & 3 \\ \midrule
 & $\edgelen$ & {\SI{16.1}{\kilo\metre}} & {\SI{16.5}{\kilo\metre}} \\ \midrule
\multirow{3}{*}{\textbf{Mesh nodes}} & Level 1 & 50075 & 35426 \\
 & Level 2 & 5569 & 8858 \\
 & Level 3 & 624 & 2210 \\ \midrule
 & \textbf{Total nodes} & 56268 & 46494 \\ \midrule
\multirow{3}{*}{\textbf{Intra-level edges}} & Level 1 & 398126 & 210528 \\
 & Level 2 & 43726 & 52137 \\
 & Level 3 & 5569 & 12756 \\ \midrule
\multirow{2}{*}{\textbf{Up edges}} & Level 1 $\rightarrow$ 2 & 50075 & 61661 \\
 & Level 2 $\rightarrow$ 3 & 5569 & 15319 \\ \midrule
\multirow{2}{*}{\textbf{Down edges}} & Level 2 $\rightarrow$ 1 & 50075 & 61661 \\
 & Level 3 $\rightarrow$ 2 & 5569 & 15319 \\ \midrule
\multirow{2}{*}{\textbf{\begin{tabular}[c]{@{}l@{}}Grid-mesh\\ edges\end{tabular}}} & $\setsize{\gtmedges}$ & 535184 & 518116 \\
 & $\setsize{\mtgedges}$\rule{0pt}{.4cm} & 907920 & 680940 \\ \midrule
 & \textbf{Total edges} & 2057457 & 1705417 \\ \bottomrule
\end{tabular}
\end{table}

\fi

\section{Additional results}
In this appendix we include some additional results from our experiments.

\subsection{Boundary width}
\label{sec:extra_boundary_width}
\Cref{fig:danra_boundary_width_experiment} shows results from models on the \gls{DANRA} domain, using boundary regions of different width. 
As for the \gls{COSMO} case, we see no significant difference between the thinner or wider boundary regions.

\danraplots{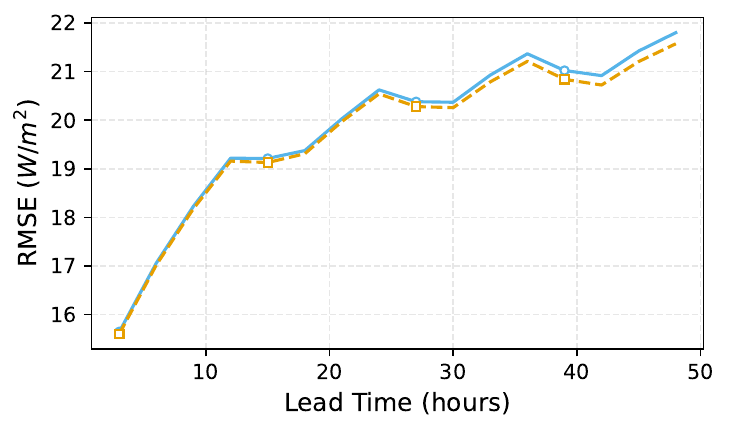}{danra_boundary_width_experiment}{
    \gls{RMSE} on validation set for \gls{DANRA} models with 400 and \SI{800}{\kilo\metre} boundary regions. The boundary forcing comes from \gls{ERA5}.
}

\subsection{Boundary forcing time}
\label{sec:extra_boundary_handling}
In \cref{fig:danra_boundary_handling_experiment} we show the same investigation around boundary forcing time for \gls{DANRA} as presented for \gls{COSMO} in \cref{fig:cosmo_boundary_handling_experiment}.
The overall results show that there is a benefit to including future boundary forcing, but the addition of an explicit time embedding does not impact the results significantly.
This aligns with the conclusions for the \gls{COSMO} setup.

\danraplots{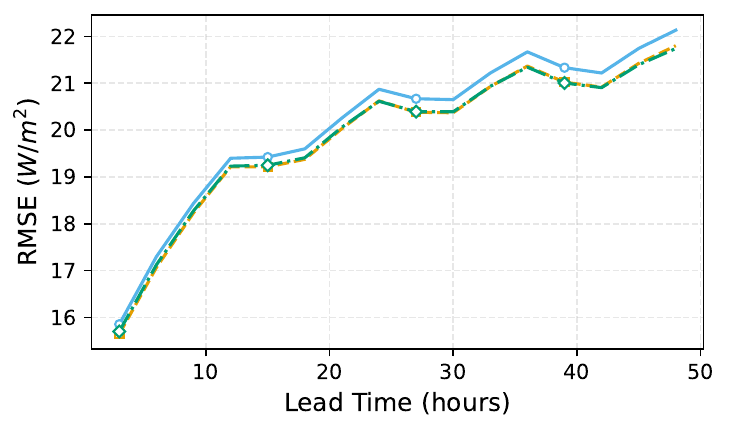}{danra_boundary_handling_experiment}{
    \gls{RMSE} on validation set for \gls{DANRA} models with different configurations for the boundary forcing.
    For the model with future boundary (our standard model), $\bforcing_{k' + 1}$ is included in its input, whereas for the model with no future boundary it is not.
    All boundary forcing here comes from \gls{ERA5}.
}

\subsection{Multi-step fine-tuning}
\label{sec:extra_finetuning}
Previous works have found it crucial to fine-tune deterministic \gls{MLWP} models by rolling out forecasts to $\fclen > 1$.
We show here that also in the \gls{LAM} setting this has a major impact on model performance through reducing error accumulation.
The focus is here mainly on the \gls{COSMO} domain, as fine-tuning is more important for this setting due to the short \SI{1}{\hour} time steps.

\paragraph{Importance of fine-tuning}
In \cref{fig:cosmo_no_finetune_experiment} we compare our standard setup model before and after the $\fclen = 4$ step fine-tuning.
Across all variables fine-tuning the model improves performance. Calculating the loss across multiple forecast steps allows the model to learn the trajectory of the atmospheric state. This is particularly important for variables with strong temporal dependencies, such as precipitation and wind fields.

\cosmoplots{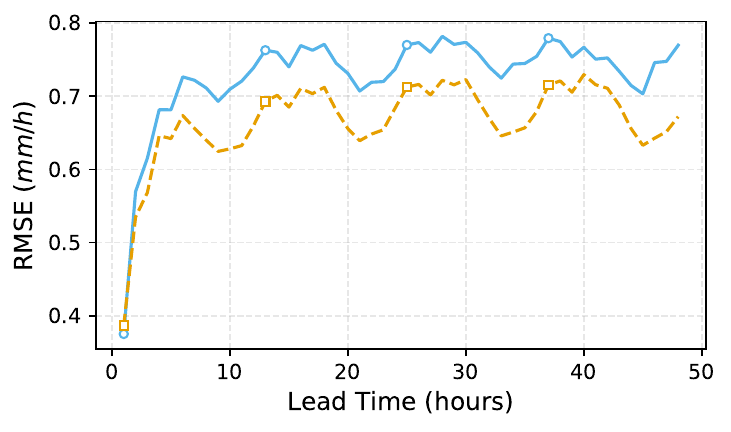}{cosmo_no_finetune_experiment}{
    \gls{RMSE} on validation set for \gls{COSMO} models with and without multi-step fine-tuning using $\fclen = 4$ autoregressive steps. 
}

\paragraph{Fine-tuning the best model}
Based on the results from our design studies the best models were chosen for more extensive verification.
For the \gls{COSMO} model we noted that when considering longer lead times (past the \SI{48}{\hour} considered in the design studies) the errors from many shorter autoregressive steps started significantly adding up.
To improve its performance on longer lead times we investigate fine-tuning this final model further, also rolling out to longer $\fclen$.

\cosmoplots{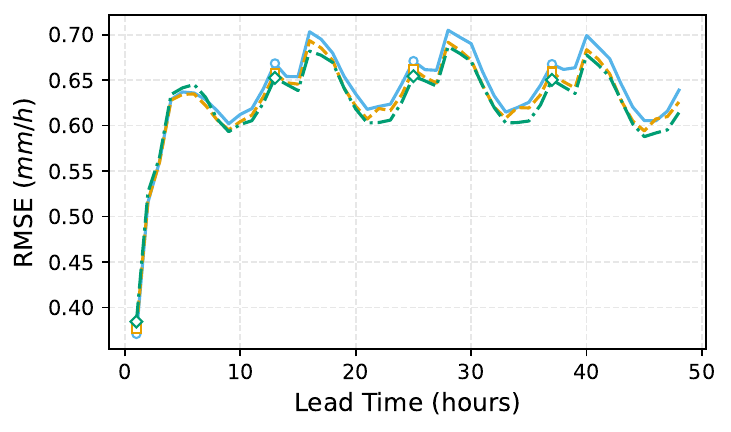}{cosmo_finetuning_strategy_experiment}{
    \gls{RMSE} on validation set for the best \gls{COSMO} model with additional fine-tuning.
    We consider further fine-tuning with $\fclen \in \set{4, 6, 12}$ steps.
    The boundary forcing comes from \gls{ERA5}.
}

In \cref{fig:cosmo_finetuning_strategy_experiment} we start from the previously best \gls{COSMO} model and then try fine-tuning it using either $\fclen=4$, 6 or 12 unrolled steps.
We note that unrolling for a higher number of autoregressive steps generally improves the model performance. With the hourly time steps in the  \gls{COSMO} setup, a 12-step rollout is beneficial for all variables. This could be related to the many variables affected by a diurnal cycle. 
The number of steps does however also affect the effective model resolution, as discussed in \citet{Selz_2025}.
The model error does not seem to suffer during the first few hours of forecasts when trained using a 12-step rollout. 
Therefore this represents the best additional fine-tuning configuration for the \gls{COSMO} setting, and is the model used for further verification.
This 12-step training does however have a stronger smoothing effect, giving a lower effective resolution, which will be observable during verification.
It should also be noted that the computational cost increases drastically from training with more steps.
While our use of gradient checkpointing means that the \gls{GPU} memory required is constant, the computational time scales approximately linearly with the choice of $\fclen$.

\subsection{Graph design}
\label{sec:extra_graph_design}
\cosmoplots{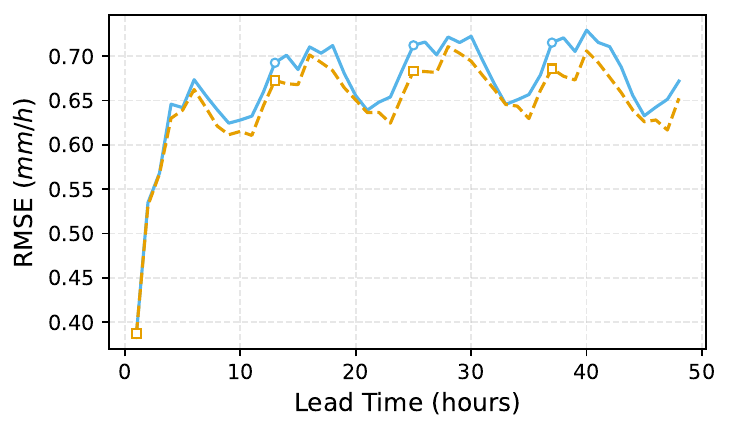}{cosmo_graph_design_experiment}{
    \gls{RMSE} on validation set for \gls{COSMO} models with triangular and rectangular graphs.
    The boundary forcing from both models comes from \gls{ERA5}.
}

For the \gls{COSMO} setting we carry out a smaller comparison of the different graph configurations.
Based on the \gls{DANRA} results we limit this experiments to 3 level hierarchical mesh graphs.
\Cref{fig:cosmo_graph_design_experiment} shows a comparison of models trained using rectangular and triangular graphs on the \gls{COSMO} data.
While the two configurations largely show similar performance, for variables like wind some improved forecasting skill is observed for the triangular graph at longer lead times.
This can be related to better capturing orography, as the Alps make up a significant portion of the \gls{COSMO}-domain.
For many variables there are however no clear differences.

\subsection{Evaluation against gridded data}
\label{sec:extra_eval_gridded}

\begin{figure}[tbp]
    \centering
    \begin{subfigure}[b]{0.5\textwidth}
        \centering
        \includegraphics[width=\textwidth]{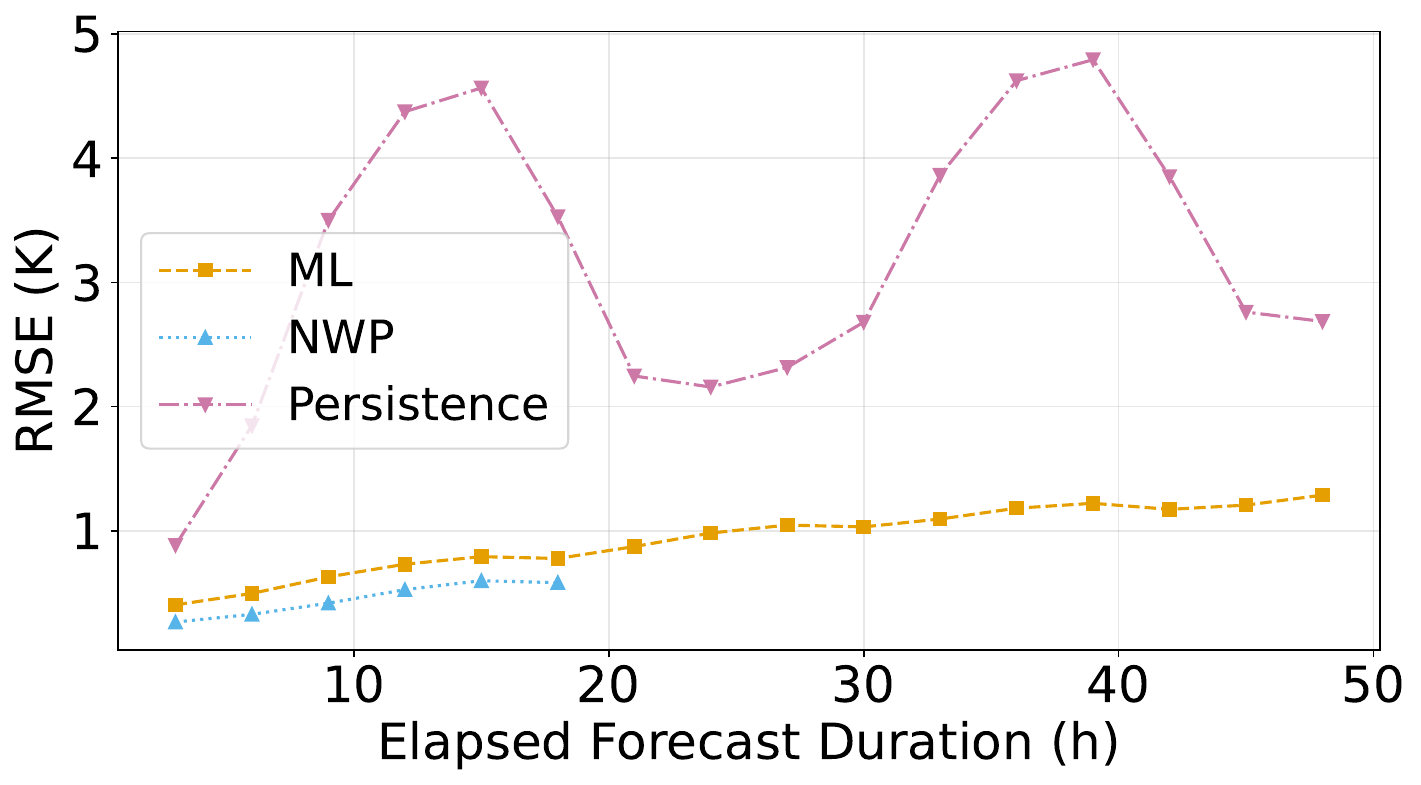}
        \caption{\SI{2}{m} temperature (\wvar{2t})}
    \end{subfigure}%
    \hfill%
    \begin{subfigure}[b]{0.5\textwidth}
        \centering
        \includegraphics[width=\textwidth]{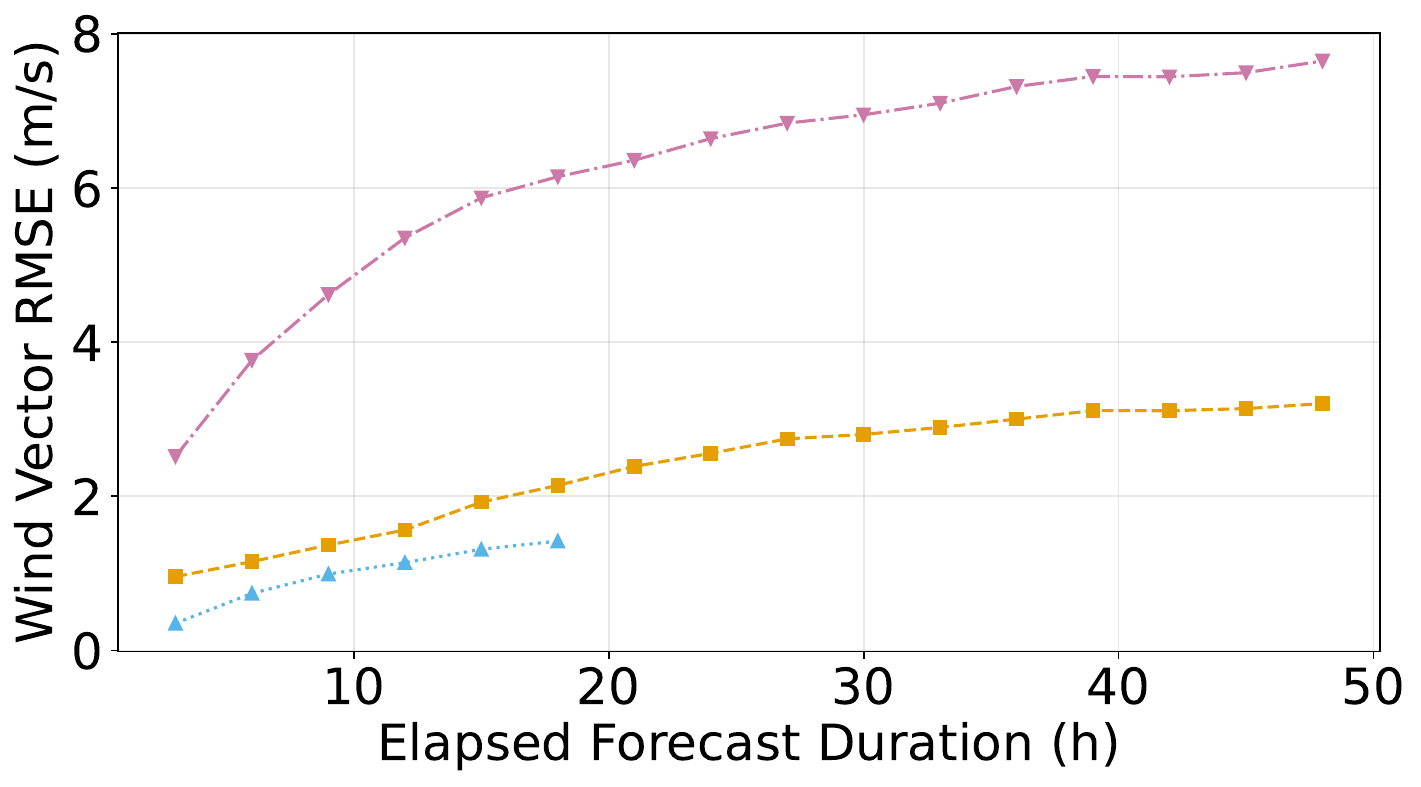}
        \caption{\SI{10}{m} wind}
    \end{subfigure}%
    \caption{
        \gls{RMSE} over Denmark, for \gls{DANRA} models compared to the gridded reanalysis.
    }
    \label{fig:danra_verif_gridded_rmse_denmark}
\end{figure}

\paragraph{\glspl{RMSE}}
To better understand the performance of the \gls{DANRA} \gls{ML} model, we compute \glspl{RMSE} against the gridded reanalysis also for a smaller region covering mainly Denmark.
This is a rectangular domain similar to what is shown in \cref{fig:danra_verif_sparse_error_map_temp}.
The \gls{RMSE} values are shown in \cref{fig:danra_verif_gridded_rmse_denmark}.
While the \gls{NWP} model still achieves lower errors, our \gls{ML} \gls{LAM} is highly competitive for forecasting of key surface variables over Denmark.

\paragraph{Mean error maps}
\begin{figure}[tbp]
    \centering
    \begin{subfigure}[b]{.5\textwidth}
        \centering
        \includegraphics[width=\textwidth]{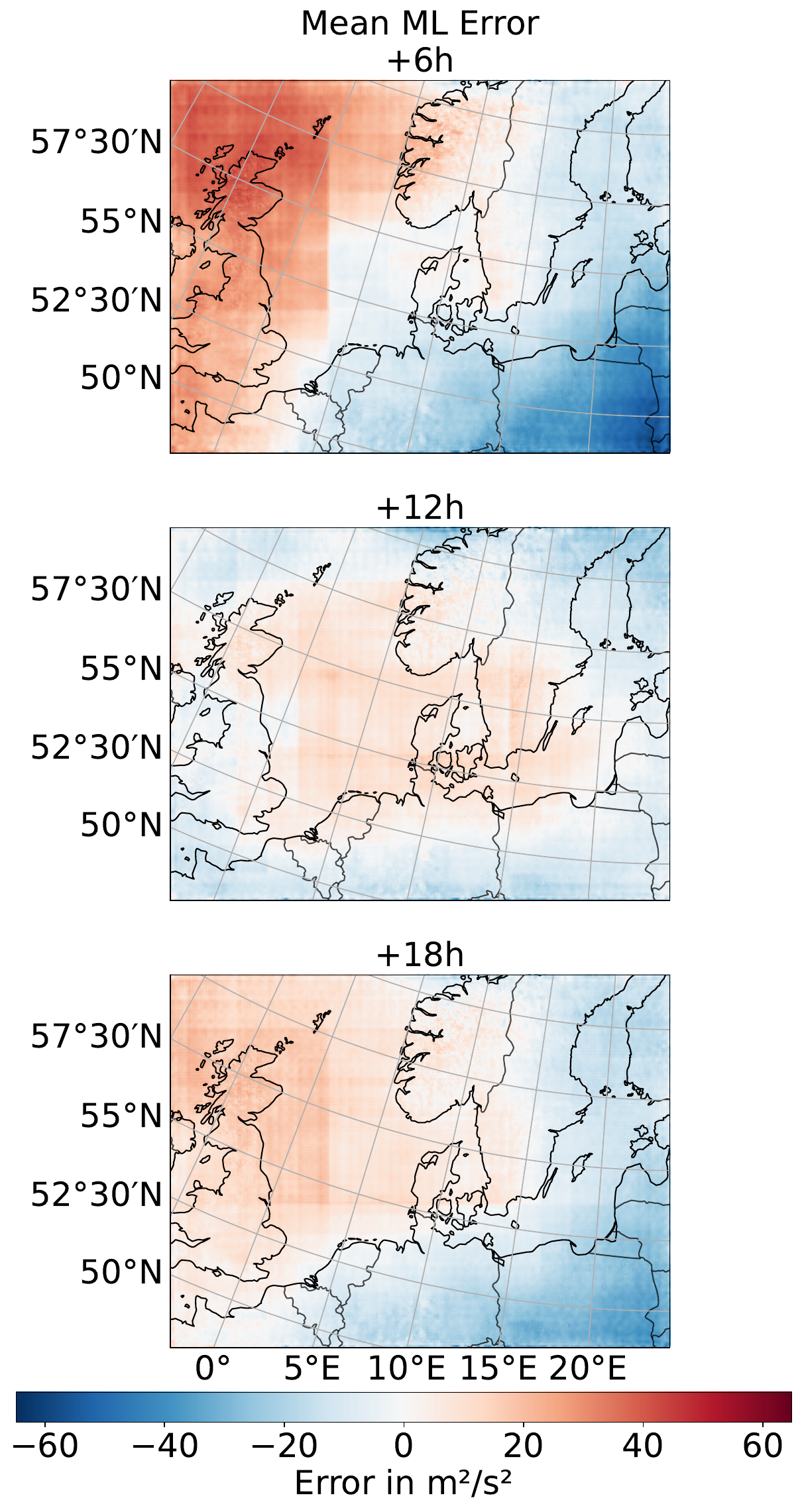}
        \caption{Geopotential at \SI{700}{\hecto\pascal} (\wvar{z700})}
        \label{fig:danra_verif_gridded_error_map_extra_geopotential}
    \end{subfigure}%
    \hfill%
    \begin{subfigure}[b]{.5\textwidth}
        \centering
        \includegraphics[width=\textwidth]{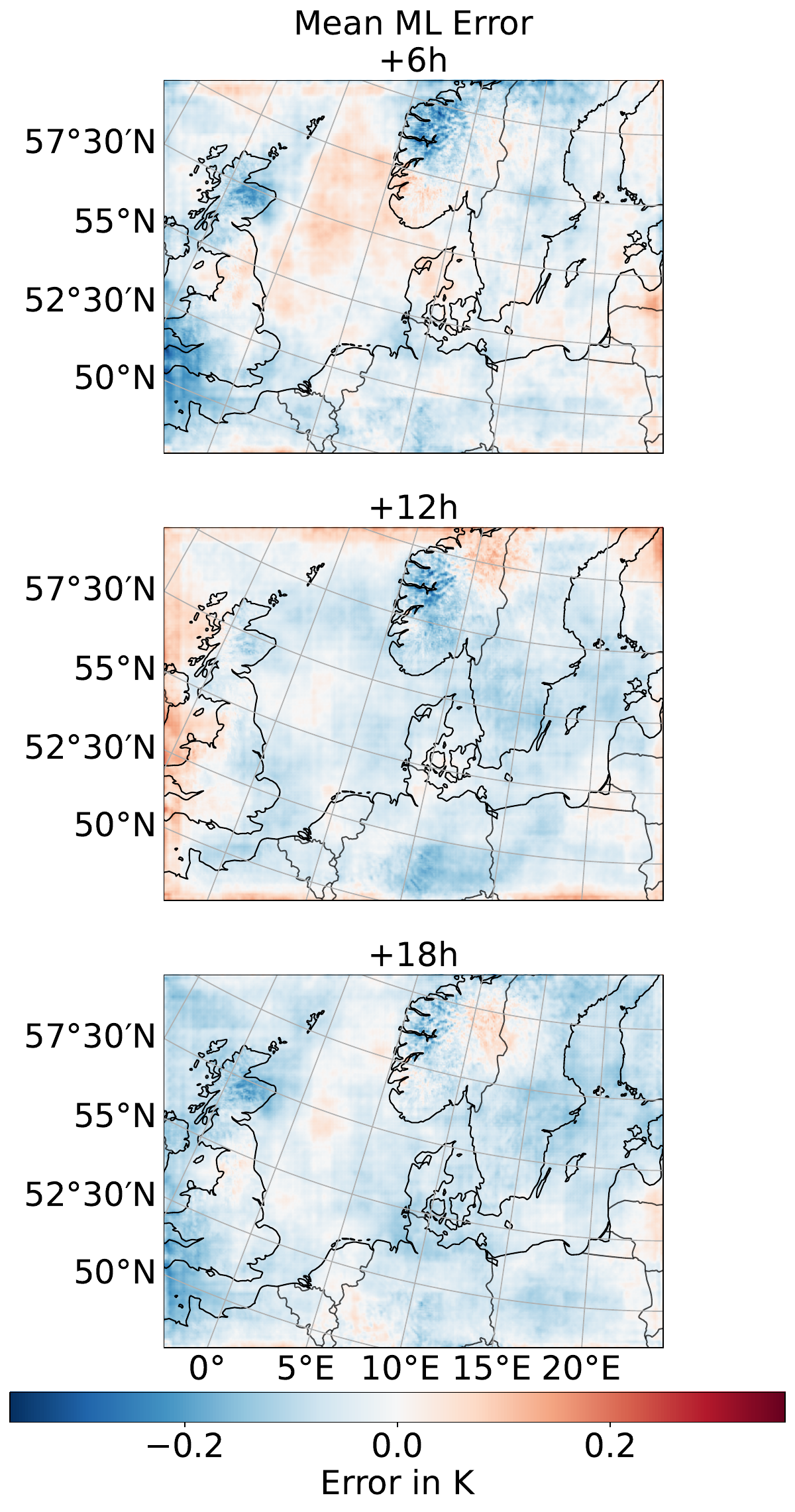}
        \caption{Temperature at \SI{700}{\hecto\pascal} (\wvar{t700})}
    \end{subfigure}%
    \caption{
        Mean error maps for variables at \SI{700}{\hecto\pascal}, for the \gls{DANRA} \gls{ML} model over the full test year period.
    }
    \label{fig:danra_verif_gridded_error_map_extra}
\end{figure}

In \cref{fig:danra_verif_gridded_error_map_extra} we show mean error maps of the \gls{DANRA} \gls{ML} model for a few vertical variables. 
The geopotential biases in \cref{fig:danra_verif_gridded_error_map_extra_geopotential} show some artifacts that can be traced back to the graph configuration used.  
Such artifacts are not clearly visible in the temperature bias, although some traces of similar patterns can be found.
In general these seem to mainly impact the smoother fields.

\paragraph{Energy spectra analysis}
To further understand the spectra of the \gls{COSMO} \gls{ML} model forecasts, we consider how the energy density evolves throughout the forecast.
\Cref{fig:cosmo_verif_gridded_wavenumber_evolution} shows the energy density of the \gls{ML} and \gls{NWP} model for three selected wavenumbers, representing different scales of the meteorological spectrum.
Note that the energy in the ground truth data does not change with lead time.
Looking at these specific wavenumbers, we note that the spectral distance between the \gls{ML} model and data is often increasing with lead time.
For some variables, such as wind and precipitation (\cref{fig:cosmo_verif_gridded_wavenumber_evolution_tp,fig:cosmo_verif_gridded_wavenumber_evolution_u10m}), we see a quick decline in energy followed by the model stabilizing at a lower energy level.

\begin{figure}[tbp]
    \centering
    \begin{subfigure}[b]{\textwidth}
        \centering
        \includegraphics[width=0.93\textwidth]{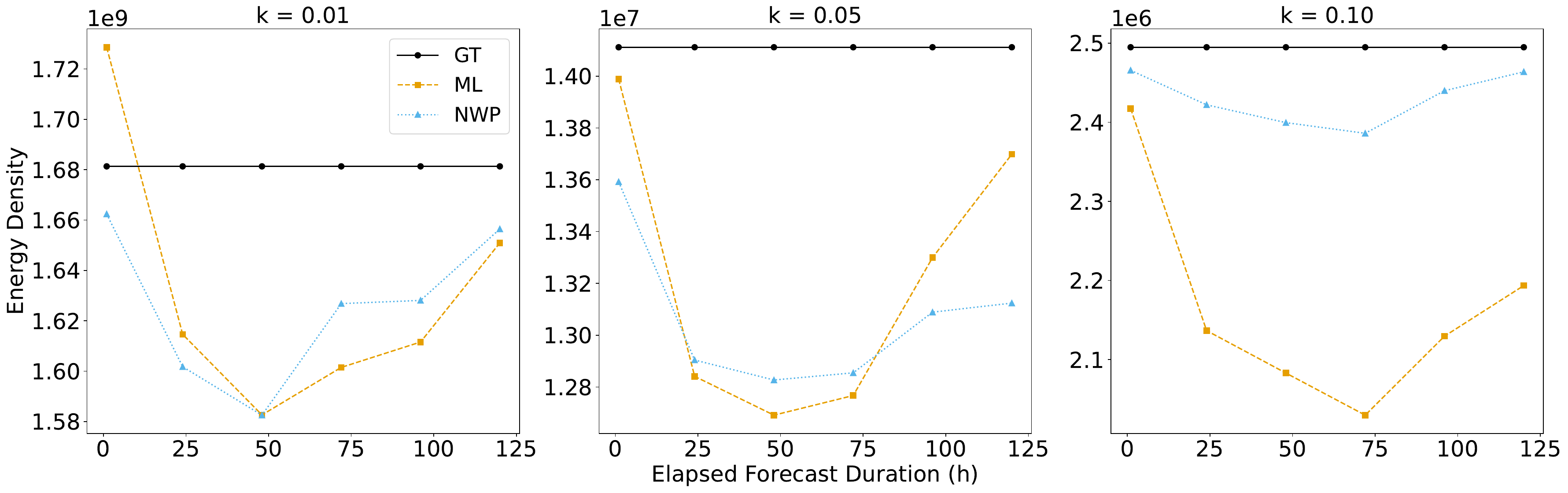}
        \caption{\SI{2}{m} temperature (\wvar{2t})}
    \end{subfigure}%
    \hfill%
    \begin{subfigure}[b]{\textwidth}
        \centering
        \includegraphics[width=0.93\textwidth]{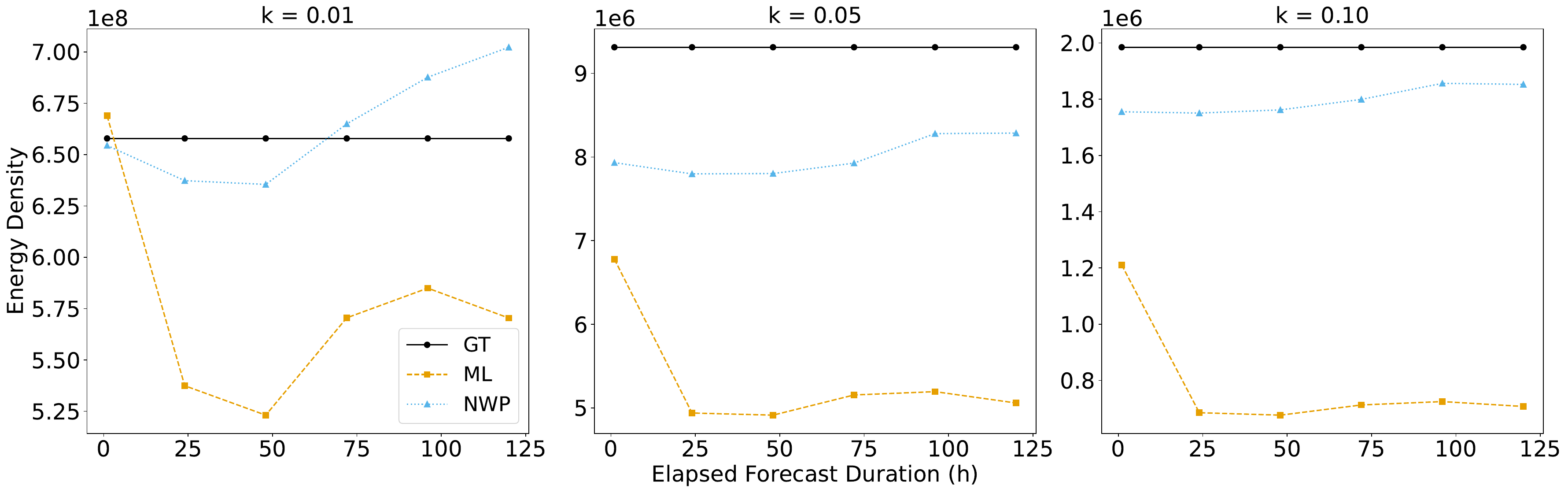}
        \caption{\SI{10}{m} wind-component \wvar{10u}}
        \label{fig:cosmo_verif_gridded_wavenumber_evolution_u10m}
    \end{subfigure}%
    \hfill%
    \begin{subfigure}[b]{\textwidth}
        \centering
        \includegraphics[width=0.93\textwidth]{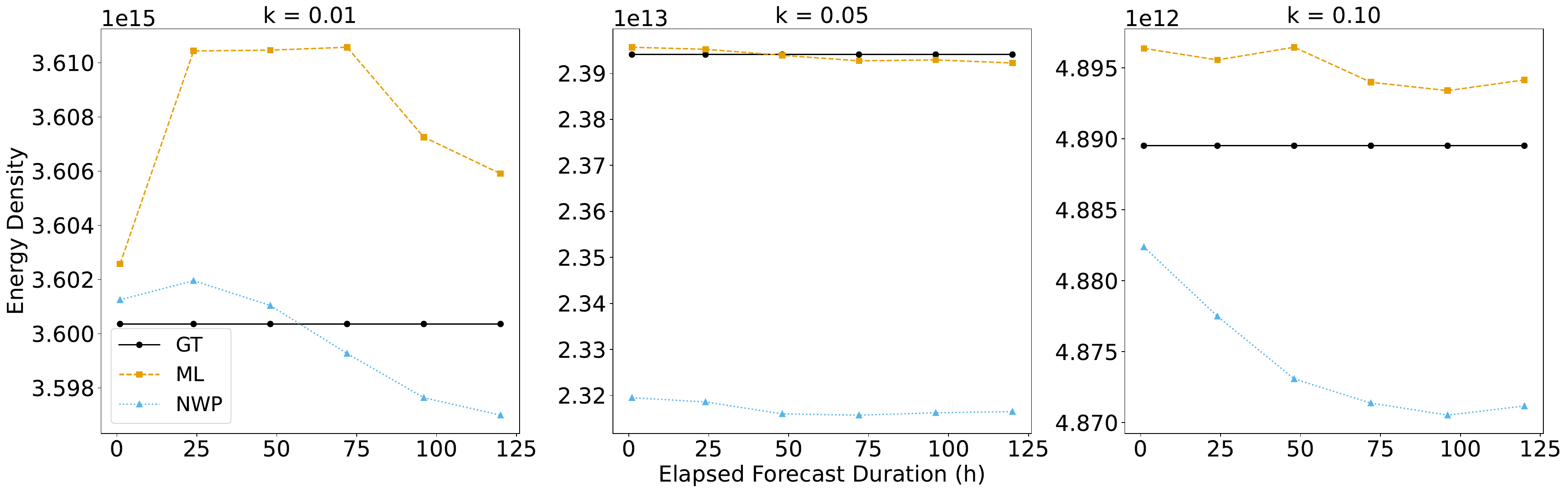}
        \caption{Surface pressure (\wvar{sp})}
    \end{subfigure}%
    \hfill%
    \begin{subfigure}[b]{\textwidth}
        \centering
        \includegraphics[width=0.93\textwidth]{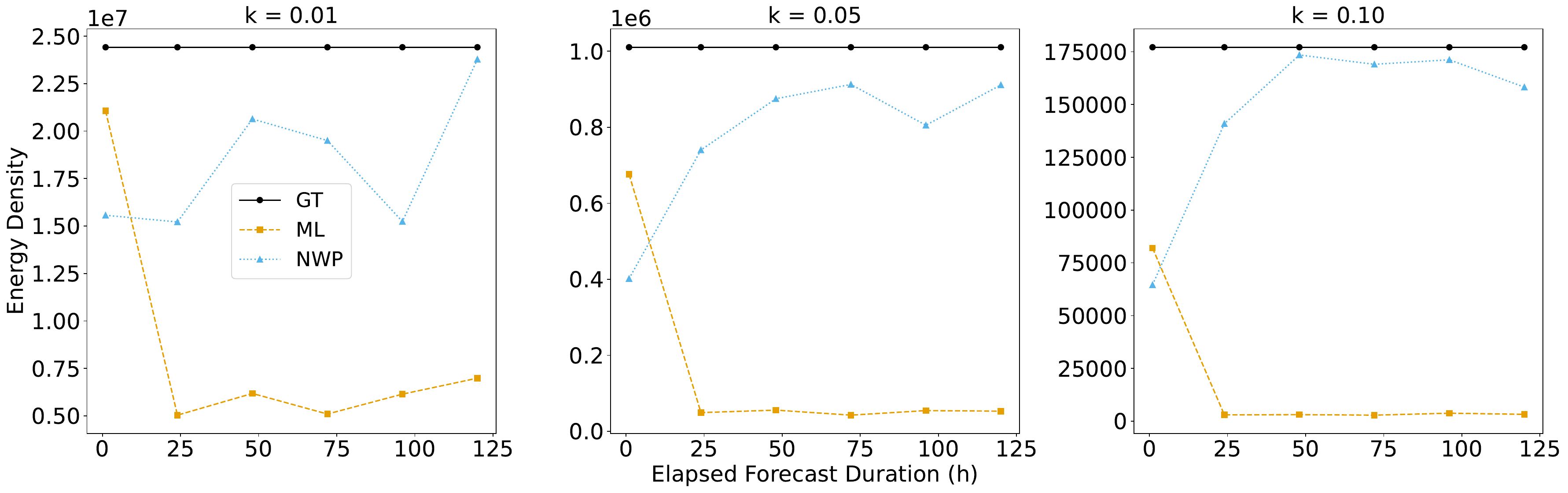}
        \caption{Precipitation (\wvar{tp01})}
        \label{fig:cosmo_verif_gridded_wavenumber_evolution_tp}
    \end{subfigure}%
    \caption{
        Evolution of energy spectral density for three characteristic wavenumbers along lead time for the \gls{COSMO} models.
    }
    \label{fig:cosmo_verif_gridded_wavenumber_evolution}
\end{figure}

For the \gls{DANRA} model we show spectra of vertical variables at pressure level \SI{700}{\hecto\pascal} in \cref{fig:danra_verif_gridded_energy_spectra_vertical}.
The spectra of the model and data generally aligns well for these smoother fields higher in the atmosphere.
However, in \cref{fig:danra_verif_gridded_energy_spectra_vertical_w} we see that the model predicts substantially less vertical movement.
This can also be observed in forecasts, where only some patterns of high vertical velocity are fully captured.

\begin{figure}[tbp]
    \centering
    \begin{subfigure}[b]{0.5\textwidth}
        \centering
        \includegraphics[width=\textwidth]{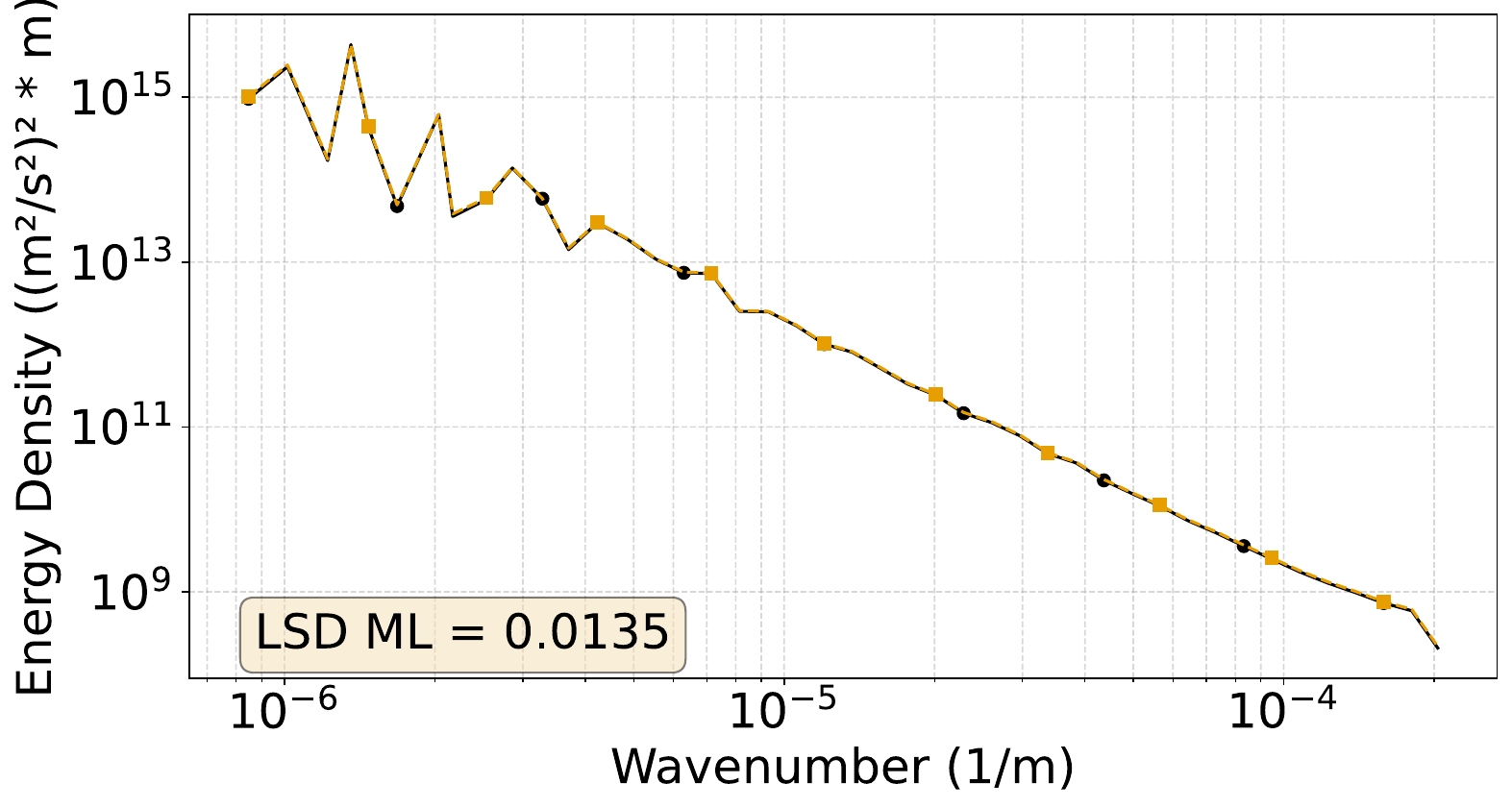}
        \caption{Geopotential at \SI{700}{\hecto\pascal} (\wvar{z700})}
    \end{subfigure}%
    \begin{subfigure}[b]{0.5\textwidth}
        \centering
        \includegraphics[width=\textwidth]{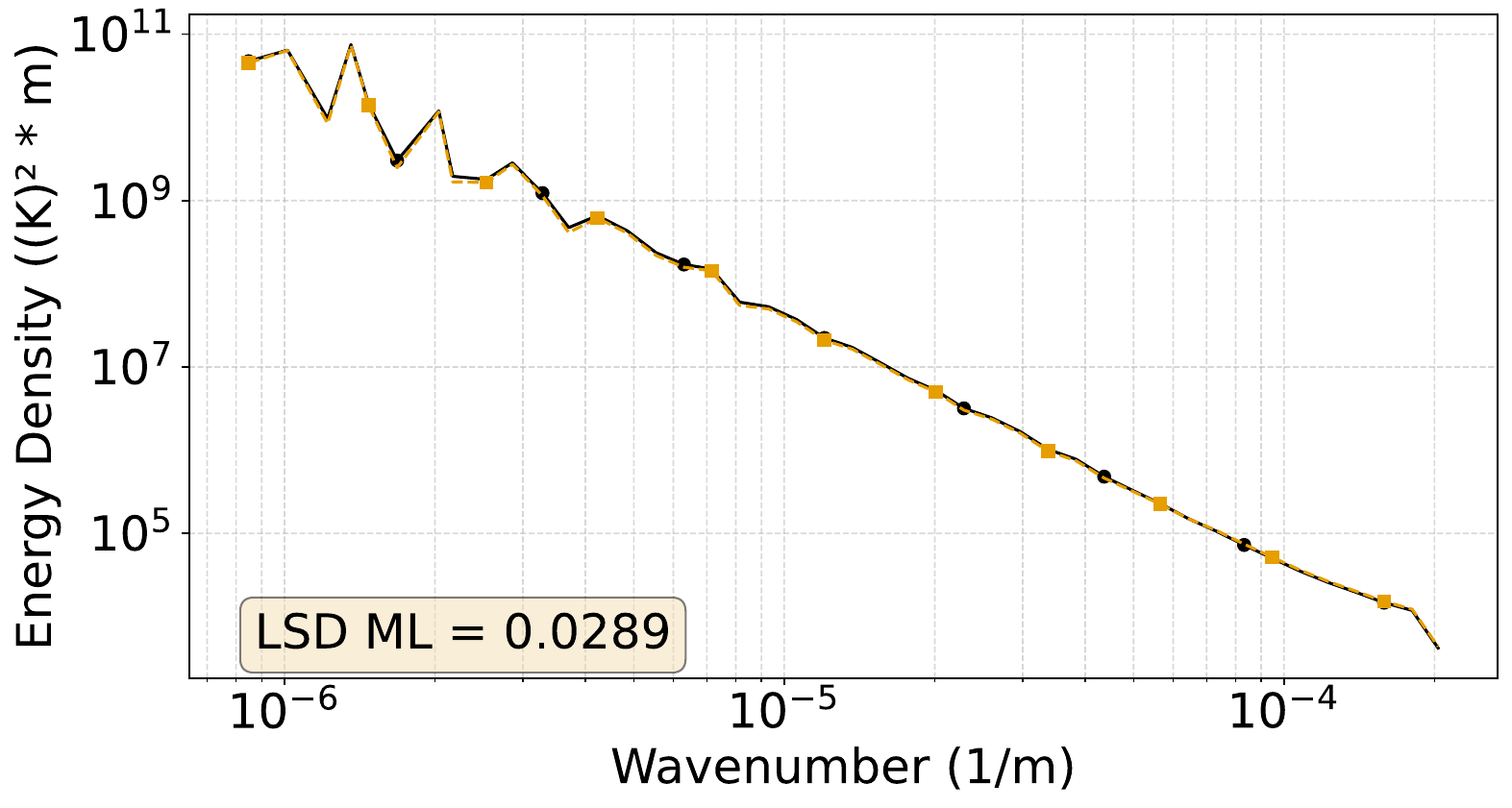}
        \caption{Temperature at \SI{700}{\hecto\pascal} (\wvar{t700})}
    \end{subfigure}%
    \hfill%
    \begin{subfigure}[b]{0.5\textwidth}
        \centering
        \includegraphics[width=\textwidth]{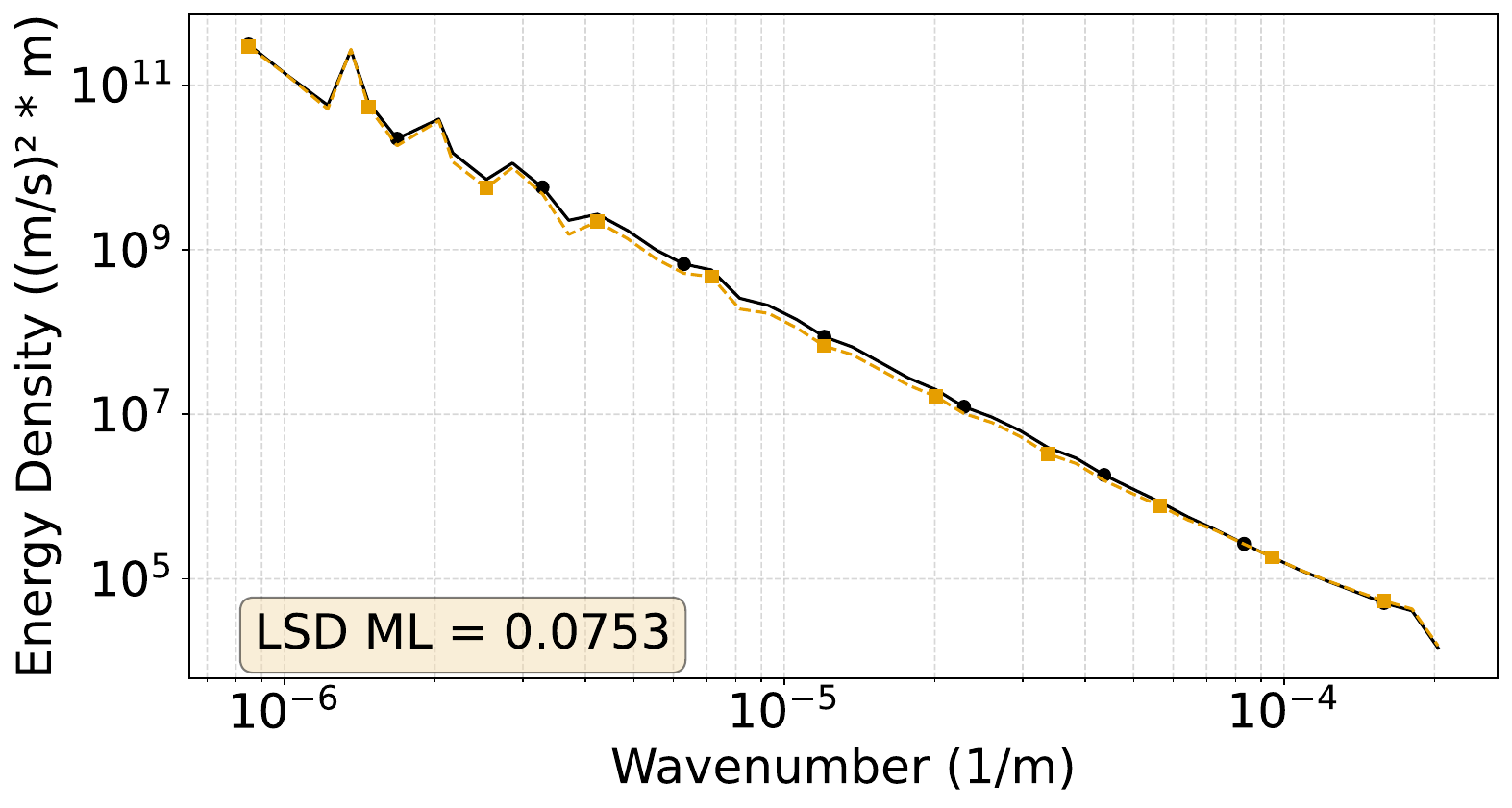}
        \caption{Wind u-component at \SI{700}{\hecto\pascal} (\wvar{u700})}
    \end{subfigure}%
    \begin{subfigure}[b]{0.5\textwidth}
        \centering
        \includegraphics[width=\textwidth]{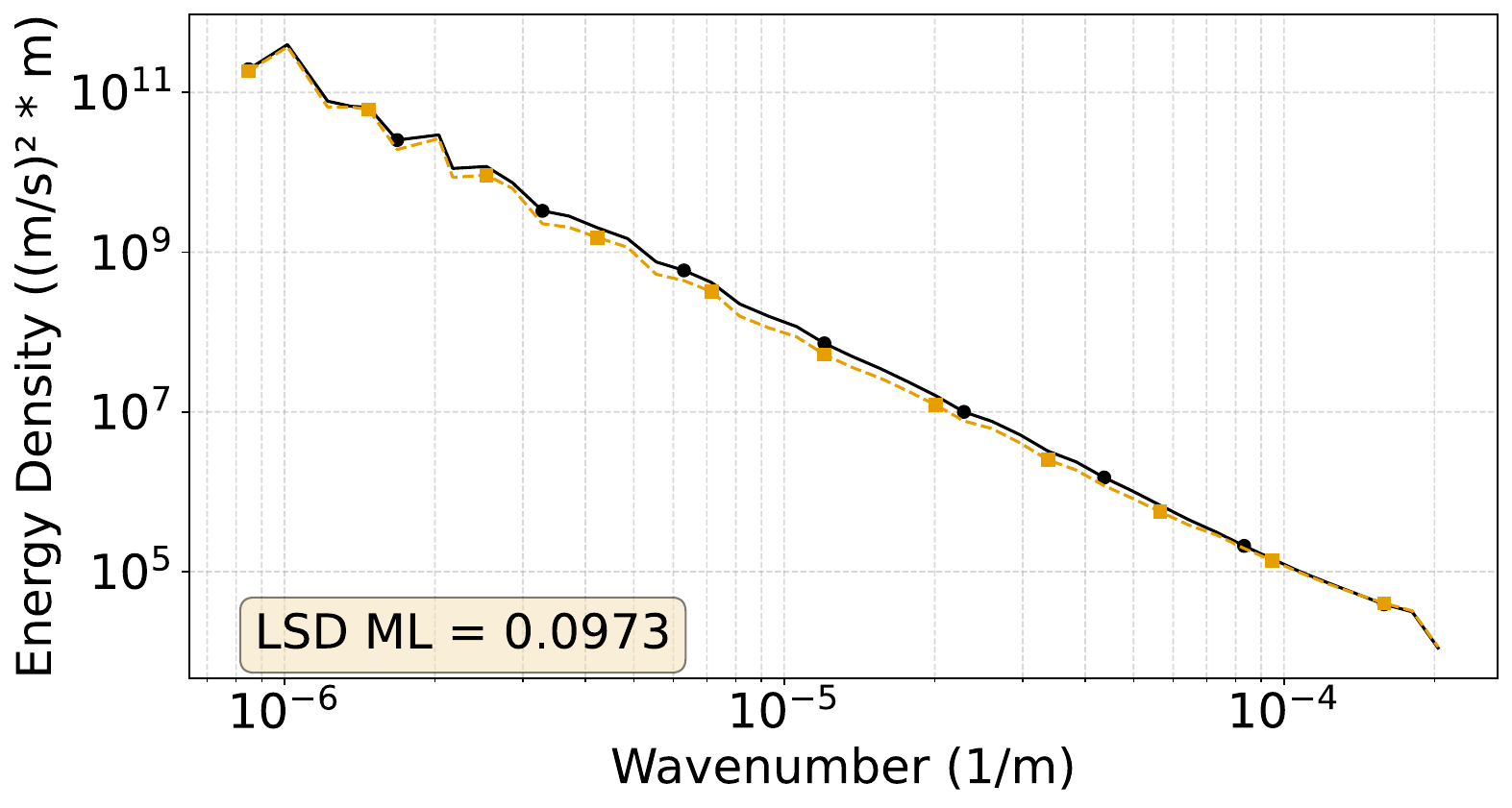}
        \caption{Wind v-component \SI{700}{\hecto\pascal} (\wvar{v700})}
    \end{subfigure}%
    \hfill%
    \begin{subfigure}[b]{0.5\textwidth}
        \centering
        \includegraphics[width=\textwidth]{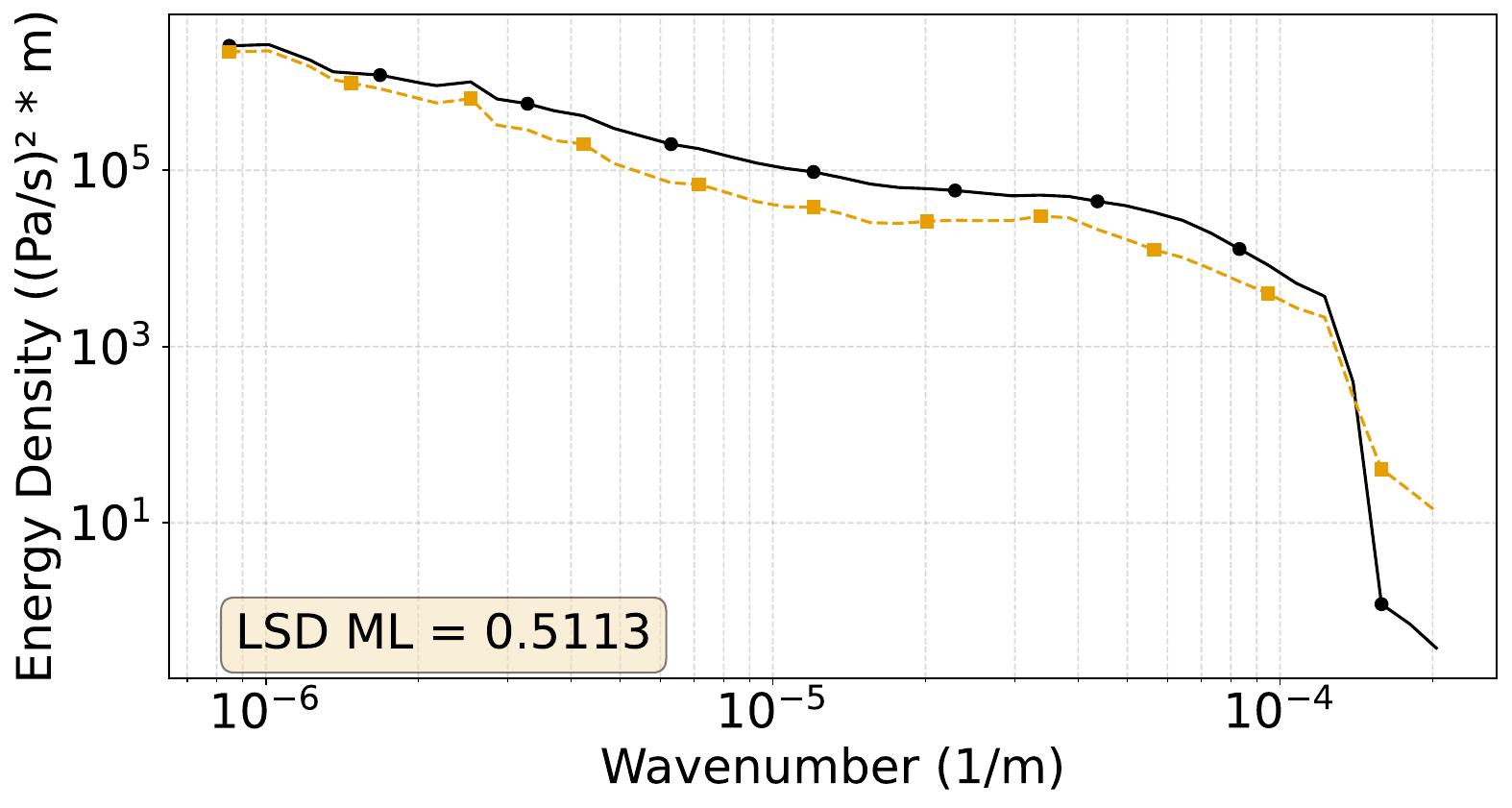}
        \caption{Vertical velocity at \SI{700}{\hecto\pascal} (\wvar{w700})}
    \label{fig:danra_verif_gridded_energy_spectra_vertical_w}
    \end{subfigure}%
    \begin{subfigure}[b]{0.5\textwidth}
        \centering
        \includegraphics[width=\textwidth]{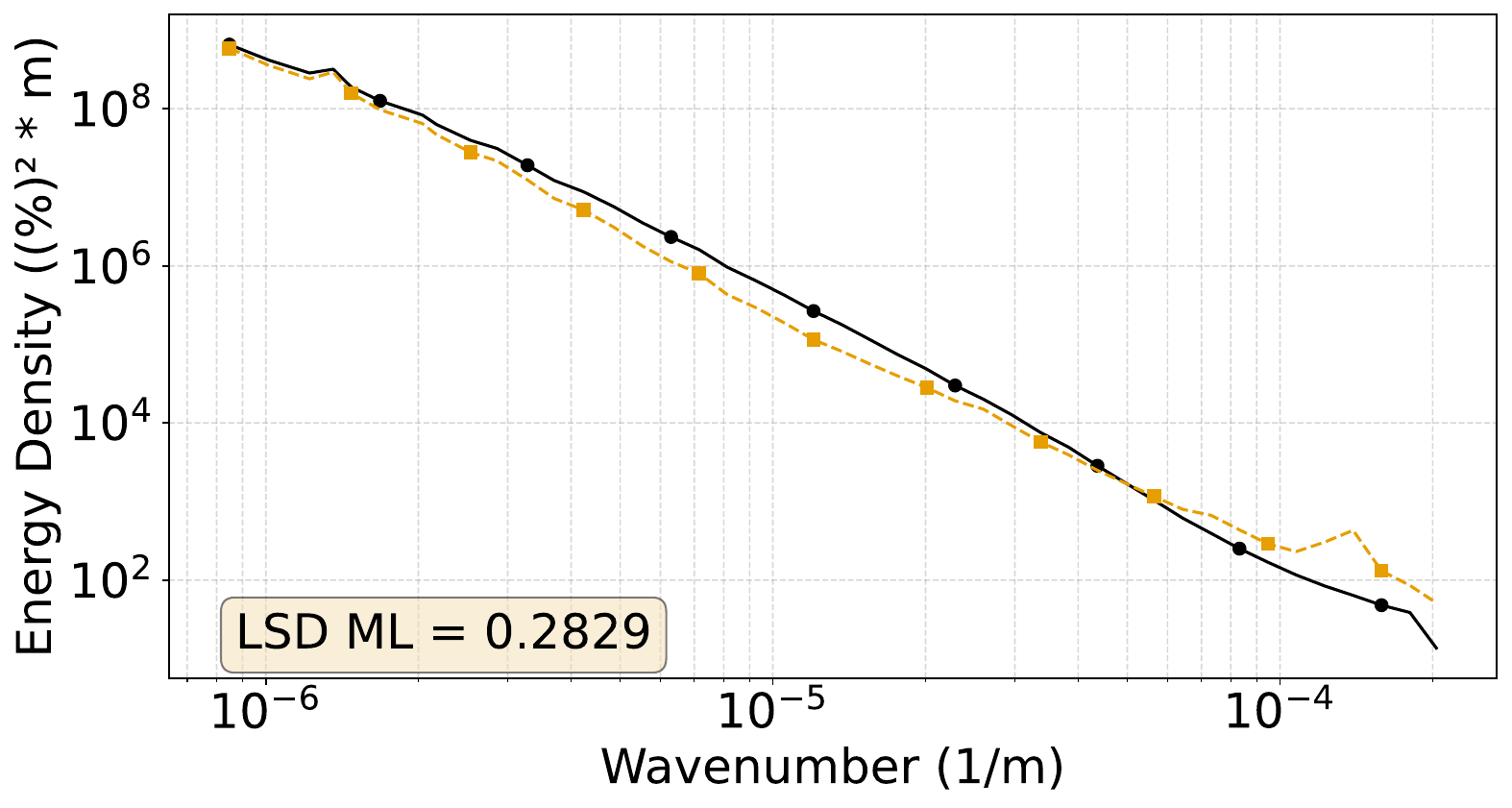}
        \caption{Relative humidity at \SI{700}{\hecto\pascal} (\wvar{r700})}
    \end{subfigure}%
    \caption{
        Energy spectra of vertical variables at \SI{700}{\hecto\pascal} for the \gls{DANRA} \gls{ML} model and ground truth data, averaged over lead times. 
    }
    \label{fig:danra_verif_gridded_energy_spectra_vertical}
\end{figure}

\paragraph{3D atmospheric variables}
We show the \gls{NMAE} along vertical profiles for all 3D variables in \cref{fig:cosmo_verif_gridded_vertical_full} for the \gls{COSMO} \gls{ML} model and in \cref{fig:danra_verif_gridded_vertical_full} for the \gls{DANRA} \gls{ML} model.

\begin{figure}[tbp]
    \centering
    \begin{subfigure}[b]{0.5\textwidth}
        \centering
        \includegraphics[width=\textwidth]{graphics/verification/cosmo/gridded/vertical_profile_wind_u.pdf}
        \caption{Wind u-component (\wvar{u}) }
    \end{subfigure}%
    \begin{subfigure}[b]{0.5\textwidth}
        \centering
        \includegraphics[width=\textwidth]{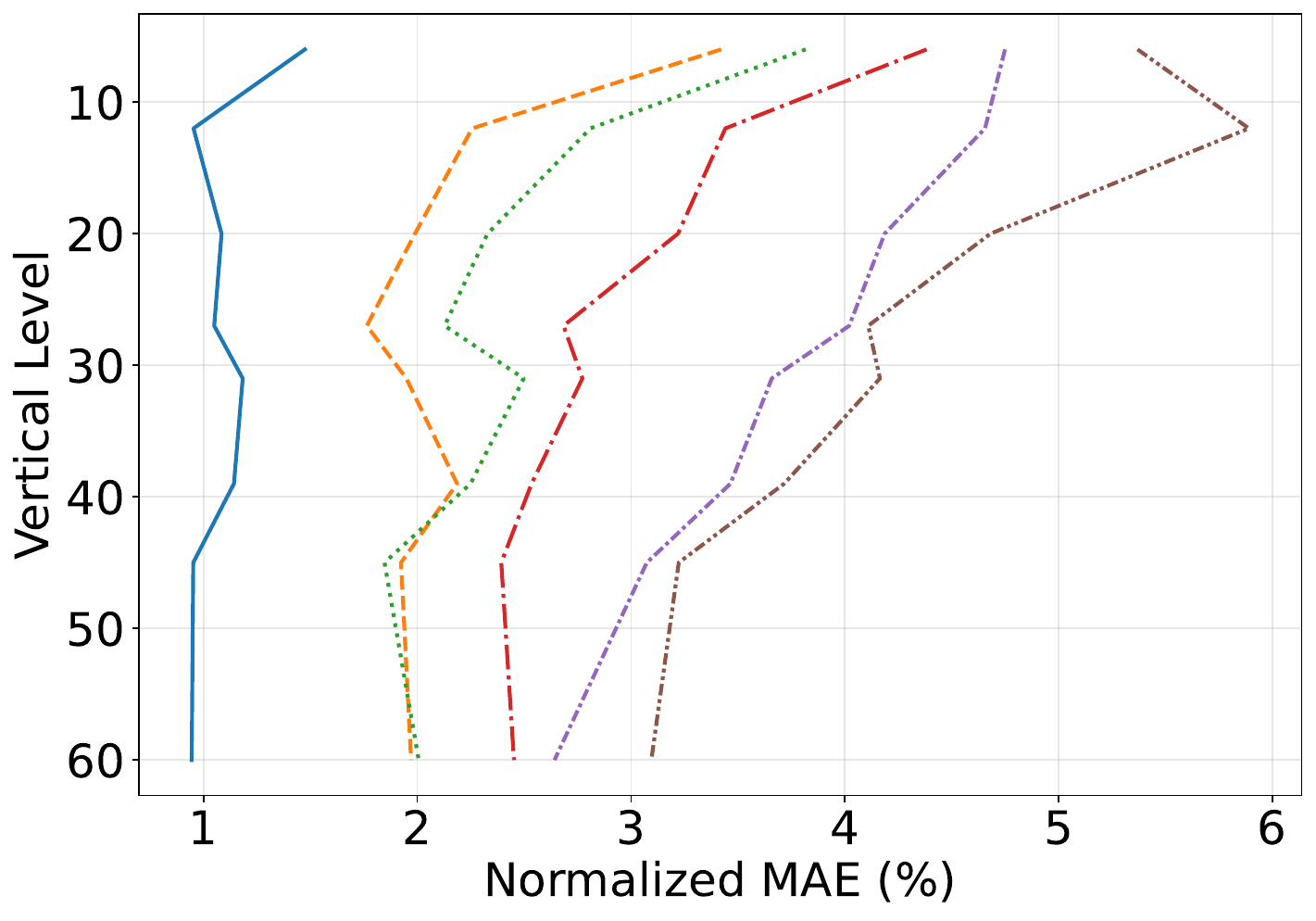}
        \caption{Wind v-component (\wvar{v})}
    \end{subfigure}%
    \hfill%
    \begin{subfigure}[b]{0.5\textwidth}
        \centering
        \includegraphics[width=\textwidth]{graphics/verification/cosmo/gridded/vertical_profile_temperature.pdf}
        \caption{Temperature (\wvar{t})}
    \end{subfigure}%
    \begin{subfigure}[b]{0.5\textwidth}
        \centering
        \includegraphics[width=\textwidth]{graphics/verification/cosmo/gridded/vertical_profile_vertical_velocity.pdf}
        \caption{Vertical velocity (\wvar{w})}
    \end{subfigure}%
    \hfill%
    \begin{subfigure}[b]{0.5\textwidth}
        \centering
        \includegraphics[width=\textwidth]{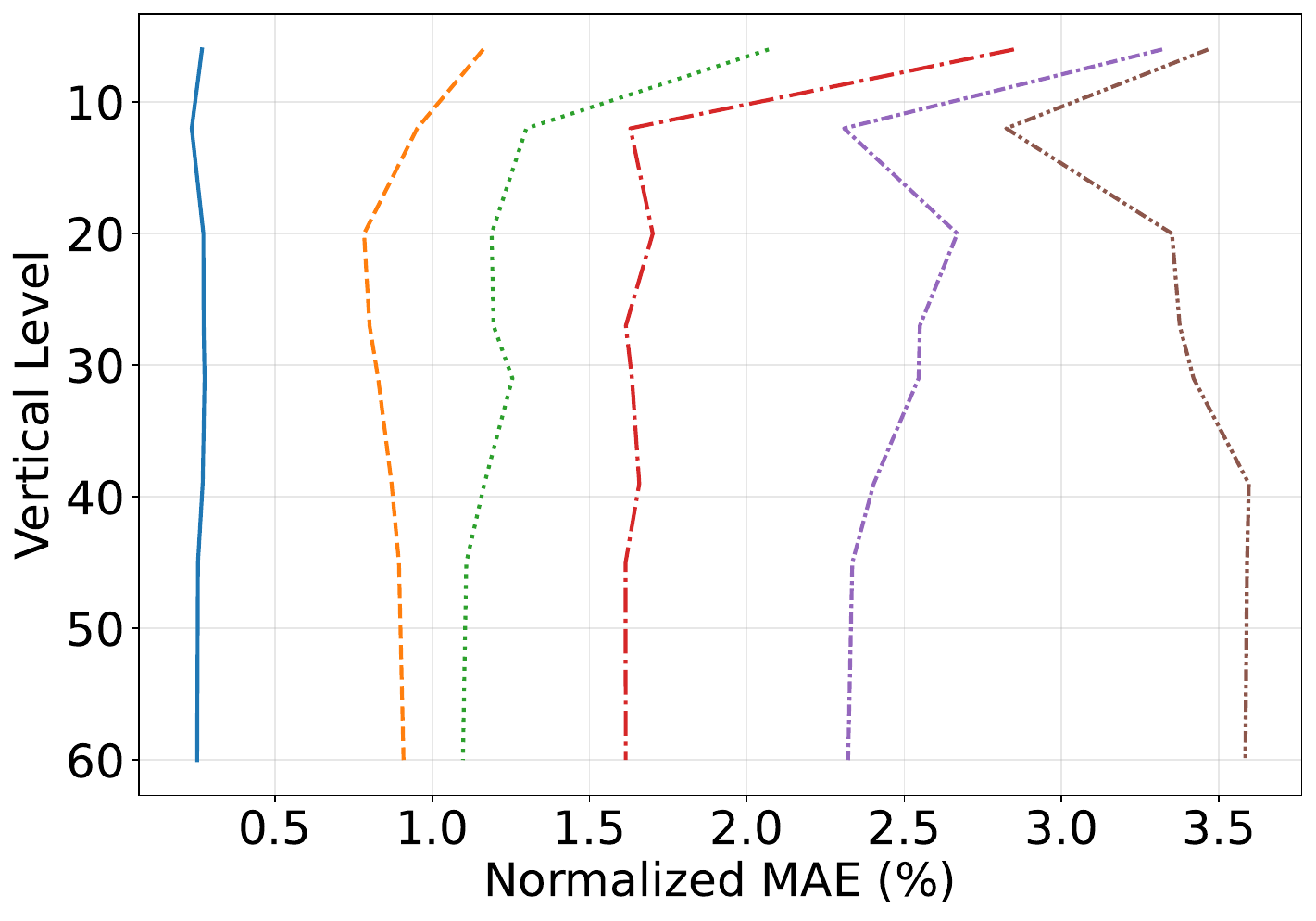}
        \caption{Pressure (\wvar{pres})}
    \end{subfigure}%
    \begin{subfigure}[b]{0.5\textwidth}
        \centering
        \includegraphics[width=\textwidth]{graphics/verification/cosmo/gridded/vertical_profile_relative_humidity.pdf}
        \caption{Relative humidity (\wvar{r})}
    \end{subfigure}%
    \caption{
        Vertical profiles of the \gls{NMAE} for the \gls{COSMO} \gls{ML} model.
    }
    \label{fig:cosmo_verif_gridded_vertical_full}
\end{figure}

\begin{figure}[tbp]
    \centering
    \begin{subfigure}[b]{0.5\textwidth}
        \centering
        \includegraphics[width=\textwidth]{graphics/verification/danra/gridded/vertical_profile_wind_u.pdf}
        \caption{Wind u-component (\wvar{u})}
    \end{subfigure}%
    \begin{subfigure}[b]{0.5\textwidth}
        \centering
        \includegraphics[width=\textwidth]{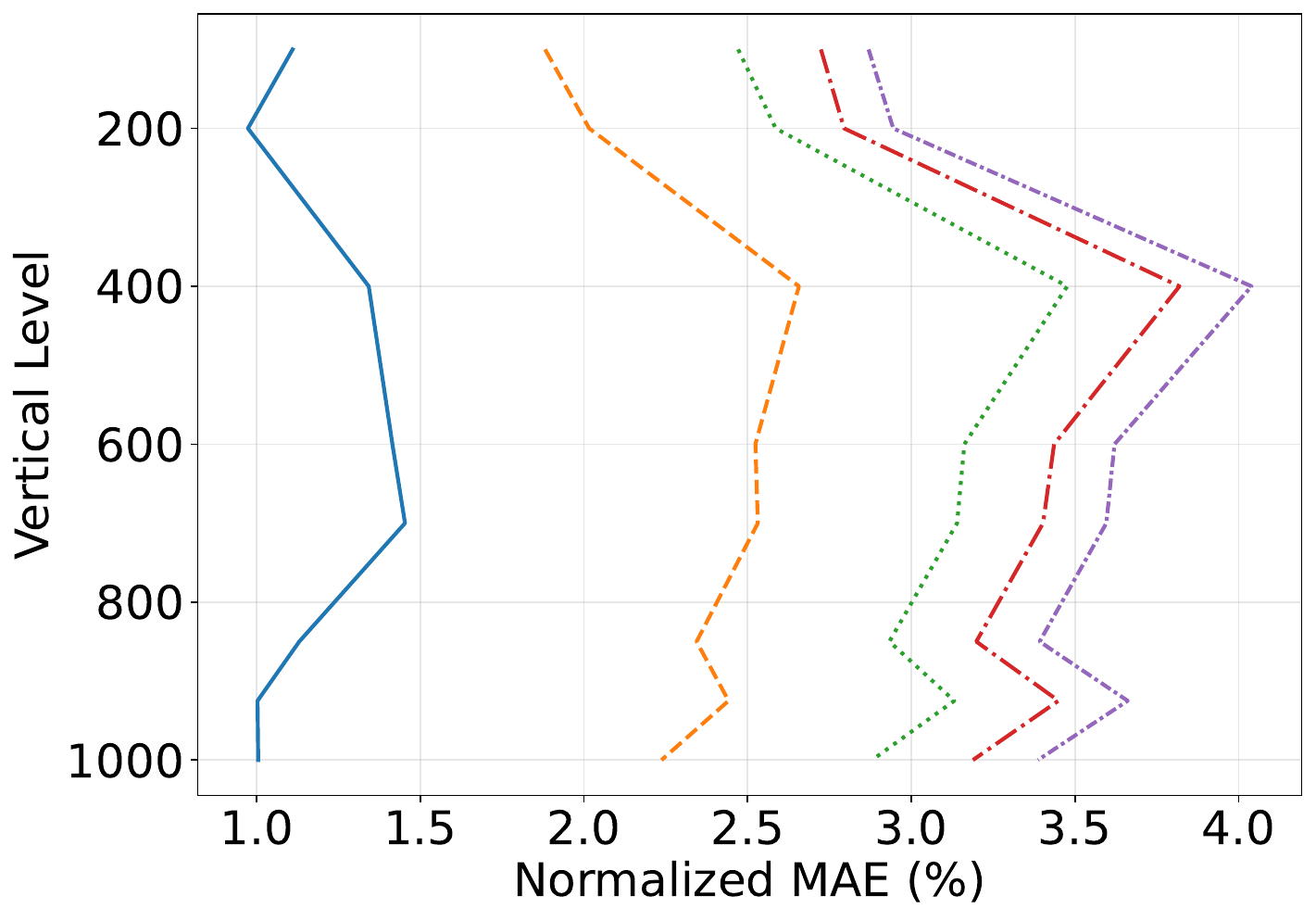}
        \caption{Wind v-component (\wvar{v})}
    \end{subfigure}%
    \hfill%
    \begin{subfigure}[b]{0.5\textwidth}
        \centering
        \includegraphics[width=\textwidth]{graphics/verification/danra/gridded/vertical_profile_temperature.pdf}
        \caption{Temperature (\wvar{t})}
    \end{subfigure}%
    \begin{subfigure}[b]{0.5\textwidth}
        \centering
        \includegraphics[width=\textwidth]{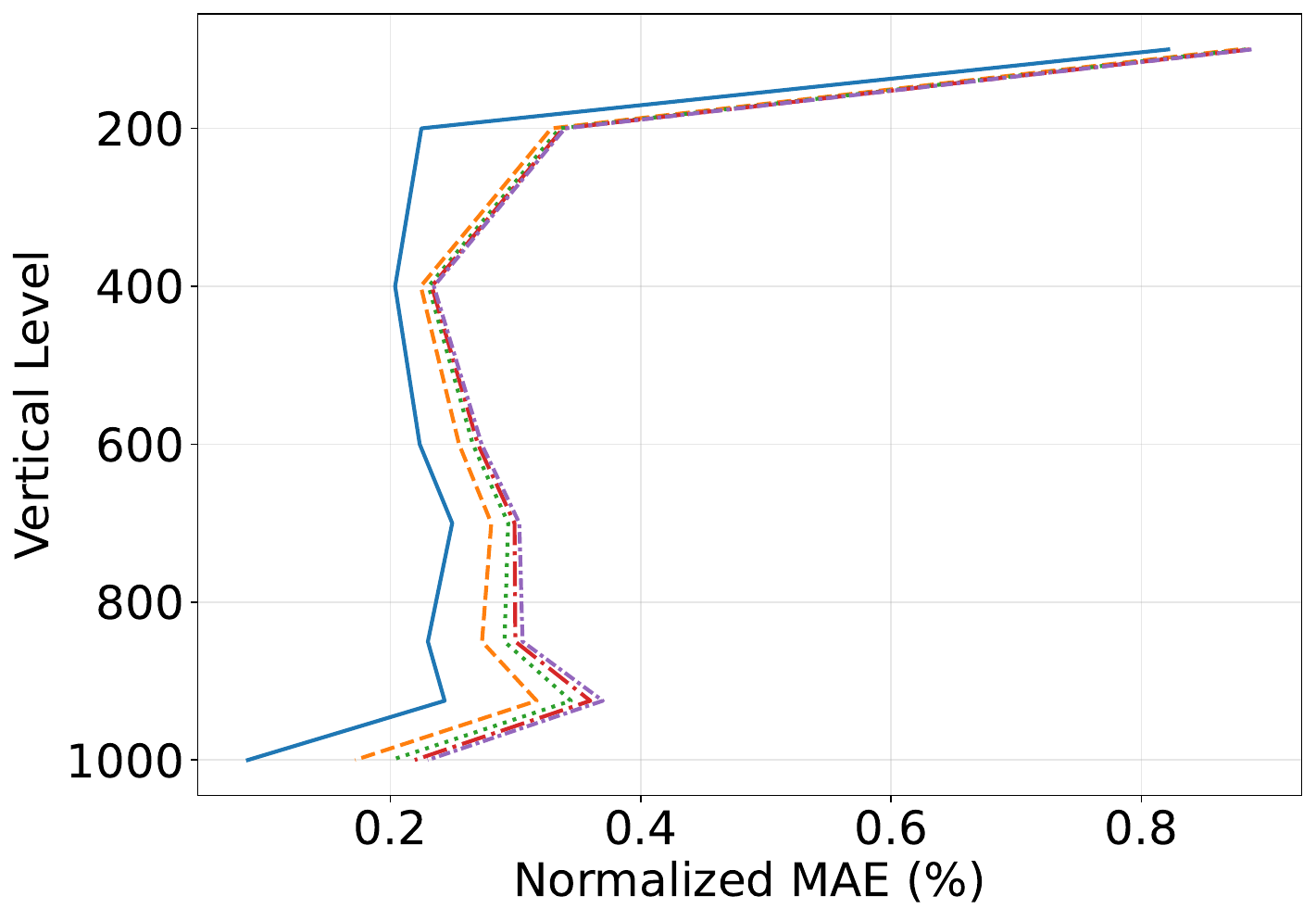}
        \caption{Vertical velocity (\wvar{w})}
    \end{subfigure}%
    \hfill%
    \begin{subfigure}[b]{0.5\textwidth}
        \centering
        \includegraphics[width=\textwidth]{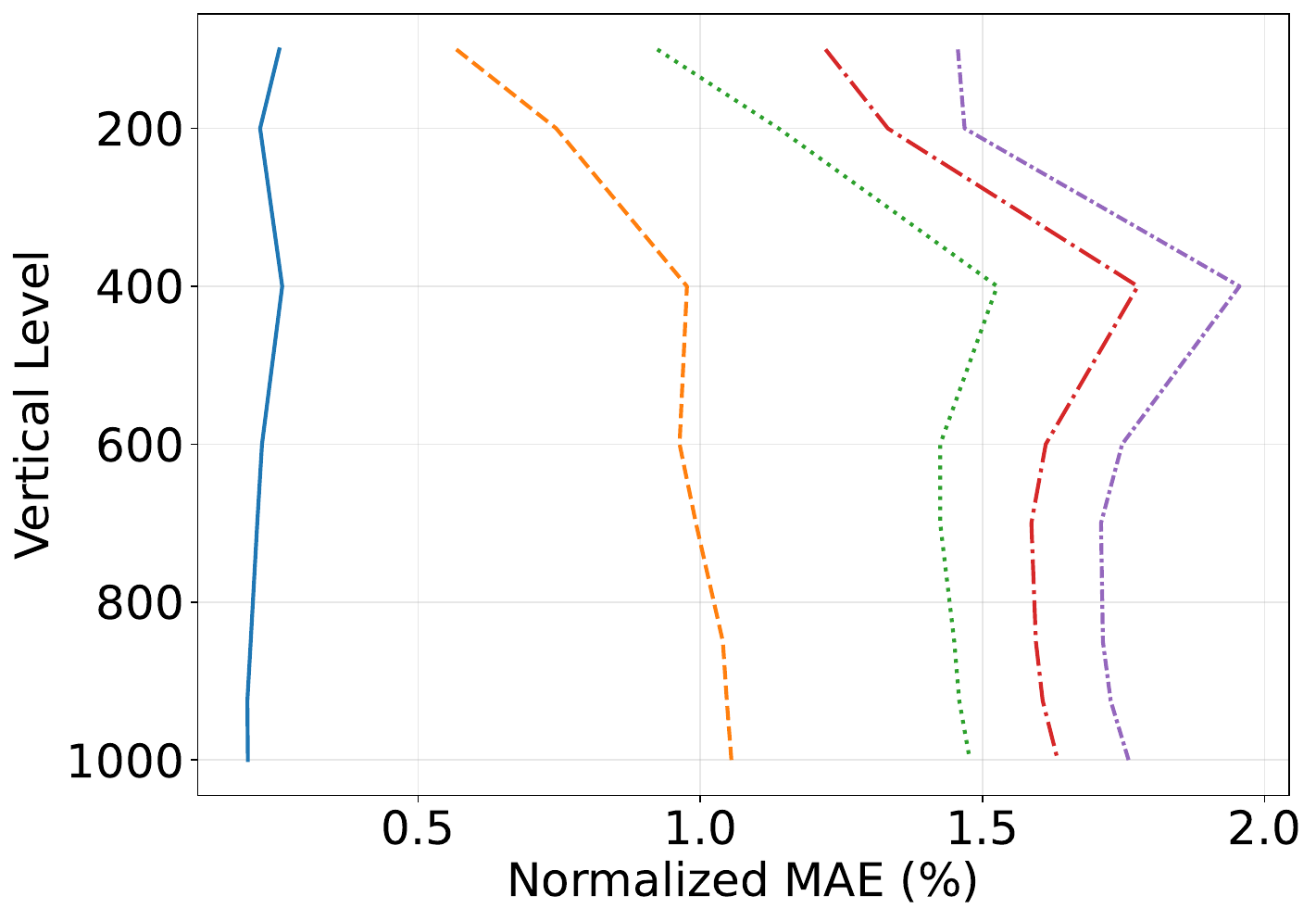}
        \caption{Geopotential (\wvar{z})}
    \end{subfigure}%
    \begin{subfigure}[b]{0.5\textwidth}
        \centering
        \includegraphics[width=\textwidth]{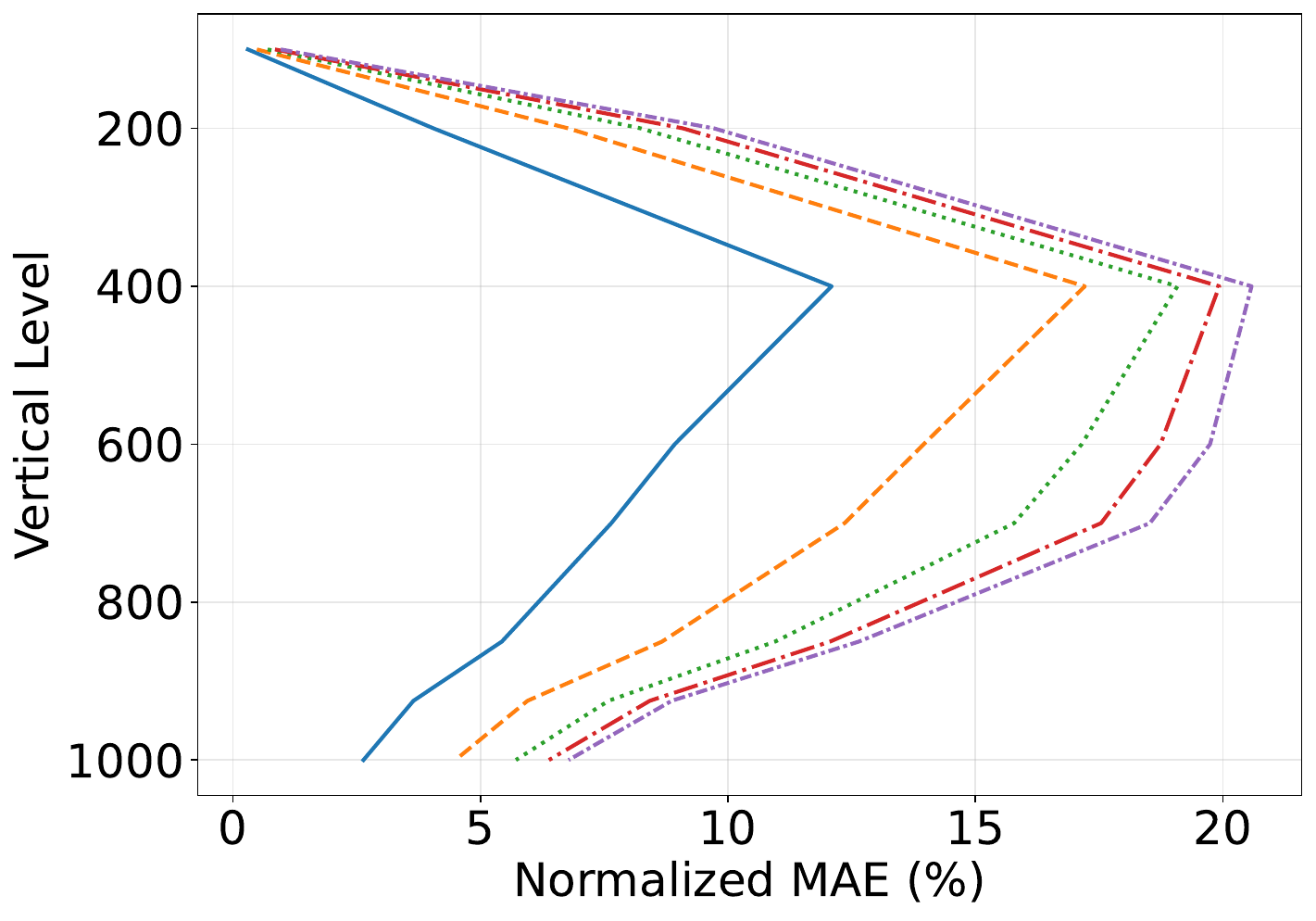}
        \caption{Relative humidity (\wvar{r})}
    \end{subfigure}%
    \caption{
        Vertical profiles of the \gls{NMAE} for the \gls{DANRA} \gls{ML} model.
    }
    \label{fig:danra_verif_gridded_vertical_full}
\end{figure}

\subsection{Evaluation against sparse observations}
\label{sec:extra_eval_station}

\paragraph{\gls{RMSE} decomposition}
For temperature and surface pressure it is common to study the error decomposition of the \gls{RMSE} into bias (\gls{ME}) and error standard deviation (\gls{STDEV_ERR}).
Looking at this decomposition in \cref{fig:cosmo_verif_sparse_me_stdev_err} for the \gls{COSMO} model, we see that the \gls{ML} model has a bias towards cold temperatures, increasingly so with lead time. 
The \gls{NWP} model is also biased towards cold temperatures, but this however decreases with lead time.
Interestingly, the \gls{STDEV_ERR} for both models remains very similar, indicating similar random forecast error.
The difference in the \gls{RMSE} can therefore be clearly attributed to model biases evolving throughout the forecast.

\begin{figure}[tbp]
    \centering
    \begin{subfigure}[b]{0.5\textwidth}
        \centering
        \includegraphics[width=\textwidth]{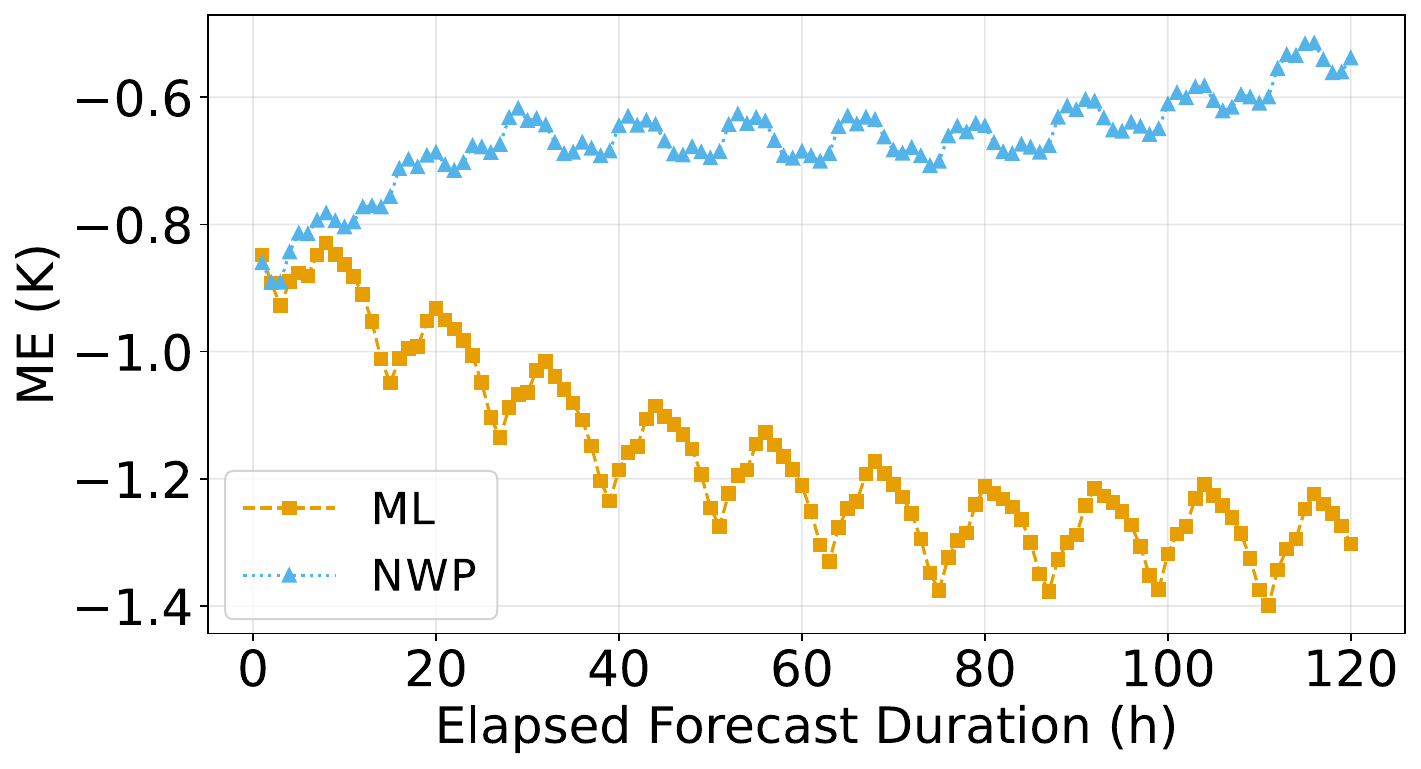}
        \caption{\gls{ME} of \SI{2}{m} temperature (\wvar{2t})}
    \end{subfigure}%
    \begin{subfigure}[b]{0.5\textwidth}
        \centering
        \includegraphics[width=\textwidth]{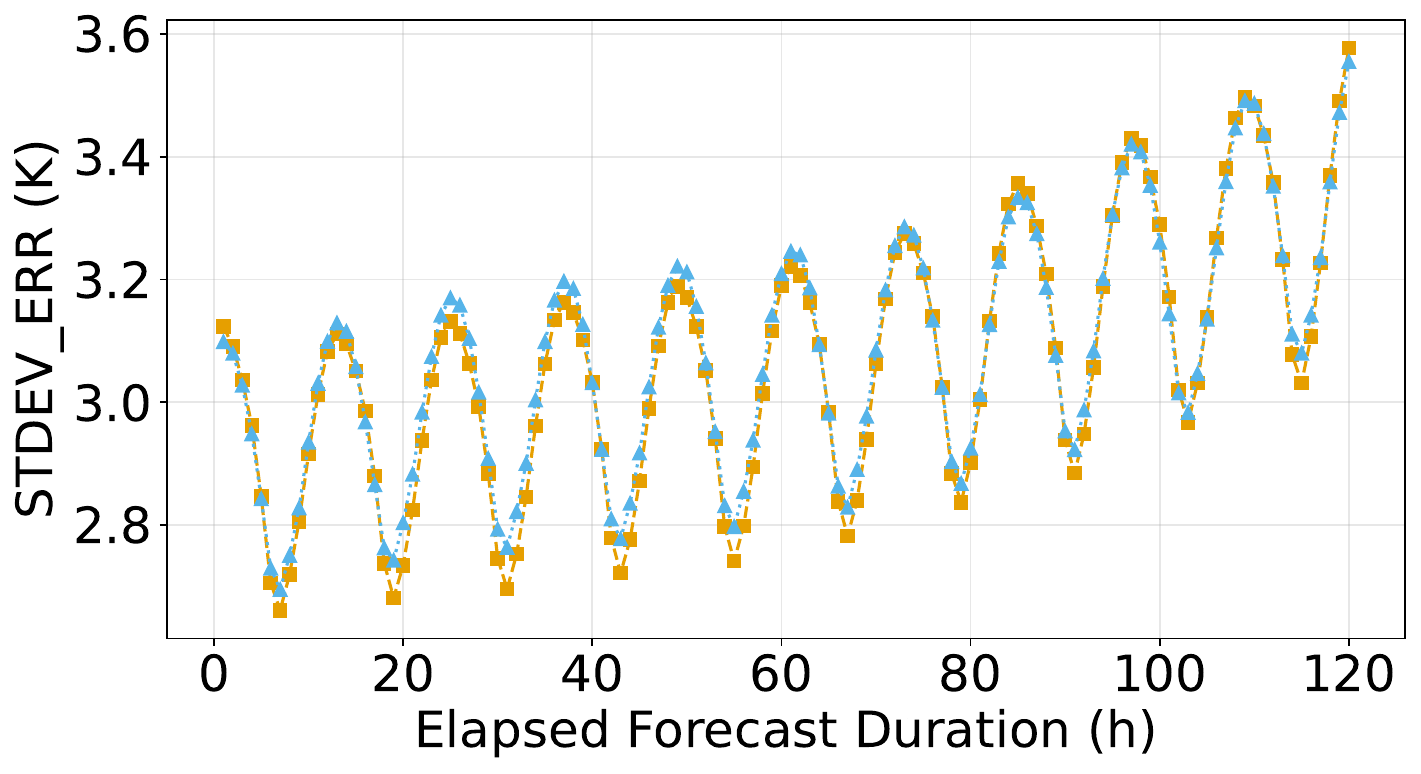}
        \caption{\gls{STDEV_ERR} of \SI{2}{m} temperature (\wvar{2t})}
    \end{subfigure}%
    \hfill%
    \begin{subfigure}[b]{0.5\textwidth}
        \centering
        \includegraphics[width=\textwidth]{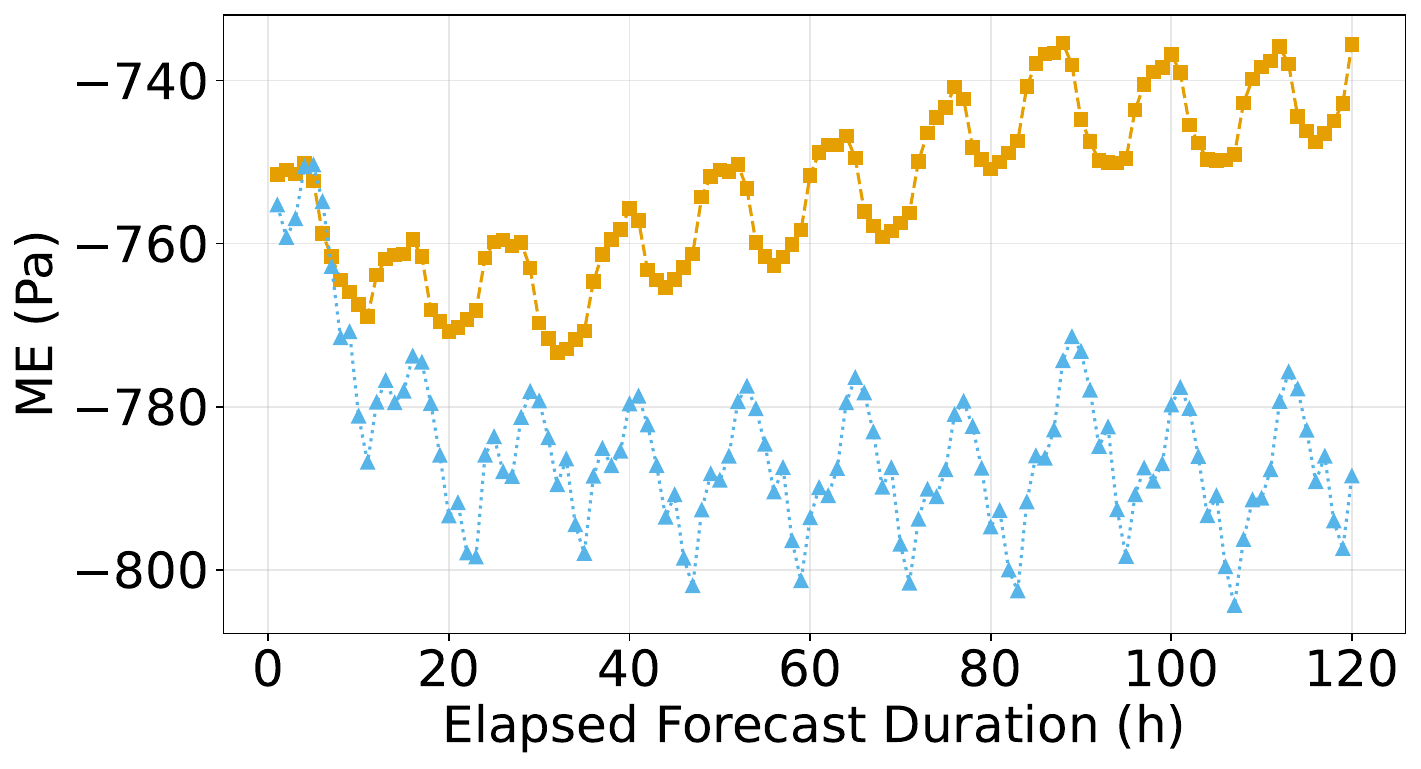}
        \caption{\gls{ME} of surface pressure (\wvar{sp})}
    \end{subfigure}%
    \begin{subfigure}[b]{0.5\textwidth}
        \centering
        \includegraphics[width=\textwidth]{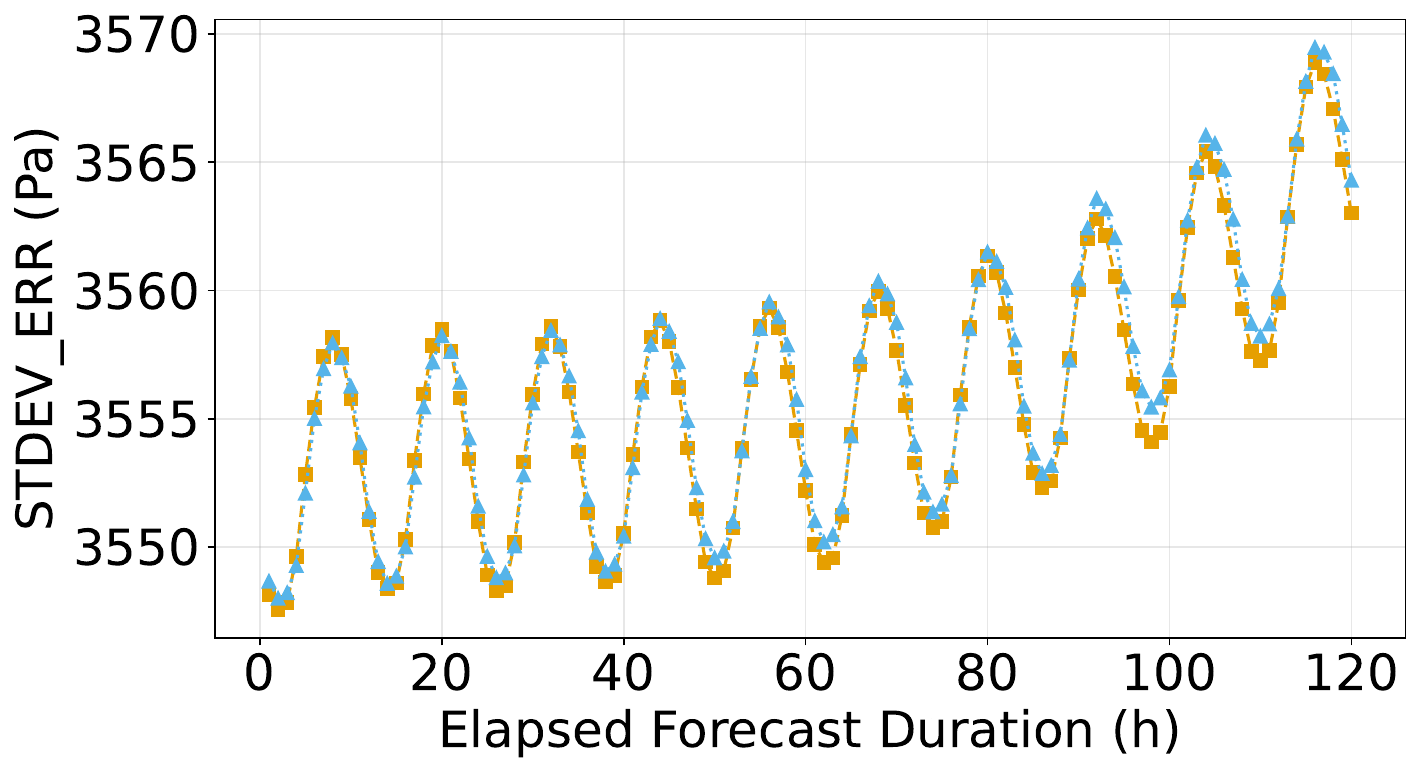}
        \caption{\gls{STDEV_ERR} of surface pressure (\wvar{sp})}
    \end{subfigure}%
    \caption{
        Bias and standard deviation of the error for the \gls{COSMO} models.
    }
    \label{fig:cosmo_verif_sparse_me_stdev_err}
\end{figure}

\paragraph{Mean error maps}
In \cref{fig:cosmo_verif_sparse_error_map_precip,fig:cosmo_verif_sparse_error_map_wind_u} we show additional mean error maps for the \gls{COSMO} models compared to station observations.
Also for these there is a general agreement between the biases of the \gls{ML} and \gls{NWP} models.
For the \gls{DANRA} models we show an additional mean error map in \cref{fig:danra_verif_sparse_error_map_wind_u}, for the wind u-component.
Both models have a clear bias to forecast higher eastward winds.

\begin{figure}[tbp]
    \centering
    \includegraphics[width=0.7\textwidth]{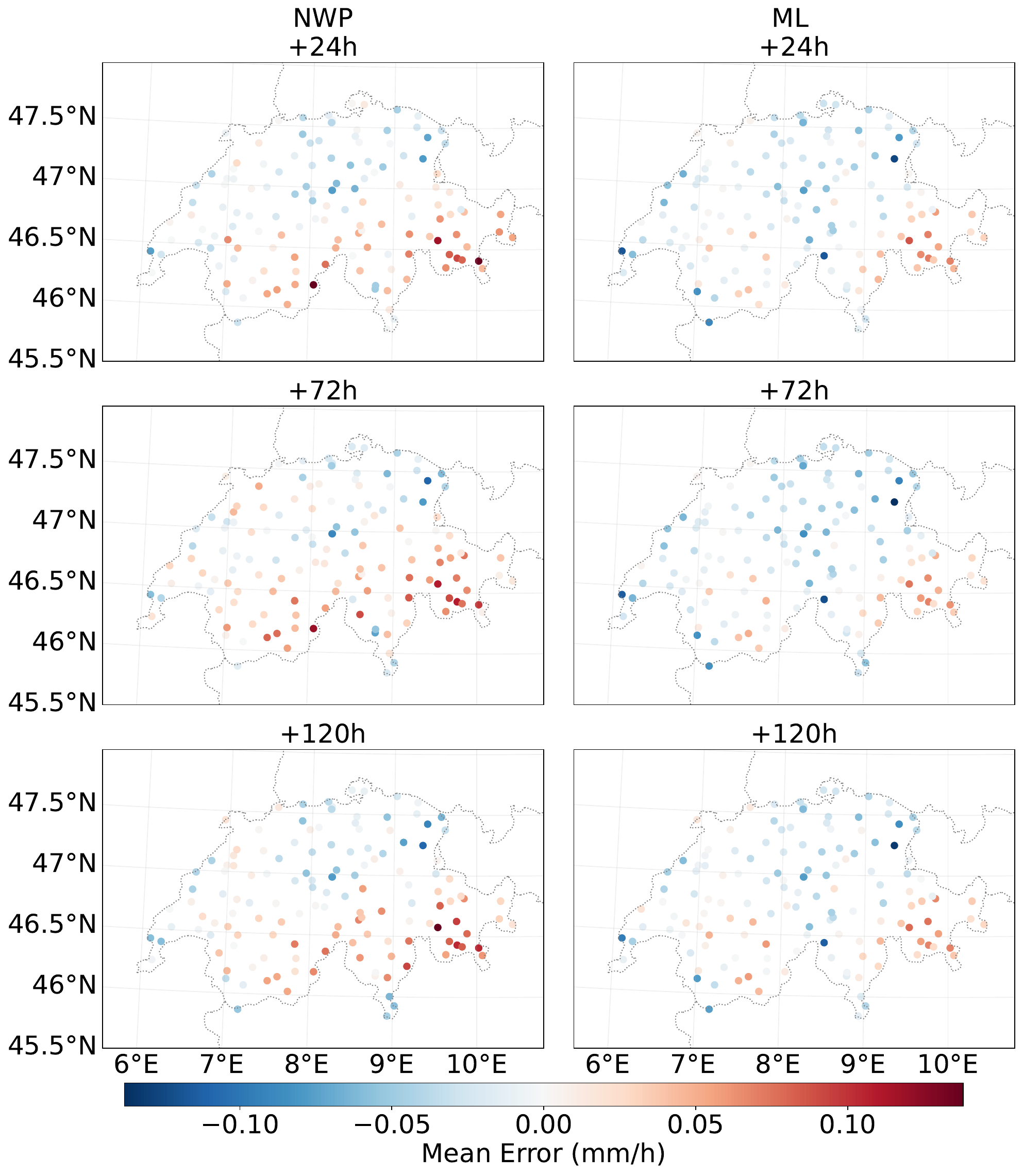}
    \caption{
        Mean error maps for precipitation (\wvar{tp01}) for the \gls{COSMO} models over the full test year period.
    }
    \label{fig:cosmo_verif_sparse_error_map_precip}
\end{figure}

\begin{figure}[tbp]
    \centering
    \includegraphics[width=0.7\textwidth]{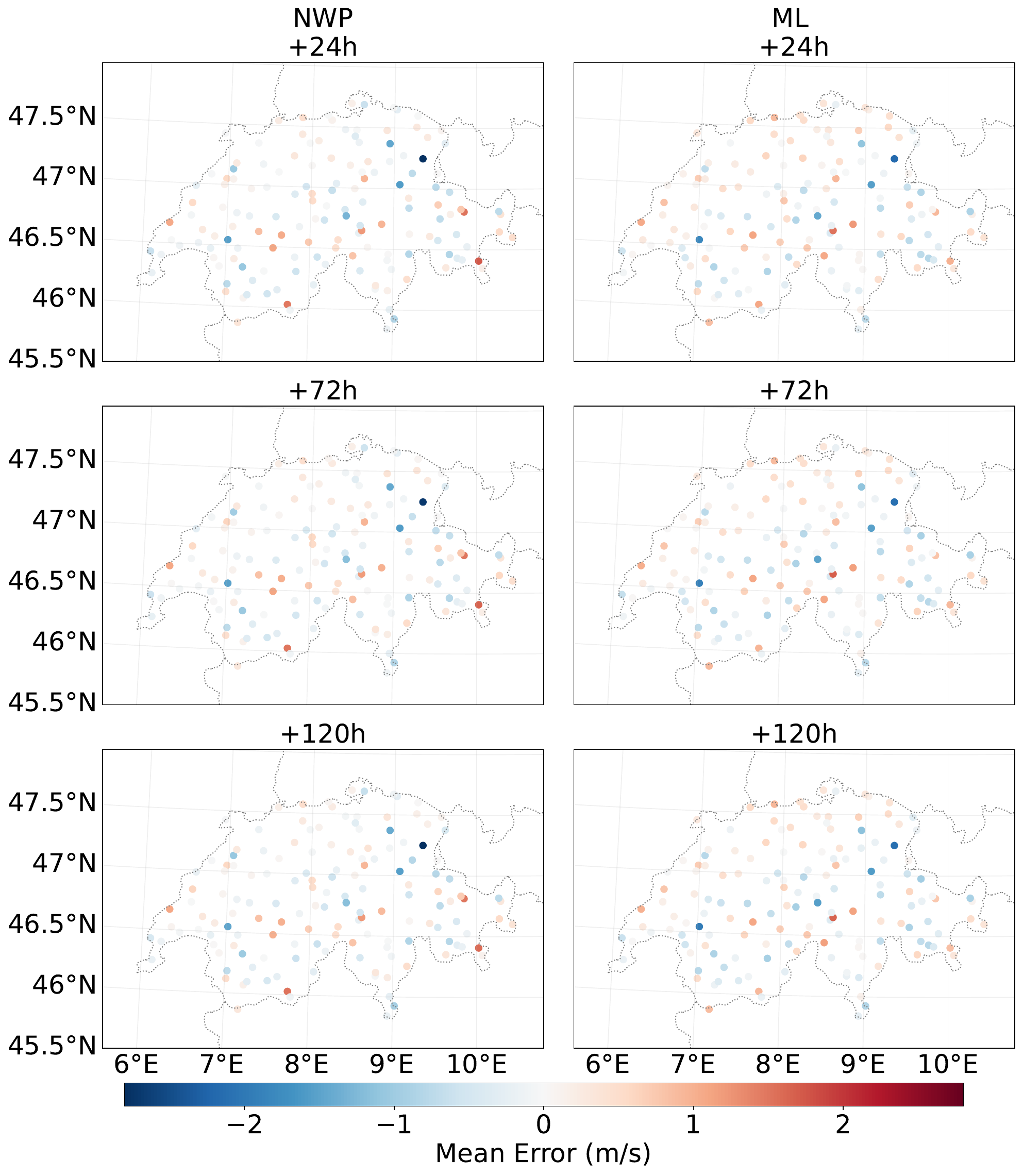}
    \caption{
        Mean error maps for the u-component of \SI{10}{m} wind (\wvar{10u}) for the \gls{COSMO} models over the full test year period.
    }
    \label{fig:cosmo_verif_sparse_error_map_wind_u}
\end{figure}

\begin{figure}[tbp]
    \centering
    \includegraphics[width=0.7\textwidth]{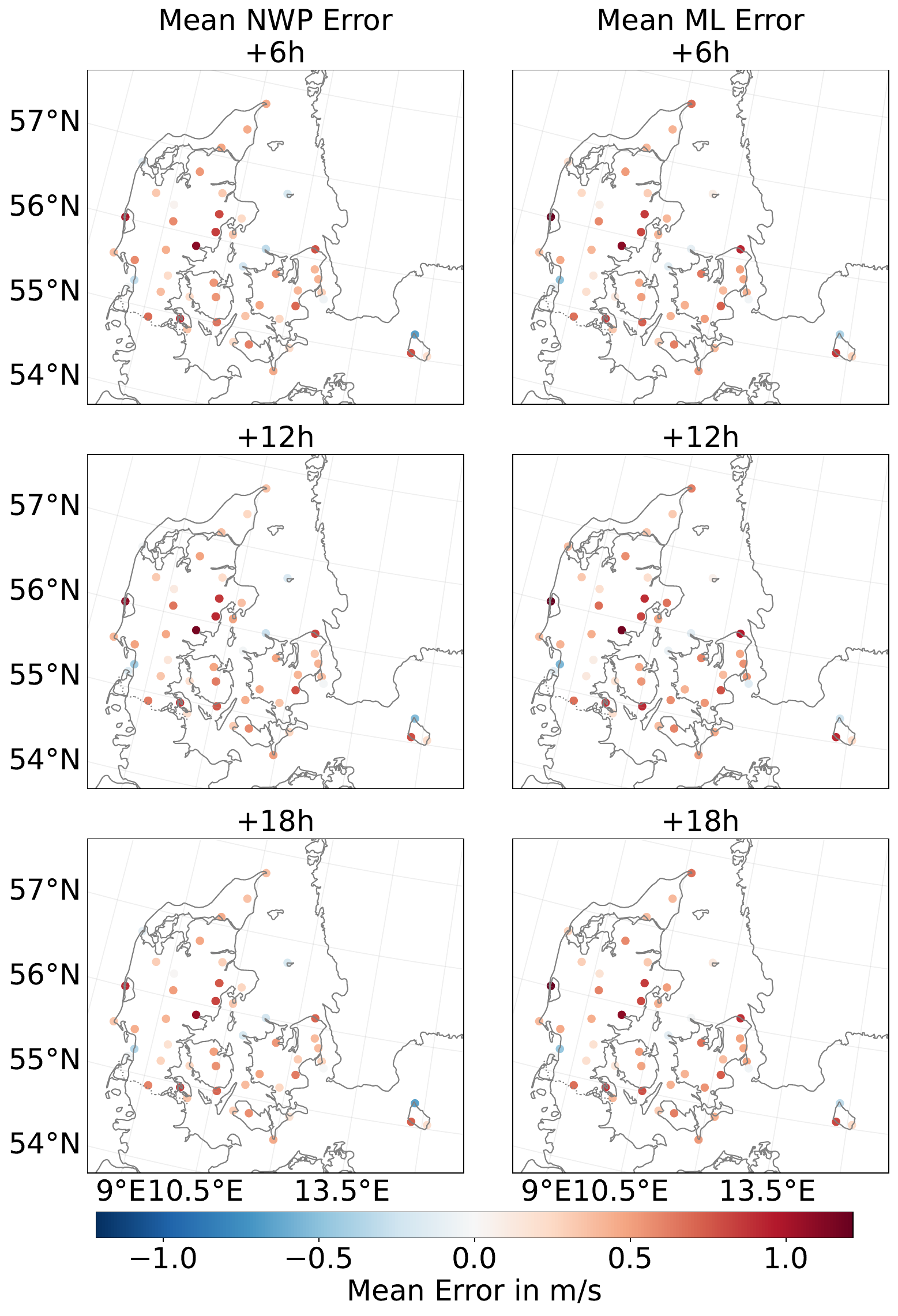}
    \caption{
        Mean error maps for the u-component of \SI{10}{m} wind (\wvar{10u}) for the \gls{DANRA} models over the full test year period.
    }
    \label{fig:danra_verif_sparse_error_map_wind_u}
\end{figure}

\paragraph{Distributional alignment}
To understand if the models are capable of representing the full distribution of values for different variables we inspect histograms of predictions from the models.
We compare such histograms to data both for the sparse station observations and the gridded datasets.
As seen in \cref{fig:cosmo_verif_sparse_histogram_sp_gridded}, the \gls{COSMO} \gls{ML} model learns the marginal distribution of the analysis data almost perfectly for surface pressure \wvar{sp}.
However, compared to the distribution of station values in \cref{fig:cosmo_verif_sparse_histogram_sp_sparse} there is a clear distributional shift between both models and the observations.
As the surface pressure is sensitive to the height above sea level, this shift can be be attributed to the naive interpolation method used.
Since we have not applied any height adjustments, the interpolated model points are often biased. 
However, as we focus on model vs. model comparison, we need to make sure that both models are affected similarly by the simplistic interpolation. 
The overlapping dynamic ranges after interpolation for both models clearly illustrate that this is the case.
For the wind component in \cref{fig:cosmo_verif_sparse_histogram_10u} the \gls{ML} model slightly underestimates the dynamic range of station observations, while closely matching the distribution of gridded data.

\begin{figure}[tbp]
    \centering
    \begin{subfigure}[b]{\textwidth}
        \centering
        \includegraphics[width=\textwidth]{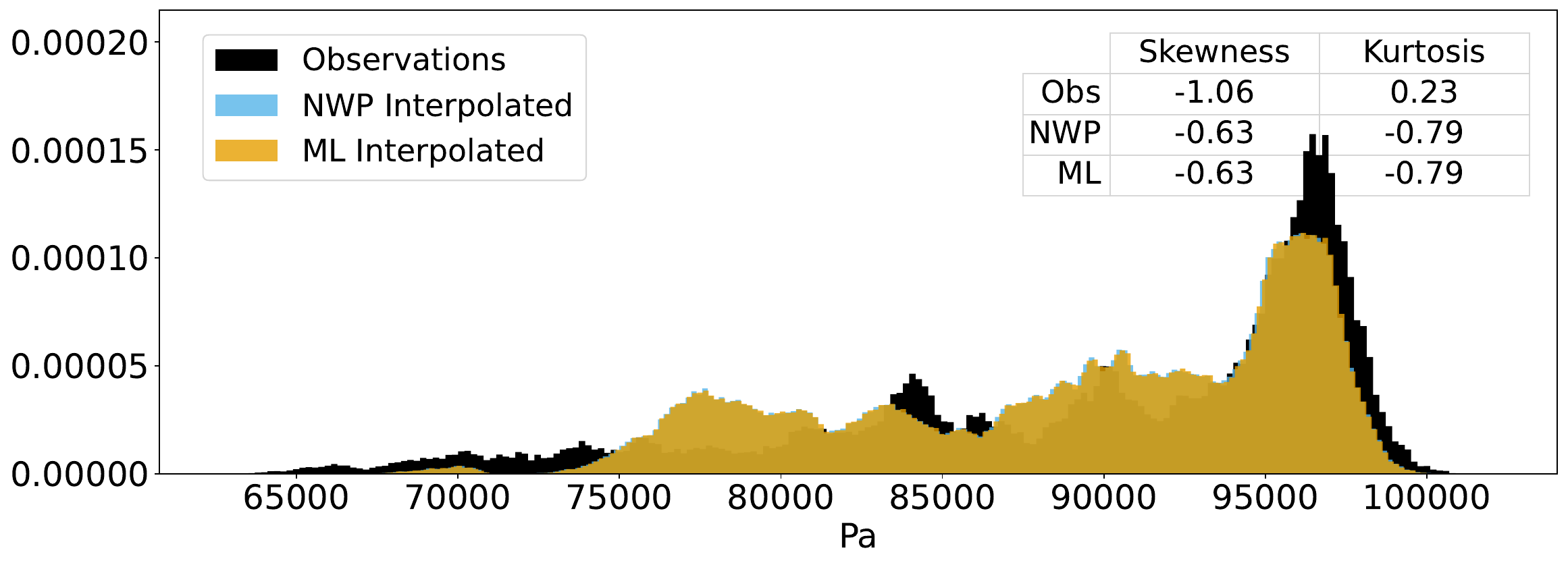}
        \caption{Compared against station observations}
    \label{fig:cosmo_verif_sparse_histogram_sp_sparse}
    \end{subfigure}%
    \hfill%
    \begin{subfigure}[b]{\textwidth}
        \centering
        \includegraphics[width=\textwidth]{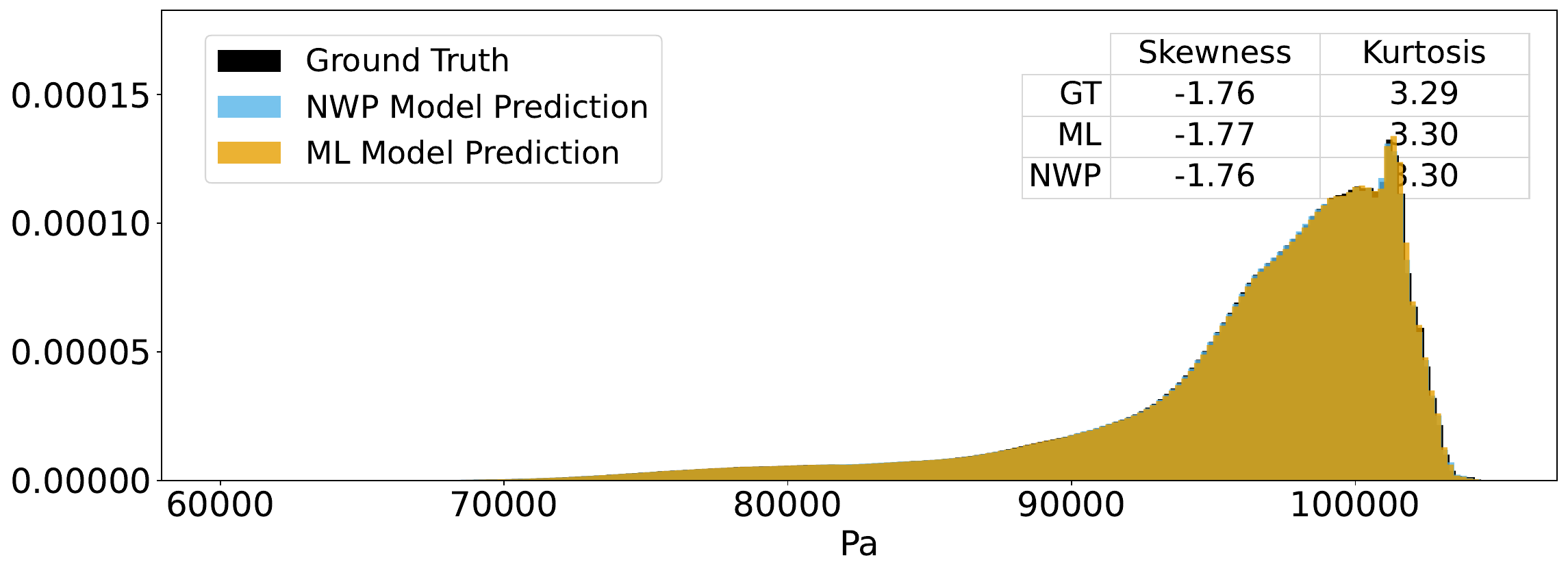}
        \caption{Compared against gridded analysis}
    \label{fig:cosmo_verif_sparse_histogram_sp_gridded}
    \end{subfigure}
    \caption{
        Histogram of the surface pressure (\wvar{sp}) values for the \gls{COSMO} models and data. The model distributions were estimated jointly across all lead times and spatial locations (interpolated station points or analysis grid points). 
        As a reference we tabulate skewness and kurtosis of the different distributions.
    }
    \label{fig:cosmo_verif_sparse_histogram_sp}
\end{figure}

\begin{figure}[tbp]
    \centering
    \begin{subfigure}[b]{\textwidth}
        \centering
        \includegraphics[width=\textwidth]{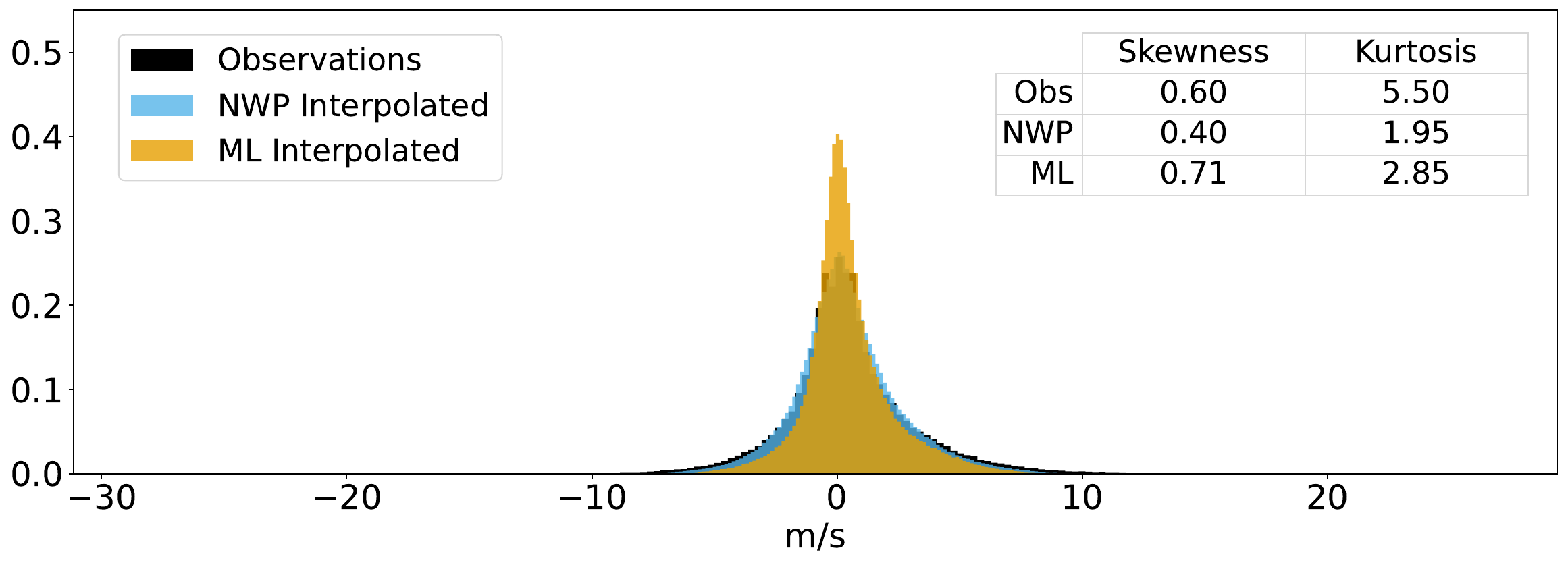}
        \caption{Compared against station observations}
    \end{subfigure}%
    \hfill%
    \begin{subfigure}[b]{\textwidth}
        \centering
        \includegraphics[width=\textwidth]{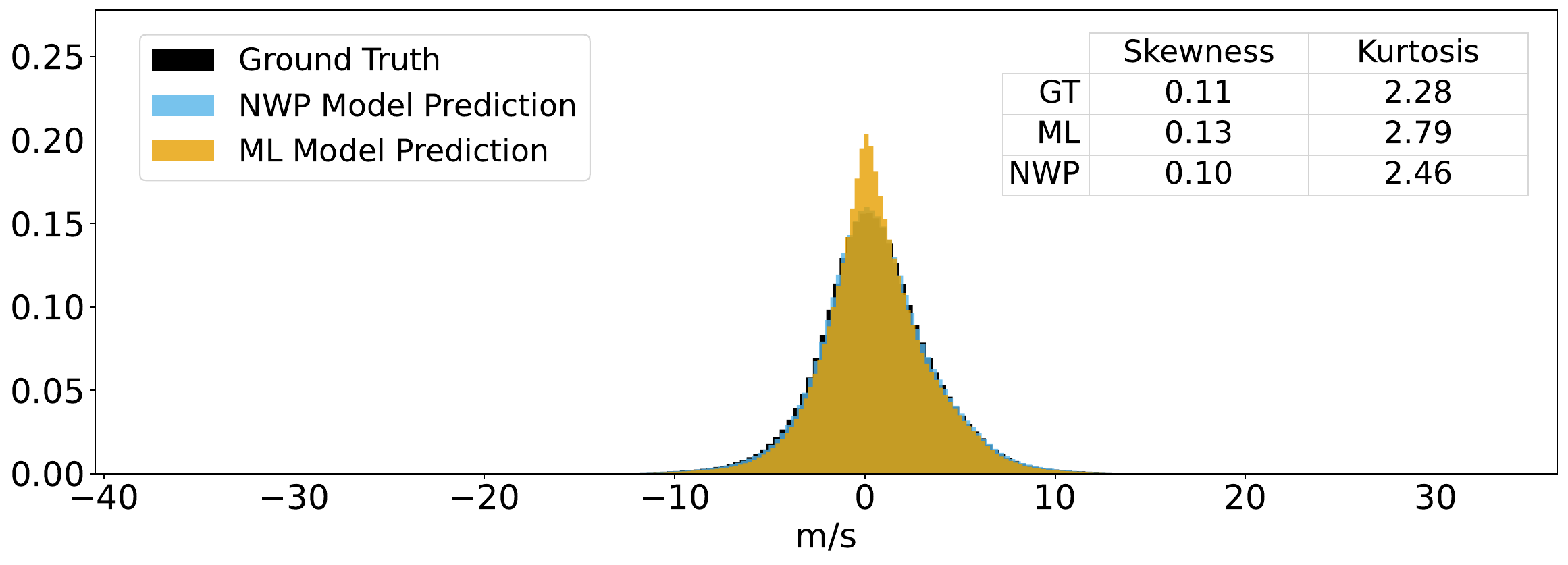}
        \caption{Compared against gridded analysis}
    \end{subfigure}
    \caption{
        Histogram of the wind u-component (\wvar{10u}) values for the \gls{COSMO} models and data. The model distributions were estimated jointly across all lead times and spatial locations (interpolated station points or analysis grid points).
        As a reference we tabulate skewness and kurtosis of the different distributions.
    }
    \label{fig:cosmo_verif_sparse_histogram_10u}
\end{figure}

To better understand these distributional shifts in the \gls{COSMO} models we measure the Wasserstein distance between the histogram of station observations and model forecasts at specific lead times.
In \cref{fig:cosmo_verif_sparse_wasserstein} we see that the \gls{ML} model experiences an increasing distributional shift for all surface variables except surface pressure.
Surface pressure being a well behaved, smooth field without strong local extremes, is easier to capture for the \gls{ML} model. 
The \gls{NWP} model has a flat Wasserstein distance for both wind components, increasing for surface pressure and even decreasing for temperature.
Precipitation is omitted here, as its distribution is dominated by a large peak at \SI{0}{\milli\metre\per\hour}.

\begin{figure}[tbp]
    \centering
    \begin{subfigure}[b]{0.5\textwidth}
        \centering
        \includegraphics[width=\textwidth]{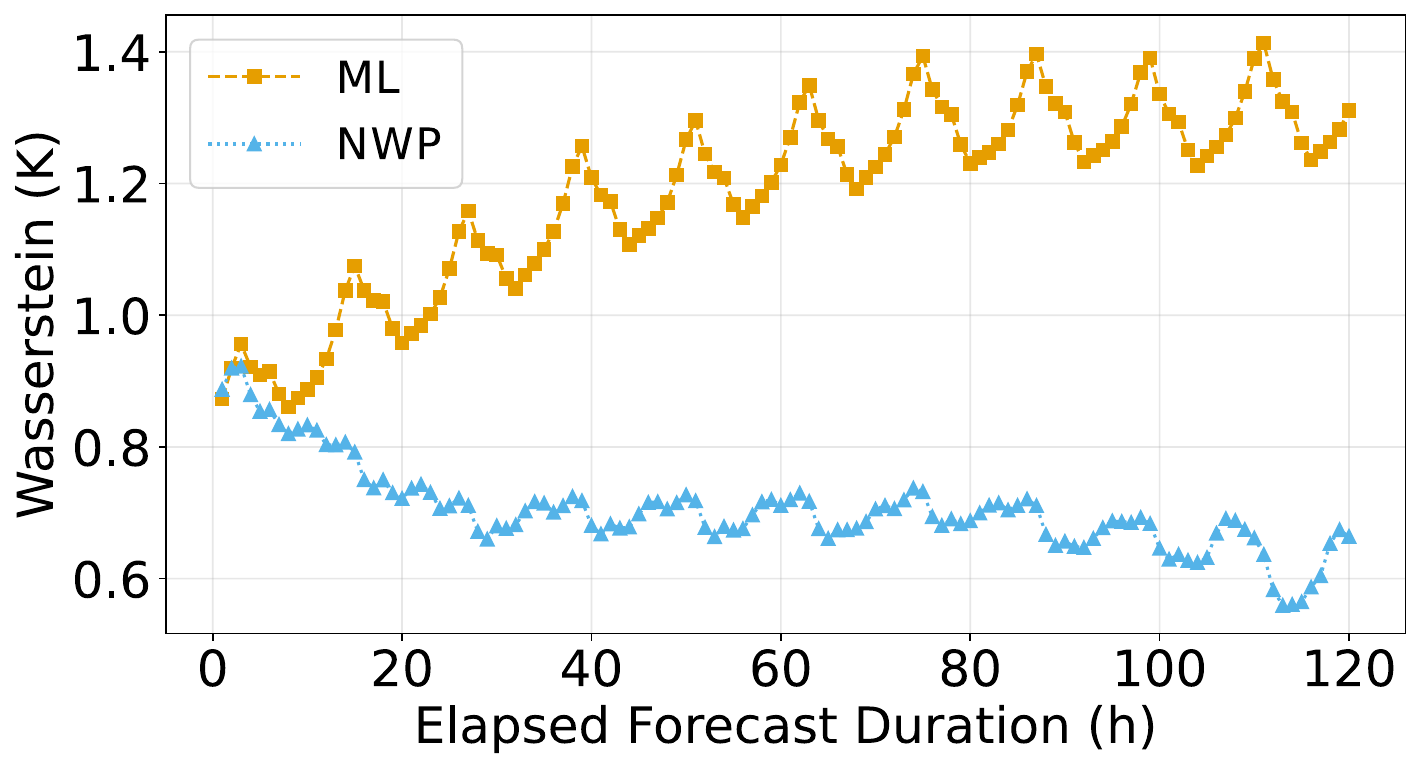}
        \caption{\SI{2}{m} temperature (\wvar{2t})}
    \end{subfigure}%
    \begin{subfigure}[b]{0.5\textwidth}
        \centering
        \includegraphics[width=\textwidth]{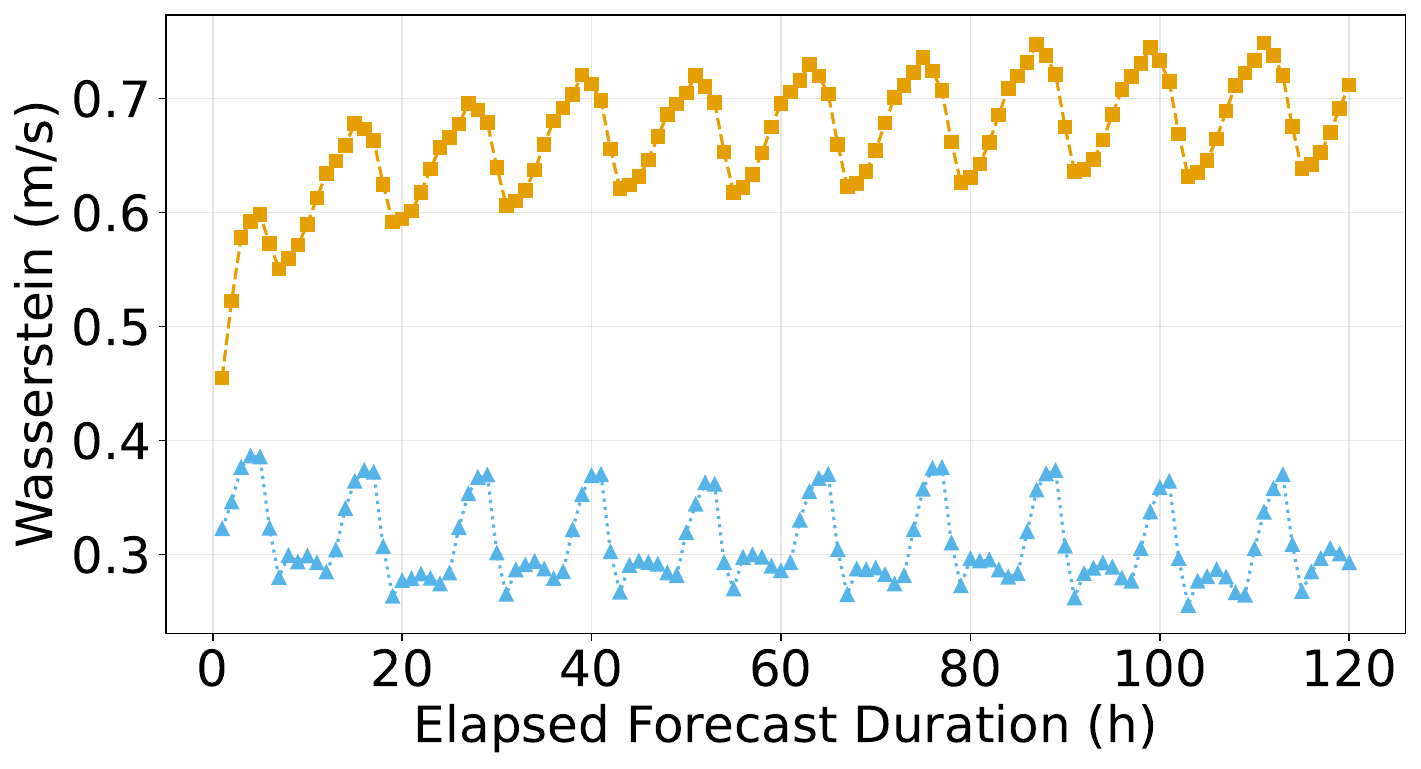}
        \caption{Wind u-component (\wvar{10u})}
    \end{subfigure}%
    \hfill%
    \begin{subfigure}[b]{0.5\textwidth}
        \centering
        \includegraphics[width=\textwidth]{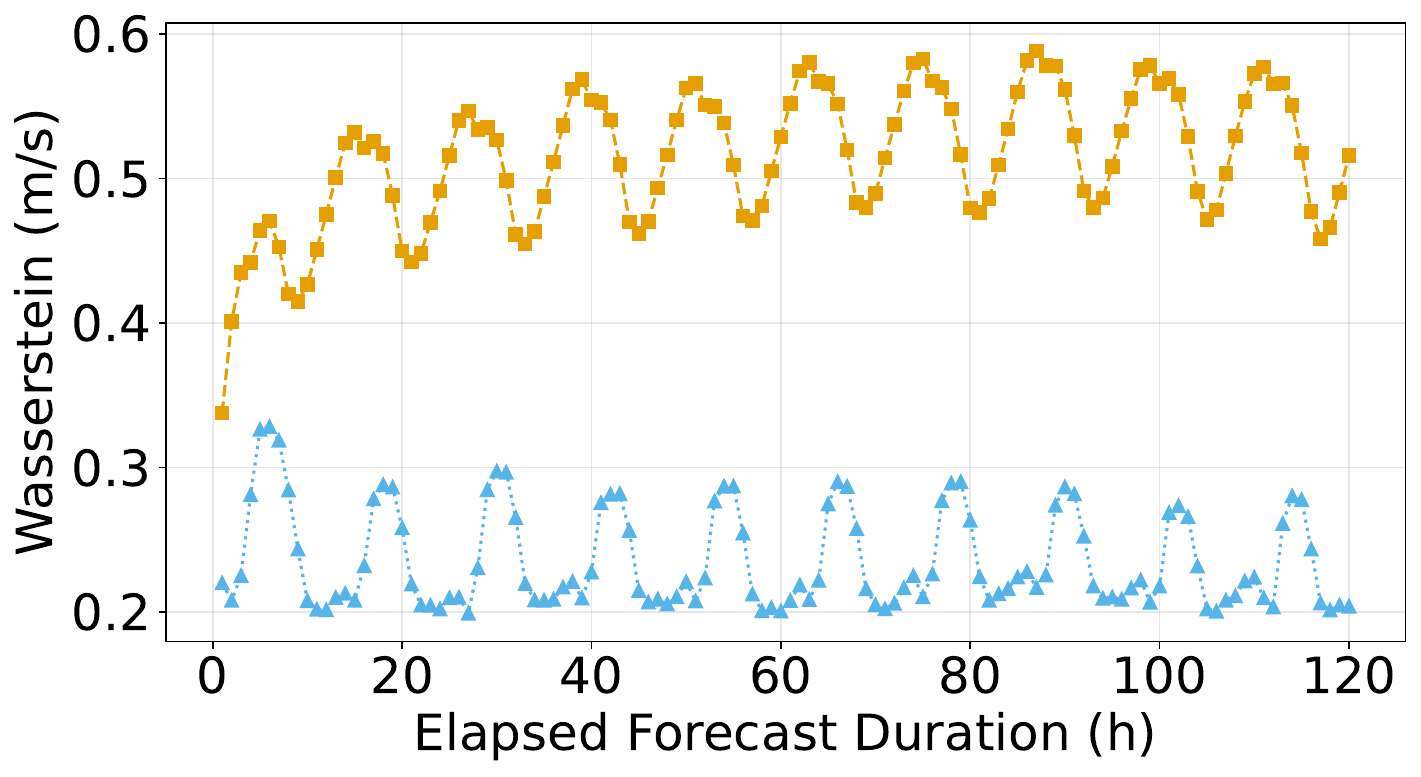}
        \caption{Wind v-component (\wvar{10u})}
    \end{subfigure}%
    \begin{subfigure}[b]{0.5\textwidth}
        \centering
        \includegraphics[width=\textwidth]{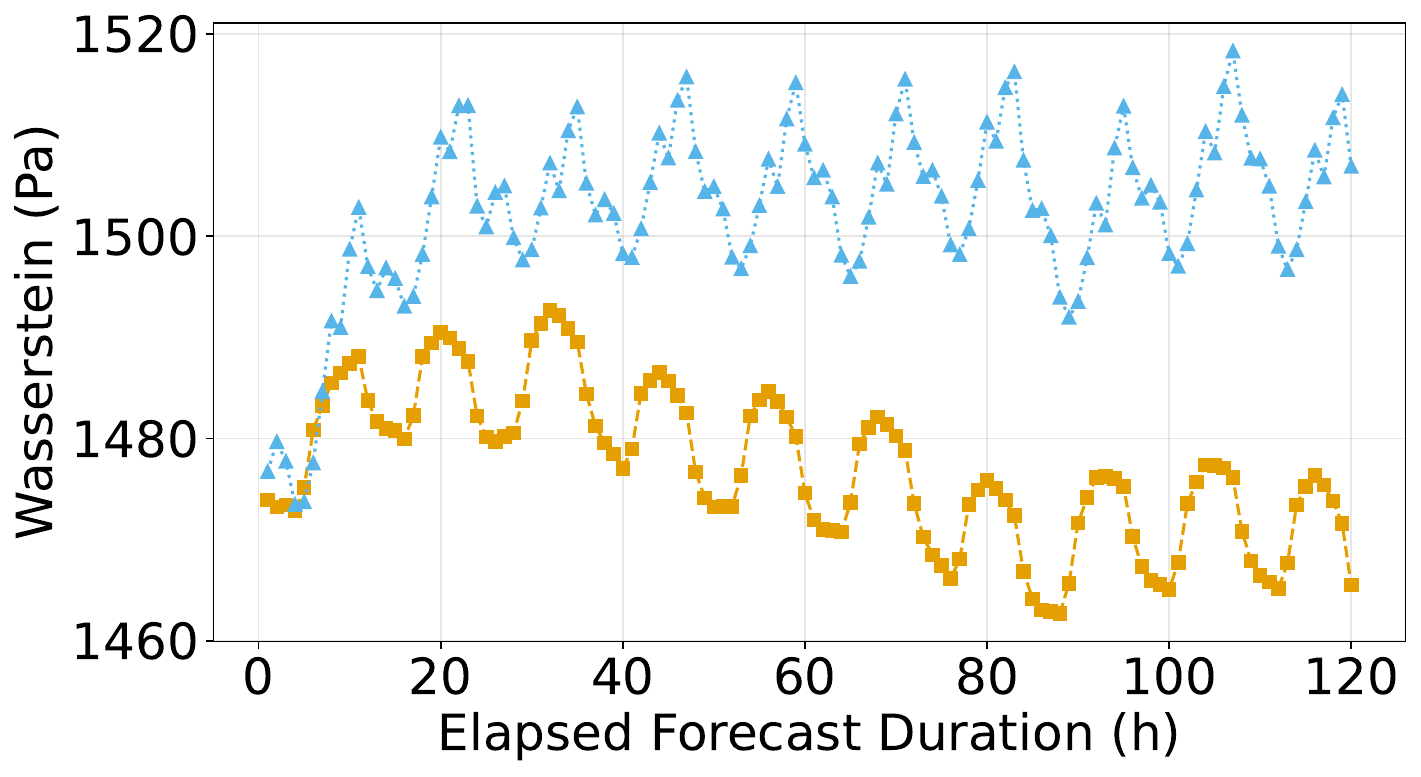}
        \caption{Surface pressure (\wvar{sp})}
    \end{subfigure}%
    \caption{
        Wasserstein distance between the histogram of station observations and model forecasts at different lead times for the \gls{COSMO} models.
    }
    \label{fig:cosmo_verif_sparse_wasserstein}
\end{figure}

\begin{figure}[tbp]
    \centering
    \begin{subfigure}[b]{\textwidth}
        \centering
        \includegraphics[width=\textwidth]{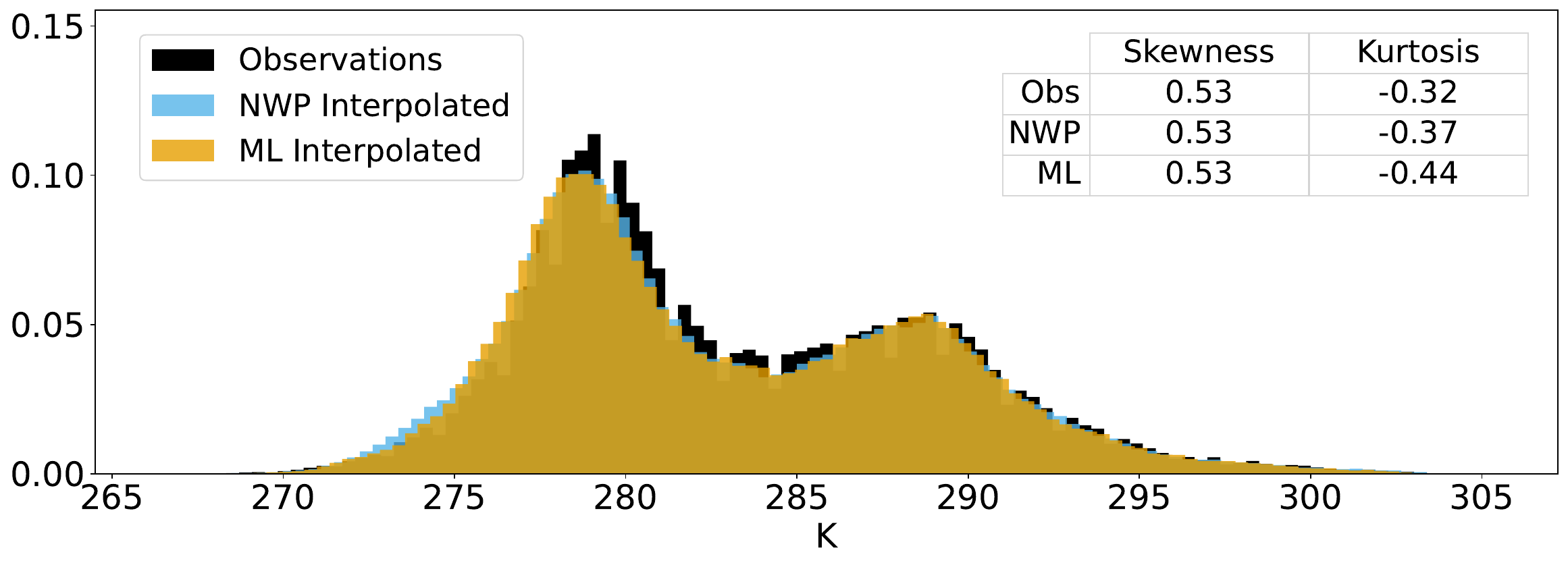}
        \caption{Compared against station observations}
    \end{subfigure}%
    \hfill%
    \begin{subfigure}[b]{\textwidth}
        \centering
        \includegraphics[width=\textwidth]{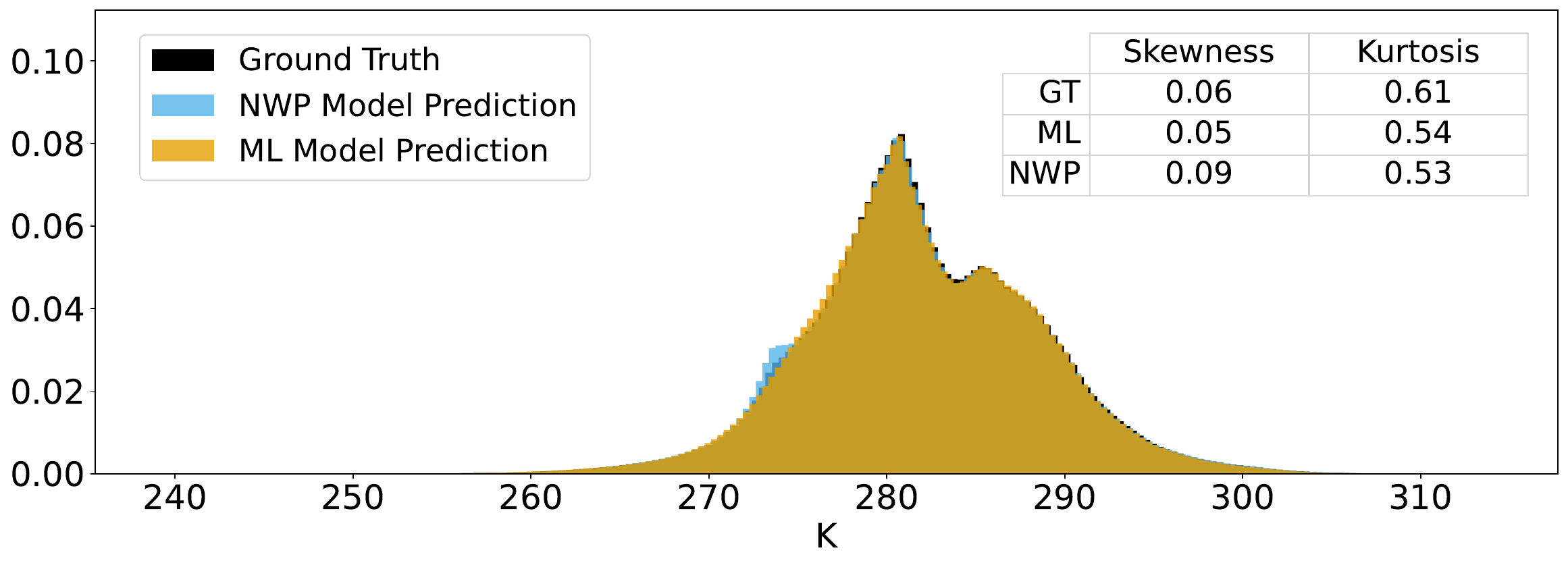}
        \caption{Compared against gridded analysis}
    \end{subfigure}
    \caption{
        Histogram of the \SI{2}{m} temperature (\wvar{2t}) values for the \gls{DANRA} models and data. The model distributions were estimated jointly across all lead times and spatial locations (interpolated station points or analysis grid points).
        As a reference we tabulate skewness and kurtosis of the different distributions.
    }
    \label{fig:danra_verif_sparse_histogram_2t}
\end{figure}

We look at histograms also to understand the range of values represented by the \gls{DANRA} \gls{ML} model.
\Cref{fig:danra_verif_sparse_histogram_2t} shows the distribution of values for \SI{2}{m} temperature for models and data at station locations and in the gridded reanalysis.
For the \gls{DANRA} \gls{ML} model, and the \gls{NWP} one, the distributions align well both for the station observations and gridded data.
Also for wind in  \cref{fig:danra_verif_sparse_histogram_10v}.
the distributions of model predictions align well with the values in the data, both for stations observations and the full gridded analysis.
While there are some differences in the modes of the distributions, we note that the \gls{ML} model accurately reproduces the tails of the data distribution.
The Danish domain has less orographic complexity, especially around the stations located in Denmark.
This again supports the fact that the distributional shift in the \gls{COSMO} domain relates to orography, either through the simple interpolation method or direct modeling challenges.

It should be noted here that what we are considering is a summary of the marginal distributions of values produced by the models.
This should be differentiated from any conditional distribution $p\left(\wstate_{k+1} \middle|\wstate_{k}, \dots\right)$ that the model predicts.
As we work with deterministic models, it is of little interest to directly reason about such conditional distributions.
What is showed through the distributional alignment here is that the model is capable of representing the full dynamical range of the data.
The large magnitude values occur equally often in the forecasts and data, but not necessarily at the same time points.
This type of distributional alignment is a necessary, but not sufficient requirement for alignment of conditional predictive distributions.

\begin{figure}[tbp]
    \centering
    \begin{subfigure}[b]{\textwidth}
        \centering
        \includegraphics[width=\textwidth]{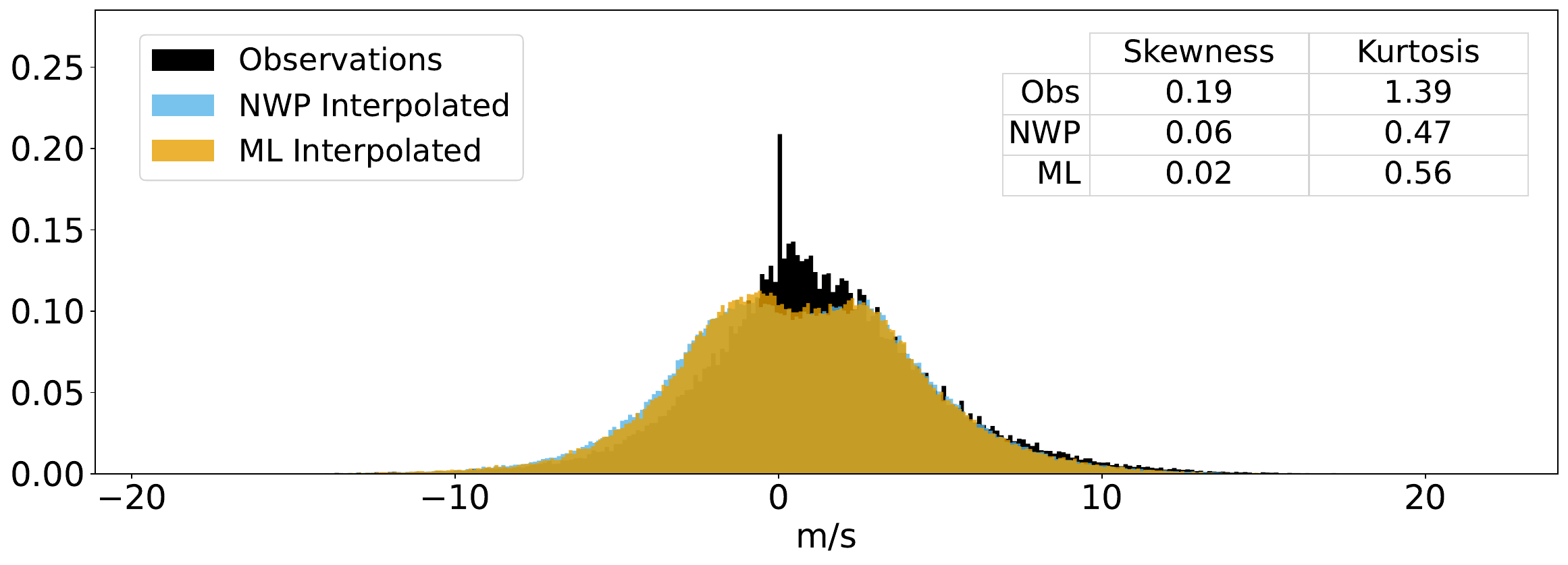}
        \caption{Compared against station observations}
    \end{subfigure}%
    \hfill%
    \begin{subfigure}[b]{\textwidth}
        \centering
        \includegraphics[width=\textwidth]{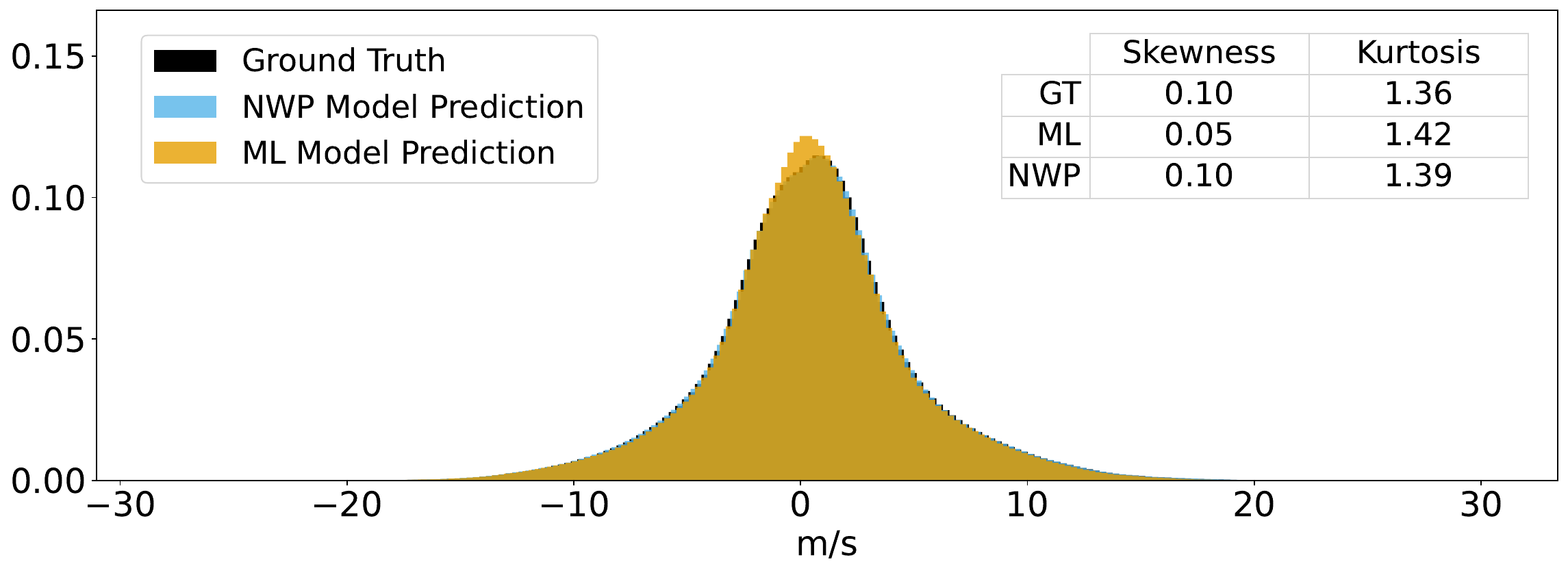}
        \caption{Compared against gridded analysis}
    \end{subfigure}
    \caption{
        Histogram of the wind v-component (\wvar{10v}) values for the \gls{DANRA} models and data. The model distributions were estimated jointly across all lead times and spatial locations (interpolated station points or analysis grid points).
        As a reference we tabulate skewness and kurtosis of the different distributions.
    }
    \label{fig:danra_verif_sparse_histogram_10v}
\end{figure}

\fi

\end{document}